\documentclass[12pt,german,english,titlepage,a4paper,twoside]{report}
     \usepackage{babel}
     \usepackage{amssymb}
     \usepackage{epsfig}
     \usepackage[usenames]{color}
     
 \usepackage{cite}

     \usepackage[]{a4}
     \usepackage{fancyhdr,thesis,slashed}
     \usepackage{rotating}     
\newcommand{\dcauthorpre}{Herr Dipl.-Phys. } 
\newcommand{\dcauthorsurname}{Hoffmann} 
\newcommand{\dcauthorname}{Roland } 
\newcommand{\dcauthoradd}{geboren am 06.06.1977 in M\"unchen} 
\newcommand{\dctitle}{Chiral properties of dynamical Wilson fermions} 
\newcommand{\dcsubtitle}{~}

\newcommand{\dcapprovala}{Prof. Dr. Ulrich Wolff}
\newcommand{\dcapprovalb}{Dr. Rainer Sommer}
\newcommand{\dcapprovalc}{Prof. Dr. Sinya Aoki}

\newcommand{\dcdegree}{doctor rerum naturalium\\ (Dr. rer. nat.)} 
\newcommand{\dcsubject}{Physik} 
\newcommand{\dcfaculty}{Mathematisch-Naturwissenschaftlichen Fakult\"at I}
\newcommand{\dcuniversity}{der Humboldt-Universit\"at zu Berlin}
\newcommand{\dcdean}{Prof. Thomas Buckhout, PhD}
\newcommand{\dcpresident}{Prof. Dr. J\"urgen Mlynek}
\newcommand{\dcdatesubmitted}{18. Mai 2005} 
\newcommand{\dcdateexam}{12. August 2005}

\newcommand{\dckeydea}{Gitter QCD}
\newcommand{\dckeydeb}{Chirale Symmetrie}
\newcommand{\dckeydec}{Renormierung}
\newcommand{\dckeyded}{Verbesserung}

\newcommand{\dckeywordsde}{\vspace*{2cm} \\{\bf{Schlagw\"orter:}}\\ \dckeydea, \dckeydeb, \dckeydec, \dckeyded \\}

\newcommand{\dckeyena}{Lattice QCD}
\newcommand{\dckeyenb}{chiral symmetry}
\newcommand{\dckeyenc}{renormalization}
\newcommand{\dckeyend}{improvement}
\newcommand{\dckeywordsen}{\vspace*{2cm} \\{\bf{Keywords:}}\\ \dckeyena, \dckeyenb, \dckeyenc, \dckeyend \\}

\author{von \\ \dcauthorpre  \dcauthorname  \dcauthorsurname  \\ \dcauthoradd}

\title{
\vspace{-3cm} \begin{flushright}
   \normalsize HU-EP-05/71\\
 \end{flushright}
\vspace{1cm}
\dctitle \\ 
\vspace{0.5cm}
\large{\dcsubtitle} \\ 
\vspace{0.5cm} {\Large{D I S S E R T A T I O N }}\\ 
\vspace{0.5cm} \large{zur Erlangung des akademischen Grades \\ 
\dcdegree\\ im Fach \dcsubject \\ 
\vspace{0.5cm} eingereicht an der \\ 
\dcfaculty \\ 
\dcuniversity \\}}
\date{\vspace{0.5cm}
\raggedright{
Pr\"asident der Humboldt-Universit\"at zu Berlin:\\
\dcpresident \vspace{-0.3cm}
}\vspace{0.5cm}\\
\raggedright{
Dekan der \dcfaculty:\\
\dcdean \vspace{-0.3cm}
}\vspace{0.5cm}\\
\raggedright{
Gutachter:
\begin{enumerate} 
\item{\dcapprovala} \vspace{-0.3cm}
\item{\dcapprovalb} \vspace{-0.3cm}
\item{\dcapprovalc} \vspace{-0.3cm}
\end{enumerate}} \vspace{0.5cm}
\raggedright{
\begin{tabular}{lll}
eingereicht am: &  &\dcdatesubmitted\\
Tag der m\"undlichen Pr\"ufung: & & \dcdateexam
\end{tabular}}\\ 
}

\usepackage{exscale}                    
\begin{document}
\pagenumbering{roman}
\maketitle
\selectlanguage{english}
\abstract
\renewcommand{\baselinestretch}{1.2}\normalsize
Quantum Chromodynamics with two light quark flavors is considered
in the lattice regularization with improved Wilson fermions.
In this formulation chiral symmetry is explicitly broken
by cutoff effects linear in the lattice spacing $a$.
As a consequence the isovector axial currents require
improvement (in the Symanzik sense) as well as a finite renormalization
if they are to satisfy the continuum Ward--Takahashi
identities associated with the isovector chiral symmetries
up to small lattice corrections of $\rmO(a^2)$.

In exploratory numerical simulations of the lattice theory
algorithmic difficulties were encountered at coarse
lattice spacings.
There the hybrid Monte Carlo algorithm used suffers from a
distorted Dirac spectrum in the form of unphysically small
eigenvalues. This is shown to be a cutoff
effect, which disappears rapidly as the lattice spacing
is decreased. An alternative algorithm, the polynomial hybrid
Monte Carlo algorithm, is found to perform significantly better
in the presence of exceptionally small eigenvalues.

Extending previously used methods both the improvement and
the renormalization of the axial current are implemented
non--perturbatively in terms of correlation functions
formulated in the framework of the Schr\"odinger functional.
In both cases this is achieved by
enforcing continuum Ward identities at finite lattice
spacing. Together, this restores the isovector chiral symmetry
to quadratic order in the lattice spacing. With little additional
effort the normalization factor of the local vector current is
also obtained.

The methods developed and implemented here can easily be applied
to other actions formulated in the Schr\"odinger functional
framework. This includes improved gauge actions as well as
theories with more than two dynamical quark flavors.
\\
\dckeywordsen
\selectlanguage{german}
\abstract \setcounter{page}{2}  
Quantenchromodynamik mit zwei leichten Quarks wird in
der Gitterregularisierung
mit verbesserten Wilson Fermionen betrachtet.
Die chirale Symmetrie in dieser Formulierung wird von
Gitterartefakten, die linear
im Gitterabstand $a$ sind, explizit gebrochen.
Daher erfordern die axialen Isospin Str\"ome Verbesserung
(im Symanzik Sinn),
sowie eine endliche Renormierung, wenn sie die Ward--Takahashi
Identit\"aten
des Kontinuums bis auf kleine Gitterkorrekturen proportional
zu $a^2$ erf\"ullen
sollen.

Algorithmische  Probleme bei gro{\ss}en
Gitterabst\"anden machen die numerischen Simulationen der
Gittertheorie
schwierig. Der Hybrid Monte Carlo Algorithmus leidet unter
einem verformten
Dirac Spektrum in Form unphysikalisch kleiner Eigenwerte.
Es wird gezeigt, da{\ss} dies ein Gitterartefakt ist,
welches schnell verschwindet,
wenn der Gitterabstand verringert wird.
Ein alternativer Algorithmus, der polynomische Hybrid Monte
Carlo Algorithmus,
zeigt erheblich bessere Eigenschaften im Umgang mit den
au{\ss}ergew\"ohnlich
kleinen Eigenwerten.

Durch Erweiterung und Verbesserung vorher verwendeter Methoden
wird die nicht--perturbative Verbesserung und Renormierung des
Axialstroms
durch  Korrelationsfunktionen im Schr\"odinger Funktional
implementiert.
In beiden F\"allen wird dies erzielt, indem man Ward
Identit\"aten
des Kontinuums bei endlichem Gitterabstand erzwingt.
Zusammen stellt dies
die chirale Symmetrie bis zur quadratischen Ordnung im
Gitterabstand wieder her.
Mit wenig zus\"atzlichem Aufwand wird auch der
Normierungsfaktor des lokalen
Vektorstroms berechnet.

Die Methoden, die hier entwickelt und implementiert wurden,
k\"onnen leicht
auch f\"ur andere Wirkungen verwendet werden, die im
Schr\"odinger Funktional
formuliert werden k\"onnen.
Dies umfa{\ss}t verbesserte Eichwirkungen sowie Theorien
mit mehr als
zwei dynamischen Quarks.\\
\renewcommand{\baselinestretch}{1.12}\normalsize
\dckeywordsde
\selectlanguage{english}

$ $
\thispagestyle{empty}
\newpage
\setcounter{page}{1}

\tableofcontents
\renewcommand{\baselinestretch}{1.0}\normalsize
\newpage
\pagenumbering{arabic}

\pagestyle{fancy}
\renewcommand{\chaptermark}[1]{\markboth{#1}{}}
\renewcommand{\sectionmark}[1]{\markright{\thesection\ #1}{}}
\fancyhf{}
\fancyhead[LE,RO]{\bfseries\thepage}
\fancyhead[LO]{\bfseries\rightmark}
\fancyhead[RE]{\bfseries\leftmark}
\renewcommand{\headrulewidth}{0.5pt}
\renewcommand{\footrulewidth}{0.0pt}
\addtolength{\headheight}{2.5pt}                

\chapter{Introduction}
	Compared to the electroweak sector of the standard model of particle
physics, quantum chromodynamics (QCD) with its few parameters and
extensive symmetries seems to be a rather simple theory. Still, it is
expected to describe the whole spectrum of strong interaction phenomena
from high--energy hadron collisions to the decays of heavy--quark bound
states and of course the hadron masses themselves.
Since its birth in the 1960s QCD has been confronted with experiment
in innumerable cases 
and is now a well--established part of the standard model.

At high energies the relevant degrees of freedom (quarks and gluons)
are found to be only weakly coupled and the non--interacting theory
can serve as a starting point for a perturbative treatment.
The breaking of Bjorken scaling \cite{Bjorken:1968dy} in deep
inelastic lepton--hadron scattering, which is the original
and still one of the most powerful
quantitative tests of (perturbative) QCD, is associated with
these energy scales.

At low energies the strong interactions show a very different behavior.
The coupling becomes strong such that a description in terms of
weakly interacting quarks and gluons is no longer appropriate.
Instead, the relevant degrees of freedom seem to be the light
mesons (pions).
Also here precise experimental data are available, in particular for
the masses and other properties of the light
hadrons.
If one wants to establish that the same Lagrangian describes both
regimes, vast energy differences need to be bridged
\cite{DellaMorte:2004bc}.

As an additional complication, the transition in the effective degrees
of freedom makes it much more difficult to work out the QCD predictions
at low energies, because perturbation theory, the ''standard tool'' of
particle physics, fails here. While effective theories are useful at this
point, they do not constitute a first--principle method. Also, some
non--perturbative predictions can be made on the basis of symmetry
considerations or QCD sum rules, but these can not be applied to all
problems one might be interested in and often also require additional
assumptions.

With the lattice regularization Wilson \cite{Wilson:1974sk} proposed a
radically different approach. Lattice QCD allows for first--principle
predictions without any additional assumptions and can be applied to a
variety of problems, from the hadron spectrum to meson decay constants
and the running of the coupling or even the topological structure
of the QCD vacuum.

In this formulation one discretizes the fields and the
action using a Euclidean space--time lattice. Thus, before
any computation is made, the high frequencies are removed
from the theory, rendering the latter ultraviolet finite.
The infrared divergencies related to zero modes can also be
be cured by either considering a non--zero quark mass or
a finite volume with fixed boundary conditions in at least one
direction (see \chap{chap:SF}). At this point the theory is
mathematically well--defined without reference to perturbation
theory.

Moreover, if the lattice is small enough, a numerical
evaluation on the computer becomes feasible. In this way the
lattice serves two purposes: it regularizes the theory with
a momentum cutoff proportional to the inverse lattice spacing
and at the same time it is a tool to evaluate observables
non--perturbatively. While today the main aspect of lattice
QCD is clearly in numerically obtaining phenomenological
predictions from Monte Carlo simulations, one should not forget
that
in many cases the lattice is the only way to define a
quantum field theory beyond perturbation theory.

As with any other regularization, the regulator has to be
removed before results can be compared to experiment. 
On the lattice this means that the \emph{continuum limit}
has to be taken by making the lattice spacing $a$
smaller and smaller. Here one encounters the usual
ultraviolet divergencies, which require renormalization
of the bare parameters and operators.

In numerical simulations the accessible lattice spacings
are strongly constrained by the computer resources and the
scaling behavior of the available algorithms.
In fact, for a long time after the invention of lattice QCD,
simulations were only possible in the \emph{quenched
approximation}. Here the observables are evaluated on a
gauge background generated with the gauge action \emph{only},
which in terms of Feynman diagrams amounts to removing all
virtual quark lines. While this makes the simulations
significantly easier, it does not represent a controlled
approximation and thus (in principle) calls into question
all physical predictions obtained in this way. Still,
impressive results were obtained from quenched simulations
and in particular the hadron spectrum is in rather good
agreement with experiment \cite{Aoki:1999yr}.

With new computer generations and algorithmic improvements
unquenched (or ''dynamical'') simulations are now possible
at reasonable lattice spacings and volumes. However, these
algorithms typically slow down proportional to $a^7$
(see e.g. \cite{Ukawa:2002pc}) or even
worse, such that a factor two in lattice spacing can change the
computational effort by more than two orders of magnitude.
As a result $a$ is usually varied only in a very limited
range, say from $0.1\fm$ to $0.05\fm$ and hence close
attention should be paid to how the observables of interest
approach the continuum. Naturally, this is strongly influenced
by the details of how the continuum theory was discretized.

Today there exists a large variety of discretizations of
lattice QCD in addition to Wilson's original formulation, on
which this work is based. These differ in both the
discretization of the gauge and -- more importantly -- the
fermionic action. A brief overview with emphasis on their
different properties concerning chirality will be given in
\chap{chap:lqcd}.

A systematic description of the approach to the continuum
limit of a lattice theory was found by Symanzik
\cite{impr:Sym1,impr:Sym2,Symanzik:1981hc}. The symmetries
of the lattice action define a set of operators, which, if
inserted in correlation functions of the continuum theory,
asymptotically reproduce the discretization errors of the
lattice theory. In other words, the lattice theory is
described in terms of an effective low--energy ($\ll a^{-1}$)
continuum theory, whose Lagrangian contains $(4+k)$--dimensional
interaction terms with couplings proportional to $a^k$.
On this basis one exploits the freedom to change the lattice
discretization  by irrelevant operators to remove the cutoff
effects, i.e. the lattice artifacts, order by order in the
lattice spacing.

This procedure is known as the Symanzik \emph{improvement}
programme and concerns both the action and the composite fields,
i.e. for QCD in particular the quark bilinears. In this way
the continuum limit can be ''accelerated'' and a smaller range
of lattice spacings might be sufficient for a reliable
extrapolation. The application of the improvement programme to
Wilson fermions is a central part of this work and a first
(general) discussion will be given in \chap{chap:lattice}.

Renormalization is commonly discussed in the framework of
perturbation theory, where divergencies in Feynman
diagrams are removed e.g. through a subtraction prescription
based on dimensional regularization. Due to the
non--perturbative nature of the low--energy sector
of QCD this approach is no longer sufficient in this case.
As will be detailed in \chap{chap:renorm}, a perturbative
calculation of renormalization factors does not result in
a controlled estimate of systematic errors of renormalized
correlation functions at low energies.

When approaching the continuum, the bare parameters in the
action have to be tuned such that a set of renormalized
quantities is kept fixed. These can be mass ratios or
otherwise defined renormalized couplings and their specific
choice defines the renormalization scheme. In this context
as well as for the  additional renormalizations required
for the composite fields we employ a finite volume scheme
based on the Schr\"odinger functional
first discussed in \cite{Luscher:1992an,Sint:1993un}.
\chap{chap:SF} will provide the reader with the necessary
background.

Despite its rich structure, some of the phenomenologically
most relevant questions are not concerned with effects
within QCD alone. In particular, one often considers QCD observables,
which can be interpreted as matrix elements of 
the effective weak Hamiltonian between QCD bound states.
A prominent example is the pion decay constant $F_\pi$,
defined through the matrix element
\be
\langle0|A_\mu(0)|\pi\rangle=ip_\mu F_\pi\label{fpi}
\ee
of the axial current between the vacuum and a pion state
with momentum $p$. Regarded as an insertion of the
effective weak Hamiltonian, it parameterizes the amplitude
for a pion to decay into a lepton anti--neutrino pair
through a virtual $W$ boson. Here it is most evident how the
renormalization and improvement of
the axial current directly affect the physical result of
a lattice estimate of $F_\pi$.

For quenched lattice QCD with Wilson fermions the
renormalization factor \cite{Luscher:1996jn} and improvement
constant \cite{Luscher:1996ug} of the axial current are
known non--perturbatively. Both are obtained by enforcing
the chiral symmetry of the underlying continuum
theory at finite lattice spacing. The non--perturbative
implementation for the case of lattice QCD with two
degenerate flavors of Wilson quarks is the central part
and the resulting renormalization factor and improvement
coefficient are the main result of this work.

Another important application of the renormalized improved
axial current is in the context of a programme to calculate
the fundamental parameters of QCD from hadronic input
parameters. In fact, the work presented here is an integral
part of an effort to calculate renormalized quark masses
from first principles. The former also
constitute an essential ingredient in tests of the 
standard model.

With
the same method as in the quenched case \cite{Capitani:1998mq},
the starting point for such a calculation is the current quark
mass derived from the non--conservation of the axial current
(PCAC relation). The renormalization of the pseudo--scalar
density, which is also required in this context, has already
been performed \cite{Knechtli:2002vp,DellaMorte:2005kg} such that together
with the results from this work all necessary tools are
available.

Starting from a review of the most important properties
of QCD (\chap{chap:cont}), which includes a derivation of
the Ward--Takahashi
identities associated with the chiral symmetry, we will
turn to the peculiarities of the lattice formulation
(\chap{chap:lattice}) and
discuss the continuum limit, renormalization and Symanzik
improvement. This is followed by a chapter concerned with
the status of chiral symmetry in different lattice
discretizations (\chap{chap:lqcd}) with particular emphasis
on Wilson fermions.
The introductory part concludes with a presentation of the
Schr\"odinger functional (SF) as a renormalization scheme
and our chosen method to compute renormalization factors
and improvement constants (\chap{chap:SF}). The necessary
correlation
functions and notation to be used are also introduced there.

The first chapter of the main part (\chap{chap:algo})
discusses Monte Carlo algorithms,
data analysis and other technical aspects of our simulations.
Algorithmic issues, which we faced in the numerical
evaluation of the axial current normalization condition,
are also reported. These problems are traced back to a
distortion of the spectrum of the Wilson--Dirac
operator, which in turn can be interpreted as a cutoff effect.

The remaining chapters are devoted to the
non--perturbative axial current improvement
(\chap{chap:impr}) and renormalization (\chap{chap:renorm}).
The former is implemented by requiring a current quark mass
derived from a Ward identity to be independent of the external
states, which are varied using projection techniques. The
integrated axial Ward identity with operator insertions is
used to formulate a normalization condition for the axial
current on the lattice. Numerical results of our
implementation are presented and summarized in interpolating
formulae for future use.

Finally, the last chapter summarizes all results
and discusses their possible application. In abbreviated
form the findings of the present work have been published in
\cite{Hoffmann:2003mm,DellaMorte:2004hs,DellaMorte:2004sz,
DellaMorte:2005se,DellaMorte:2005rd}.

\chapter{Continuum QCD}
\label{chap:cont}
	\section{History and properties}

\label{sect:propQCD}

Already in the 1960s it was conjectured that the observed
large number of strongly interacting particles, the hadrons,
are composite objects made from supposedly fundamental building
blocks called \emph{partons}. In the spirit of Rutherford's
scattering experiments, which revealed the structure of the atom,
hadrons were probed with beams of highly energetic
leptons in order to learn about the
underlying dynamics that govern the formation of
partons into hadrons.

The most interesting
experimental results came from the kinematic region of deep
inelastic
scattering, where both the momentum transfer $q^2$ and the
energy transfer $\nu$
from the leptons are very large with the ratio $q^2/\nu$ fixed.
As proposed by Bjorken in 1969 \cite{Bjorken:1968dy},
in this region the structure
functions, which
parameterize the momentum distribution within the hadron,
were found to depend only on the ratio $q^2/\nu$ (Bjorken scaling).

The easiest way to understand this behavior is to assume
that the leptons scatter off almost-free pointlike particles,
the constituents of the hadrons.
To accommodate Bjorken scaling, the theory describing the
dynamics of the partons should therefore
have the feature that the interaction becomes
weak at high energies (or small distances). The only
known quantum field theories\footnote
{More precisely: the only renormalizable quantum field
theories in four dimension.} with this property, which
is now called \emph{asymptotic freedom}, are
the non--Abelian gauge theories introduced by Yang and Mills
\cite{Yang:1954ek}.

The working hypothesis is now that the interaction
of the fundamental degrees of freedom -- at this point
renamed \emph{quarks} by Gell-Mann --
is described by a non--Abelian gauge theory.

At the same time experimental observations
required the quarks to have an additional unobserved
quantum number (''color'') in order to avoid conflict
with the Pauli exclusion principle. Most of the difficulties
could be resolved by identifying the symmetry corresponding
to the new quantum numbers with the non--Abelian gauge
symmetry. Further experimental input and theoretical arguments
uniquely fixed the gauge group to be $SU(3)$.
This theory was named quantum chromodynamics, QCD,
and its gauge quanta are called gluons.

At this point one should note that it is the property
of being asymptotically free that ensures the applicability
of perturbation theory at high energies. One of the successes
of perturbative QCD is the correct prediction of the (logarithmic)
corrections to Bjorken scaling, which is now known to
be strictly valid only in the limit of infinite energy.

In contrast to Abelian gauge theories like quantum
electrodynamics, QED, the field quanta of non--Abelian
theories also carry charge and thus interact among
themselves. Indeed, perturbation theory shows that it
is precisely this self--interaction, which is responsible
for asymptotic freedom. The situation is opposite
to the one found in QED, where perturbation theory predicts
the effective
charge of the electron to \emph{increase} with energy.

We can now write down the Lagrangian density for the gluonic
($\LL_g$) and fermionic ($\LL_f$) part of
QCD with $\nf$ flavors of quarks.
\be
\begin{array}{rcl}
\LL_g&=&-\frac12\,\tr F_{\mu\nu}F^{\mu\nu}\;,\\
\textrm{with the field strength }\
F_{\mu\nu}&=&\partial_\mu A_\nu-\partial_\nu A_\mu
+g_0[A_\mu,A_\nu]\;.\label{gaugelagrangian}
\end{array}
\ee
Here $g_0$ is the bare gauge coupling and $A_\mu$ the Lie--algebra
valued anti--Hermitian $SU(3)$ gauge field. For the fermionic part we have
\bea\LL_f&=&\sum_{f=1}^{\nf}\psibar_f(i\slashed D-m_f)\psi_f\;,
\label{llf}
\eea
with the gauge covariant derivative
$\slashed D=\dirac\mu(\partial_\mu\pl g_0A_\mu)$ and the bare
quark masses $m_f$. The two
color indices of $A_\mu$ as well as the color and
Dirac indices of $\psibar$ and $\psi$ have been suppressed.

To account for the observed particle spectrum, hadron states
and in general all physical observables are postulated
to be color singlets. In this way an unobserved proliferation
of states due to the color symmetry is prevented.
This non--perturbative phenomenon (known
as \emph{confinement} of color)
is ascribed to an increase of the force between
color sources at long distances ($\simeq1\fm$). Since, as a
consequence, even at high interaction energies the initial
and final states are subject
to confinement, also perturbation theory is affected and hadronic
matrix elements cannot be obtained using a purely perturbative
treatment.
This problem is usually addressed by factorizing them
into a ''hard'' (perturbative) and a ''soft'' (non--perturbative)
part using e.g. the operator product expansion (OPE). The soft
part is then parameterized by effective couplings.

Lattice simulations of pure non--Abelian gauge theories show that
for large distances the energy of two static color sources grows
linearly with the so--called string tension $\sigma\simeq1\GeV/\fm$
(see e.g. \cite{Gattringer:2001jf}). Hence
there is strong numerical evidence in support of the confinement
hypothesis. However, an
analytical first--principle explanation from the underlying
dynamics is still missing although many attempts have been made,
e.g. in terms of condensation of topological excitations
\cite{'tHooft:1977hy}.

Finally, the most important ingredient to understanding
the strong interactions at low energies is
\emph{chiral symmetry}.
For degenerate quarks the QCD Lagrangian (\ref{llf}) is
flavor--blind and hence invariant under (global, i.e.
space--time independent) unitary
transformations
of the $\nf$--component fermion fields. In addition, for
massless quarks there is also no coupling between the left-- and
right--handed field components
\be
\psi_L=\half(1\mi\dirac5)\psi\quad\textrm{and}\quad\psi_R
=\half(1\pl\dirac5)\psi\;.
\ee
Therefore separate
$U(\nf)$ transformations of these fields according to
\be
\left.
\begin{array}[c]{cc}
\psi_L\rightarrow U_L\psi_L\\
\psi_R\rightarrow U_R\psi_R
\end{array}\right\}
\ \ \Rightarrow\ \ \psi\rightarrow U_R\psi_R+U_L\psi_L
\ee
are also a symmetry
of the massless QCD Lagrangian. These are usually written in terms of the
vector and axial vector transformations
\be\begin{array}{rcll}
U_R=U_L=U&\Rightarrow&\psi\rightarrow U\psi=\rme^{i\omega^a\lambda^a}\psi
&\quad\textrm{\small (vector)}\\[1mm]
U_R=U_L^\dagger=U&\Rightarrow&\psi\rightarrow U_5\psi=\rme^{i\omega^a\lambda^a\dirac5}\psi
&\quad\textrm{\small (axial)}\;.
\end{array}
\label{chiral_trafo}
\ee
At the classical
level this
$U(\nf)_{\rm V}\times U(\nf)_{\rm A}$ symmetry gives
rise to two conserved currents (the vector and the
axial current) for each generator $\l^a$ of $U(\nf)$,
\be
\begin{array}{rcl}
A_\mu^a&=&\psibar\dirac\mu\dirac5\l^a\psi\;,\\
V_\mu^a&=&\psibar\dirac\mu\l^a\psi\;,
\end{array}\ee
where the $\l^a$ act on the flavor indices of the
fermion fields. To summarize, the concrete requirements
for the $U(\nf)_{\rm V}\times U(\nf)_{\rm A}$ flavor chiral symmetry
are that the Dirac operator is diagonal in flavor space and that it
anti--commutes with $\dirac5$.

Since the  masses of the up and down quarks are much below typical
hadronic energy scales, massless
QCD with $\nf\!=\!2$ seems to be a reasonable approximation of nature.
In this case, the vector symmetry $SU(2)_{\rm V}\times U(1)_{\rm V}$
implies isospin symmetry and quark number conservation,
which are indeed experimentally confirmed to a high precision.

The fact that the corresponding axial symmetries are not
found in the strong interactions is believed to be due
to a condensation of quark--antiquark pairs, such that these
symmetries are spontaneously
broken by the QCD vacuum. In this picture the isospin triplet
of light pseudo--scalar mesons, the pions, become the quasi--Goldstone bosons
of the spontaneously broken axial symmetry.\footnote
{A direct proof of the
Goldstone theorem for this case will be given in \sect{int_ward_later}.}
Their non--vanishing
mass is explained
by the fact that for small but non--zero quark mass the
$SU(2)_{\rm A}$ is only an approximate symmetry and the pions
are hence not true Goldstone bosons.

The identification of the pions as the Goldstone bosons
of spontaneous chiral symmetry breaking leads to
various relations between matrix elements, whose
experimental verification helped in establishing this picture.
A systematic study of the low--energy limit of QCD
can be performed in the framework of \emph{chiral perturbation
theory} \cite{Gasser:1982ap}. This is a low--energy effective theory
of the strong interactions, where pions are introduced as fundamental
degrees of freedom. The quark masses are treated
as a perturbation to the chirally invariant Lagrangian.

In a somewhat worse approximation of nature this picture can be
extended to three flavors if the strange quark mass is also neglected.
The Ward--Takahashi identities resulting from the
$SU(2)_{\rm V}\times SU(2)_{\rm A}$ symmetry of massless
two--flavor QCD are derived in \sect{contward}.

The conservation of the singlet axial current corresponding
to the remaining $U(1)_{\rm A}$ symmetry is violated by quantum
corrections, which are related to the topological structure
of the QCD vacuum. This is known as the axial (or chiral)
anomaly and is crucial in explaining the absence of an
isosinglet pseudo--scalar meson with mass comparable to that of
the pions. In the path integral formulation (see below) the
anomaly can be understood from the fact that the $U(1)_{\rm A}$
transformation does not leave the integral measure invariant
\cite{Fujikawa:1979ay}. This is particularly transparent
\cite{Luscher:1998pq} in
a lattice regularization using Ginsparg--Wilson fermions,
which are invariant under a chiral symmetry, differing from
(\ref{chiral_trafo}) only at finite lattice spacing.

\section{Euclidean path integral}

While scalar field theories can be quantized efficiently in
the operator language, for gauge theories it is more
convenient to employ the
functional integral (or path integral) formalism. 
In this approach the fundamental quantity is the generating
functional
\be
\Z[J]=\int_{\rm fields}\!\!\exp\left\{i\!\int\!\rmd^4
x (\LL+J\!\cdot\textrm{fields})\right\}\;.\label{partfunc}
\ee
Correlation functions are generated by taking functional
derivatives with respect to the sources $J(x)$ and perturbation
theory is set up by expanding the (gauge fixed) action
$S=\int\rmd x^4 \LL$ around a saddle point. Feynman rules can
essentially be read directly from the Lagrangian.

However, the functional integral is also the starting point
for the so far most successful non--perturbative treatment
of QCD, the lattice regularization.
There the partition function $\Z\!=\!\Z[0]$ is evaluated directly
using Monte Carlo methods. This approach requires
a statistical interpretation of the path integral
(\ref{partfunc}), where the phase factor $\exp(iS)$ becomes
a Boltzmann weight. To achieve this the theory is transcribed
from Minkowski to Euclidean space by introducing an imaginary
time coordinate (''Wick rotation'').
This formulation also
reveals the deep connection between quantum field theories
and statistical systems.

With a purely imaginary time coordinate the space--time metric
becomes Euclidean. Under certain condition the Minkowskian
correlation functions (the so--called Wightman functions)
can be analytically
continued into this region to obtain Euclidean correlation
functions (Schwinger functions). From the more accessible
Schwinger functions the physical Wightman functions in
Minkowski space and thus the quantum field theory generated by
(\ref{partfunc}) can be reconstructed.

In the path integral formulation the fermionic anti--commutation
properties are accommodated through a representation of the 
fields $\psi$ and $\psibar$ as Grassmann variables. The integral
over the fields is therefore a Grassmann integral for the fermionic part.
On the lattice the gauge fields are represented by
compact link variables (see \chap{chap:lattice}) such that
the path integral is then defined in terms of the Haar measure on
the gauge group.

The QCD fermion action for one quark flavor in its Euclidean
form is given by

\vspace*{-4.5mm}

\bea
S&=&\int\!\!\rmd^4x\, \psibar(\dirac\mu D_\mu\pl m_0)\psi,
\label{cont_act}\;,\\
\textrm{where }\quad D_\mu&=&\partial_\mu\pl g_0A_\mu\;.
\eea
Before the properties of the Euclidean lattice formulation
as a non--perturbative regulator are discussed in \chap{chap:lattice},
where for simplicity the case of a pure gauge theory is considered,
we will derive the Ward identities associated with the \emph{local}
versions of the transformations (\ref{chiral_trafo}).

\section[Current algebra and continuum Ward identities]
{Current algebra and continuum\\[-0.5mm] Ward identities}

\label{contward}
For the two--flavor case ($\lambda^a=\tau^a/2$ with the Pauli matrices
$\tau^a$) the infinitesimal chiral transformations are 
written as
\bea
\textrm{vector}&&
\left\{\begin{array}{rcl}
\psi&\rightarrow&\psi+i\omega^a\cdot\half\tau^a\psi\\[1mm]
\psibar&\rightarrow&\psibar-i\omega^a\cdot\psibar\half\tau^a\;,
\label{vtrafo}
\end{array}\right.\\[2mm]
\textrm{axial}&&
\left\{\begin{array}{rcl}
\psi&\rightarrow&\psi+i\omega^a\cdot\half\tau^a\dirac5\psi\\[1mm]
\psibar&\rightarrow&\psibar+i\omega^a\cdot\psibar\dirac5\half\tau^a\;.
\end{array}\right.\label{atrafo}
\eea
We define the vector and axial variations of (composite) fields $\op$
through\linebreak $\op\rightarrow\op+i\delta\op$, with
\be
\dv\op=\omega^a\dv^a\op\quad
\textrm{and}
\quad \da\op=\omega^a\da^a\op\;,
\ee
where (in case of a local transformation) $\omega^a$ is evaluated
at the space--time point where $\op$ lives.
For the quark fields we obtain
\begin{eqnarray}
\dv^a\psi(x)=\half\tau^a\psi(x), & \dv^a\psibar(x)=
  -\psibar(x)\half\tau^a\;,\label{varV}\\
\da^a\psi(x)=\half\tau^a\dirac{5}\psi(x), & \da^a\psibar(x)=
  \psibar(x)\dirac{5}\half
\label{varA}\tau^a\;.
\end{eqnarray}
These can be used to calculate the variation of arbitrary expressions 
$\op$ built from the basic fields by treating $\dv^a$ and $\da^a$
as first order differential operators. In particular, for the
variations of the isospin vector and axial vector currents,
\begin{equation}
 V^a_{\mu}(x)=
\psibar(x)\dirac{\mu}\half\tau^a\psi(x),
\qquad\quad
A^a_{\mu}(x)=
\psibar(x)\dirac{\mu}\dirac{5}\half\tau^a\psi(x),
\label{currents}
\end{equation}
one obtains for example
\begin{eqnarray}
\da^aV_\mu^b&=&\ts\pbruch {V_\mu^b}{\psi}\da^a\psi+\da^a\psibar\,
\pbruch{V_\mu^b}{\psibar}\nonumber\\
\ts
&\use{currents}&\textstyle\psibar\dirac\mu\half\tau^b\da^a\psi+
\da^a\psibar\dirac\mu\half\tau^b\psi\nonumber\\
\ts
&\use{varA}&\textstyle\psibar\dirac\mu\half\tau^b\half
\tau^a\dirac5\psi+\psibar\dirac5\half\tau^a\dirac\mu\half
\tau^b\psi\nonumber\\[1mm]\ts
&=&\textstyle\psibar\dirac\mu[\frac{\tau^b}2,\frac{\tau^a}2]
\dirac5\psi=-i\epsilon^{abc}A_\mu^c\nonumber\;,
\end{eqnarray}
since the generators of $SU(2)$ satisfy the algebra
\begin{displaymath}\ts
[\frac{\tau^a}2,\frac{\tau^b}2]=\frac{\tau^a}2\frac{\tau^b}2-
\frac{\tau^b}2\frac{\tau^a}2=i\epsilon^{abc}\frac{\tau^c}2\;.
\end{displaymath}
Calculating all variations using $\{\dirac\mu,\dirac5\}=0$ it is
easily verified that the currents form a closed algebra
\begin{eqnarray}
 \dv^aV^b_{\mu}(x)=-i\epsilon^{abc}V^c_{\mu}(x), &&  \da^aV^b_{\mu}(x)=
 -i\epsilon^{abc}A^c_{\mu}(x)\;,\label{dV}\\[1mm]
  \dv^aA^b_{\mu}(x)=-i\epsilon^{abc}A^c_{\mu}(x), && \da^aA^b_{\mu}(x)=
  -i\epsilon^{abc}V^c_{\mu}(x)\;.\label{dA}
\end{eqnarray}
The Ward identities associated with
the chiral symmetry of the action
are now derived by performing local infinitesimal transformations
of the quark and anti-quark fields in the Euclidean functional
integral.

\subsection{Variation of the action}

The starting point is the Euclidean fermion action (\ref{cont_act}) for
two quark flavors,
where now $m_0$ is a diagonal $2\!\times\!2$ matrix in isospin space.
We perform \emph{local} infinitesimal variations of the fermionic
fields and study their effect on the classical QCD action.
These variations are parameterized by
$\omega^a(x)$
with support in a space-time region $\mathcal R$.
The isospin vector variation of the fermionic action is
\begin{eqnarray}
\dv S&=&\int_\R\!\!\rmd^4x\, \Big[(\dv\psibar)(\dirac\mu D_\mu\pl m_0)\psi+
    \psibar(\dirac\mu D_\mu\pl m_0)\dv\psi\Big]\nonumber\\
&\use{varV}&\int_\R\textstyle\!\!\rmd^4x\, \Big[-\omega^a\psibar\half
\tau^a(\dirac\mu\partial_\mu\pl m_0)\psi+
    \psibar(\dirac\mu\partial_\mu\pl m_0)\omega^a\half\tau^a\psi\Big]
    \nonumber\;,
\end{eqnarray}
where the terms
containing the gauge fields $A_\mu$ canceled. Performing a partial integration
of the second expression, which does not give boundary terms since the variation
vanishes outside the region $\mathcal R$, yields
\begin{eqnarray}
\dv S&=&\int_\R\!\!\ts\rmd^4x\, \omega^a\Big[-\psibar\half\tau^a[\dirac\mu
\partial_\mu+m_0]\psi-
    \Big[\dirac\mu(\partial_\mu\psibar)-m_0\Big]\half\tau^a\psi\Big]\nonumber\\
&=&\int_\R\!\!\ts\rmd^4x\, \omega^a\Big[-\partial_\mu\left(
\psibar\half\tau^a\dirac\mu\psi\right)-\psibar\,[\half\tau^a,m_0]\,\psi\Big]
\nonumber\\
&\use{currents}&\int_\R\!\!\ts\rmd^4x\, \omega^a(x)\Big[-\partial_\mu
V_\mu^a(x)-\psibar(x)\,[\half\tau^a,m_0]\,\psi(x)\Big]\label{dVS}\;.
\end{eqnarray}
In the case of degenerate quarks $m_0$ is proportional to the unit matrix and
hence the commutator with $\tau^a$ vanishes.

In the same way the isospin
axial vector variation of the action can be computed
using the anti-commutator $\{\dirac\mu,\dirac5\}=0$. Since in (\ref{varA})
the variations of $\psi$ and $\psibar$ appear with the same sign,
the mass now results in an anti--commutator term. Again performing
a partial integration of the second term results in
\begin{eqnarray}
\da S&\use{varA}&\int_\R\ts\!\!\rmd^4x\, \Big[\omega^a\psibar\dirac5\half
\tau^a\left(\dirac\mu\partial_\mu+m_0\right)\psi+
    \psibar\left(\dirac\mu\partial_\mu+m_0\right)\omega^a\half
    \tau^a\dirac5\psi\Big]\nonumber\\
    &\!=\!&\!\int_\R\!\!\textstyle\rmd^4x\, \omega^a(x)\Big[-\partial_\mu
A_\mu^a(x)+\psibar(x)\dirac5\{\half\tau^a,m_0\}\psi(x)\Big]\;.\nonumber
\end{eqnarray}
For two degenerate quarks this expression becomes
\begin{eqnarray}
\da S&\!=\!&\!\int_\R\!\!\textstyle\rmd^4x\, \omega^a(x)\Big[-\partial_\mu
A_\mu^a(x)+2m_0P(x)\Big]\;,\label{dAS}
\end{eqnarray}
with the pseudo-scalar density
\be
P^a(x)=\psibar(x)\dirac5\half\tau^a\psi(x)\;.\label{density}
\ee

\subsection{Ward identities}

Through a formal
manipulation of the functional integral the expectation value
of the variation of an operator can be related to that of the action.
Later we will discuss how such relations are realized in the regularized
quantum theory.
If the fermion integration measure is denoted by $\DD\psi$,
we have for a linear transformation of the Grassmann fields
\footnote{The rules of Grassmann integration imply that the Jacobian is the
\emph{inverse} of what one would expect from bosonic integrals, hence
the appearance of $J^{-1}$ in (\ref{jacob}).}
\bea
\psi'=A\psi\!\!&&\!\!\psibar'=\psibar\bar A\nonumber\\
\Rightarrow\qquad[\DD\psibar\DD\psi]&=&
\det\bar A\det A[\DD\psibar'\DD\psi']
=J^{-1}[\DD\psibar'\DD\psi']\;.\label{jacob}
\eea
Using $\det(1\pl\omega X)=1\pl\omega\tr X\pl\rmO(\omega^2)$, one obtains
for the infinitesimal versions of the isovector transformations
(\ref{vtrafo}) and (\ref{atrafo})
\bea
\textrm{vector}\qquad J&=&1+\rmO(\omega^2)\label{jvec}\\
\textrm{axial}\qquad  J&=&1+i\label{jaxial}
\omega^a\tr(\tau^a\dirac5)+\rmO(\omega^2)\;.
\eea
Since the Pauli matrices are traceless, the Jacobian is unchanged 
for both isovector transformations
and one can perform the infinitesimal variable
transformation in the path integral according to
\begin{eqnarray}
\Z=\int_{\textrm{\tiny fields}}\!\!\!\! e^{-S}\quad\Rightarrow\quad
\langle\op\rangle&\!\!\!=\!\!\!&\Z^{-1}\int_{\textrm{\tiny fields}}\!\!\!\!\op\, e^{-S}
=\Z^{-1}\int_{\textrm{\tiny fields}}\!\!\!\!(\op+\delta \op)\, e^{-S}
(1-\delta S)\nonumber\\[1mm]\nonumber
&\!\!\!=\!\!\!&\langle\op\rangle+\langle\delta\op\rangle-\langle\op\delta S\rangle\\[2mm]
\Rightarrow\quad\langle\delta\op\rangle&=&\langle\op\delta S\rangle\label{deltaOS}\;.
\end{eqnarray}
We can now derive
the integrated Ward identities associated with the flavor chiral symmetry
of the theory with two degenerate quarks.
To this end we insert the vector (\ref{dVS}) and axial (\ref{dAS})
variations into (\ref{deltaOS}).
Choosing there an operator $\op_{\rm ext}$ (for \emph{ext}ernal)
with support only
outside the region $\R$, where we perform the transformations,
the left--hand side of \eq{deltaOS} vanishes and we have
\bea
0&=&\int_\R\!\!\textstyle\rmd^4x\, \omega^a(x)\Big\langle-\partial_\mu
V_\mu^a(x)\,\op_{\rm ext}\Big\rangle\\
\textrm{and}\quad
0&=&\int_\R\!\!\textstyle\rmd^4x\, \omega^a(x)\Big\langle[-\partial_\mu
A_\mu^a(x)+2m_0P(x)]\,\op_{\rm ext}\Big\rangle\;.
\eea
These hold for any variation $\omega^a(x)$ and one can thus conclude that
the expectation value multiplying it vanishes.
This gives the vector current conservation in the
form
\be
\langle\partial_\mu V_\mu^a(x)\op_{\rm ext}\rangle=0\;,
\label{cvc}
\ee
and for the isovector axial current one has the PCAC
(\emph{p}artial \emph{c}onservation
of the \emph{a}xial \emph{c}urrents) relation
\be
\langle\partial_\mu A_\mu^a(x)\op_{\rm ext}\rangle=2m_0\langle P^a(x)
\op_{\rm ext}\rangle\;.\label{pcac}
\ee
Both are valid for operators $\op_{\rm ext}$ not located at the point
$x$. Integrated Ward identities with non--trivial operator insertions
will be derived in \sect{int_ward_later}
to be used in the normalization conditions for the vector and axial
currents in the lattice regularization with Wilson fermions.

\subsection{Anomalous symmetries}

Since the quantized theory requires a regularization, it is not
immediately obvious how e.g. the above relations have to be
interpreted. If a symmetry is preserved in the
quantum theory, this implies that it is either unbroken
by the regulator or is recovered in the limit where the latter
is removed. Perturbation theory shows that this is the case for
the isovector symmetries discussed above.
In \chap{chap:renorm} we will see that in a specific lattice
formulation of the theory the above relations can be made to hold
up to (small) lattice corrections. This can be interpreted as showing
that even beyond perturbation theory the isovector symmetries are
preserved in the quantized theory.

A transformation is called \emph{anomalous} if it is a symmetry of the
classical theory (i.e. the unregulated action), but not of
the quantum theory. Depending on the chosen regularization,
an anomaly can arise in different ways. One possibility is that
the integration measure of the functional integral is not invariant
under the transformation.
If such a transformation is parameterized by an infinitesimal $\omega$,
we can write the Jacobian as $J=1+i\omega\delta J$ and instead of
(\ref{deltaOS})
one then has
\be
\langle\delta\op\rangle=\langle\op\delta S\rangle+\ev{\op\delta J}\;.
\label{anomaly}
\ee
It can also happen that a symmetry is broken
by the regulator in such a way that it is not recovered in
the renormalized theory.
To illustrate these points we consider the iso\emph{singlet} transformations
\be
\begin{array}{rr@{\,=\,}lr@{\,=\,}l}
\textrm{vector}\ \ &
\dv\psi&\psi\;,&
\dv\psibar&-\psibar\;,\\[1mm]
\textrm{axial}\ \ &
\da\psi&\dirac5\psi\;,&
\da\psibar&\psibar\dirac5\;.
\label{anotrafo}
\end{array}
\ee
For those \eq{jvec} is still true and in fact the $U(1)_{\rm V}$
(quark number conservation) is preserved in the
quantized theory. For the axial transformation, instead
of (\ref{jaxial}) we obtain for the Jacobian $\delta J=2\tr\dirac5$.

The axial anomaly mentioned in the introduction has the non--conservation
of the isosinglet axial current
$A_\mu\!=\!\psibar\dirac\mu\dirac5\psi$ even at vanishing quark mass
as a \emph{physical} consequence. Therefore any admissible regularization
of QCD needs to be able to reproduce it correctly. This can happen
through a non--trivial redefinition of $\dirac5$, i.e. for (\ref{anotrafo})
to be a symmetry of the regularized action one has to replace $\dirac5$
with some $\hat\dirac5$. This is the case for dimensional
regularization or Ginsparg--Wilson fermions (see \sect{sec:chirallat}), where
the anomaly comes from $\delta J$ in (\ref{anomaly}) according to ($m=0$)
\be
\langle\partial_\mu A_\mu(x)\op_{\rm ext}\rangle=2\langle 
\op_{\rm ext}\tr\hat\dirac5\rangle\;.\label{isoaxial}
\ee
Alternatively, the anomaly can arise from an explicit chiral symmetry breaking
by the regulator, such that the iso\emph{vector} axial symmetry is recovered in the
renormalized theory, but the $U(1)_{\rm A}$ is not.
This the case in a lattice regularization with Wilson fermions, where the anomaly
can be understood 
as coming from the chiral variation of the Wilson
term (see \sect{sec:wilson}).

\chapter{Gauge theories on the lattice}
	\label{chap:lattice}

Consider a field theory with a matter field $\psi^i(x)$, where 
$x$ is the
space--time coordinate and $i$ refers to some internal degree of freedom.
The requirement that one should be able
to choose the basis to describe this internal degree of freedom
independently on every space--time point results in the
principle of \emph{local gauge invariance}. To achieve this,
the matter fields need to be coupled to a \emph{gauge field}
$A_\mu^{ij}(x)$, which takes care of the basis transformation
between different (but infinitesimally close) space--time points
(''parallel transporter'').
This is done via the ''minimal coupling prescription'', which
replaces the derivative $\partial_\mu$ with the covariant
derivative $D_\mu=(\partial_\mu\pl g_0A_\mu)$ such that
$D_\mu\psi$ has the same properties under local basis
transformations (\emph{gauge transformations}) as $\psi$ itself. 
For the case of QCD the internal degree of freedom of
the quark fields $\psi$ is that of
''color'' and $i$ is the index of the fundamental (3)
representation of $SU(3)$ ($\psibar$ is in the $\bar3$
representation). The anti--Hermitian gauge fields are
conventionally parameterized as
\be
A_\mu^{ij}=\sum_ai{\rm A}_\mu^a(\lambda_a)_{ij}\;,
\label{contgaugefield}
\ee
where $\lambda_a$ are the eight generators of $SU(3)$
(Gell-Mann matrices). The trace of the square of the local field
strength
(\ref{gaugelagrangian}) is the only term that can be made from
the gauge fields that is gauge--invariant, has mass--dimension
four or less
and is compatible with time--reversal, charge--conjugation and
parity. The geometric interpretation of the field strength tensor
$F_{\mu\nu}$ is the local curvature in color space as obtained
by connecting color bases around an infinitesimal square in
the $\mu$--$\nu$--plane.

In the discrete theory the matter fields are defined on the
sites of the lattice and to gauge connect those we always need
parallel transporters for \emph{finite} distances. It is
therefore more natural to describe the gauge degrees
of freedom in terms of \emph{link variables}
$U_\mu(x)\in SU(3)$. These are the path--ordered exponentials
of the integral of (\ref{contgaugefield}) from $x$ along the link
of length $a$
to the next lattice site in the $\mu$ direction.
The variable $U_\mu(x)$
is thus associated with this link and by
definition the link
in the opposite direction is its inverse
$U_{-\mu}(x\pl \hat\mu)=U_\mu^\dagger(x)$, where $\hat\mu$ is
a vector of length $a$ in the $\mu$ direction.
%
%
Under a local gauge transformation $\L(x)\in SU(3)$ the links
and quark fields transform as
\bea
U_\mu(x)&\stackrel \L\rightarrow&\L(x)U_\mu(x)\L(x\pl\hat \mu)^\dagger\nonumber\\
\psi(x)&\stackrel \L\rightarrow&\L(x)\psi(x)\\
\psibar(x)&\stackrel \L\rightarrow&\psibar(x)\L(x)^\dagger\;,\nonumber
\eea
such that
\begin{eqnarray}
\nabla_\mu\psi(x)&=&{\ts\frac1a}[U_\mu(x)\psi(x\pl\hat\mu)-\psi(x)]\nonumber\\
\nabla^\star_\mu\psi(x)&=&{\ts\frac1a}[\psi(x)-
U_\mu^\dagger(x\mi\hat\mu)\psi(x\mi\hat\mu)]\label{covder}\\
\widetilde\nabla_\mu\psi(x)&=&\half(\nabla_\mu\pl\nabla^\star_\mu)\psi(x)\nonumber
\end{eqnarray}
define covariant derivatives on the lattice. Normal
lattice derivatives $\partial_\mu$, $\partial_\mu^\star$ and $\tilde\partial_\mu$ 
are defined by setting $U_\mu\equiv1$ in eqs.~(\ref{covder}).

On the lattice the field strength is measured by tracing the
gauge links around closed
loops and hence the simplest admissible lattice gauge action is
Wilson's plaquette action
\be
S_g[U]=\frac1{g_0^2}\sum_p\tr\left\{1-U_p\right\}
\;,\label{wilsongauge}
\ee
where $p$ runs over all oriented elementary plaquettes
($1\!\times\!1$ loops) and
$U_p$ is the product of the gauge links around a plaquette
\bea
U_p(x)&=&U_{x;\mu,\nu}
=U_\mu(x)\cd U_\nu(x\pl\hat\mu) \cd U_{\mu}^{\dagger}(x\pl\hat\nu)
\cd  U_{\nu}(x)^{\dagger}\;,\\
U_p(x)&\stackrel \Lambda\rightarrow&\Lambda(x)U_p(x)\Lambda(x)^\dagger\;.
\eea
For gauge
group $SU(N)$ the bare gauge coupling is commonly expressed through
$\beta=2N/g_0^2$.
A small $a$ expansion
of the gauge links in terms of the gauge fields gives
\be
U_{x;\mu,\nu}=1+a^2F_{\mu\nu}+\frac{a^4}2F_{\mu\nu}^2+\ldots
\ee
such that the sum over oriented plaquettes in (\ref{wilsongauge}) reduces
to the Yang--Mills action (\ref{gaugelagrangian}). In fact this is true
for any lattice action of the form (\ref{wilsongauge}) composed of closed loops
of gauge links. Improved gauge actions (see \sect{sect:impr}) can thus be defined
by adding more extended loops with suitably chosen coefficients
\cite{Luscher:1984xn,Iwasaki:1983ck} to (\ref{wilsongauge}).

\section{Monte Carlo integration}

To obtain the expectation value of a gauge--invariant observable
$F[U]$
in the quantum field theory
described by (\ref{wilsongauge}) we need to evaluate the integral
\be
\ev F=\frac1\Z\int\!\!\DD U \ F[U]\rme^{-S[U]}
\quad\textrm{with}\quad \Z=\int\!\!\DD U
\ \rme^{-S[U]}\;,\label{funcint}
\ee
where $\DD U$ denotes the product of the $SU(3)$ Haar measures for
all links on the lattice. 
For a 4--dimensional lattice of linear
extension $L$ there are $4\cd(L/a)^4$ links, which in typical
simulations is of order $10^5$ or larger. The only way to evaluate
such high--dimensional integrals is using Monte Carlo techniques.

To this end one generates gauge link configurations $U_i$ that are
distributed with a probability $P_i\propto\DD U_i\rme^{-S[U_i]}$
(importance sampling).
The integral (\ref{funcint}) is then statistically approximated
by
\be
\ev F=\frac1N \sum_{i=1}^NF[U_i]+\rmO\Big(\frac1{\sqrt N}\Big)\;.
\ee
Unless a \emph{global} heatbath algorithm is available for the
system one wants to simulate,
configurations are produced by an algorithm that
updates a given configuration to obtain the next one (Markov chain).
Configurations
generated in this way are not statistically independent and
therefore \emph{autocorrelation} effects need to be taken into
account in the data analysis.

For the case of pure gauge actions (or equivalently in quenched QCD)
the local heatbath \cite{Creutz:1980zw,Cabibbo:1982zn} is an efficient
algorithm to generate correctly
distributed ensembles of gauge configurations and lattice sizes up to
$L/a=\rmO(10^2)$ are possible today. When the correlation lengths
in the system become large, autocorrelation along the Markov chain
increases and the algorithm rapidly becomes inefficient (critical
slowing down). This can be countered by adding (microcanonical)
overrelaxation updates \cite{Creutz:1987xi} to the heatbath algorithm.

If we want to consider full QCD, the fermionic part needs to be treated
analytically because the Grassmann--valued quark fields cannot be handled
in a computer simulation. Since the fermion action is bilinear in the quark
fields, the Grassmann integral can be performed and for the partition
function one gets ($M=D+m$)
\bea
\Z&=&\int\!\!\DD U\DD\psibar\DD\psi\, \rme^{-S_g[U]-\psibar M[U]\psi}\\
  &\propto&\int\!\!\DD U\, \det\!M[U] \rme^{-S_g[U]}\;.
\eea
In the path integral for an observable $O(U,\psi,\psibar)$ this integration
of the fermionic fields results in a new observable $\widetilde O(U)$,
which depends only on the gauge field (Wick contraction). Instead
of the fermion fields $\widetilde O$
contains matrix elements of the fermion propagator $S=M^{-1}$ and as a
consequence it can
be a very complicated and non--local function of $U$.
For the ''pion--pion'' correlator $(\psibar\dirac5\psi)_x(\psibar\dirac5\psi)_y$
one obtains e.g.
\bea\nonumber
&&\ev{(\psibar\dirac5\psi)_x(\psibar\dirac5\psi)_y}=\\[2mm]
\nonumber
&&\qquad\frac1\Z
\int\!\!\DD U \,\Big(\tr\{
S_{xx}[U]\dirac5\}\tr\{S_{yy}[U]\dirac5\}
\\[-2mm]
&&\qquad\hspace*{19mm}
-\tr\{
S_{xy}[U]\dirac5S_{yx}[U]\dirac5\}
\Big)\det\!M[U] \rme^{-S_g[U]}\;,
\eea
where the trace is taken over Dirac indices.
Algorithms for simulating full QCD, i.e. the generation of a gauge
ensemble according to $\DD U \det\!M[U] \rme^{-S_g[U]}$, will be discussed in detail in
\chap{chap:algo}.

\section{Continuum limit and renormalization}

As already mentioned, the regulator has to be removed before
physical predictions can be made. In the statistical analogy the
mass (in lattice units) $a\cd m$ of the lowest--lying state
(e.g. the $0^{++}$ glueball for a pure gauge theory) corresponds
to the inverse correlation length of the statistical system.
Usually masses are obtained from the exponential decay of suitable
correlation functions.

If a continuum limit with a finite physical mass exists, it means
that by tuning the bare parameters ($\beta$ in pure gauge theory)
we can find a limit, where $a\cd m$ goes to zero, while a physical
quantity like a mass ratio ${am_1}/{am_2}$ remains finite. Thus
the continuum limit of the lattice field theory corresponds to a
critical point (second order phase transition) of a statistical
system. When approaching the continuum, one has to increase $L/a$
such that the correlators are not affected by finite--size effects.
If one is interested in the continuum limit of explicit
finite--volume quantities, $L\cd m$ has to be kept fix as $a$ goes
to zero. 

In practice one picks one observable with small variance to \emph
{set the scale}, i.e. other observables are expressed in units
of this one. For $SU(3)$ gauge theory and also QCD the most commonly
used quantity is the hadronic length scale
$r_0$ \cite{Sommer:1993ce}, defined through the force $F(r)$
between two static color sources. More precisely, $r_0$ is the distance
where\newline
\vspace*{-6mm}
\bea
\textstyle 1.65=r^2F(r)\Big|_{r=r_0}\;.\label{arenot}\\[-5mm]\nonumber
\eea
The choice of the constant $1.65$ is based on phenomenological quark
potential models, where the distance $r_0$ defined in this way corresponds 
to approximately $0.5\fm$. Assigning physical units to lattice
results in this way introduces large systematic errors due to
uncertainty in the physical value of $r_0$, but as long as results
are expressed in terms of $r_0$ this provides a well--defined way
to compare lattice results from different actions and/or lattice
spacings. For the simple plaquette gauge action the value of $r_0/a$
is known quite precisely for the relevant range of $\beta$ values
\cite{Necco:2001xg}.

While this amounts to a non--perturbative determination of $a(\beta)$,
the evolution $a\cd\partial g_0/\partial a$ can also be calculated
in perturbation theory. The result is valid for $g_0$ going to zero
and predicts an exponential decrease of the lattice spacing as $g_0$
vanishes (\emph{asymptotic scaling}). Thus the continuum limit is
obtained by sending $g_0$ to zero, or equivalently
$\beta$ to infinity. 
As will be detailed in the next section, physical observables approach
their continuum values with powers of the lattice spacing $a$ such that
in practice a small range of $\beta$ values is generally sufficient
for a reliable continuum extrapolation.

The situation in full QCD is more complicated since additional dimensionful
parameters are present in the form of the quark masses. These have to be tuned
such that in the $\beta\rightarrow\infty$ limit a corresponding number
of renormalized quantities is kept constant. If a
lattice regularization with (at least remnant) chiral symmetry is employed,
a massless continuum limit
is obtained by simply setting the bare quark masses to zero at each value
of $\beta$.
In other cases (e.g. Wilson fermions, see \sect{sec:wilson}) additive mass
renormalization requires that one finds a line in bare parameter space, where
a suitable defined quark mass (i.e. through
the PCAC relation) vanishes.

\section{Symanzik improvement}

\label{sect:impr}

A very insightful way of looking at the continuum limit of a lattice
field theory is due to Symanzik \cite{impr:Sym1,impr:Sym2,Symanzik:1981hc}.
Instead of
taking the lattice as something that approximates a continuum theory,
he turns the tables and declares the discretized theory \emph{at finite 
cutoff} to be the main object of interest. One can then construct a
new theory in the continuum that asymptotically
describes the lattice theory.

More precisely, the discretization effects are modeled
\cite{Luscher:1998pe} by adding higher--dimensional interaction terms
(accompanied by powers of the lattice spacing) to the original continuum
action
\be
S_{\rm eff}=\int\!\rmd^4x\Big\{
\LL_0(x)+a\LL_1(x)+a^2\LL_2(x)+\ldots\Big\}\label{symeff}\;.
\ee
Here $\LL_0$ is the continuum QCD Lagrangian and the terms
$\LL_1,\LL_2,\ldots$ are linear combinations of those local
operators, which The coefficients $a$, $a^2$, \ldots are additional couplings
with negative mass dimension, which renders the theory described
by (\ref{symeff}) non--renormalizable by power--counting.
However, we are not interested in the renormalization of these
''new couplings''. Instead (\ref{symeff}) is treated as an effective
theory to finite order in the latter (but to all orders in $g_0^2$)
and is thus renormalizable order by order in $a$. It is in this sense
that renormalized correlation functions are matched between the lattice
at finite cutoff and the effective continuum theory.

This is analogous to the phenomenological approach of describing yet
undetected substructures or the effects of heavy particles through
higher--dimensional interaction terms. The most prominent example is
probably Fermi's current--current interaction that approximates two
weak--interaction vertices connected by the propagator of a $W$ boson.
In the low--energy limit $E\ll m_W$, the $W$-propagator can be
neglected and Fermi's description becomes exact.

In this sense the additional terms in Symanzik's low--energy effective
theory (\ref{symeff}) represent ''new physics'' (namely the lattice
artifacts)
entering at a scale of $a^{-1}$.
Naturally, this description can be valid only for energies small compared
to the cutoff, i.e. lattice spacings fine compared to hadronic length
scales.

In order to study how correlation functions approach their continuum limit,
one also has to represent renormalized (composite) lattice fields in the
effective theory. These effective fields
\be
\phi_{\rm eff}=\phi_0+a\phi_1+a^2\phi_2+\ldots\label{phieff}
\ee
are given as a sum of terms of increasing mass dimension that have the
same lattice
symmetries as the (composite) lattice field to be represented.

With the help of the effective continuum theory the connected renormalized
$n$--point
lattice correlation function
\be
G_n(x_1,\ldots,x_n)=(Z_\phi)^n\langle\phi(x_1)\ldots\phi(x_n)\rangle_{\rm con}
\ee
can be written in terms of expectation values with the QCD Lagrangian only
\bea
  G_n(x_1,\ldots,x_n)&=&\langle\phi_0(x_1)\ldots\phi_0(x_n)\rangle_{\rm con}
  \nonumber\\
  &&-a\int\!\!\rmd^4y\langle\phi_0(x_1)\ldots\phi_0(x_n)\LL_1(y)
  \rangle_{\rm con}\nonumber\\                                  
  &&+a\sum_{k=1}^n\langle\phi_1(x_0)\ldots\phi_1(x_k)\ldots\phi_0(x_n)
  \rangle_{\rm   con}+\rmO(a^2)\;.\label{symanzik}
\eea
The higher--dimensional terms from the action and the composite field now
appear as operator insertions, thus explicitly showing the origin of the
lattice artifacts.
If the symmetries of the lattice action in question are such that the
terms $\LL_1$ and $\phi_1$ are forbidden, correlation functions will
approach the continuum with a rate proportional to $a^2$. This is the
case for the lattice chiral symmetry of Ginsparg--Wilson fermions,
which will be discussed in \sect{sec:chirallat}.

The Symanzik improvement programme aims at a removal of the lattice
artifacts of a given action order by order in the lattice spacing.
This is done by modifying the lattice action and composite fields
such that the terms $\LL_{k}$ and $\phi_k$ in the effective continuum
theory vanish up to a fixed order $k$. If e.g. $\LL_1$ is a linear
combination of operators $\op_i$, improvement (of the action) to
order $a$ can be achieved by adding a term $\sum c_i\widehat \op_i$
to the action, where $\widehat \op_i$ is a lattice representation
of $\op_i$ and $c_i$ are suitably chosen chosen improvement 
coefficients. The same procedure is also applied to the fields $\phi$.

While in principle the improvement programme can be applied to
arbitrary orders in the lattice spacing, in practice the increasing
number of possible improvement terms limits its application
to a cancellation of the cutoff effects linear in the lattice spacing.
For Wilson fermions $\rmO(a)$ improvement of the action and the
isovector currents and densities will be discussed at the end of
\sect{sec:wilson}.

\chapter{Lattice QCD}
\label{chap:lqcd}
	In the path integral formulation fermions are represented by
two Grassmann--valued fields, $\psi$ and $\psibar$. On the lattice
these become an assignment of
$\psibar$ and $\psi$
to each lattice site $x_\mu=an_\mu$, $n_\mu\in\mathbb N^4$. They
carry color ($\alpha,\beta,\ldots$) and Dirac ($A,B,\ldots$) indices.
A lattice action $S$ combines the two lattice vectors into a scalar and
is hence defined in terms of a matrix $M\!=\!D+m_0$, where
$D$ is the lattice Dirac operator and $m_0$ the bare quark mass
\be
S=\!\!\sum_{x,y;\alpha,\beta;A,B}\psibar(x)^\alpha_A
M(x,y)^{\alpha\beta}_{AB}\psi(y)^\beta_B\;.
\ee
In the following several choices for $D$ will be discussed and in
particular we will see which properties of the continuum Dirac
operator $\dirac\mu D_\mu$ can be preserved on the lattice.

\section{Chiral symmetry on the lattice}

\label{sec:chirallat}

As we have seen in \chap{chap:cont},
the global chiral symmetry is a central aspect of continuum QCD. Unfortunately
it also seems to be the symmetry that is most difficult to obtain in a lattice
formulation. In their 1981 paper \cite{Nielsen:1981hk} Nielson and Ninomiya
gave a mathematical foundation to this finding. If we denote the Fourier
space representation of the translation--invariant Dirac operator $D$ by
$\widetilde D(p)$, the famous Nielson--Ninomiya no--go theorem states
\cite{Luscher:1998pq} that there is no lattice Dirac operator that simultaneously
has \emph{all} of the following properties.

\begin{itemize}
\item[$\circ$] {\bf [locality]} $\widetilde D(p)$ is analytic in the momenta
				$p_\mu$	with period $2\pi/a$,
				\vspace*{-1.5mm}
\item[$\circ$] {\bf [small $p$ limit]} For small momenta $\widetilde D(p)$
						behaves like $i\dirac\mu p_\mu+\rmO(ap^2),$
						\vspace*{-1.5mm}
\item[$\circ$] {\bf [no doublers]} $\widetilde D(p)$ is invertible at all
				non--zero momenta (mod $2\pi/a$),
				\vspace*{-1.5mm}
\item[$\circ$] {\bf [chirality]} $D$ anti--commutes with $\dirac5$.
\end{itemize}

The first condition ensures that $D$ is an essentially local operator,
i.e. that in position space the coupling between sites decreases exponentially
with their distance. Obviously the small momentum and invertibility conditions
are required to obtain a continuum limit with the right number
of particles and the correct dispersion relation. Finally, the
last condition guarantees the invariance under the chiral transformations
discussed at the end of \sect{sect:propQCD}. The violations of these properties
are useful to classify the various fermion discretizations.

Directly transcribing the continuum action (\ref{cont_act})
on a space--time lattice using the covariant lattice derivatives
(\ref{covder}) leads to the na{\"\i}ve lattice fermion action
\be
S_n=a^4\mysum{x}\psibar(x)\dirac\mu\widetilde\nabla_\mu\psi(x)+m_0\psibar(x)\psi(x).
\label{naiveaction}
\ee
This action is ultra--local and due to its Dirac structure it also inherited all
the (isovector) chiral properties of its continuum counterpart. By construction,
the behavior for small momenta is also correct.
However, in the
chiral limit $m_0\rightarrow 0$ the inverse momentum--space propagator
\be
S^{-1}(p) = m_0 + \frac{i}{a}\mysum{\mu}\gamma_{\mu}\sin(p_{\mu}a).
\ee
of this action
has zeros on all corners
of the Brillouin zone in addition to the one at $p\!=\!(0,0,0,0)$. 
These spurious poles of the propagator cannot be ignored, since they survive
in the continuum limit. They describe
additional unwanted particles, the so--called ''doublers'' and hence the na{\"\i}ve
lattice fermions violate the third property mentioned above.
To make things worse, the doublers come in pairs with opposite chirality,
thus spoiling the $U(1)_{\rm A}$ axial anomaly \cite{Karsten:1980wd}.\footnote{
One can define a different axial transformation to remove this cancellation
\cite{Sharatchandra:1981si}, but this still produces the anomaly
for 16 flavors.}

To achieve a correct continuum limit, Wilson added a second derivative term to the
na{\"\i}ve action (\ref{naiveaction}). This gives a mass of order $a^{-1}$ to all
doublers such that they decouple
in the continuum. The locality of the action is not affected and also the small
momentum behavior is only modified at $\rmO(ap^2)$. The Wilson fermion action with
Wilson parameter $r$ is given by
\bea
S_f&=&a^4\mysum{x}\psibar(x)(D_{\rm w}+m_0)\psi(x)\nonumber\\[-1mm]
&=&a^4\mysum{x}\psibar(x)\dirac\mu\widetilde\nabla_\mu\psi(x)
-ar\psibar(x)\half\nabla_\mu\!\nabla^\star_\mu\psi(x)+m_0\psibar(x)\psi(x)\;.
\label{wilsonaction}
\eea
The Wilson term proportional to $r$ ($0<r\leq1$, \cite{Montvay}) is diagonal
in Dirac space and hence violates
the last point in the no--go theorem. The properties of Wilson fermions and
in particular the consequences of this explicit breaking of chiral
symmetry (also with respect to the axial anomaly) will be discussed
in more detail in the next section.

In 1982 Ginsparg and Wilson \cite{Ginsparg:1981bj} suggested a way to evade
the no--go theorem and preserve the consequences of chiral symmetry.
They proposed to replace the condition $\{D,\dirac5\}=0$ with the milder
condition
\be
D\dirac5+\dirac5D=a D\dirac5D\;\qquad\textrm{(GW relation)}\footnote{
The original and more general form is actually
$D\dirac5+\dirac5D=2a D R \dirac5D$, where $R$ is some local operator.
},
\label{gwrelation}
\ee
An action defined with a Dirac operator satisfying this relation
is invariant under a continuous chiral symmetry, which becomes the
continuum
chiral symmetry as the lattice spacing $a$ goes to zero. Indeed,
it is easily verified that (\ref{gwrelation}) implies the invariance
of $\psibar D\psi$ under the transformation
\be
\delta\psi=\dirac5(1-\half aD)\psi\quad\textrm{and}\quad
\delta\psibar=\psibar(1-\half aD)\dirac5
\ee
and also its isovector counterpart. Comparing this to the axial
transformation in (\ref{anotrafo}), it can be seen as a
redefinition of $\dirac5$ to $\hat\dirac5\!=\!\dirac5(1\mi aD/2)$.

Since one now has an exact chiral symmetry at \emph{finite} lattice
spacing, there exist exactly conserved (albeit rather
complicated) isovector axial currents \cite{Hasenfratz:2002rp}.
Also the axial anomaly with GW fermions appears not only in the
continuum limit. Inserting $\hat\dirac5$ into
(\ref{isoaxial}) one immediately obtains the anomaly in the
form 
\be
\langle\partial_\mu A_\mu(x)\op_{\rm ext}\rangle=-a\langle 
\op_{\rm ext}\tr[\dirac5D]\rangle\;.
\ee
In \cite{Luscher:1998pq} it is shown that the GW relation also
implies that for any gauge background $-a\tr[\dirac5D]=2\nf\times
\rm{index}(D)$, where $\nf$ is the number of quark flavors and
the index is the difference of the right-- and left--handed zero
modes of $D$. The topological interpretation of the anomaly
is then obtained via the Atiyah--Singer index theorem
\cite{Atiyah:1971rm}, which (for smooth gauge configurations)
states that
\be
{\rm index} (D)=\frac{g_0^2}{64\pi^2}\epsilon_{\mu\nu\rho\sigma}
\!\!\int\rmd^4x\,
F_{\mu\nu}^aF_{\rho\sigma}^a\;.
\ee
It took 15 years to find solutions to the GW relation
for the interacting case of QCD. These now include the (classically
perfect) fixed point action \cite{Hasenfratz:1997ft}, domain wall fermions
\cite{Kaplan:1992bt} and the related overlap formalism \cite{Neuberger:1997fp}.
Tempted by the beautiful properties implied by the GW relation, attempts have
been made at dynamical QCD simulations using overlap fermions
\cite{Fodor:2003bh,Cundy:2004xf,DeGrand:2004nq}. However, currently available
algorithms and computer resources essentially forbid a physical application
except in the quenched case.

While already the \emph{global} chiral symmetry of QCD is a very important
aspect of the theory, its r{\^o}le in the electroweak theory is even more
fundamental. There the flavor (or weak isospin) $SU(2)$ is promoted to a local
gauge symmetry and its presence is therefore necessary to provide a
renormalizable theory. The existence of solutions to the GW--relation is
thus crucial to define a chiral gauge theory non--perturbatively
(see \cite{Luscher:2000hn} for a review).
Hence they help in establishing the renormalizability of the gauge sector
of the standard model beyond perturbation theory.

The ''staggered'' or Kogut--Susskind (KS) fermions
\cite{Kogut:1974ag,Susskind:1976jm} represent a very different approach 
by trying to directly address the
problem of fermion doubling. Through a unitary transformation of the fermionic
fields the action is brought to a spin--diagonal form, which is used to reduce
the doubling problem to a four--fold ''taste'' degeneracy. While KS fermions
keep a remnant chiral symmetry, they do not completely solve the doubling
problem and also the construction of operators with correct quantum numbers
is far from trivial. An additional theoretical problem \cite{Bunk:2004br}
with the locality of the staggered operator might lie in the
''square--root--trick'' \cite{Marinari:1981qf}, which is used
to further reduce the number of fermion flavors. With respect to the anomaly,
staggered fermions have the same properties as the na{\"\i}ve ones, i.e.
the species doubling spoils an anomaly for any number of flavors different from
four (or multiples thereof).

The last lattice fermion formulation we mention here is again based on the Wilson Dirac
operator. The (in practice) most immediate consequence of its lack  of chiral symmetry
is the fact that it is not protected against zero--modes even at non--vanishing
quark mass. In the so--called twisted mass formulation \cite{Frezzotti:2000nk} of lattice
QCD a term proportional
to $i\dirac5\tau^3$ (with the Pauli matrix $\tau^3$ acting in flavor space)
is added to the Wilson Dirac operator for two quark flavors. As a result this operator
has a manifestly positive determinant. Also here part of the continuum chiral symmetry
is recovered at the cost of breaking the vector flavor symmetry such that the total
number of conserved currents remains the same as for (untwisted) Wilson fermions.

One particularity of this lattice fermion formulation is related to a spurionic symmetry
of the terms appearing at $\rmO(a)$ in the Symanzik expansion (\ref{symanzik}). Using
this symmetry it is possible to set up a twisted mass formulation such that $\rmO(a)$
lattice artifacts cancel in most physical observables without the need to tune improvement
coefficients \cite{Frezzotti:2003ni}.

	\section{Wilson fermions}

\label{sec:wilson}

We will now proceed with a more detailed discussion of Wilson's
fermion action (\ref{wilsonaction}). Inserting the expressions
for the covariant derivatives (\ref{covder}) and fixing the
Wilson parameter $r$ to unity, it can be written
as
\bea
&&\hspace*{-6mm}S_f=a^4\!\sum_x\bigg\{\frac1{2a}\psibar(x)\dirac\mu\Big[U_\mu(x)\psi(x\pl\hat\mu)
-U_\mu^\dagger(x\mi\hat\mu)\psi(x\mi\hat\mu)\Big]\\
&&\quad\quad -\frac 1{2a}\psibar(x)\Big[U_\mu(x)\psi(x\pl\hat\mu)-2\psi(x)+
U_\mu^\dagger(x\mi\hat\mu)\psi(x\mi\hat\mu)\Big]\pl m_0\psibar(x)\psi(x)\bigg\}
\;,\nonumber
\eea
where a sum over $\mu$ is still implied in all but the last expression. The
action can be rearranged into a diagonal and off--diagonal part
\begin{eqnarray}
&&\hspace*{-6mm}S_f=
a^4\!\sum_x\bigg\{-\frac1{2a}\sum_\mu\Big[\psibar(x)(1\mi\dirac\mu)
U_\mu(x)\psi(x\pl\hat\mu)+\psibar(x\pl\hat\mu)(1\pl\dirac\mu)\nonumber
U_\mu^\dagger(x)\psi(x)\Big]\\
&&\qquad\quad\ \ +\psibar(x)\Big(m_0+\frac{4}a\Big)\psi(x)\bigg\}\;.
\end{eqnarray}
In practice the \emph{hopping parameter} $\kappa=(2am_0+8)^{-1}$ is used
instead of the bare mass $m_0$. If one rescales the fields $\psi$ and $\psibar$
by a factor $\sqrt{2\kappa}/a^{3/2}$, the Wilson action assumes the simple
form
\begin{eqnarray}
&&\hspace*{-6mm}S_f=
\!\sum_x\bigg\{-\kappa\sum_\mu\Big[\psibar(x)(1\mi\dirac\mu)
U_\mu(x)\psi(x\pl\hat\mu)+\psibar(x\pl\hat\mu)(1\pl\dirac\mu)\nonumber
U_\mu^\dagger(x)\psi(x)\Big]\\
&&\qquad\quad\ \ +\psibar(x)\psi(x)\bigg\}.
\end{eqnarray}
The transformations of charge conjugation $C$, parity $P$ and time
reversal $T$ listed in \tab{cpt} are all discrete symmetries of the
Wilson fermion action (this is of course also true for the plaquette
gauge action). In addition it also inherited the $\dirac5$--Hermiticity
($\dirac5 D\dirac5=D^\dagger$ due to the use
of the symmetric derivative $\widetilde\nabla_\mu$) from the continuum.

\def\arraystretch{1.25}
\begin{table}[!b]
\center
\begin{tabular}{|@{$\hspace*{-1mm}$}c@{$\hspace*{-1mm}$}|@{$\hspace*{-0.2mm}$}
l@{$\hspace*{-1mm}$}|@{$\hspace*{-0.2mm}$}l@{$\hspace*{-1mm}$}|}\hline
$C$       &       
$\begin{array}{rcl}
U_\mu(x)&\Ctrafo&U_\mu^*(x)
\end{array}$
&
$\begin{array}{rcl}
\psi(x)&\Ctrafo&\phantom{-}C\psibar^T(x)\\
\psibar(x)&\Ctrafo&-\psi^T(x)C^{-1}\\
\multicolumn{3}{c}{
C=i\dirac0\dirac2}
\end{array}$\\\hline
\begin{tabular}{c}
$P$\\
$x\Ptrafo x_P=(x_0,-\bx)$
\end{tabular}
&
$\begin{array}{rcl}
U_0(x)&\Ptrafo&U_0(x_P)\\
U_k(x)&\Ptrafo&U_k^\dag(x_p\mi \hat k)
\end{array}$
& 
$\begin{array}{rcl}
\psi(x)&\Ptrafo&\dirac0\psi(x_P)\\
\psibar(x)&\Ptrafo&\psibar(x_P)\dirac0
\end{array}$\\\hline
\begin{tabular}{c}
$T$\\
$x\Ttrafo x_T=(-x_0,\bx)$
\end{tabular}
&
$\begin{array}{ccc}
U_0(x)&\Ttrafo&U_0^\dag(x_T\mi\hat0)\\
U_k(x)&\Ttrafo&U_k(x_p)
\end{array}$
& 
$\begin{array}{rcl}
\psi(x)&\Ttrafo&T\psi(x_T)\\
\psibar(x)&\Ttrafo&\psibar(x_T)T^{-1}\\
\multicolumn{3}{c}{
T=i\dirac0\dirac5}
\end{array}$\\\hline
\end{tabular}
\caption{$C$,$P$ and $T$ transformations on the lattice.
\label{cpt}}
\end{table}

As already mentioned, chiral symmetry is broken at order $a$ by the
Wilson term and recovered only in the continuum. As a consequence
the point of vanishing bare quark mass is no longer endowed with an
enhanced symmetry (leading to the conservation of the axial current)
and protected from renormalization. However,
by tuning the bare quark mass $m_0$ to a non--zero
value one can still find a point, 
where the axial current is conserved
(up to cutoff effects). At this \emph{critical mass} $\mc(g_0^2)$ a
quark mass derived from the PCAC relation
(see \sect{sect:latticeSF}) vanishes and thus the theory can be parameterized
in terms of the bare subtracted quark mass $\mq=m_0-\mc$.
In perturbation theory the critical
mass if found to be \cite{Panagopoulos:2001fn}
\be
\mc(g_0^2)=-0.2701\,g_0^2-0.0430\,g_0^4+\rmO(g_0^6)\label{mcrit}\;.
\ee
In addition to this \emph{additive quark mass
renormalization} the axial current itself requires a finite renormalization,
which is discussed in \sect{currentnorm}.
In analogy to \sect{contward} we now study
the lattice currents associated with the vector and axial flavor
transformations in more detail.

\subsection{Vector currents}

The isospin symmetry, which the continuum action possesses in the case of
degenerate quarks, is not broken by the Wilson
term. Hence a global isospin vector variation of the Wilson action vanishes and
a local one, parameterized by $\omega^a(x)$, can be written as a divergence
\be
\dv S_f=a^4\sum_{x}-\omega^a(x)\drvstar\mu \widetilde V_\mu^a(x)\;,
\ee
where $\drvstar\mu$ is the backward lattice derivative and $\widetilde V_\mu^a(x)$
the split--point vector current
\be
\widetilde V_\mu^a(x)\!=\!
\half\Big\{\psibar(x\pl\hat\mu)\half\tau^a(1\pl\dirac\mu)
U_\mu^\dagger(x)\psi(x)-
\psibar(x)\half\tau^a(1\mi\dirac\mu)U_\mu(x)\psi(x\pl\hat\mu)\Big\}\;.
\label{splitvector}
\ee
Although one now has an exactly conserved vector current for Wilson fermions,
in practice the local current
\be
V_\mu^a(x)=\psibar(x)\dirac\mu\half\tau^a\psi(x)
\ee
is generally used for convenience since sources for
correlation functions of this current are easier to construct.
Not being a Noether current, it is not protected from
renormalization. However, its normalization factor $Z_{\rm V}(g_0^2)$
is calculated essentially as a by--product of the axial current renormalization
discussed in \chap{chap:renorm}.

\subsection{Axial currents}

Of more interest are the axial transformations. Let us first consider
the case of na{\"\i}ve fermions, where also chiral symmetry is preserved on the
lattice. The associated current $\widetilde A_\mu^a(x)$ is conserved
at zero quark mass due to
\begin{eqnarray*}
\da S_n&=&a^4\sum_{x}\omega(x)\Big[-\drvstar\mu \widetilde A_\mu^a(x)+2m_0P^a(x)\Big]\;,\\
\textrm{with }\ \ \widetilde A_\mu^a(x)&=&
\half\Big\{\psibar(x\pl\hat\mu)\half\tau^a\dirac\mu\dirac5
U_\mu^\dagger(x)\psi(x)+
\psibar(x)\half\tau^a\dirac\mu\dirac5U_\mu(x)\psi(x\pl\hat\mu)\Big\}\;.
\end{eqnarray*}
At non--vanishing bare quark mass the divergence of this axial current is
proportional to the pseudo--scalar density
\be
P^a(x)=\psibar(x)\dirac5\half\tau^a\psi(x)\;.
\ee
In this way na{\"\i}ve lattice fermions have the chiral symmetry of
the continuum if the local axial current
\be
A_\mu^a(x)=\psibar(x)\dirac\mu\dirac5\half\tau^a\psi(x)
\ee
is replaced by the split--point axial current $\widetilde A_\mu^a(x)$.
However, this symmetry is broken by the Wilson term such that even at
vanishing bare quark mass no conserved axial current can be defined.
Denoting the axial variation of the Wilson term by $X^a(x)$ we obtain
\cite{Bochicchio:1985xa}
\bea
\da S_f&\!\!=\!\!&a^4\sum_x\omega^a(x)\Big[-\drvstar\mu
\widetilde A_\mu^a(x)+2m_0P^a(x)+X^a(x)\Big]\;,\nonumber
\qquad\textrm{ with}\\
 X^a(x)&\!\!=\!\!&-{\ts\frac1{2a}}\sum_\mu\Big\{
\psibar(x)\half\tau^a\dirac5U_\mu(x)\psi(x\pl\hat\mu)\nonumber
+\psibar(x\pl\hat\mu)\half\tau^a\dirac5U_\mu^\dagger(x)\psi(x)\\[-4mm]
&&\qquad\qquad+\Big[x\rightarrow(x\mi\hat\mu)\Big]-4P^a(x)\big\}
\;.\label{wilsonvari}
\eea
Here the explicit breaking of chiral symmetry by the Wilson term can be
seen from the fact that $X^a$ can not be written as the divergence of a current.

If Wilson fermions are an admissible regularization of QCD, it must be
possible to reproduce the correct axial anomaly in the continuum
limit. Performing an iso\emph{singlet} axial transformation of the Wilson action
one obtains \eq{wilsonvari} with the replacement $(\half\tau^a\rightarrow1)$.
It is
precisely the isosinglet axial variation $X$ of the Wilson term, which
(under certain assumptions \cite{Kerler:1981tb}) reproduces the correct
axial anomaly
\be
\lim_{a\rightarrow 0}X=\frac{\nf g_0^2}{32\pi^2}\epsilon_{\mu\nu\rho\sigma}
F_{\mu\nu}^aF_{\rho\sigma}^a\;.
\ee

\subsection{Improvement}

\label{sect:wilsonimpr}
For the Wilson fermion action the $\LL_1$ term in Symanzik's effective theory
can contain the five expressions ($\sigma_{\mu\nu}\!=\!\frac i2[\dirac\mu,\dirac\nu]$)
\be
\begin{array}{r@{\,=\;}l@{\quad}r@{\,=\;}l}
\op_1 & \psibar i\sigma_{\mu\nu}F_{\mu\nu}\psi\;, &
\op_2 & \psibar D_\mu D_\mu\psi+\psibar \lvec D_\mu\lvec D_\mu\psi\;,\\
\op_3 & \mq\tr\{F_{\mu\nu}F_{\mu\nu}\} \;,&
\op_4 & \mq\{\psibar\dirac\mu D_\mu\psi-\psibar\dirac\mu\lvec D_\mu\psi\}\;,\\
\op_5 & \mq^2\psibar\psi\;,\\
\end{array}
\ee
since these are compatible with the symmetries ($C$,$P$,$T$ and
$\dirac5$--Hermiticity) of the Wilson--Dirac operator. Note that
none of these terms can appear in the effective action
for a Dirac operator that satisfies the GW--relation (\ref{gwrelation}).

Through formal manipulations of the Euclidean functional integral
(which amount to an application of the classical field equations)
two of these (e.g. $\op_2$ and $\op_4$) can be eliminated
\cite{Luscher:1998pe}. While this
is necessary from a practical point of view, it restricts the
improvement to on--shell quantities, i.e. correlation functions over
finite distances. However, since in practice all physically relevant matrix
elements are on--shell, this does not represent any loss of applicability

On the other hand, adding $\op_3$ and $\op_5$ merely amounts to a rescaling
of the bare coupling and quark mass by factors $1\pl\rmO(a\mq)$. We will
return to this point in the discussion of normalization factors.
What remains is the Pauli term $\op_1$. On the lattice the field
strength tensor can be represented by

\vspace*{-6mm}

\bea
\widehat F_{\mu\nu}&=&\frac1{8a^2}[Q_{\mu\nu}-Q_{\nu\mu}]\;,\\[2mm]
\textrm{where }\ Q_{\mu\nu}(x)&=&U_{x;\mu,\nu}+U_{x;-\nu,\mu}
+U_{x;-\mu,-\nu}+U_{x;\nu,-\mu}\;,
\eea
to arrive at the action \cite{Sheikholeslami:1985ij}
\be
S_{\rm SW}=\sum_x\psibar(x)\Big[D_w+m_0+a\csw{\ts\frac i4}\sigma_{\mu\nu}
\widehat F_{\mu\nu}\Big]\psi(x)\;.\label{cloveraction}
\ee
The normalization of the improvement term is chosen such that
the improvement coefficient $\csw=1$ at tree--level in perturbation
theory. In general however, $\csw$ depends on $g_0^2$ and a non--perturbative
determination for the
two--flavor case \cite{Jansen:1998mx,Yamada:2004ja} is summarized by the
interpolating formula
\be
\csw(g_0^2)=\frac{1-0.454g_0^2-0.175g_0^4+0.012g_0^6+0.045g_0^8}
{1-0.720g_0^2}\;.
\ee
For the gauge action $\LL_1$ vanishes and thus at $\rmO(a)$ no
improvement is needed.

The number of possible improvement terms
for the isovector axial and vector currents can be reduced using the
same arguments as for the Wilson action. The on--shell improved currents
are given by
\begin{eqnarray}\label{AI}
(\aimpr)_{\mu}^a&=& A_{\mu}^a
+a\cA \drvtilde\mu{}P^a\;,\\[1mm]
(\vimpr)_{\mu}^a&=& V_{\mu}^a
+a\cv\drvtilde\mu{}T_{\mu\nu}^a,\label{VI}\\[1.8mm]
\textrm{where }\quad
  T_{\mu\nu}^a(x)&=&i\psibar(x)\sigma_{\mu\nu}\half\tau^a\psi(x)\;.\label{Tmunu}
\eea
The pseudo--scalar density $P^a(x)$ is already improved, i.e. no $\rmO(a)$
counter term with the correct symmetries exists. For the vector current
correlation functions we will consider, the term proportional to $\cv$ does not
contribute.
The 1--loop result for
$\cA$ is \cite{Luscher:1996vw}
\be
\cA(g_0^2)=-0.00756g_0^2+\rmO(g_0^4)\label{capt}
\ee
and a non--perturbative determination will be presented in \chap{chap:impr}.

\section{Current renormalization}

\label{currentnorm}

A scheme, in which the renormalization conditions are imposed on
a set of correlation functions at a point $(g_0,am_0)$ in bare
parameter space, would result in renormalization factors,
which depend on both the bare coupling and the bare mass as
well as a renormalization scale $a\mu$.
Inherently simpler are mass--independent schemes, where one
imposes the normalization condition at zero quark mass with the
consequence that no dependence of the renormalization factors on
the quark mass is introduced \cite{Weinberg:1973ss}.

Neglecting $\rmO(a)$ improvement, in such a scheme one would introduce renormalized
parameters through
e.g. $\mr=Z_m\mq$, where the subtracted mass $\mq=m_0-\mc$ parameterizes
deviations from the critical line and the renormalization factor depends
on the bare coupling $g_0^2$ as well as the renormalization scale $a\mu$.
As is discussed in \cite{Luscher:1996sc}, such a definition leads
to uncanceled $\rmO(am)$ effects and is thus not compatible
with the improvement programme. This can be understood from
the fact that in \sect{sect:wilsonimpr} we ignored $\rmO(a)$
counterterms, which merely
amounted to a rescaling of the bare parameters.

If we want to employ a massless renormalization scheme and
at the same time
have on--shell $\rmO(a)$ improvement, we need to allow
for a more general form of renormalized quantities.
These are expressed in terms of the modified bare coupling
and quark mass
\bea
\tilde g_0^2&=&g_0^2(1+\bg a\mq)\;,\\
\tilde \mq&=&\mq(1+\bm a\mq)\;,
\eea
where the $b$--coefficients are functions of $g_0^2$.
The renormalized coupling and quark mass can then
be written as
\bea
\gr^2&=&\tilde g_0^2 Z_g(\tilde g_0^2,a\mu)\\
\mr  &=&\tilde \mq Z_m(\tilde g_0^2,a\mu)\;.
\eea
Also for local composite fields $\phi$ the $1+\rmO(a\mq)$
counterterm is conventionally not included in the
definition of the improved field $\phi_{\rm I}$, but rather
in the renormalized field according to
\be
\phi_{\rm R}(x)=Z_\phi(\tilde g_0^2,a\mu)(1+b_\phi a\mq)\phi_{\rm I}(x)\;.
\ee
In this context the isovector axial and vector currents are
special since in the (massless) continuum limit they become
the Noether currents of the flavor chiral symmetry and
at finite lattice spacing their normalization can thus be fixed by
imposing a continuum
Ward identity.
The latter is a local identity and hence the renormalization factors
of the isovector axial and vector currents do not depend
on a scale. In addition they approach unity for $g_0^2$
going to zero, i.e. when the flavor chiral symmetry is recovered
\cite{Bochicchio:1985xa,Testa:1998ez}.
\bea
(\ar)_{\mu}^a&=&\za(\tilde g_0^2)\,(1+\ba
a\mq)(\aimpr)_{\mu}^a\label{ar}\;,\\
(\vr)_{\mu}^a&=&\zv(\tilde g_0^2)\,(1+b_{\rm V}
a\mq)(\vimpr)_{\mu}^a\;.\label{vr}
\eea
To one loop in perturbation theory \cite{Gabrielli:1990us,
Gockeler:1996gu,sint:notes}
one finds that
\bea
\za&=&1-0.116458\,g_0^2+\rmO(g_0^4)\;,\label{pta}\\
\zv&=&1-0.129430\,g_0^2+\rmO(g_0^4)\;.\label{ptv}
\eea
One should point out again that we have ignored the existence
of an exactly conserved vector current (\ref{splitvector}) for
Wilson fermions and treat $V_\mu^a$ and $A_\mu^a$ on exactly the
footing. The normalization of the split--point vector current
(\ref{splitvector}) is given naturally, i.e. it does not renormalize.

The renormalization of the pseudo--scalar density
\be
(P_{\rm R})^a(x)=\zp(1\pl \bp am_q) P^a(x)\label{zp}
\ee
on the other hand will depend on the scale, at which
the normalization condition is imposed, i.e. $\zp=
\zp(\tilde g_0^2,a\mu)$.

A massless scheme requires that the normalization conditions
are defined at vanishing quark mass. In a non--perturbative
implementation this implies that numerical simulations have
to be performed at zero (or at least very small) quark mass.
With periodic boundary conditions such simulations are
technically very
difficult or even impossible since there is no lower bound
on the spectrum of the Dirac operator.
This problem is cured in the Schr\"odinger functional setup,
which will be discussed in the following chapter, at
the end of which we will define a renormalized quark mass
in terms of correlation functions of renormalized composite
fields.

\chapter{The Schr\"odinger functional}
\label{chap:SF}
	So far no boundary conditions have been specified
since they would have been of no importance
in the discussion of the local symmetries.
It was thus implicitly assumed that we are working either
in infinite volume or in finite volume with periodic
boundary conditions for the fields.
All numerical results report here were obtained in the
Schr\"odinger functional (SF) setup, which employs
periodic boundary conditions in the three spatial directions
and fixed (Dirichlet) boundary conditions in time.

There are several advantages of this method compared to
the torus, where all four directions are periodic. The Dirichlet
boundary conditions provide an infrared cutoff (inversely
proportional to the time extension $T$) to the Dirac operator
\cite{Sint:1993un}.\footnote{This can be rigorously shown only
in the
$g_0^2\rightarrow0$
limit.}
The gluonic boundary conditions can be used to induce
a color background field, which was employed e.g. in the
definition of a running coupling
\cite{Luscher:1993gh,DellaMorte:2004bc}.
The fermionic boundary fields can serve as sources
for correlation functions in an elegant way, particularly
if a zero--momentum projection  or
wave--functions are required. This results in mesonic
correlation functions with smaller statistical error than
on the torus.

A vast literature (see e.g.
\cite{Luscher:1992an,Sint:1993un,Sommer:1997xw,Luscher:1998pe}) exists on the SF and we will therefore
restrict the discussion here to the aspects required
in the following chapters, in particular
the ingredients required for the definition
of the lattice SF with Wilson fermions.
Some SF correlation functions needed in the axial current
improvement (\chap{chap:impr}) and renormalization
(\chap{chap:renorm}) are also introduced.

The QCD Schr\"odinger functional is the propagation kernel
of a field configuration $C,\rhobar,\rho$ at Euclidean time
$0$ to a configuration $C',\rhobarprime,\rhoprime$ at time $T$.
Following Feynman, this is written as an integration over all
interpolating field configurations
\be
\Z[C,C';\rhobar,\rho;\rhobarprime,\rhoprime]=\int_{A,\psibar,\psi}
\rme^{-S[A,\psibar,\psi]}\;.\label{donotuseme}
\ee

\section{Lattice SF with Wilson fermions}

\label{sect:latticeSF}

We consider an $L^3\times T$ space--time cylinder, i.e. the
set of sites ($k=1,2,3$)
\be
\big\{x\,\big|\,x/a\in\mathbb Z^4, 0\le x_0\le T, 0\le x_k<L \big\}
\ee
and the associated links $U_\mu(x)$.\footnote
{Except for $U_0(x)$ when $x_0=T$.}
The gauge links are spatially strictly periodic, while the fermionic
fields are allowed to pick up a phase $\theta_k$
at the boundary
\bea
U_\mu(x\pl L\hat k/a)&=&U_\mu(x)\nonumber\\
\psi(x\pl L\hat k/a)&=&\rme^{i\theta_k}\psi(x)\\
\psibar(x\pl L\hat k/a)&=&\rme^{-i\theta_k}\psibar(x)\nonumber
\;.
\eea
The Dirichlet boundary conditions are implemented via
\be
U_k(x)|_{x_0=0}=\exp(C_k)\quad \textrm{and}\quad U_k(x)|_{x_0=T}=\exp(C_k')\;,
\ee
where $C_k$ is a traceless anti--Hermitian $3\times 3$ matrix.
Throughout this work we use $C_k=C_k'=0$, i.e. no background field.
From the lattice action it will be obvious that only
half of the fermion fields components on the boundaries actually
couple to the interior of the SF cylinder. The SF therefore becomes
a well--defined boundary value problem by prescribing
($P_\pm=(1\pm\dirac0)/2$)
\be
\begin{array}{r@{\,=\,}lr@{\,=\,}l}
P_+\psi(x)\big|_{x_0=0}&\rho(\bx)\;,  &
\psibar(x)P_-\big|_{x_0=0}&\rhobar(\bx)\;,\\[1mm]
P_-\psi(x)\big|_{x_0=T}&\rho'(\bx)\;, &
\psibar(x)P_+\big|_{x_0=T}&\rhobarprime(\bx)\;.
\end{array}
\ee
While the fermionic boundary fields ($\rho,\ldots$) are always
set to zero, functional derivatives of the effective action
with respect to them are used to define correlation functions
involving the SF boundary.
Thus, expectation values of composite fields $\op$ 
in the lattice Schr\"odinger functional
are obtained
from
\be
\langle\op\rangle=\left\{
\frac1\Z\int\!\!\DD U\DD\psibar\DD\psi\, \op\,\rme^{-S[U,\psibar,\psi]}
\right\}_{\rhobarprime=\rhoprime=\rhobar=\rho=0}\;.\label{ev}
\ee
In addition to the dynamical variables $U,\psibar$ and $\psi$,
the composite fields may then also involve the ''boundary fields''
(see \app{app:SF} for details)
\be
\begin{array}{r@{\ =\ }l@{\;,\quad}r@{\ =\ }l}
\ds\zeta(\bx)&\ds\frac{\delta}{\delta\rhobar(\bx)}	&
\ds\zetabar(\bx)&\ds-\frac{\delta}{\delta\rho(\bx)}\;,\\[4mm]
\ds\zetaprime(\bx)&\ds\frac{\delta}{\delta\rhobarprime(\bx)}	&
\ds\zetabarprime(\bx)&\ds-\frac{\delta}{\delta\rhoprime(\bx)}\;.
\end{array}\label{zeta}
\ee
When acting on the Boltzmann factor in (\ref{ev}),
these have the effect of inserting certain combinations
of $\psi$ and $\psibar$ close to the boundary, together with the
appropriate gauge field variables to ensure gauge covariance.
While the lattice action $S[U,\psibar,\psi]$ is in principle given by
(\ref{wilsongauge}) and (\ref{cloveraction}) together with the above boundary
conditions, the issue of $\rmO(a)$ improvement needs to be re--addressed
due to boundary effects.

In the presence of the Dirichlet boundary conditions, surface terms
have to be added to the effective action (\ref{symeff}). At
$\rmO(a)$ this is \cite{Luscher:1996sc}
\be
\lim_{\epsilon\rightarrow0}\int\rmd^3\bx\{\mathcal B_1(\bx)|_{x_0=\epsilon}+
\mathcal B_1'(\bx)|_{x_0=T-\epsilon}\}\;,
\ee
where $\mathcal B_1$ and $\mathcal B_1'$ are linear combinations of composite fields
of mass dimension four. For the gauge action the only fields that can contribute
are $\tr\{F_{kl}F_{kl}\}$ and $\tr\{F_{0k}F_{0k}\}$. On the lattice these are
implemented by modifying the weight of the plaquettes \emph{in} the
boundaries ($\cs/2$) and those \emph{touching} the boundaries ($\ct$).

On--shell $\rmO(a)$ improvement of the fermion action can be achieved by e.g.
adding the field products
\be
\begin{array}{r@{\,=\;}l}
\op_{11} & \psibar P_+D_0\psi+\psibar\lvec D_0P_-\psi \;,\\
\op_{14} & \psibar P_-\dirac kD_k\psi-\psibar\lvec D_k\dirac kP_+\psi\;,
\end{array}\label{SFimp1}
\ee
at $x_0=0$ and
\be
\begin{array}{r@{\,=\;}l}
\op_{12} & \psibar P_-D_0\psi+\psibar\lvec D_0P_+\psi \;,\\
\op_{13} & \psibar P_+\dirac kD_k\psi-\psibar\lvec D_k\dirac kP_-\psi\;,
\end{array}\label{SFimp2}
\ee

\noindent at $x_0=T$. The corresponding lattice discretizations of these terms are
added to the action with coefficients $\ctildet$ for the temporal derivatives
and $\ctildes$ for the spatial derivatives.

To keep the discussion transparent, the explicit form of the improved action
is given in \app{app:SF}. Here we want to note only that for our specific
choice of boundary conditions the terms proportional to $\cs$ and $\ctildes$
do not contribute, while $\ct$ and $\ctildet$ are known perturbatively
\cite{Bode:1999sm,Sint:1997jx}
\bea
\ct&=&1-0.089\,g_0^2-0.03\,g_0^4\;,\\
\ctildet&=&1-0.018\,g_0^2\;.
\eea
Unless $T/a$ is very small, a non--perturbative knowledge of these
coefficients is not necessary since the lattice artifacts they modify
are surface effects and thus suppressed by $a/T$ compared to the
cutoff--effects from the bulk. Moreover, in \cite{Luscher:1996sc}
it is shown explicitly that these term can not contribute
$\rmO(a)$ terms to matrix elements of the PCAC relation (since the
latter holds exactly in the continuum theory).

The two--point functions are derived by taking functional derivatives
of the generating functional and all the fermionic propagators
involving the boundary and bulk
quark fields are also given in \app{app:SF}.

\section{Fermionic correlation functions}

\label{addcorr}

We are interested in mesonic correlation functions, i.e. we want to excite
states with the quantum numbers of isovector pseudo--scalar particles. In terms
of the boundary fields (\ref{zeta}) possible choices are
\begin{eqnarray}\label{O}
\op^a&\!\!=\!\!&a^6\sum_{\bf u,v}
\zetabar(\bu)\dirac{5}\half
\tau^a\zeta(\bv)\;,\\
\oprime^a&\!\!=\!\!&a^6\sum_{\bf u,v}
\zetabarprime(\bu)\dirac{5}\half
\tau^a\zetaprime(\bv)\;.\label{Oprime}
\end{eqnarray}
With these operators both constituent quarks are projected
to zero momentum separately. 
In \chap{chap:impr} we will also consider
generalizations of (\ref{O}, \ref{Oprime}), which make use
of spatial trial wave functions to influence the excitations
of pseudo--scalar states.

To obtain non--trivial expectation values in (\ref{ev}) the operator
appearing there needs to be compatible with the lattice symmetries,
in particular it must be invariant under parity and an isoscalar.
Restricting ourselves
to two quark bilinears, there are five possible combinations
involving (\ref{O}, \ref{Oprime})
\bea 
\fa(x_0)&=&-\frac{a^3}{3L^3}\sum_\vecx\langle A_0^a(x)\,\op^a
\rangle\;,\label{fA}\\[-1mm]
\fp(x_0)&=&-\frac{a^3}{3L^3}\sum_\vecx\langle P^a(x)\,\op^a
\rangle\;,\label{fP}\\[-1mm]
\ga(x_0)&=&\phantom{-}\frac{a^3}{3L^3}\sum_\vecx\langle \oprime^a\,A_0^a(x)
\rangle\;,\label{gA}\\[-1mm]
\gp(x_0)&=&-\frac{a^3}{3L^3}\sum_\vecx\langle \oprime^a\,P^a(x)
\rangle\;,\label{gP}\\[-1mm]
\textrm{and }\ \ f_1&=&- \frac1{3L^6}\langle\op'^a \,\label{f1def}
\op^a\rangle\;.
\eea
Their explicit form in terms of quark propagators is given in
\app{app:corr}.
From these correlation functions one can define
a current quark mass $m$ through the PCAC relation (\ref{pcac}).
If the transformations (\ref{dA}) are applied in the interior,
the boundary fields $\op$ and $\oprime$ are ''external'' and
we get
\bea
\tilde\partial_\mu \langle A_\mu^a\op^a\rangle&=&2m\langle P^a\op^a\rangle
\eea
and thus e.g.
\be
\frac{\tilde\partial_0\fa(T/2)}{2\fp(T/2)}\label{mass}
\ee
defines a bare current quark mass since the spatial derivatives vanish
under periodic boundary conditions. The lattice artifacts,
by which \eq{mass} is affected can be reduced by using the
improved axial current
(\ref{AI}) to arrive at
\be
m_{\rm I}=\frac{\tilde\partial_0\fa(T/2)+a\cA\partial_0\partial_0^\star
\fp(T/2)}{2\fp(T/2)}\label{massimp}
\ee
as a possible estimator for an improved current quark mass.
Of course, also $\ga$ and $\gp$ could be used instead.
In fact, for vanishing background field ($C_k=C_k'=0$)
time--reflection is a symmetry of the SF and
$f_{\rm X}(x_0)=g_{\rm X}(T\mi x_0)$. An average of $f$ and $g$
can then be used to improve the statistical precision.

Since their derivation is based on local symmetry transformations
of the continuum action, the quark masses defined in this way
are independent of the lattice extensions $L$ and $T$ up to
cutoff effects of $\rmO(a^2)$ in the improved theory. This is
explicitly shown in \cite{Luscher:1996vw}.

The Schr\"odinger functional also defines a (finite volume)
renormalization scheme, which has been used e.g. in the calculation
of the scale dependence of the SF coupling $\bar g^2$
\cite{DellaMorte:2004bc}. In the SF scheme the renormalization
scale $\mu$ is identified with the inverse box length $L^{-1}$.
In the context of a project to determine the scale evolution of
quark masses, the renormalization constant of the
pseudo--scalar density has been determined non--perturbatively in
this scheme \cite{Capitani:1998mq,Knechtli:2002vp}. To this end one
requires that a certain matrix
element of $(P_{\rm R})^a$ is equal to its tree--level value $c$.
The concrete requirement is that
\be
\zp(g_0,L/a)=c\frac{\sqrt{3f_1}}{\fp(T/2)}
\ee
at vanishing quark mass. One can then proceed and insert the
renormalized currents into
(\ref{massimp}) to define a renormalized improved quark mass
\bea
\overline m&=&\frac{\za(1\pl\ba a\mq)}{\zp(1\pl\bp a\mq)}
\,m_{\rm I}+\rmO(a^2)\\
&=&\frac\za\zp(1\pl(\ba\mi\bp)\, a\mq)
\,m_{\rm I}+\rmO(a^2)\;.
\eea
Since $\za$ depends only on the (modified) bare coupling,
the entire scale (and scheme) dependence of the renormalized quark mass
enters through the normalization of the pseudo--scalar density.

The evolution of $\overline m$ is then traced to very high energies, where
the SF scheme can be related to a more common renormalization scheme
like $\overline{\rm MS}$ by means of a 1--loop calculation.
The application of this programme to the strange quark mass
\cite{DellaMorte:2005kg} is one
of the immediate applications of the axial current improvement
and renormalization presented here.
In the quenched approximation this has been done in
\cite{Capitani:1998mq,Garden:1999fg}

\chapter{Algorithmic issues}
\label{chap:algo}
The basics of our lattice theory as well as the framework, in which we
implement the axial current improvement and renormalization, have now been
laid out. However, before proceeding to a detailed discussion of the latter,
it is necessary to detour into the more technical aspects of full QCD
simulations.

\section{Hybrid Monte Carlo algorithms}
\label{sect:hmc}
Since the fermionic action is bilinear in the quark fields,
the Grassmann integral can be performed analytically, 
resulting in the determinant of the Dirac operator. This step
is even necessary unless one can find an efficient way of
dealing with
the Grassmann fields directly. For two quark flavors one
has (writing $M=D+m$)
\bea
\Z&=&\int\!\!\DD U\DD\psibar\DD\psi\, \exp\Big\{-S_g[U]-\!\!
\sum_{\rm flavors}\!\psibar M[U]\psi\Big\}\\
&=&\int\!\!\DD U\ \rme^{-S_g[U]}
\det M[U]^2\;.
\eea
Due to the $\dirac5$--Hermiticity of $M$ this can be expressed
in terms of the positive matrix $Q^2=Q^\dagger Q$ with $Q=\dirac5M$,
such that $\det M^2=\det QQ^\dagger$.
With the introduction of a so--called pseudo--fermion field $\phi$
the determinant can be expressed as a \emph{bosonic} Gaussian integral
\be
\Z=\int\!\!\DD U\DD\phi\ \exp\left\{
-S_g[U]-\phi^\dagger (Q Q^\dagger)^{-1} \phi\right\}
=\int\!\!\DD U\DD\phi\ \rme^{-S_{\rm eff}}\;.\label{seff}
\ee
To generate gauge field configurations distributed according
to the effective action, the hybrid Monte Carlo (HMC)
\cite{Duane:1987de}
algorithm evolves the gauge field in a fictitious time
$\tau$ (molecular dynamics). This evolution is described
by anti--Hermitian traceless momentum matrices $\pi_\mu(x)$
\be
\frac\partial{\partial\tau}U_\mu(x)=\pi_\mu(x)U_\mu(x)\;.
\ee
Adding a Gaussian action for the momenta to $S_{\rm eff}$,
we obtain the Hamiltonian of the molecular dynamics (MD) evolution
\be
H=S_{\rm eff}-\half\tr\pi^2\;.\label{hamilt}
\ee
The evolution equation of the momenta $\pi$ is derived
by requiring the Hamiltonian $H$ to be a constant of the
MD motion. The update of the momenta
requires the computation of the ''force'', i.e.
$\delta S_{\rm eff}/\delta U$, which involves an inversion
of $Q^\dagger Q$ for the fermionic contribution. For this inversion
the conjugate gradient algorithm is used (see \sect{sect:invert}).
Since the momenta don't couple to the gauge or pseudo--fermion
fields they will have no effects on correlation functions involving
$U$ and $\phi$. An HMC update step (trajectory) now proceeds from
a starting configuration $U(0)$
according to

\vspace*{1.5mm}

\begin{minipage}[c]{12.5cm}
\begin{enumerate}
\item Generate a complex Gaussian random vector and apply $Q$ to it, to obtain
		$\phi$ distributed according to (\ref{seff});
		\vspace*{-1.8mm}
\item Generate momenta according to (\ref{hamilt}) by drawing Gaussian
		components in the Lie algebra generators;
		\vspace*{-1.8mm}
\item For fixed $\phi$, numerically evolve $U$ and $\pi$ for a MD time
		$\tau_0$ using a numerical approximation of Hamilton's equations
		of motion;
		\vspace*{-1.8mm}
\item Accept $U(\tau_0)$ with a probability $\rme^{-\Delta H}$,
		where\\ $\Delta H=H(\tau_0)-H(0)$ (Metropolis step).
\end{enumerate}
\end{minipage}

\vspace*{4mm}

In case of rejection $U(0)$ becomes also the next
configuration in the Markov chain.
In the next trajectory new pseudo--fermion fields are used to estimate the
determinant contribution and also the direction of evolution is again
chosen randomly by using new Gaussian momenta (\emph{refreshed}
molecular dynamics). Several aspects of the HMC are worth mentioning
at this points.

For the algorithm to satisfy detailed balance with respect to the
effective action, the numerical integration scheme needs to be
reversible and area preserving (in phase space).
The molecular dynamics evolution is the computationally most expensive part
of the HMC algorithm.

The simplest reversible and area preserving integration scheme, is the
\emph{leapfrog}. With $\dtau=\tau_0/\nstep$ this is defined through
\beann
\hspace*{29.1mm}\pi&\rightarrow&\pi-
\delta S_{\rm eff}/\delta U\cd\delta\tau/2\\[-6.4mm]
\eeann
\be
(\nstep\mi1)\ \times\ \left\{
\begin{array}{rcl}
U&\rightarrow&\exp(\pi\dtau)U\\[-1mm]
\pi&\rightarrow&\pi-
\delta S_{\rm eff}/\delta U\cd\delta\tau
\end{array}
\right\}
\ee
\vspace*{-4.4mm}
\beann
U&\rightarrow&\exp(\pi\dtau)U\\[-0.5mm]
\hspace*{31.4mm}\pi&\rightarrow&\pi-
\delta S_{\rm eff}/\delta U\cd\delta\tau/2\;.
\eeann
Throughout this work we always consider molecular dynamics trajectories 
of unit length, i.e. $\tau_0=1$ and $\dtau=1/\nstep$.
The half--steps differ from the exact integration of Hamilton's equation by
errors of order $\dtau^2$, whereas the $\nstep=\rmO(1/\dtau)$ intermediate
steps each have errors
of $\dtau^3$.
More elaborate (and hence more expensive) schemes can increase the
acceptance rate by lowering the violation of energy (Hamiltonian)
conservation.
If the accept/reject step at the end of the HMC trajectory is omitted,
one generates a systematically wrong ensemble and an extrapolation to
$\dtau=0$ becomes necessary ($\phi$--algorithm \cite{Gottlieb:1987mq}).
In this case observables are expected
to differ by $\rmO(\dtau^2)$ from their correct value
if a leapfrog integrator is employed.

While most of the computer time is spent in evaluating the fermionic
force, one should note that already for pure gauge theory, the HMC
itself is a rather inefficient algorithm. Although it globally moves the
field configurations, in practice the autocorrelation times are
large.

To illustrate these points, results from three--dimensional
$SU(2)$ gauge theory are shown in \fig{toy}. At this statistical
precision the $\rmO(\dtau^2)$ errors from the leapfrog integration
are clearly visible for the $\phi$--algorithm (left plot).
The HMC results for the average plaquette agree
with the (much more precise) heatbath data for all values of the leapfrog
step--size. A comparison of the
statistical error as a function of the CPU time (right plot) shows that the
heatbath gives a significantly more precise result for a given
computational effort.

\EPSFIGURE[t]{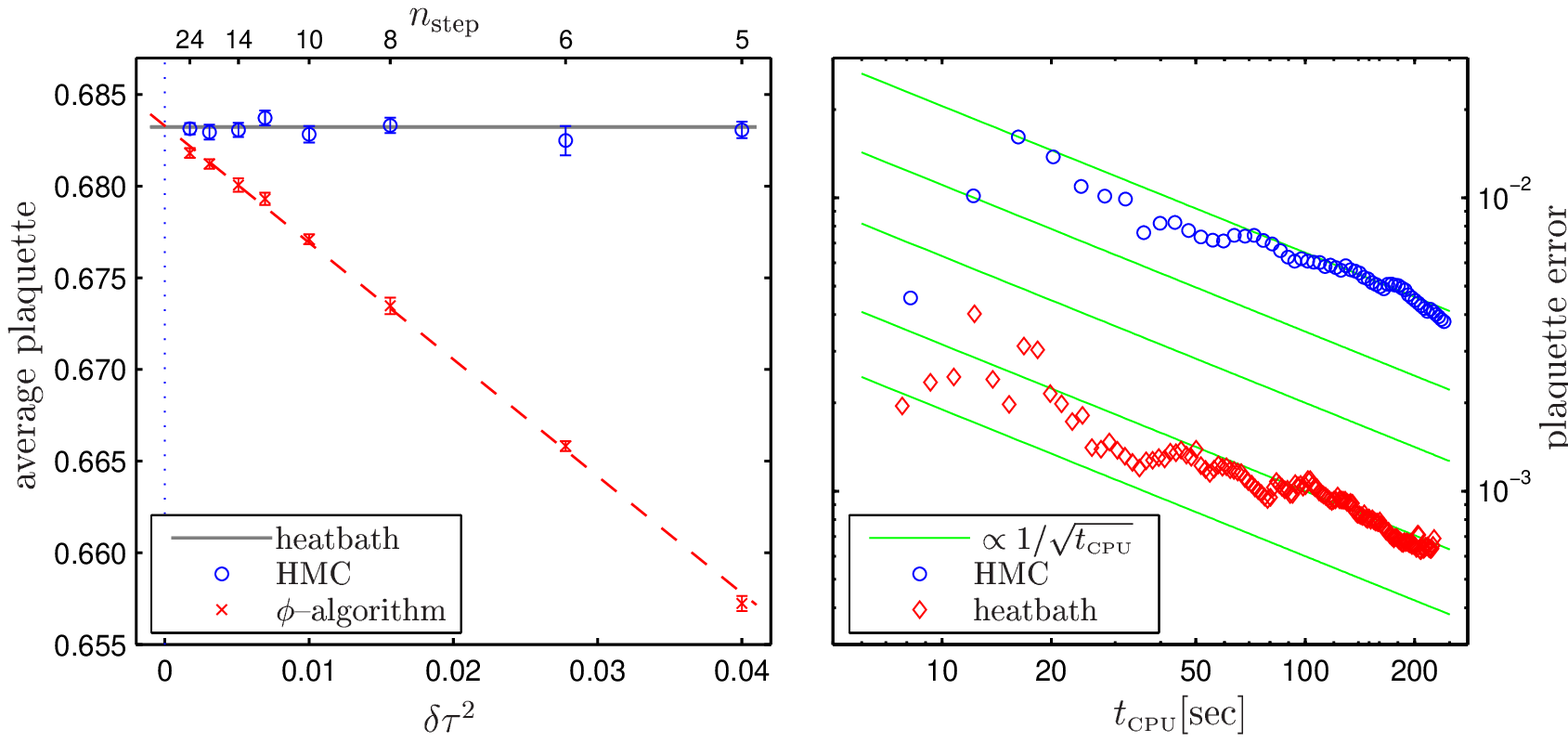,width=144mm}{
Left: Average plaquette from different algorithms in $SU(2)$ gauge
theory on a $6^3$ lattice with periodic boundary conditions
at $\beta=3.5$. Right: scaling of the
error with CPU time $t_{\rm\scriptscriptstyle CPU}$ for heatbath and HMC
in this setup.
\label{toy}}

Many improvements to the ''basic'' HMC described here have been
proposed and tested to speed up full QCD simulations. This includes
preconditioning and different pseudo--fermion
representations of the fermion matrix as well as higher--order integration
schemes for the MD evolution.
However, so far none of these represent a real breakthrough in
algorithmic development since the performance gain is in general
quite moderate (i.e. factors of $\lesssim\!2$ at most).
The situation might be different for a recent promising proposal
\cite{Luscher:2003vf}, which relies on a domain--decomposition
of the fermion matrix. Here significant advantages in the region of
small quark masses are expected.

Another possibility is to use in the MD evolution a different action
(guidance Hamiltonian)
than one really wants to simulate. This can be corrected either through
an adjustment of the acceptance probability in the
Metropolis step or by calculating a \emph{reweighting factor} to be included
in the data analysis. The latter
possibility is employed in the polynomial hybrid Monte Carlo (PHMC)
algorithm, which will be discussed in \sect{sect:sampling}.

All of the simulations presented here are done using even--odd
preconditioning \cite{DeGrand:1990dk}, where
the lattice sites are divided into an even and an odd part according
to $\sum_\mu x_\mu$. Since the hopping term
in the Wilson Dirac operator couples only even to odd sites, it
is possible to rewrite $Q$ in a form, where the non--trivial part
acts only on the odd sites. As a result the effective condition
number is reduced and inversions become computationally cheaper.
We will denote the
even--odd--preconditioned
Hermitian Dirac operator by $\hat Q$.

Another improvement is
the introduction of a second pseudo--fermion field as proposed
in \cite{Hasenbusch:2001ne}. More precisely, the implementation
we employed is the one discussed in \cite{DellaMorte:2003jj}, where
the pseudo--fermion action from (\ref{seff}) is split according to

\vspace*{-6mm}

\bea
S_{F_1}&=&\phi_1^\dagger(Q^2\pl\rho^2)^{-1}\phi_1\\
\textrm{and}\quad
S_{F_2}&=&\phi_2^\dagger(\rho^{-2}+Q^{-2})\phi_2\;,
\eea
such that
\be
\det Q^2=\int\!\!\DD\phi_1\DD\phi_2\ \rme^{-S_{F_1}-S_{F_2}}\;.
\ee
With this split--up the fermionic forces in the MD evolution
become smaller, such that larger step--sizes can be used in the
numerical integration.

Finally, an improved Sexton--Weingarten \cite{Sexton:1992nu}
integrator is used in the molecular dynamics. It uses different
time scales for the gauge and the fermionic part of the forces.
A performance improvement results since the computationally
cheaper gauge
forces are evaluated more often. This scheme partially removes
the integration errors of $\rmO(\dtau^3)$. In practice it is
found that the Metropolis acceptance rate $P_{\rm acc}$ behaves as
\be
-\log P_{\rm acc}\simeq 1-P_{\rm acc}\propto \dtau^4\;,
\ee
thus indicating that the errors of $\rmO(\dtau^4)$ dominate.

An important quantity to monitor in an HMC simulation is
the Hamiltonian violation $\Delta H$ during one molecular
dynamics trajectory. In the large volume limit
its distribution is Gaussian with mean and width related through
\cite{Gupta:1990ka}
\be
2\ev{\Delta H}=\ev{\Delta H^2}-\ev{\Delta H}^2\;.
\ee
This relation holds for the simulation described in \tab{t_simpar},
while for the simulations in smaller volume (Tables \ref{t_simpar_algo} and
\ref{longtable}) deviations were observed.
Only in the limit of an exact integration, $\dtau\rightarrow0$,
the mean (and hence also the variance) of $\Delta H$ vanish.
However, from the detailed balance condition it is possible to
show that even at finite step--size \cite{Creutz:1988wv}
\be
\ev{\rme^{-\Delta H}}=1
\ee
on the generated ensemble (including of course the rejected
proposals). A violation of this condition could indicate
problems with reversibility in the numerical integration
of the equations of motion.

\section{Inverting the Dirac operator}

\label{sect:invert}
To evaluate the fermionic contribution to the MD force and to
calculate fermionic correlation functions, matrix
elements of the quark propagator $S=M^{-1}$ need to
be calculated on the gauge background
produced in the Monte Carlo run. Depending on the specific
correlator we need different linear combinations
of matrix elements. These are obtained by inverting
M on different sources $q$. Writing the Dirac indices $A,B,\ldots$ and color
indices $\alpha,\beta,\ldots$ explicitly, we need to solve
\be
M(x,y)_{AB}^{\alpha\beta}\cdot \widetilde S(y,z)_{BC}^{\beta\gamma}=
q(x,z)^{\alpha\gamma}_{AC}\;.\label{invertme}
\ee
Since the source for the inversion can only be a vector, we need
one inversion for each combination of the ''column'' indices
$z,\gamma$ and $C$ of $q$. Exploiting the linearity of (\ref{invertme}),
one can add different sources to obtain the corresponding linear
combination of columns of $S$ with a single inversion.
For a given source $y$ the solution $x$ of $Mx\!=\!y$ is found by
minimizing the residue
\be
r=\frac{||Mx-y||^2}{||y||^2}\;. \label{residue}
\ee
Since $M$ is a large sparse matrix, iterative conjugate gradient methods
provide an efficient tool for this task. The most commonly
used version is the stabilized biconjugate gradient (BiCGstab)
\cite{131930}. However, we found that in cases where the matrix
$M$ is rather ill--conditioned, it can happen that the BiCGstab
does not converge to a solution.

This poses a problem in particular because during the iteration,
the residue $r$ is not computed according to (\ref{residue}).
To save an additional application of $M$, the current residue
is computed from its value during the previous iteration
(iterated residue). Thus, due to accumulation of roundoff the iterated
residue might become small, indicating that a solution has been found,
while the real residue is still of $\rmO(1)$.

\EPSFIGURE[!ht]{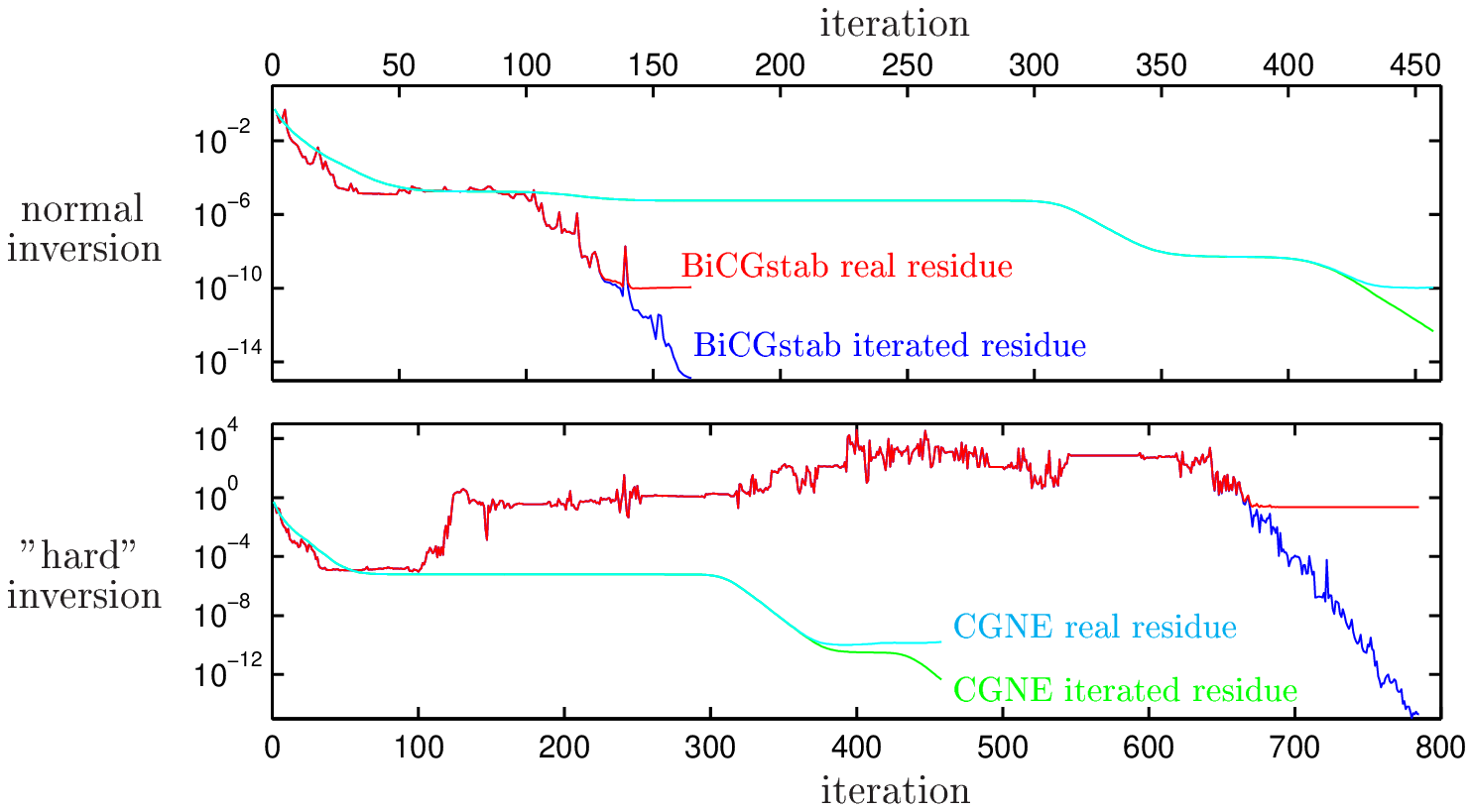,width=14cm}{
Solver residues as a function of the iteration number. For difficult inversions
(lower plot) the BiCGstab might indicate a convergence in the iterated
residue although no solution has been found.
\label{solvfig}}

The simple conjugate gradient (CGNE)\footnote
{NE stands for normal equation, $M^\dagger Mx=M^\dagger y$.
CG itself applies only to positive matrices.} solver has
a much smoother convergence behavior and is very stable
even when applied to ill--conditioned problems.
\fig{solvfig} shows the real and the iterated residue for
BiCGstab and CGNE during the iterations of a normal inversion and
one, where BiCGstab did not converge to a solution. For CGNE the real
and iterated residues start to differ only when the
single--precision limit is approached.\footnote{
Apart from global sums all our calculations are
carried out in single--precision arithmetics.}

To account for the known insufficiencies of the BiCGstab while
still making use of its superior performance, we perform the
inversions for the calculation of correlation functions in the
following way. We start by first running the BiCGstab solver
until the required precision or the maximum number of iterations,
$\rmO(100)$, is reached. This is followed by a CGNE inversion starting
from the BiCGstab result with
a maximum number of $\rmO(1000)$ iterations, although for a
''normal'' inversion only $\rmO(1)$ CGNE iterations are performed
until the required precision is obtained.
To compensate for
precision loss due to roundoff, the CGNE is restarted up to ten
times with the previously found solution. The real residue is
always recomputed after the entire inversion and monitored
in the data analysis program.

\section{Data analysis}

As mentioned earlier, the field configurations in a Markov chain
produced by a hybrid Monte Carlo algorithm can be strongly
(auto)correlated in Monte Carlo time. This needs to be taken
into account in order to obtain reliable error estimates for
quantities measured in such simulations.

Traditionally this is done by combining a number of consecutive
measurements into \emph{bins} and then performing a na{\"\i}ve
error analysis on the binned data (jackknife analysis). However,
it has been shown \cite{Wolff:2003sm} that this procedure is
only suboptimal concerning the reliability of the error estimate.

For all our data analysis we use the explicit numerical integration
of the autocorrelation function for primary and derived observables
from \cite{Wolff:2003sm}. The integrated autocorrelation time
$\tau_{\rm int}$ enters  the error estimate $\sigma$ in the form
that the number of estimates $N$ is ''effectively'' reduced
according to
\be
\sigma^2=\frac{2\tau_{\rm int}}{N}\nu\;,
\ee
where $\nu$ is the variance of the quantity in question.
While the latter is a property of the observable and hence of the
theory one is simulating, the size of the statistical error is
also influenced by $\tau_{\rm int}$, which depends on the
algorithm used to generate the Markov chain. This method also
provides us with an error of $\tau_{\rm int}$ and hence we can
directly assess the reliability of our error estimate, i.e. the 
error of the error.
\section{Sampling problems on coarse lattices}

\label{intro_algo}

In order to motivate the choice of simulation algorithms
in the evaluation of the axial current normalization
condition (\chap{chap:renorm}) we now proceed to a more
detailed discussion of the algorithmic difficulties
we faced.
This concerns the simulations at the coarsest lattice spacing
we consider, $a\!\simeq\!0.1\fm$ at a bare gauge coupling of
$\beta=5.2$.

It has long been established
that in the corresponding quenched situation cutoff--effects
are compatible with Symanzik's description (\sect{sect:impr})
and a continuum extrapolation can be started there.
However, over the last years more and more evidence has been
accumulated that for
\emph{dynamical} improved Wilson fermions at this lattice spacing
the cutoff--effects are much larger than expected. As an extreme
example, for three flavors the existence of a phase transition in the
$\beta$--$\kappa$--plane has been numerically conjectured and is 
interpreted as a lattice artifact \cite{Aoki:2001xq}.
A summary of large scaling violations in the two--flavor--theory
is given in ref.~\cite{Sommer:2003ne}.

In the following we will establish that the algorithmic
problems we encountered in
the MD integration
and the efficient simulation of the canonical ensemble can also
be interpreted as cutoff--effects. More precisely, at the infrared
end of the spectrum of the Dirac operator we find a behavior different
from what one would expect close to the continuum.

One of the important results will be our finding that the distortion
of the Dirac spectrum makes it advantageous to deviate from importance
sampling. In this context
we study the behavior of the 
polynomial hybrid Monte Carlo (PHMC) algorithm
\cite{deForcrand:1996ck,Frezzotti:1997ym}
in this situation and find it a very useful tool for a detailed
investigation of the properties of the small eigenvalues of the
Dirac operator.

In \tab{t_simpar_algo} we list the lattice sizes and bare parameters for
the simulations discussed in this chapter.
In all cases we have $T\!=\!9/4\;L$ for the time extension $T$.
The bare quark mass $m$ is defined through (\ref{massimp}), where
we used the perturbative value of $\cA$ from \eq{capt}.
In the algorithm column '$\rm H_2$' refers to the HMC with
two pseudo--fermion fields described at the end of \sect{sect:hmc} and
'$\rm P_n$' stands for PHMC with a polynomial of degree $n$. This algorithm
will be introduced in \sect{sect:sampling}. To gain more
statistics, each simulation was performed with 16 independent replica.

\TABULARSMALL[!ht]{r|cccllccc}{
    \hline
run & $L/a$ &  $\beta$ & $\kappa$ & $\ \ Lm$ & algo. & 
    $N_{\rm traj}$    &   $\dtau$ & $P_{\rm acc}$\\
    \hline
[A1] & 8 & 5.2 & 0.13550   & 0.205(10)  & $\, \rm H_2$ & $16\!\cdot\!500$ &    1/16 & 91\% \\[0mm]
[A2] & 8 & 5.2 & 0.13515   & 0.307(9) & $\, \rm H_2$ & $16\!\cdot\!520$ &  1/25 & 97\% \\[0mm]
[A3] & 8 & 5.2 & 0.13515   & 0.314(8) & $\, \rm P_{140}$  & $16\!\cdot\!500$ & 1/26 & 87\% \\[0mm]\hline
[A4] & 8 & 5.2 & 0.13550   & 0.195(7) & $\, \rm P_{140}$  & $16\!\cdot\!400$ & 1/25 & 85\% \\[0mm]
[A5] & 8 & 6.0 & 0.13421   & 0.193(3) & \multicolumn{4}{c}{--- quenched ---} \\[0mm]\hline 
[A6] &12 & 5.5 & 0.13606   & 0.287(3)& $\, \rm H_{2}$ & $16\!\cdot\!240$ & 1/20 & 91\% \\[0mm]
[A7] &12 & 6.26& 0.13495   & 0.295(3)& \multicolumn{4}{c}{--- quenched ---} \\     \hline}
{Summary of simulation parameters.\label{t_simpar_algo}}

\section[Instabilities in the molecular dynamics integration]
{Instabilities in the molecular dynamics\\[-0.5mm] integration}
\label{instab}

We have seen that in the numerical approximation to the molecular
dynamics evolution the Hamiltonian is conserved up to powers of the
step--size $\dtau$ employed in the integration. Apart from these
(usually small) deviations, under certain conditions
the currently used integration schemes can become unstable and produce
very large Hamiltonian violations $\Delta H$. For a more detailed
discussion see ref.~\cite{Joo:2000dh}, where a connection between
these instabilities and large driving forces in the MD is proposed
in analogy to a harmonic oscillator model.
In this model the integrator becomes unstable when the product of the
force and the integration step--size exceeds a certain threshold.

Large (positive) values of $\Delta H$ result in a rejection of the
configuration proposed by the MD. In a histogram of $e^{-\Delta H}$
these contribute to bins close to zero
while the distribution is peaked around one, indicating that in most
cases the numerical integrator performs as expected.
They can also lead to an unusual autocorrelation of
this quantity, making the Monte Carlo error estimate
difficult.
In particular this applies also to the
integrated autocorrelation time of $e^{-\Delta H}$ itself. 
This is due to the long periods of rejection in the Metropolis step,
which sometimes follow large $\Delta H$ values.

\EPSFIGURE[t]{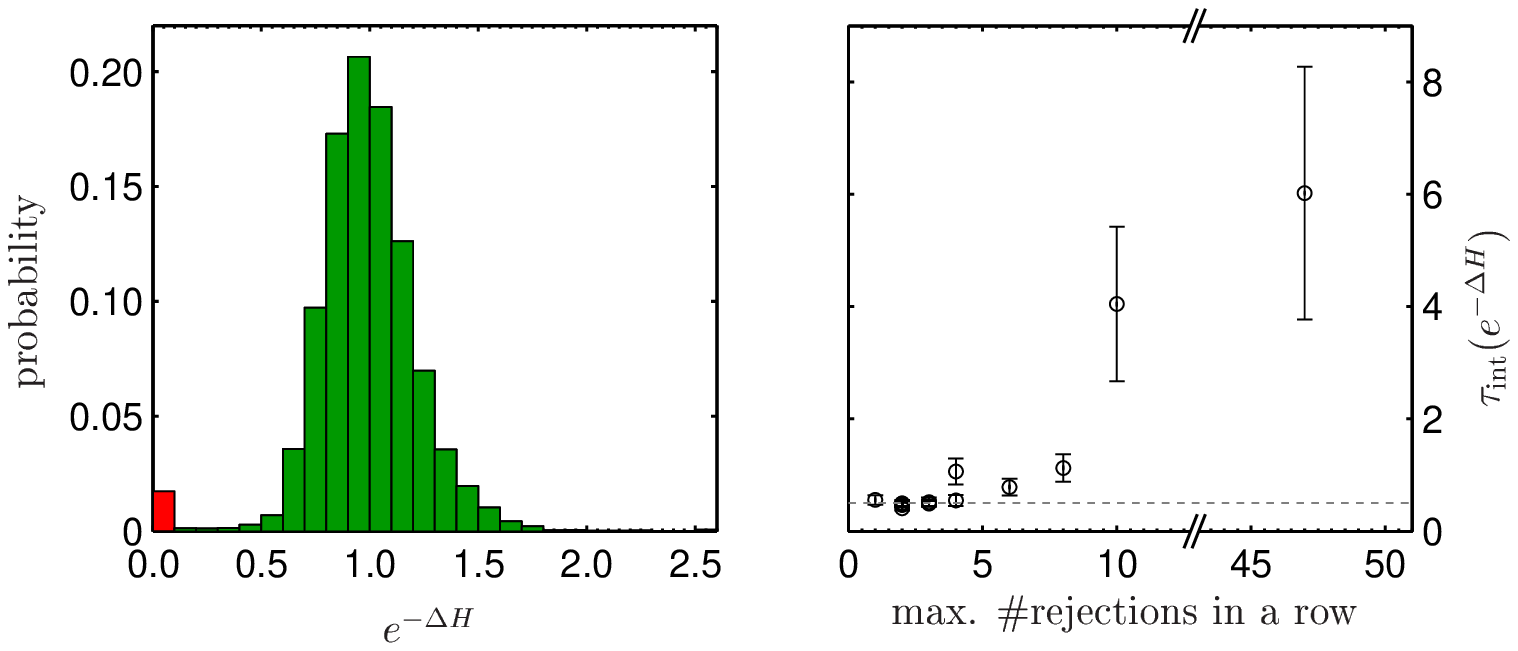,angle=0,width=13cm}
{$e^{-\Delta H}$ from run [A1]. Left plot: histogram.
Right plot: our estimates of $\tau_{\rm int}$ in units
of MD time separately for the 16 replica. In our normalization
$\tau_{\rm int}=0.5$ means no autocorrelation (dotted line).\label{tau}}

\fig{tau} shows a histogram of $e^{-\Delta H}$ and also its integrated
autocorrelation time from one of our simulations.
In this data set there are several series of large $\Delta H$ values, during
which the proposed configurations were rejected.
In the distribution of $e^{-\Delta H}$ these lead to an additional peak close
to zero.
One also sees from the right--hand plot that $e^{-\Delta H}$ is noticeably
autocorrelated only when a large number of proposals were rejected in a row.
As argued above in these cases the error of $\tau_{\rm int}$ could be
underestimated. These two effects might cause some concerns when using
$\langle e^{-\Delta H}\rangle-1$ as an indicator for the absence of
reversibility violations \cite{DellaMorte:2003jj}.

Spikes in $\Delta H$ have been observed by
several collaborations using (improved) Wilson fermions in various setups
(e.g.~different gauge actions and volumes)
at relatively large lattice spacings
\cite{Joo:2000dh,Jansen:1998mx,Namekawa:2002pv,Allton:2004qq}.
There these spikes have been traced back to large values of the driving force in the
MD evolution and also their dependence on the quark mass has been investigated.

Here we want to clarify a point, which is essentially implied by the
previous observations \cite{Namekawa:2002pv,Frezzotti:2002iv}, namely
the strong correlation
between spikes in $\Delta H$ and small eigenvalues of the Dirac
operator.\footnote{Here and in the following we will always refer to the eigenvalues
of the square of the Hermitian even--odd preconditioned Dirac
operator $\hat Q^2$ in the Schr\"odinger functional. For its precise definition
see ref.~\cite{DellaMorte:2003jj}.}
In this way we hope to be able to separate physical
effects from cutoff--effects, i.e. the occurrence of unphysically
small eigenvalues. The low--lying eigenvalues are computed using the
method described in \cite{Kalkreuter:1995mm}.
In \fig{lmin} we clearly see a long period of rejection
(corresponding to the rightmost data point in \fig{tau})
caused by the presence of a very small eigenvalue. Although we did
not measure them, this is expected to
produce large fermionic contributions to the driving forces since
they involve an inverse power of the Dirac operator.

\EPSFIGURE[t]{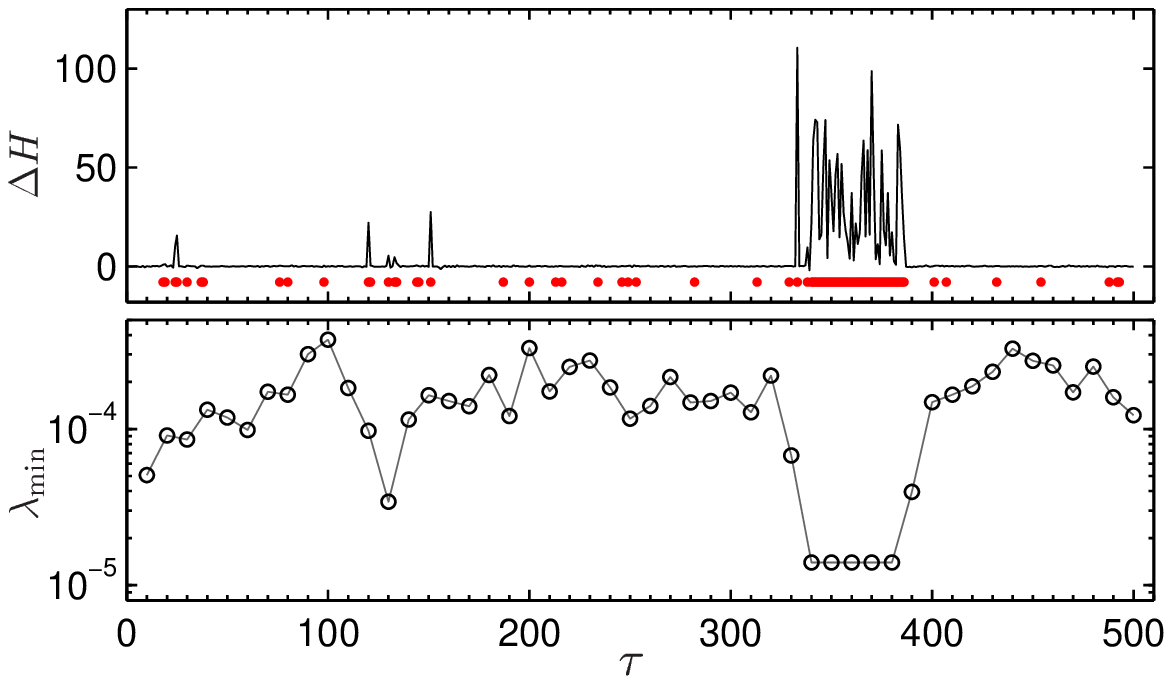,angle=0,width=11cm}
{Monte Carlo history for one replicum of run [A1] with a long
period of rejection. Configurations where the new proposal was
rejected are marked by a dot. At $\tau\!\!\simeq\!340$
the algorithm gets stuck with a configuration carrying an exceptionally low
smallest eigenvalue $\lambda_{\rm min}$ of $\hat Q^2$.\label{lmin}}

We found the observed \emph{average} $\lambda_{\rm min}$
to be close to its tree--level estimate with Schr\"odinger functional
boundary conditions~\cite{Sint:1993un}.
However, the \emph{smallest} $\lambda_{\rm min}$ is an order of magnitude below that and
we therefore consider these eigenvalues unphysical and will later
establish their nature as cutoff--effects.

Finally, following the procedure of ref.~\cite{DellaMorte:2003jj}, the
absence of global reversibility violations is explicitly verified even
for trajectories resulting in large values of $\Delta H$. Nevertheless
our experience shows that the increased cost of using a smaller
$\dtau$ such that no long periods of rejection occur is
more than compensated by the reduction in autocorrelation time of
all observables. The reason is that already a small decrease of the
integration step--size greatly reduces the Hamiltonian violations.
For example, repeating run [A1] with a step--size of $1/20$ instead
of $1/16$, the longest period of rejection was $4$ (instead of
$47$) consecutive trajectories. Such a rapidly changing behavior
also supports the picture of integrator instabilities from
\cite{Joo:2000dh}, where an abrupt increase of Hamiltonian violations
is predicted when a certain $\dtau$ threshold is exceeded.

\section{MC estimates of fermionic observables}

\label{sect:sampling}

We concluded in the previous section that unphysically small eigenvalues
of $\hat Q^2$ produce algorithmic problems only on a practical
and not on a theoretical level. But apart from slowing down the algorithm
these small eigenvalues also cause problems in the MC evaluation of
fermionic Green's functions. 

\EPSFIGURE[t]
{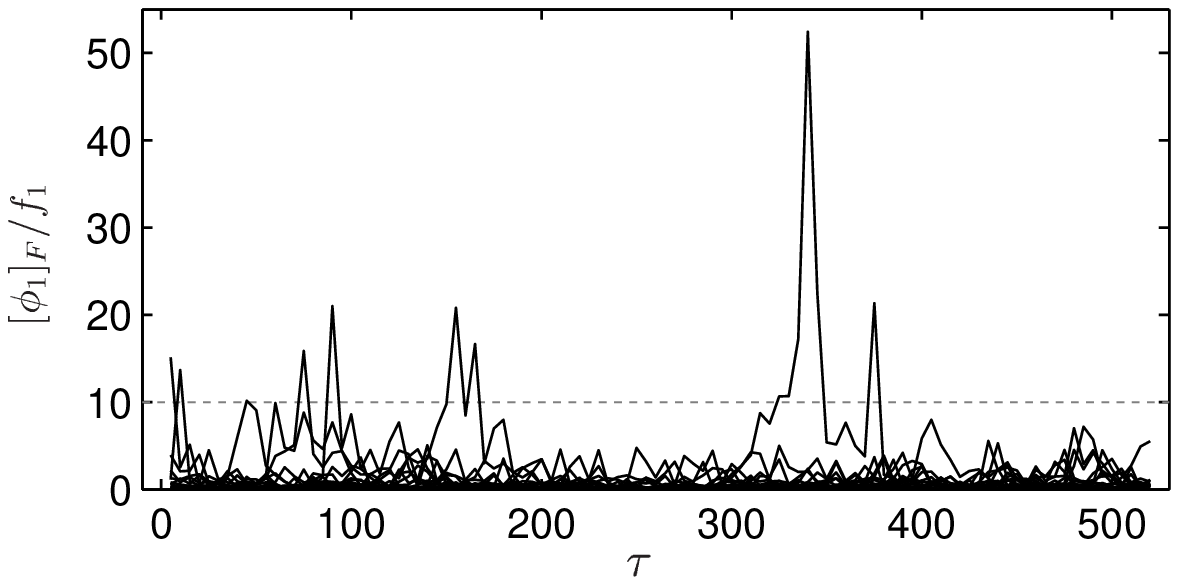,angle=0,width=11cm}
{Monte Carlo history for the $N_{\rm rep}\!=\!16$ replica of run [A2] showing
the normalized $\ff$.\label{f1_algo}}

As an example we consider the Schr\"odinger functional correlation function 
$f_1$ as defined in \eq{f1def}. It is the correlation between pseudo--scalar
composite fields at the first and last time--slice, respectively. We will
denote its value on a given gauge field configuration by
$\ff$.\footnote{
From \eq{f1prelast} it follows that $\ff=\tr\overline S_T
\overline S_T^\dagger/2L^3$.}
\fig{f1_algo} shows the MC history of the normalized $\ff$ for the 16 replica of run [A2].
Here $\tau$ refers to the molecular dynamics time for each replicum.
While on this scale the bulk of the data are below one and hence not visible,
there are several peaks, which give a significant contribution to the mean value.
These spikes also affect the error estimate $\sigma(f_1)$ through both the variance
and the integrated autocorrelation time.
For statistically accessible quantities the error should
approach a $1/\sqrt{\tau}$ behavior in the limit
$\tau\!\rightarrow\!\infty$.
In this respect we found $f_1$ and all other fermionic correlation functions we
considered to be very hard to measure. Even when averaging over $16$ replica,
this asymptotic behavior does not set in after $\tau\!\simeq\!500$. 

The reason is the rare occurrence of very large values of $\ff$,
which appear to be correlated with small eigenvalues of $\hat Q^2$.
However, this  effect is washed out by using several replica.
We therefore show in \fig{f1sick} the MC history
of $\ff$, $\lambda_{\rm min}$ and our error estimate for $f_1$ 
 for one replicum of run [A2] with such a spike in $\ff$.
Indeed, for each spike in $\ff$ the smallest eigenvalue drops below its
average. That the converse is not true could be ascribed to a lack of overlap 
of the eigenvector corresponding to $\lambda_{\rm min}$ with the source needed
to compute the quark propagator.
Quantitatively, for the correlation between
$\ff$ and $\lambda_{\rm min}$ we measure a value of
$C_{\ff,\lambda_{\rm min}}=-0.33(4)$ if we use all replica and
$-0.46(6)$ from the replicum shown in \fig{f1sick} alone.
Here we used as a definition of the correlation $C_{A,B}$ between two
(primary) observables $A$ and $B$

\be
C_{A,B}=\frac{\ev{AB}-\ev A\ev B}
{\sqrt{\Big\langle A^2-\ev A^2\Big\rangle\Big\langle
B^2-\ev B^2\Big\rangle}}\;,\ \ \textrm{ so that }\ \
-1\leq C_{A,B}\leq1\;.
\ee

\EPSFIGURE[t]
{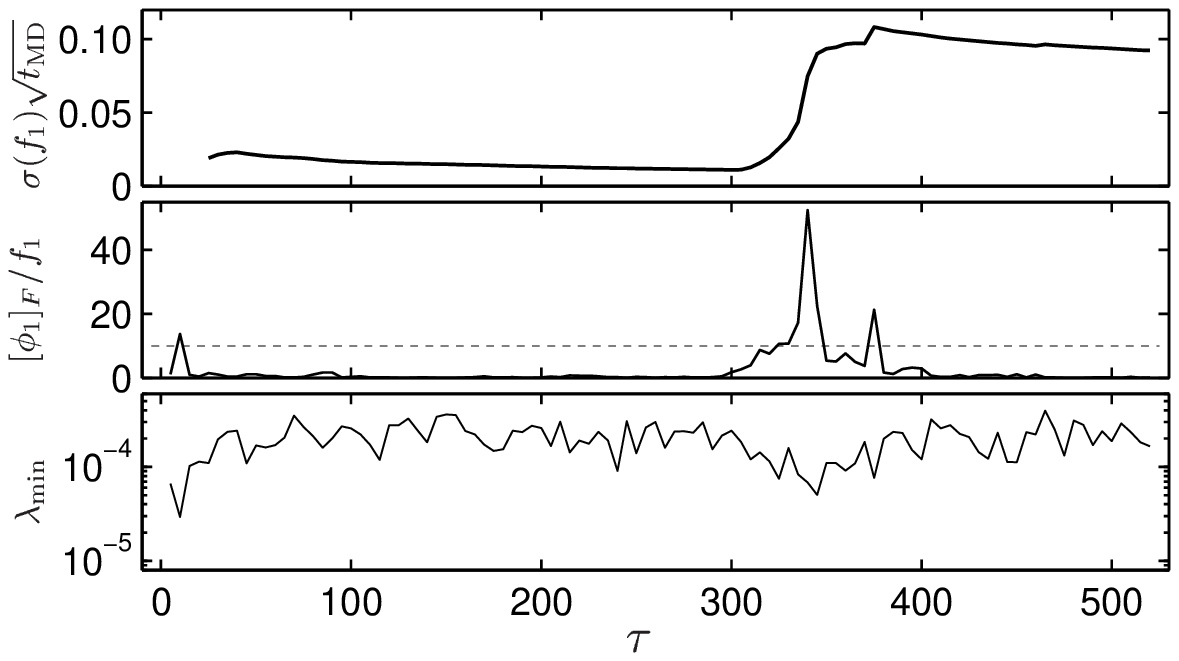,angle=0,width=11cm}
{Normalized $\ff$ and  smallest eigenvalue from one ''sick'' replicum
of run [A2]. Evidently the spike in $\ff$ is dominating the
statistical error $\sigma(f_1)$.\label{f1sick}}

Even though in the limit of infinite statistics configurations carrying very
small eigenvalues are given the correct weight, depending on the algorithm this
might be badly approximated for a typical ensemble size. 
Similar arguments referring in particular to the HMC algorithm motivated the
introduction of the polynomial hybrid Monte Carlo (PHMC) algorithm in
refs.~\cite{Frezzotti:1997ym,Frezzotti:1998eu,Frezzotti:1998yp}.

Hence the difficulty in measuring fermionic correlation functions
might be an efficiency problem related to the choice of the algorithm.
To check this conjecture we employ a second algorithm and compare ensembles
generated by HMC (with two pseudo--fermion fields) with PHMC ensembles.
Indeed, PHMC can be tuned in such a way that it enhances the occurrence of
configurations carrying small eigenvalues, thus resulting in a better sampling
of this region of configuration space. A reweighting step is introduced to
render the algorithm exact. As a preparation for the following discussions
we want to recall some properties and introduce the notations concerning
the PHMC.

\subsection{The PHMC algorithm}

One of the main ideas of the PHMC algorithm is to deliberately move away from
importance sampling
by using an approximation to the fermionic part of the lattice QCD action.
More precisely, in an HMC algorithm the inverse of $\hat Q^2$ is replaced
by a polynomial $P_{n,\epsilon}(\hat Q^2)$ of degree $n$.
Here $P_{n,\epsilon}(x)$ approximates
$1/x$ in the range $\epsilon\leq x\leq1$. 
As a consequence this algorithm stochastically implements the weight
$\DD U\det P^{-1}_{n,\epsilon}(\hat Q^2)e^{-S_g}$,
whereas standard HMC generates ensembles according to $\DD U\det \hat Q^2e^{-S_g}$.
Denoting averages over the PHMC ensemble by $\langle\dots\rangle_P$,
the correct sample average of an observable $\langle O\rangle$ can then be
written as
\be \label{QCD_ave}
\langle { O} \rangle = \langle O\omega \rangle_P\;
,\textrm{ where}\ \  \omega=\frac W{\ev W_P}\;,
\ee
and we introduce the reweighting factor $W$ as a (partially)
\footnote{Through the separate treatment of the lowest eigenvalues of $\hat Q^2$
the infrared part of $W$ is evaluated exactly.}
stochastic estimate of $\det\{\hat Q^2 P_{n,\epsilon}(\hat Q^2)\}$.
When using Chebyshev polynomials the relative approximation error of
$P_{n,\epsilon}$ is bounded by $\delta \simeq 2 \exp (-2\sqrt{\epsilon}n)$
in the range
$\epsilon\leq x\leq1$ .

\EPSFIGURE[t]
{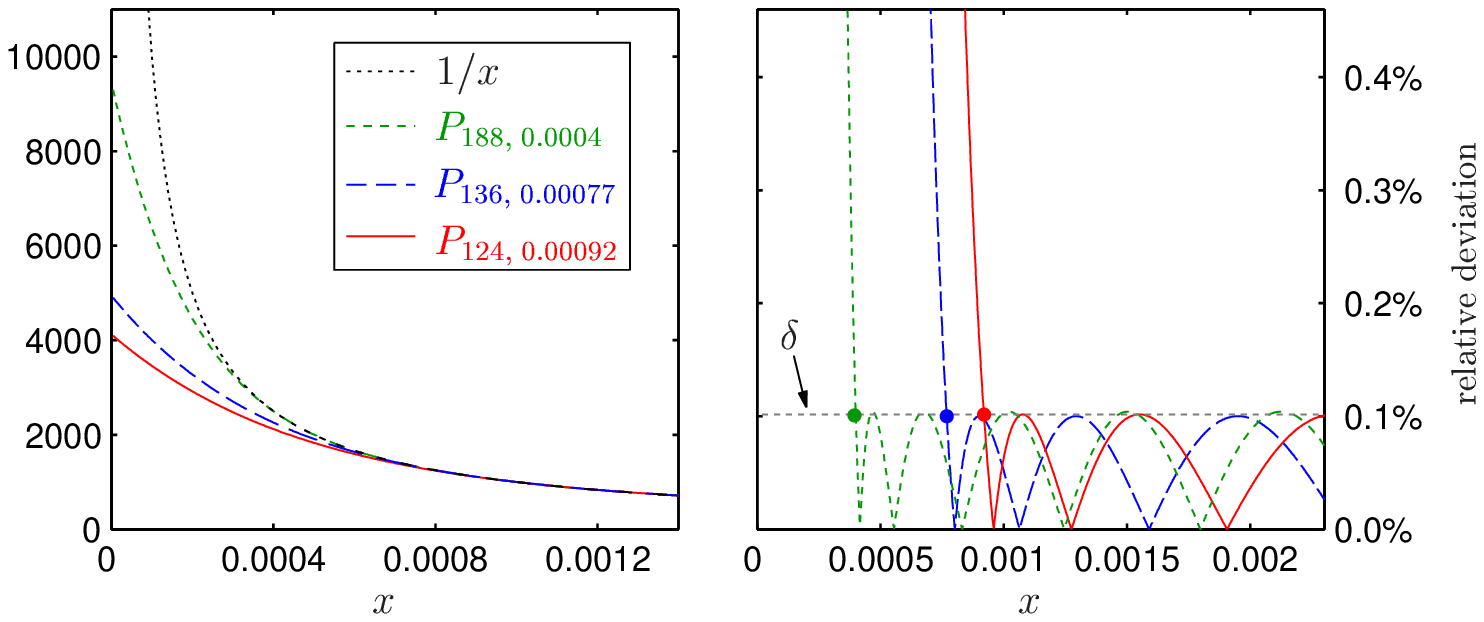,angle=0,width=13cm}
{Three different Chebyshev polynomials approximating $1/x$,
all with $\delta=0.001$. The right--hand plot shows the relative
deviation from $1/x$ as a function of $x$. There the points
$(\epsilon,\delta)$ are marked by dots.\label{polynomial}}

To give an impression of the r{\^o}le of $\epsilon$ and $\delta$ we plot in
\fig{polynomial} a set of polynomials $P_{n,\epsilon}(x)$ for typical
(in our simulations) values of these parameters and compare them with
$1/x$ in the region of small $x$. Depending on the smallest eigenvalue
of $\hat Q^2$ the parameters $\epsilon$ and $n$ have to be tuned
such that the reweighting factor does not fluctuate too much.
The authors of ref.~\cite{Frezzotti:1997ym} suggested to take $\epsilon$
of the same order as $\langle\lambda_{\rm min}\rangle$ and in practice
used $\epsilon\simeq 2\langle\lambda_{\rm min}\rangle$ and
$\delta\lesssim 0.01$.

Recalling that PHMC replaces  $\det \hat Q^2$ in the  HMC weight with
$\det P^{-1}_{n,\epsilon}(\hat Q^2)$ and observing from \fig{polynomial} that
$P_{n,\epsilon}(x)$ is smaller than $1/x$ for $x\leq\epsilon$,
the aforementioned property of enhancing the occurrence
of small eigenvalues is evident. At this point we would like to
note that the fermionic contribution to the driving force in the PHMC
is bounded from above since $P_{n,\epsilon}(x)$ is finite even at
$x=0$.
In this way the polynomial provides a regularized inversion of
$\hat Q^2$, thus also addressing the problems mentioned in \sect{instab}.

\subsection{HMC vs. PHMC}

Coming back to the comparison of samples from HMC and PHMC, we repeated
run [A2] with PHMC using a polynomial of degree $140$ and
$\epsilon=6\cdot\!10^{-4}$, resulting in $\delta\simeq0.002$. The ratio
$\epsilon/\ev{\lambda_{\rm min}}$ turned out to be around $2.7$.
\EPSFIGURE[t]
{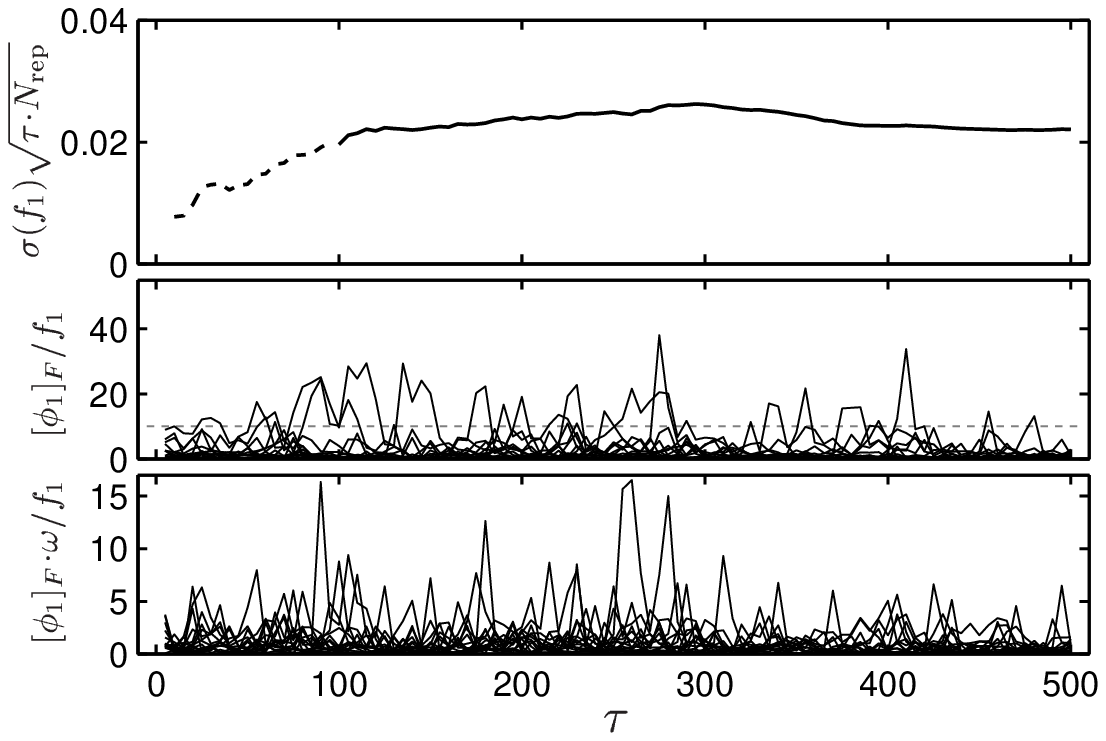,angle=0,width=11cm}
{Monte Carlo history for the 16 replica of run [A3] showing
the correlation function $\ff$ and the product $\ff\cdot \omega$,
where $\omega$ is the normalized reweighting factor. Our error
estimate of $f_1$ shows the expected scaling behavior as soon
as the run is long enough for a reliable extraction of $\tau_{\rm int}$.\label{reweight}}
In \fig{reweight} we plot for this run the MC history of $\ff$ and of
$\ff\cdot \omega$,
which enters into \eq{QCD_ave} if we consider $O=\ff$, i.e.
\be
f_1=\ev \ff=\ev{\ff\cdot \omega}_P=\frac{\ev{\ff\cdot W}_P}{\ev W_P}\;.
\ee
We first observe that apart from 
removing the largest spikes the inclusion of the reweighting factor does not
seem to significantly change the relative fluctuations. This means that
the parameters of the polynomial have been chosen properly.
Events where $\ff$ assumes a value $\rmO(10)$ times larger than
$f_1$ are no longer isolated as in \fig{f1_algo} but happen frequently,
which means that the PHMC algorithm can more easily explore the
associated regions in configuration space.
This is what allows a reliable error estimate as shown in the upper
part of \fig{reweight}, i.e. with 16 replica the asymptotic behavior of
the error sets in after $\tau\!\simeq\!100$.

The advantage of using PHMC instead of HMC can be clearly seen by
considering the spread of $\sigma(f_{1})\sqrt{\tau}$ among
the replica. We analyzed this quantity in extensions of runs [A2]
and [A3]. The result is shown in figure \fig{scaling}, where the
shaded areas represent the range of values covered by the 16 replica
as a function of the MD time.
In the limit of infinite statistics all replica should converge to the
same value, which need not be the same for the two algorithms because of
reweighting and different autocorrelation times. We see
that the spread for the HMC data is more than twice as large as for
PHMC, i.e. the error on $f_1$ is significantly harder to estimate with
HMC.

\EPSFIGURE[t]
{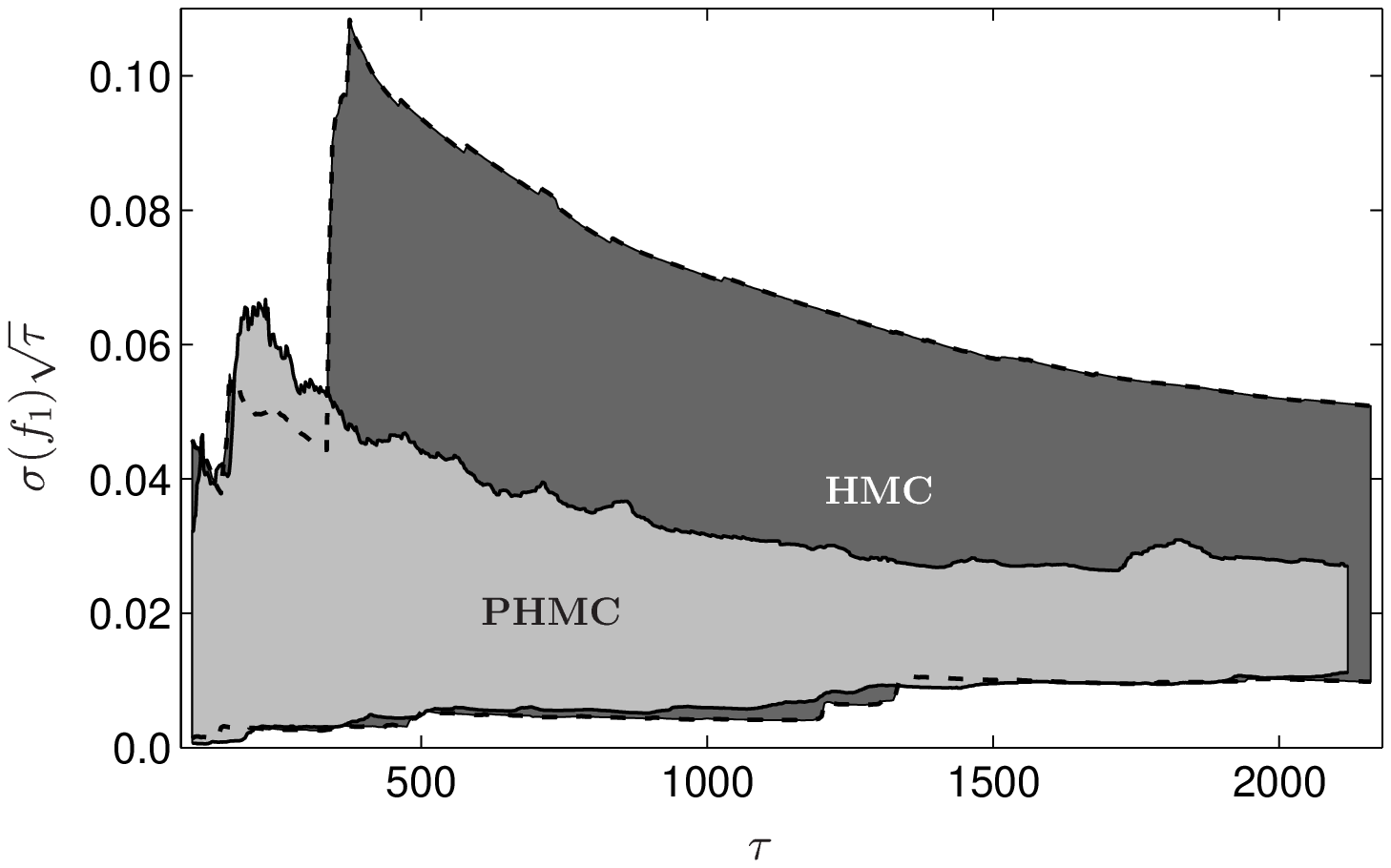,angle=0,width=11cm}
{Monte Carlo history of $\sigma(f_1)\sqrt{\tau}$ from
extensions of runs [A2] and [A3]. For the two algorithms we show
the ranges covered by the 16 replica.\label{scaling}}

What we are suggesting here is that the algorithm should be chosen
depending on the type of observables and the parameter values.
From our experience we conclude that PHMC sampling might just be more
effective than HMC when computing fermionic 
quantities that are sensitive to small eigenvalues. 

To gain some more insight into the difference in sampling we consider the
distribution of $\lambda_{\rm min}$ since this is where we expect the largest
effect. The distributions are analyzed by treating
$\Lambda_{\rm bin}\!=\!\chi_{\rm bin}(\lambda_{\rm min})$ as a primary observable. Here
$\chi_{\rm bin}$ denotes the characteristic function of each given bin in the
histogram.
We then perform our normal error analysis for
$\ev{\Lambda_{\rm bin}}$,
where \eq{QCD_ave} has to be used if it is a
PHMC sample. For comparison $\ev{\Lambda_{\rm bin}}_P$ is also analyzed in
this case.

The histograms in the upper part of \fig{compsample} compare the results
from 200 independent measurements produced by HMC and PHMC (runs [A2] and [A3],
respectively). As expected the distributions agree within errors.
For the PHMC run we also plot the unreweighted histogram,
i.e. $\ev{\Lambda_{\rm bin}}_{\rm P}$.
Here we again confirm that with the parameters we chose for the polynomial
the PHMC produces more configurations with small eigenvalues than HMC.
As a consequence
of the reweighting the errors at the infrared end of the spectrum should be
smaller for the PHMC data. This is explicitly verified in the lower part
of the plot where we show the ratio of the errors on $\ev{\Lambda_{\rm bin}}$
from the two algorithms. One
can see that below $\approx7\cd10^{-5}$ the error from PHMC is at least a factor
two smaller than from HMC, with a more pronounced difference towards even
smaller eigenvalues. 
The advantage in using PHMC to sample this part of the spectrum is
significant and we will make use of this in the following discussion.

\EPSFIGURE[t]
{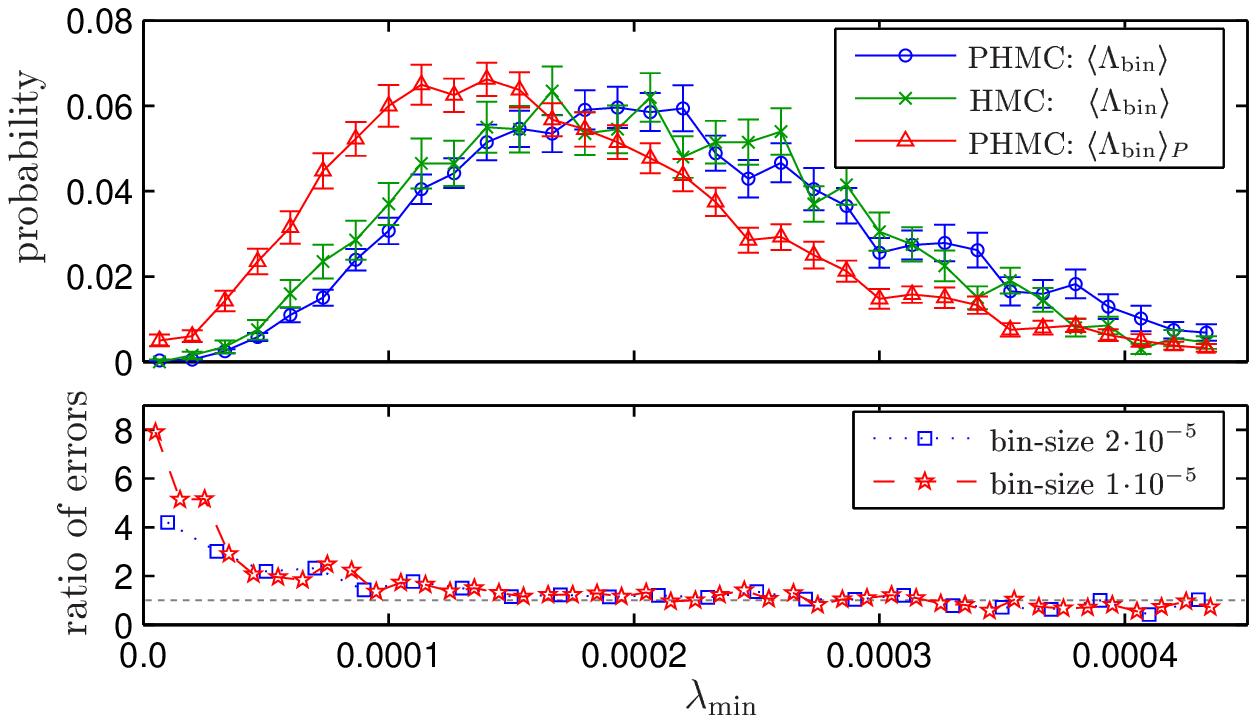,angle=0,width=12cm}
{Upper panel: histograms of $\lambda_{\rm min}$, i.e. 
$\ev{\Lambda_{\rm bin}}$ vs. 'bin', from runs [A2] and [A3]. For run
[A3] we also show $\ev{\Lambda_{\rm bin}}_P$.
Lower panel (from the same data): ratio of the error on $\ev{\Lambda_{\rm bin}}$
from HMC to that from PHMC.\label{compsample}}

\section{Comparison to the quenched case}\label{sect:three_algo}

In the previous section we studied various problems related
to the occurrence of small eigenvalues. All the data presented
there were produced at bare parameter values, which correspond
to relatively large quark masses and small volumes.
These small eigenvalues might therefore have a different nature from
the ''physical'' ones expected to show up in large volumes and/or
close to the chiral limit. Here and in the next section we will establish
them as cutoff--effects.

To this end we made an additional simulation at the parameters of run [A2] and
calculated the ten lowest--lying eigenvalues
$\lambda_i$, $i=1\ldots10$.
In \fig{bulk} the smallest eigenvalue, $\lambda_1$, is denoted by an open
symbol. It seems that while $\lambda_2$ through $\lambda_{10}$ form a
rather compact band, the lowest eigenvalue fluctuates to very small
values quite independently of the others. It is expected and has been
shown numerically \cite{Hernandez:1998et} that the spectrum of the
Dirac operator depends quite strongly on the bare gauge coupling.
A well--defined lower bound should be recovered close to the continuum
limit only. Therefore we take the strong fluctuations of $\lambda_{\rm min}$
as an indication for the presence of large cutoff--effects. Here we should
point out that the eigenvalues of the Dirac operator are not on--shell 
quantities and hence the Symanzik improvement programme does not necessarily
reduce cutoff--effects here. Quenched experience even suggests 
that the opposite might be true \cite{DeGrand:1998mn}.

\EPSFIGURE[t]
{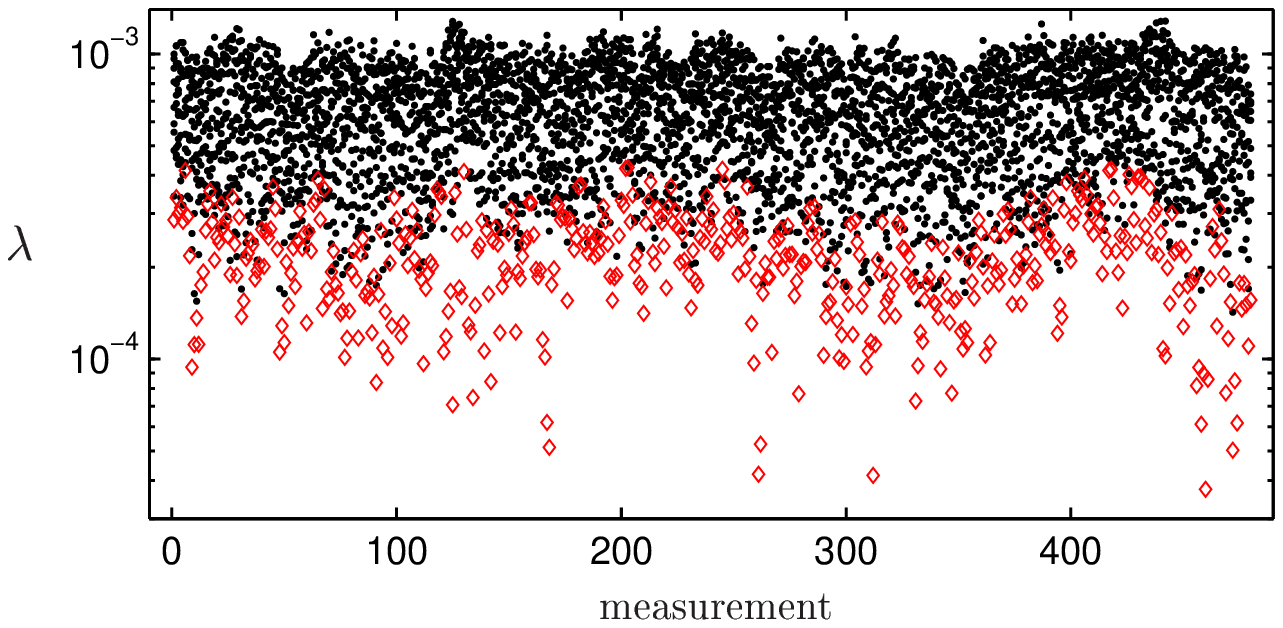,angle=0,width=12.4cm}
{Monte Carlo history of the ten lowest eigenvalues 
at the parameters of run [A2]. The open symbols denote $\lambda_{\rm min}$.\label{bulk}} 

The occurrence of small eigenvalues at these bare parameters
poses a somewhat unexpected problem in dynamical simulations.
Comparing the quenched situation to the $N_f\!=\!2$ dynamical case, the 
na{\"\i}ve expectation is that at fixed bare parameters the
probability of finding configurations with small eigenvalues should be reduced
by the determinant.
To us the more relevant question seems to be whether small eigenvalues are suppressed
in a situation where the physical parameters (e.g. volume and
pseudo--scalar mass) are kept constant.

Using the quenched data from ref.~\cite{Garden:1999fg} and the dynamical data
from refs.~\cite{Aoki:2002uc} and \cite{Allton:2001sk}
(where an estimate of $r_0/a\!=\!5.21(6)$ for $\beta\!=\!5.2$ can be found)
we chose the parameters of the quenched run [A5]
such that the lattice spacing and the (large volume) pseudo--scalar mass
are matched to run [A4].
This was found to occur at almost equal bare current quark mass
(see $Lm$ in \tab{t_simpar_algo}).
In \fig{cutoff} we compare the distributions of
$\lambda_{\rm min}$ for these two runs. Two comments are in order here:
\begin{itemize}
\item
For the dynamical run the mean value is shifted
up from $1.44(1)\!\cdot\!10^{-4}$ to $1.72(5)\!\cdot\!10^{-4}$. 
This agrees with the na{\"\i}ve expectation but in a physically matched 
comparison it is a non--trivial observation.
\item
The distribution itself is
significantly broader compared to the quenched case and in particular it is 
falling off more slowly towards zero. This means that even though
$\ev{\lambda_{\rm min}}$ is larger for $N_f\!=\!2$ the probability of finding
\emph{very} small eigenvalues is enhanced.
\end{itemize}

The second point, i.e. that the lower bound of $\lambda_{\rm min}$ is 
less well--defined, seems to imply that at a
lattice spacing of $a\approx0.1\fm$ the cutoff--effects are much larger
in the $N_f\!=\!2$ case. 
To substantiate this we will compare the
distribution of $\lambda_{\rm min}$ to that from a run at finer lattice spacing
and matched physical parameters.

\EPSFIGURE[t]
{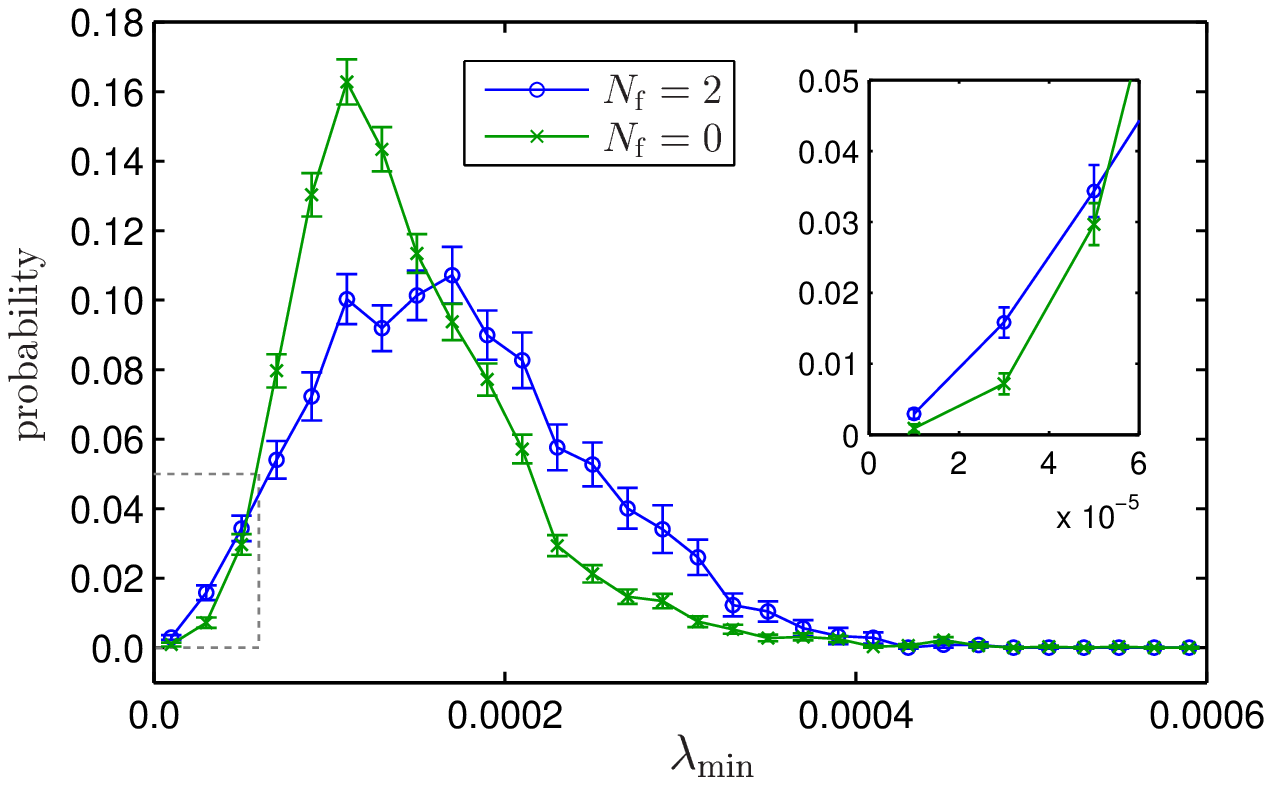,angle=0,width=12cm}
{The histograms of $\lambda_{\rm min}$ from run [A4] ($N_f\!=\!2$) and
run [A5] ($N_f\!=\!0$). Despite its higher mean value the dynamical data show
a much larger probability of finding very small eigenvalues.\label{cutoff}}

\section{Finer lattices}\label{sect:four_algo}

Apart from cutoff--effects, in the massless theory the Schr\"odinger functional
coupling $\bar g^2$ is a
function of the box size $L$ only \cite{Luscher:1992an,Sint:1993un}.
We measured it on a small lattice of extension $L/a\!=\!4$ at $\beta=5.2$,
obtaining a value of $\bar g^2\!=\!3.7(1)$. We then extrapolated to this value the
$L/a\!=\!6$ data used in ref.~\cite{Bode:2001jv} as a function of $\beta$.
Our result from the matching
is that for the two--flavor theory a bare gauge coupling of $\beta\!=\!5.5$
roughly corresponds to a lattice spacing, which is 1.5 times smaller than at
$\beta\!=\!5.2$. This estimate will be confirmed by other non--perturbative
data as well as the perturbative evolution of $a$ in \chap{chap:impr}.

Hoping that the algorithmic difficulties arising from cutoff--effects
would be much smaller in this situation, we simulated a $12^3\!\times\!27$
lattice at this value of $\beta$ (run [A6]) using the HMC algorithm.
With the $\kappa $ we chose (and ignoring the change in
renormalization factors) the bare quark mass $Lm$ is roughly
matched to the heavier runs at $\beta\!=\!5.2$.
We therefore compare run [A6] with run [A3].

Normally, a constant acceptance requires a decrease of the MD
integration step--size if ones goes to finer lattices at fixed
physical conditions. This argument is based on the scaling of
the small eigenvalues\footnote{
As the squared Dirac operator, its eigenvalues have mass dimension
two and should thus
behave like $a^2$.}, which influence the MD driving force.
We found that $\ev{\lambda_{\rm min}}$ in run [A6] is a factor two
smaller than in run [A3].
Nevertheless, at $\beta\!=\!5.5$ the step--size necessary for
a certain ($\simeq90\%$) acceptance is roughly the same as at $\beta\!=\!5.2$.
This indicates that the value of $\dtau$ we had to use in the HMC runs
at $\beta\!=\!5.2$ was dictated by the occurrence of extremely small
eigenvalues rather than by the average smallest eigenvalue.
In addition, where in run [A1] at the same average
acceptance a maximum of $47$ proposals were rejected in a row, the maximum
for run [A6] is $4$ trajectories. For this reason $e^{-\Delta H}$ shows no
autocorrelation and its distribution is well separated from zero.

Concerning fermionic observables, we have not observed spikes and hence
expect the error to scale properly. However, for an accurate estimate of
the error on e.g. $f_1$ our present statistics is not
sufficient. One should note that \emph{ratios} of correlators are easier
to estimate since usually numerator and denominator are correlated, which
reduces the impact of statistical fluctuations. This applies to essentially
all relevant quantities, including the axial current improvement and
renormalization constants discussed in the next chapters.

\EPSFIGURE[p]{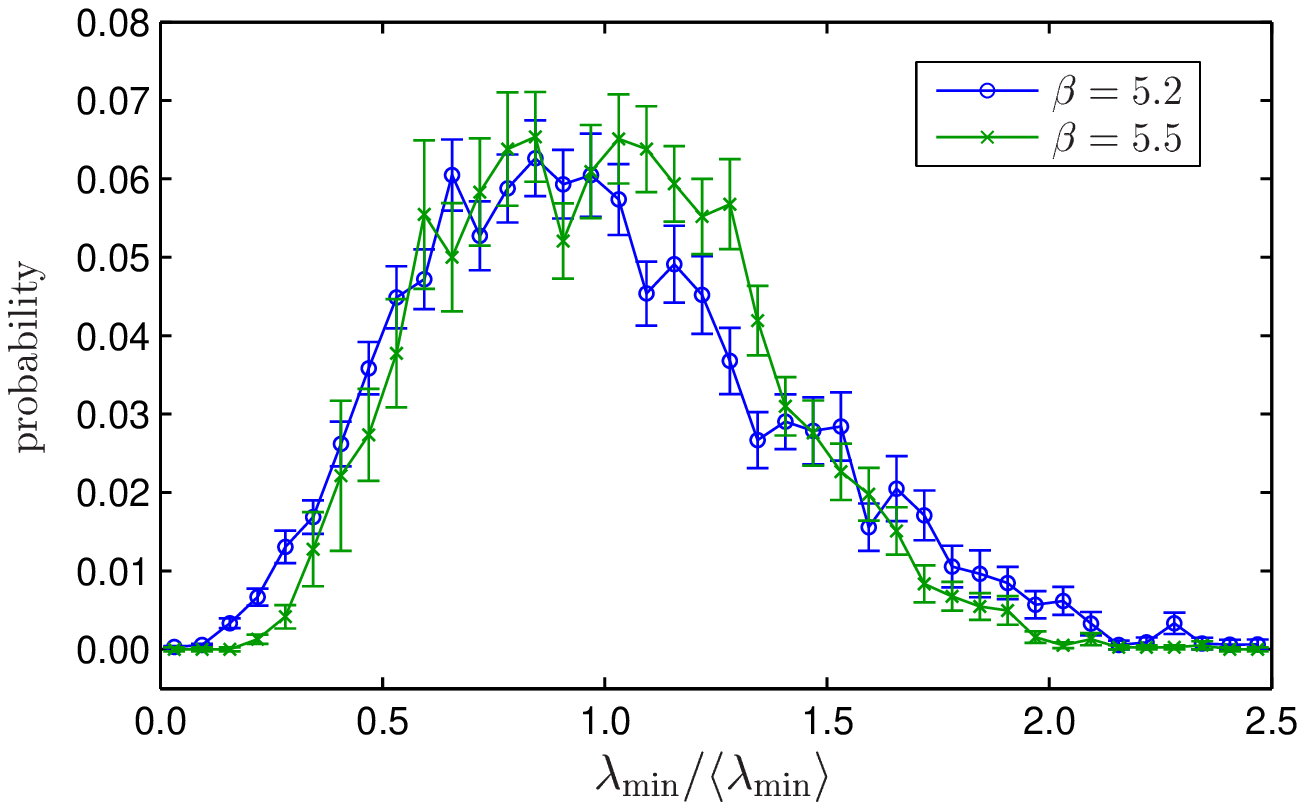,angle=0,width=12cm}
{Normalized distributions of $\lambda_{\rm min}$ from runs [A3]
($\beta\!=\!5.2$) and [A6] ($\beta\!=\!5.5$). While the data from the coarse lattice
stretch almost to zero, the $\beta\!=\!5.5$ data seem to have a more well--defined
lower bound.\label{roughmatch}}

\EPSFIGURE[p]
{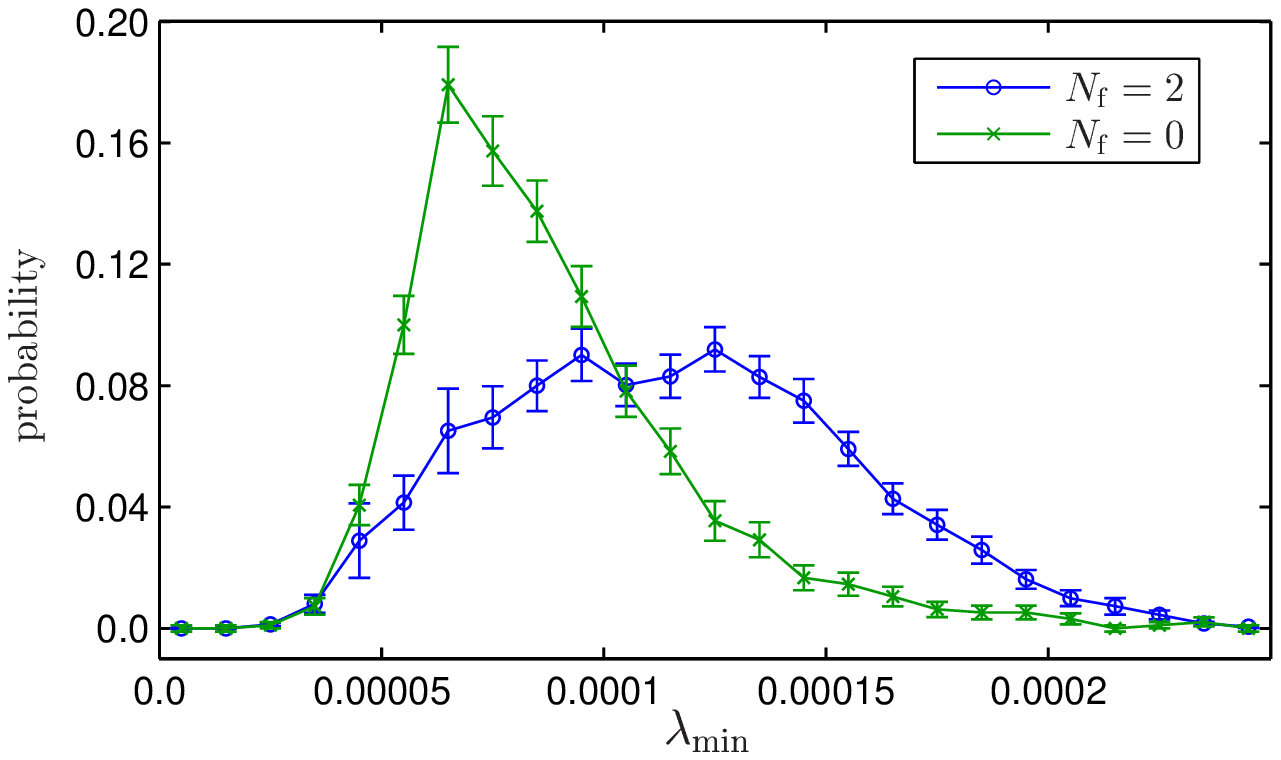,angle=0,width=12cm}
{The histograms of $\lambda_{\rm min}$ from run [A6] ($N_f\!=\!2$) and
run [A7] ($N_f\!=\!0$). At this finer lattice spacing the
lower end of the spectrum
appears to be similar in the quenched and the dynamical case.
\label{matchfine}}

The reason for these effects is the change in the distribution of
$\lambda_{\rm min}$. To compensate for the different lattice spacing,
\fig{roughmatch} compares $\lambda_{\rm min}/\ev{\lambda_{\rm min}}$
from runs [A3] and [A6]. One can clearly see that at the finer lattice spacing
the probability of finding a smallest eigenvalue less than half its average
is greatly reduced compared to $\beta\!=\!5.2$. The width of the distribution
is smaller in this case and in particular the spectrum is now clearly
separated from zero. Quantitatively, the normalized variance of
$\lambda_{\rm min}$ is reduced from $0.18(1)$ to $0.13(2)$.

This comparison explicitly shows that the long tail of the eigenvalue
distribution we observed at $a\simeq0.1\fm$, and which caused the
problems we have discussed, is a cutoff--effect. Matching also run [A6]
to a quenched simulation (run [A7]), we again found an upward shift of
$\ev{\lambda_{\rm min}}$ for the dynamical case. In addition, at this finer
lattice spacing, the tails of the distributions of $\lambda_{\rm min}$ look
already very similar to each other as shown in \fig{matchfine}.

\section{Conclusions}

At a lattice cutoff of approximately $2\;\rm{GeV}$, corresponding
to $a\!\simeq\!0.1\fm$, we have
studied the behavior and performance of HMC--type algorithms
in an intermediate size volume of $1\fm^4$.
We discussed problems related to the occurrence of
small eigenvalues in two--flavor dynamical simulations with improved
Wilson fermions. We found these small eigenvalues to be responsible
for large Hamiltonian violations in the molecular dynamics. Even for
integration step--sizes such that the acceptance is $80\sim90\%$ those
can still cause long periods of rejection, thus degrading algorithmic
performance. However, in spite of employing only single--precision
arithmetics we never observed
reversibility violations.

In addition, those eigenvalues make the estimate of fermionic
quantities very difficult. The na{\"\i}ve intuition is that the fermionic
determinant should suppress small eigenvalues compared to the quenched case.
Through a direct comparison at matched physical parameters we indeed verified
that $\ev{\lambda_{\rm min}}$ is larger with two dynamical flavors. On the other
hand there is no obvious expectation for the tail of the distribution
and we observed that it extends further towards zero than in the quenched case.
Given the infrared cutoff induced by the Schr\"odinger functional boundary
conditions and the quark mass
we interpret this as a lattice artifact. We were able to confirm this
picture with a simulation at finer lattice spacing, where the spectrum
turned out to have a much sharper lower bound.

In our study we found that the PHMC algorithm is more efficient than
HMC (with two pseudo--fermions) in incorporating the contribution to
the path integral of configurations
carrying small eigenvalues. In other words, the distortion of the spectrum by
cutoff--effects actually makes it advantageous to deviate from
importance sampling.
Also without such special problems we found PHMC at least comparable
in performance to HMC (in our implementations).

We want to emphasize that the problems discussed here do not
occur only in the Schr\"odinger functional setup. Without
this infrared regulator they are expected to show
up already at larger quark masses.

Due to the in a sense ''more robust'' nature of PHMC one might ask
why one should not use it in all simulations. The reason is that with
decreasing lattice spacing the smallest eigenvalues will also decrease,
which in turn requires a smaller $\epsilon$ in the construction of the
polynomial as otherwise the fluctuations of the reweighting factor will
render the algorithm inefficient.

However, a smaller lower bound in the polynomial approximation implies
a larger degree of the polynomial (at constant relative deviation $\delta$).
In the current implementation of the PHMC, single precision restricts the
degree to $\approx140$ since otherwise roundoff problems appear in the
construction of the polynomial.

We therefore used the PHMC only in the simulation for the axial current
renormalization at $\beta=5.2$ and $5.29$.
The simulations for the axial current improvement
condition are performed with a smaller time extension and larger quark mass.
Therefore one can safely use HMC at all lattice spacings we consider.

\chapter{Axial current improvement}
\label{chap:impr}
In the quenched approximation, the improvement
coefficients $\csw$
and $\cA$ have been determined non-perturbatively in
the relevant range of bare coupling (or lattice spacings)
in \cite{Luscher:1996ug}. The improvement conditions,
from which they have been determined, were derived from
the chiral symmetry of the theory in the continuum
limit.

More precisely, the PCAC relation (\ref{pcac}) was required
to hold at finite lattice
spacing~\cite{Luscher:1996sc,Luscher:1996vw,Luscher:1996ug}.
As will be detailed in the next section, the PCAC relation
is considered considered with different external states.
In \cite{Luscher:1996sc,Luscher:1996vw,Luscher:1996ug},
finite volume states were chosen, formulated in the framework
of the Schr\"odinger functional. Later, a determination of
$\cA$ was performed by evaluating the PCAC relation in large
volume \cite{Bhattacharya:2000pn,Collins:2001mm} at several
values of the lattice spacing. While at $a\approx0.1\fm$
these results for $\cA$ differ quite significantly
from the finite volume definition \cite{Luscher:1996ug},
at smaller lattice spacings the difference decreases.

For the interpretation of this difference,
one should keep in mind that beyond perturbation theory
the improvement coefficients themselves are affected by
$\rmO(a)$ ambiguities. In some detail this has been
discussed and demonstrated numerically in
\cite{Guagnelli:2000jw}. The $\rmO(a)$ ambiguity simply
corresponds to the fact that in the implementation of the
improvement programme the theory is treated only up to 
$\rmO(a^2)$ effects. Thus any change of $\rmO(a)$ of the
improvement coefficients only contributes to the terms
$\phi_2$ and $\LL_2$ in (\ref{symeff}, \ref{phieff}) and
does therefore not invalidate $\rmO(a)$ improvement.

While this obviously forbids a unique definition of the
improved theory, the $\rmO(a)$ ambiguities can be made to
disappear smoothly if the improvement condition is evaluated
with all physical scales kept fixed, e.g. in units of $r_0$
(\ref{arenot}), while only the lattice spacing is varied
\cite{Guagnelli:2000jw}. This is what is meant by a ''line
of constant physics'' (LCP).

Since Symanzik's effective
theory describes the lattice artifacts only asymptotically
(\sect{sect:impr}) it can only be valid for matrix elements
dominated by states with energy $E \ll a^{-1}$. It is
therefore important to make sure that the improvement conditions
are also imposed using low energy states. So far, the
methods of \cite{Bhattacharya:2000pn,Collins:2001mm} have
not yet been implemented such as to satisfy these conditions. 

In \cite{Durr:2003nc}, two improvement conditions for $\cA$, 
were studied, for which one can choose the kinematic
parameters such that the above criteria are satisfied.
They are formulated in finite volume
with Schr\"odinger functional boundary conditions, which
furthermore  helps to render  the numerical evaluation feasible
in full QCD. These improvement conditions will be discussed in
the next section, where we motivate our choice of one of them
to compute $\cA$ in the $\nf\!=\!2$ theory.

\section{Strategy and techniques}
\label{sect:two}

Before going into the details of our strategy and
techniques let us comment again on the constant 
physics condition.
In order to keep the physical volume constant,
we need to know how the lattice spacing depends on
$\beta$. With this knowledge we can tune $\beta$
such that a certain $L/a$ corresponds to a prescribed
value of $L$ (or $L/r_0$). However, this tuning of the
volume is not critical in the evaluation of $\cA$,
since the latter depends on the volume only through
effects of order $a/L$.\footnote{Effects of order $am$
will be discussed later.} A relative uncertainty
$\Delta$ in the physical volume $L$ (or equivalently
in $r_0$) thus implies a relative uncertainty in $\cA$,
which is proportional to $a/L\cd\Delta$. As a consequence,
even if $\Delta$ varies a bit in the considered range
of  lattice spacings, this is quite irrelevant, in
particular if $\Delta$ is a smooth function of the
lattice spacing. Therefore the constant physics condition
with respect to the volume has to be enforced with only
moderate precision.

In the remainder of this section we will state in more
detail how the constant physics condition is implemented. 
We will also discuss the methods in
\cite{Luscher:1996ug,Durr:2003nc} to determine $\cA$,
with emphasis on the one we finally used.

\subsection{Constant physics condition}

With two degenerate flavors of light quarks, the
theory has two bare parameters, $\beta$ and $m_0$. Roughly
speaking, the bare quark mass 
$m_0$ controls the physical quark mass and the bare
coupling determines the lattice spacing, defined
at vanishing quark mass (for a more precise discussion
see \cite{Luscher:1996sc,Sommer:2003ne}).
Non--perturbative estimates of
\be
t_{r_0}(\beta)=\frac{[r_0/a](5.2)}{[r_0/a](\beta)}\;,
\ee
are available in a limited range of $\beta$ 
\cite{Gockeler:2004rp,Aoki:2002uc}.
In \cite{DellaMorte:2004bc}, the results of
\cite{Gockeler:2004rp,Aoki:2002uc} were extrapolated
to zero quark mass. Taking directly these values for
$[r_0/a](\beta)$, we have the points with error bars in
\fig{pert_evol}. From those we roughly estimated the location of
the filled points, using the perturbative dependence of the
lattice spacing on $\beta$ as a guideline. For our action
this is given to three loops in \cite{Bode:2001uz}, which
builds on various steps carried out in
\cite{Hasenfratz:1980kn,Kawai:1980ja,Weisz:1980pu,
Sint:1995ch,Luscher:1995np,Alles:1996cy,Christou:1998ws}. 
Applied as a pure expansion in the bare coupling (no tadpole
improvement), one has
\begin{eqnarray} \nonumber
\frac{a(g_0^2)}{a((g_0')^2)}&=&e^{ -[g_0^{-2}-(g_0')^{-2}]/ 2b_0}
[g_0^2/(g_0')^2]^{-{b_1/2b_0^2}}\,
\left[\, 1 + q\, [g_0^2 -(g_0')^2] + {\rm O}\left((g_0')^4\right)\,\right], 
\\
&& q = 0.4529(1)\,,\quad g_0<g_0'\;. \label{e:L3l}
\end{eqnarray}
The evolution of the lattice spacing relative to our
reference point at  $(g_0')^2\!=\!6/5.2$ is then
expressed by the function $t(\beta)=a(6/\beta)/a(6/5.2)$, 
which is plotted as a thick line in the graph. It confirms
that the filled points are very reasonable choices.
Note that other forms of applying bare perturbation theory 
(differing from \eq{e:L3l} in the $g_0^4$-term)
would give somewhat different results, but since we are 
interested in a rather limited range in
$g_0^2$ this is of no great importance. 
Later we will show that
systematic uncertainties in $\cA$ introduced by this
approximate scale
setting are negligible.
\EPSFIGURE[t]{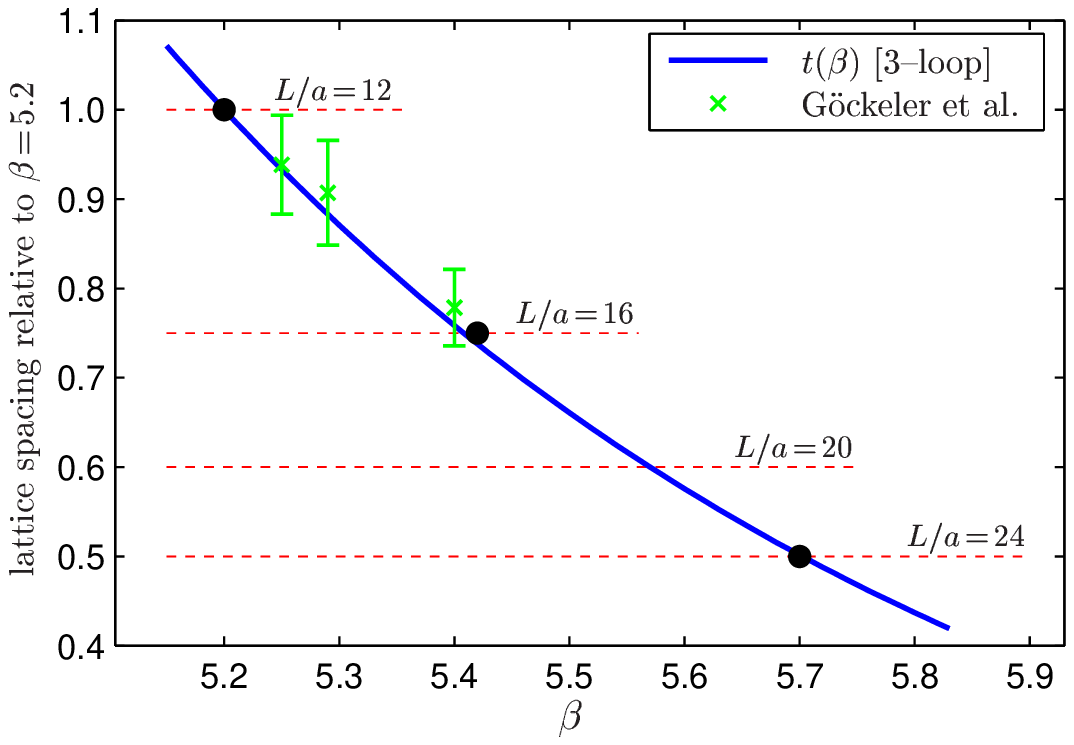,width=10cm}{
The evolution of the lattice spacing $a$ with the inverse
bare gauge coupling $\beta$ from perturbation theory in the
lattice scheme and large volume data \cite{Gockeler:2004rp}
for the scale $r_0$. The filled points correspond to our
``scaled'' simulations.
\label{pert_evol}}
Finally, we keep the PCAC mass (\ref{massimp}), evaluated
with an average of $f_{\rm X}$ and $g_{\rm X}$
approximately constant.

\subsection{Improvement conditions for the axial current}

In this section we discuss criteria for the choice of the
improvement condition. Quark masses derived from the PCAC
relation are of the form
\be
m(x;\alpha,\beta)=\frac{\langle\alpha|\tilde\partial_\mu
(A_{\rm I})_\mu^a(x)|\beta\rangle}
{2\langle\alpha|P^a(x)|\beta\rangle}\;.
\ee
Since this mass is obtained from an operator identity, it
is independent of the states $|\alpha\rangle$ and
$|\beta\rangle$ as well as the insertion point
$x$ up to cutoff--effects. Enforcing this independence at
\emph{finite} lattice
spacing leads to possible definitions of improvement
conditions \cite{Durr:2003nc}. Inserting the
expression for the improved current (\ref{AI}) in the
previous equation, the
quark mass can be written as $m=r+a\cA s+\rmO(a^2)$ with
\bea
r(x;\alpha,\beta)&=&\frac{\langle\alpha|\tilde\partial_\mu
A_\mu^a(x)|\beta\rangle}
{2\langle\alpha|P^a(x)|\beta\rangle}\label{rgeneral}\\
\textrm{and}\qquad s(x;\alpha,\beta)&=&
\frac{\langle\alpha|\partial_\mu\partial^*_\mu P^a(x)
|\beta\rangle}
{2\langle\alpha|P^a(x)|\beta\rangle}\;.\qquad\label{sgeneral}
\eea
If we now consider two sets of external states and two
insertion points, the improvement condition
$m(x;\alpha,\beta)=m(y;\gamma,\delta)$
yields
\be
-\cA=\frac{\Delta r}{a\Delta s}=\frac1a\!\cdot\!
\frac{r(x;\alpha,\beta)
-r(y;\gamma,\delta)}
{s(x;\alpha,\beta)-s(y;\gamma,\delta)}\;.
\ee
The ''sensitivity'' to $\cA$ of an improvement condition is
therefore given by $a\Delta s$, since it is this term, which
is multiplied by $\cA$ in the current quark masses.

Once a reasonably large sensitivity is achieved, 
all improvement conditions at constant physics
are equally valid in the sense
that $\rmO(a)$ effects are removed in on-shell quantities.
However, the way in which \emph{higher--order} lattice
artifacts are modified will depend on the concrete choice
of the improvement condition. In particular, if states with
energy not so far from the cutoff are involved, large
$\rmO(a^2)$ effects might be introduced.

To discuss the different improvement conditions we need to
generalize the Schr\"odinger function correlators introduced
in \sect{addcorr}. Instead of projecting both boundary quarks
to zero momentum separately (\ref{O},\ref{Oprime}), we now use
boundary fields products, which depend on spatial trial
''wave functions'' $\omega(\bx)$ \cite{Molke:thesis}. In the
following we will use
\bea 
 \fa(x_0;\omega)&=&- \frac{a^3}{3L^{3/2}} \sum_\vecx\langle
 A_0^a(x) \, \op^a(\omega)
\rangle\;,\\
 \fp(x_0;\omega)&=&- \frac{a^3}{3L^{3/2}} \sum_\vecx\langle
 P^a(x) \, \op^a(\omega)
\rangle\;,\\ 
\textrm{and }\ \     f_1(
\omega',\omega)&=&- \frac1{3L^3} \langle\op'^a (\omega') \,
\op^a(\omega) \rangle\;,
\eea
which depend on the pseudo--scalar operator
\be
\quad \op^a(\omega) =a^6 
\sum_{\vecx,\vecy} \zetabar(\vecx) \gamma_5 \tau^a\half 
\omega(\vecx-\vecy) \zeta(\vecy) \label{op}
\ee
at the $x_0\!=\!0$ boundary and the corresponding operator
$\op'^a (\omega')$ at the upper boundary of the SF cylinder.

The Schr\"odinger functional version of eqs. (\ref{rgeneral})
and (\ref{sgeneral}) is then given by
\bea
r(x_0;\omega)&=&\frac{\tilde\partial_0\fa(x_0;\omega)}
{2\fp(x_0;\omega)}\label{rSF}\\
\textrm{and}\qquad s(x_0;\omega)&=&\frac
{\partial_0\partial^*_0 \fp(x_0;\omega)}
{2\fp(x_0;\omega)}\;.\label{sSF}
\eea
To determine $\cA$ in the $\nf\!=\!0$ theory
\cite{Luscher:1996ug}, $\Delta r$ and $\Delta s$ were
originally defined through a variation of the periodicity
angle $\theta$ of the fermion fields, while keeping
$x_0\!=\!T/2$ and $\omega\!=\!const$ fixed. For this method
the sensitivity $a\Delta s$ is quite low when $L\gtrsim0.8\fm$,
$T=2L$. In addition, with dynamical fermions different values
of $\theta$ would require separate simulations. We therefore
consider this method as too expensive and disregard it in the
following. In the quenched approximation two alternatives have
been explored in \cite{Durr:2003nc}.

Requiring the quark mass to be independent of $x_0$ (for
fixed $\theta$ and $\omega\!=\!const$) is technically easy
to implement. However, also in this case the sensitivity
is small unless large values of $\theta$ are used. Moreover,
the contribution of excited states is not well controlled,
because  one insertion point must be rather close to a
boundary to achieve a sufficiently large sensitivity. Thus,
energies which are not far removed from the cutoff may contribute.

The simultaneous requirements of large sensitivity and control
of excited--state contribution can be fulfilled more easily
with the second method, where variations of the wave function
$\omega$ have been considered. Ideally, one would like to use
two wave functions $\omega_{\pi,0}$ and $\omega_{\pi,1}$,
such that the corresponding operator $\op^a(\omega)$ couples
only to the ground and first excited state in the pseudo--scalar
channel, respectively. As one can easily see from \eq{sSF} the
sensitivity to $\cA$ is then proportional to
$m_{\pi,1}^2-m_{\pi,0}^2$, where $m_{\pi,n}$ denotes the mass
of the $n$th excited state in the pseudo--scalar channel.
Higher excited states are
(by definition) not contributing and in principle the method
can be used for rather small $T$. Hence, we find this the most
attractive method both from a theoretical and practical point of
view. In the next section we will detail our approximation
to this ideal situation.

\subsection{Wave functions}

We will now proceed to the more technical aspects of our
method. In order to approximate  $\omega_{\pi,0}$ and
$\omega_{\pi,1}$ consider
a set of $N$ wave functions. Given a vector $u$ in this
$N$--dimensional space, projected correlation functions are
defined as $(u,\fa)$ and $(u,f_1u)$, i.e. $\fa$ is regarded as
a vector and $f_1$ as a matrix in this space. It is useful to
represent $f_{\rm X}$ ($\rm X=A,P$) and $f_1$ as
\cite{Guagnelli:1999zf}
\bea
f_{\rm X}(x_0;\omega_i)&=&\sum_{n=0}^{M-1} F_{\rm X}^{n} v_i^{n}
e^{-m_{\pi,n}x_0}
+\rmO(e^{-m_{\pi,M}x_0})  +\rmO(e^{-m_\mrm{G}(T-x_0)})\;,\qquad\\
f_1(\omega_i,\omega_j)&=&\sum_{n=0}^{M-1} v_i^{n} v_j^{n}
e^{-m_{\pi,n}T} 
+\rmO(e^{-m_{\pi,M}T})+ \rmO(e^{-m_\mrm{G}T})\;,\label{f1decomp}
\eea
where $n$ labels the states in the pseudo--scalar channel in
increasing energy and $v_i^{n}$ is the overlap of such a state
with the one generated by the action of $\op^a(\omega_i)$ on the
SF boundary state. The mass $m_\mrm{G}$  belongs to the lowest
excitation in the scalar channel, the $0^{++}$ glueball and the
coefficients $F_{\rm X}^{n}$ are proportional to the decay
constant of the $n$th state. Here we have suppressed the
explicit volume dependence of all quantities.

Knowledge of $v^{n}$ would allow the construction of vectors
$u^{n}$, such that -- up to corrections of order
$e^{-m_{\pi,M}T}$ -- the correlation
$(u^{n},\fa)$ receives contribution from the $n$th state only. 
These $u^{n}$ may be computed
from the $v^{n}$ by a 
Gram-Schmidt orthonormalization. Clearly,
$u^{0}$ and $u^{1}$ can then be used to approximate
$\omega_{\pi,0}$ and $\omega_{\pi,1}$.

An approximation to the $v^{n}$ 
can be obtained from the eigenvectors of the positive symmetric
matrix $f_1$. We denote the normalized eigenvectors of $f_1$
by $\eta^{0},\eta^{1},\ldots$ corresponding to eigenvalues
$\lambda^{0}>\lambda^{1}>\ldots$ and apply \eq{f1decomp}
to $\eta^0$ to obtain\footnote{
The glueball contribution will be dropped from now on.}
\be
v^0(v^0,\eta^0)\rme^{-m_{\pi,0} T}+\rmO(\rme^{-m_{\pi,1} T})=\lambda^0\eta^0\;.
\ee
Clearly $\lambda_0$ is of order $|v_0|^2\rme^{-m_{\pi,0}T}$, which
is inserted to rewrite the previous equation in terms
of the normalized vectors $\hat v^n=v^n/|v^n|$
\be
\hat v^0(\hat v^0,\eta^0)=\eta^0+\rmO(\rme^{-(m_{\pi,1}-m_{\pi,0}) T})\;.
\ee
Since $\hat v^0$ and $\eta^0$ are normalized vectors, it follows that
$(\hat v^0,\eta^0)=1$ up the error term given above. Together with
the orthogonality of the eigenvalues of $f_1$ this implies
\bea
||\hat v^{0}-\eta^{0}||^2&=&
\rmO(e^{-(m_{\pi,1}-m_{\pi,0})T})\\
\textrm{and}\quad
(\eta^{1},\hat v^{0})&=&
\rmO(e^{-(m_{\pi,1}-m_{\pi,0})T})\;.
\eea
Thus, to the order indicated above, $\hat v^{0}$ is given by $\eta^{0}$
and $\eta^{1}$ is orthogonal to the ''ground state vector''
$\hat v^{0}$. As eigenvectors of a symmetric matrix the
$\eta^{n}$ are already orthogonal and we therefore use the approximation
\bea
\omega_{\pi,0}\simeq\sum_i\eta_i^{0}\omega_i &\quad \textrm{and}\quad &
\omega_{\pi,1}\simeq\sum_i\eta_i^{1}\omega_i\label{project}
\eea
to obtain correlators, which are (for intermediate $x_0$) dominated
by the ground and first excited state, respectively.
We note in passing that the ratios $v^{n}_i/v^{n}_j$ have a continuum
limit if the wave functions are properly scaled with the lattice spacing. 

In our simulations we restrict ourselves to a basis consisting
of three (spatially periodic) wave functions defined by
\bea
\omega_i(\vecx) &=& N_i^{-1} \sum_{{\bf n}\in{\bf Z}^3}
\overline{\omega}_i(|\vecx-{\bf n}L|)\,,\; i=1,\ldots,3\,,
\nonumber \\
\overline{\omega}_1(r) &=& r_0^{-3/2}\,\rme^{-r/a_0}\,,\quad \nonumber 
\overline{\omega}_2(r) =  r_0^{-5/2}\,r\,\rme^{-r/a_0} \,,\quad\\
\overline{\omega}_3(r) &=& r_0^{-3/2}\,\rme^{-r/(2a_0)}\,\;,
\eea
where $a_0$ is some physical length scale. We thus keep
it fixed in units of $L$,  choosing $a_0\!=\!L/6$.
The sum over $\bf n$ is required to preserve the spatial
periodicity. In practice the summation is stopped when
the norm of $\omega$ no longer changes within single precision
arithmetics.
The (dimensionless) coefficients $N_i$ are fixed to normalize the
wave function via $a^3\sum_{\vecx} \omega_i^2({\bf x})=1$. In
this context the correlation functions introduced in \sect{addcorr}
can be regarded as belonging to the flat wave function
$\omega_0(\vecx)=L^{-3/2}$,
where both quarks are projected to zero momentum separately.
In this case $\vecx$ and $\vecy$ in
\eq{op} are uncorrelated and thus full translational invariance
can be used without performing additional inversions of the Dirac
operator (see \app{sectionf1}).
In the general case $\omega_{1\ldots3}$ we replace one of
the spatial sums in \eq{op} by a sum over 
eight far separated points, which means that one
performs eight times as many inversions.
This additional computational effort is still small compared
to the cost of the inversions in the HMC update.

\section{Numerical computation}

\label{sect:canum}

\subsection{Results for the improvement coefficient}

\tab{t_simpar} summarizes the parameters of our simulations for
the axial current improvement constant. The values of the
other improvement coefficients are given in
Sections \ref{sec:wilson} and \ref{sect:latticeSF}.

\TABULARSMALL[!t]{r|rrcclcll}{
    \hline
 run & $L/a$ &  $T/a$  &  $\beta$ & $\kappa$ & $N_{\rm meas}$  & $\!\!\!\tau_{\rm meas}\!\!\!$ & $\ am/t(\beta)$ &  
$\quad-\cA$ \\
    \hline
 [C1] & 12  & 12  & 5.20 & 0.135600 & $4\!\times\!80\  $&  5  &  $0.0151(9)$ & $0.0638(23)$ \\[0 mm]
 [C2] & 16  & 16  & 5.42 & 0.136300 & $1\!\times\!200\ $&  4  &$0.0171(5)$ & $0.0420(21)$ \\[0 mm]
 [C3] & 24  & 24  & 5.70 & 0.136490 & $2\!\times\!60\  $&  4  &$0.0151(4)$ & $0.0243(36)$ \\[0 mm]\hline
 [C4] & 12  & 12  & 5.20 & 0.135050 & $4\!\times\!39\  $&  5  &$0.0363(6)$ & $0.0697(31)$ \\[0 mm]
 [C5] & 16  & 20  & 5.57 & 0.136496 & $1\!\times\!285\ $&  8  &$0.0154(4)$ & $0.0366(36)$ \\[0 mm]
 [C6] & 24  & 24  & 6.12 & 0.136139 & $2\!\times\!21\  $&  12  &$0.0002(4)$ & $0.0244(21)$ \\[0 mm]\hline}
{Summary of simulation parameters and results for $\cA$. 
Runs [C1]-[C3] are at constant physics.\label{t_simpar}}

The $\beta$ values for run [C2] and [C3] have been chosen such
that $L/r_0$ is approximately the same as in
run [C1], which corresponds to $L\simeq1.2\fm$. In exploratory
quenched studies \cite{Durr:2003nc} this volume was found to be
sufficient for the described projection method to work. 
$N_{\rm meas}$ is the number of estimates
of $\cA$ (with the number of replica denoted explicitly),
separated by $\tau_{\rm meas}$ in Monte Carlo time. 
The autocorrelation of these measurements turned out 
to be negligible.
The column labeled $am/t(\beta)$ refers to the
bare quark mass $m=r(T/2;\omega_0)+a\cA s(T/2;\omega_0)$, cf.
eqs.~(\ref{rSF}, \ref{sSF}), which is equivalent
to (\ref{massimp}). The 1--loop value of $\cA$ from
\cite{Luscher:1996vw}
is used here. We tuned the hopping parameter $\kappa$ in order
to keep $am/t(\beta)$ fixed when varying $\beta$, thus ignoring
(presumably small) changes in the renormalization factors. 
Note that we have chosen a finite, but small bare quark mass
of around 30~MeV. Such a mass helps (in addition to the Dirichlet
boundary conditions) to reduce the cost of the simulations.
Results from the remaining simulations are used to
discuss systematic uncertainties in our determination of $\cA$.

We employed the hybrid Monte Carlo algorithm with two
pseudo--fermion fields as described in \sect{sect:hmc}.
For all
observables we have checked the expected scaling of the
statistical error with the sample size and thus verified the
absence of the problems described in Sections \ref{instab}
and \ref{sect:sampling} at the
volumes and masses we consider here.
\EPSFIGURE[!b]{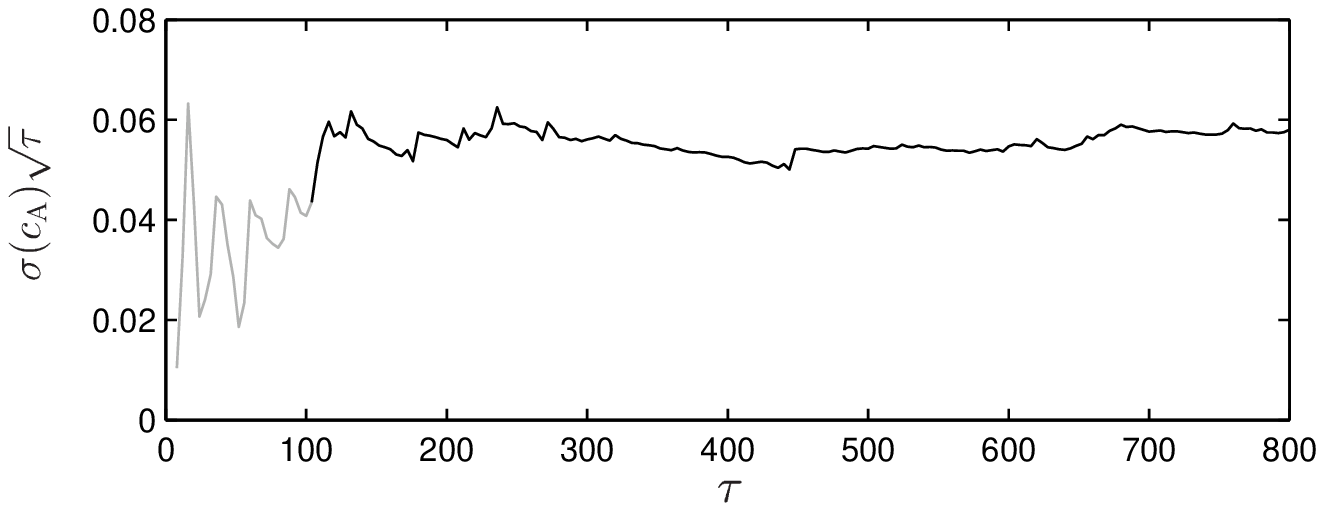,width=137mm}{
The error on $\cA$ vs. the Monte Carlo time for run [C2].
\label{cascaling}}
For run [C2] \fig{cascaling} shows how the error of our
estimate for $\cA$ (see below) behaves as a function of the Monte
Carlo time $\tau$. One can see that already after approximately
200 trajectories the region of statistical scaling is reached.

\DOUBLEFIGURE[t]{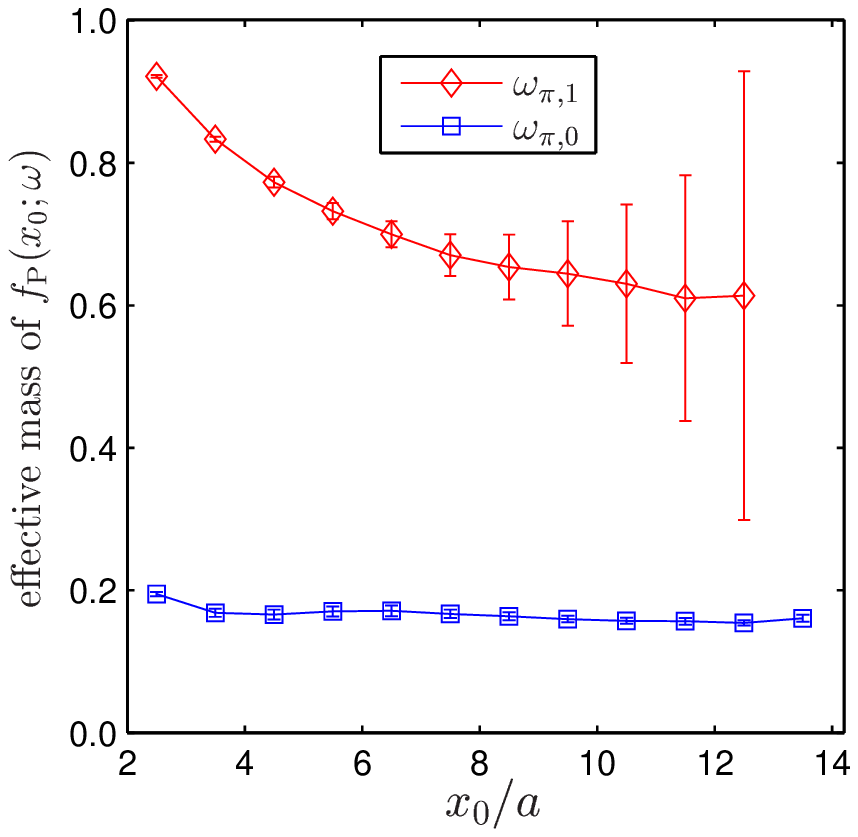,height=6.55cm}{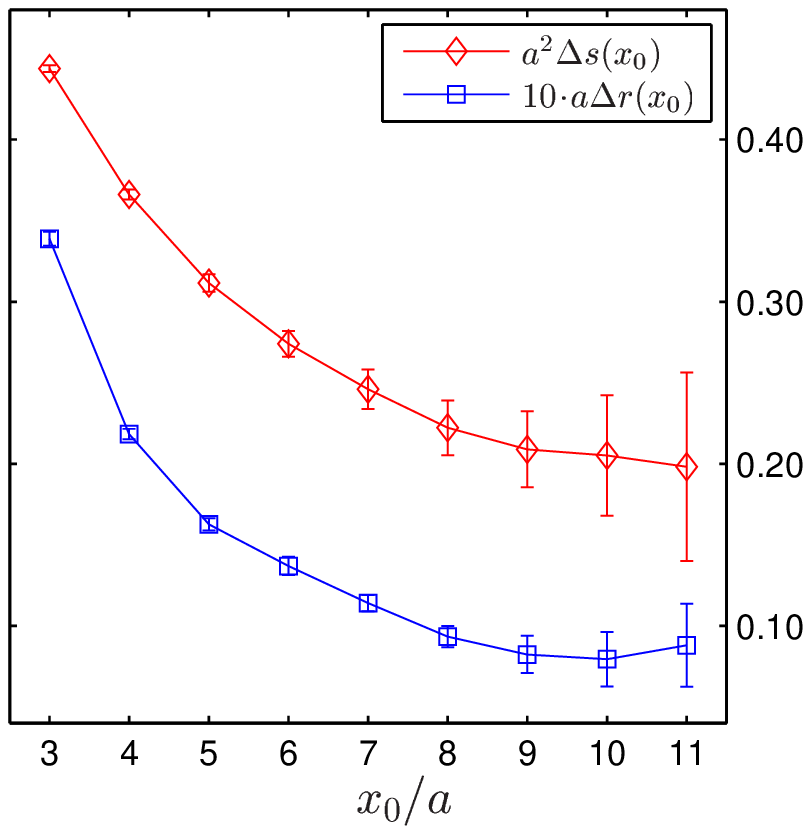,height=6.45cm}{
The effective mass in lattice units 
of the projected correlation functions $(\eta^{0},\fp)$
and $(\eta^{1},\fp)$ from run [C2].
\label{effmass}}
{$\Delta s(x_0)$ and $\Delta r(x_0)$ determined from
$\omega_{\pi,0}$ and $\omega_{\pi,1}$ in run [C2].
\label{effrs}}

In \fig{effmass} we show the effective masses from
$\fp(x_0;\omega_{\pi,0})$
and $\fp(x_0;\omega_{\pi,1})$ as obtained in run [C2].
Two distinct signals are clearly visible, which indicates
that the described approximate projection method works well
at these parameters. The energy of the first excited state is
not far away from $a^{-1}$, suggesting that in even smaller volumes
the residual $\rmO(a^2)$ effects would grow rapidly.
In the spirit of the remark after \eq{project} at the other values
of $\beta$ we used the same linear combination of wave functions to
define $\omega_{\pi,0}$ and $\omega_{\pi,1}$, namely
\be
\begin{array}{rcl}
\eta^{0}&=&[\,0.5172,\,\phantom{-}0.6023,\,\phantom{-}0.6081\,]\\
\textrm{and}\quad
\eta^{1}&=&[\,0.8545,\,-0.3233,\,-0.4066\,]\;,\label{eta}
\end{array}
\ee
which are the ones determined in run [C2].
When scaled in units of $r_0$, this yields effective masses similar
to those  shown in \fig{effmass}.
Results from a redetermination of $\eta^{(0)}$ and $\eta^{(1)}$ in
the other matched simulations agree with \eq{eta} to a high precision.
In fact, one can easily distinguish the matched and unmatched runs
using e.g. $\eta^0$. Comparing the two $24^4$ runs
\bea
\textrm{[C3]}\qquad\eta^{0}&=&[\,0.5173(2),\,0.6024(1),\,0.6079(1)\,]\;,\\
\textrm{and [C6]}\qquad\eta^{0}&=&[\,0.5126(2),\,0.6042(1),\,0.6101(1)\,]\;,
\eea
it is clear that the eigenvector $\eta^0$ from run [C3] is in good
agreement with (\ref{eta}), while a significant deviation is seen
in run [C6].

In \fig{effmass} the error on the effective mass of the first excited
state is seen to be quite large, but what actually enters the computation
of $\cA$ is the error of
\bea
\Delta r(x_0)&=&r(x_0;\omega_{\pi,1})-r(x_0;\omega_{\pi,0})\\
\textrm{ and }\Delta s(x_0)&=&s(x_0;\omega_{\pi,1})
-s(x_0;\omega_{\pi,0})
\;.
\eea
These profit from statistical correlations of the correlation functions
entering their definition and thus have smaller statistical
errors as can be seen in \fig{effrs}, where we plot $a\Delta r$ and
$a^2\Delta s$ from the same data used in \fig{effmass}.

\fig{effca} collects results
for the ''effective'' $\cA(x_0)=-\Delta r(x_0)/a\Delta s(x_0)$
from the matched runs [C1]-[C3].
We see little variation for $x_0\gtrsim 6a$, which we take as
another signal that high energy states, which could contribute large 
$\rmO(a)$ ambiguities in the improvement condition,
are reasonably suppressed in this region.
We complete
our definition of $\cA$ with the choice $x_0\!=T/2$,
which is at the same time scaled in physical units and in agreement
with the $x_0\gtrsim 6a$ bound for all our lattices.

\EPSFIGURE[!b]{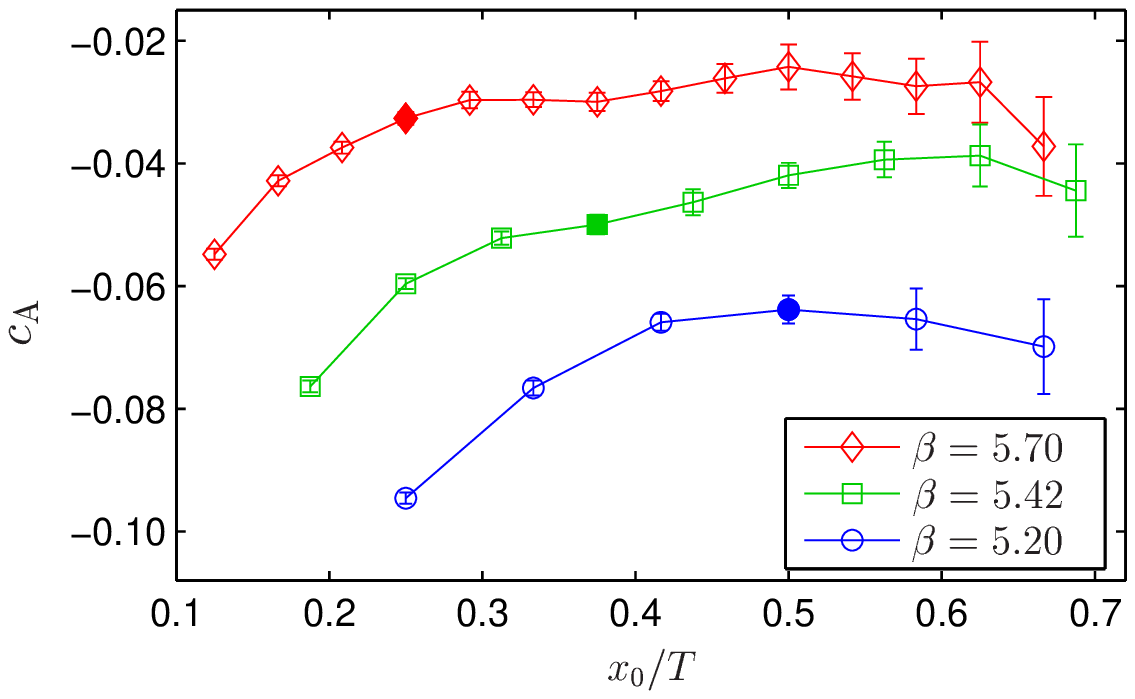,width=99mm}{
Effective $\cA$ as determined from $\omega_{\pi,0}$ and
$\omega_{\pi,1}$ for runs [C1]-[C3]. Points 
with $x_0/a=6$ are marked 
by filled symbols.
\label{effca}}

Finally, $\cA$ is plotted as a function of $g_0^2$ in \fig{figpade}. 
The solid line is a smooth interpolation of the data from
the matched simulations, constrained in addition by 1--loop perturbation
theory:
\begin{equation}
\cA(g_0^2)=-0.00756\, g_0^2\times\frac{1-0.4485\, g_0^2}{1-0.8098\, g_0^2}\;.
\label{cainterpol}
\end{equation}
It is our final result, valid in the range $0.98 \leq g_0^2 \leq1.16$  
within the errors of the data points (at most 0.004). 
\EPSFIGURE[ht]{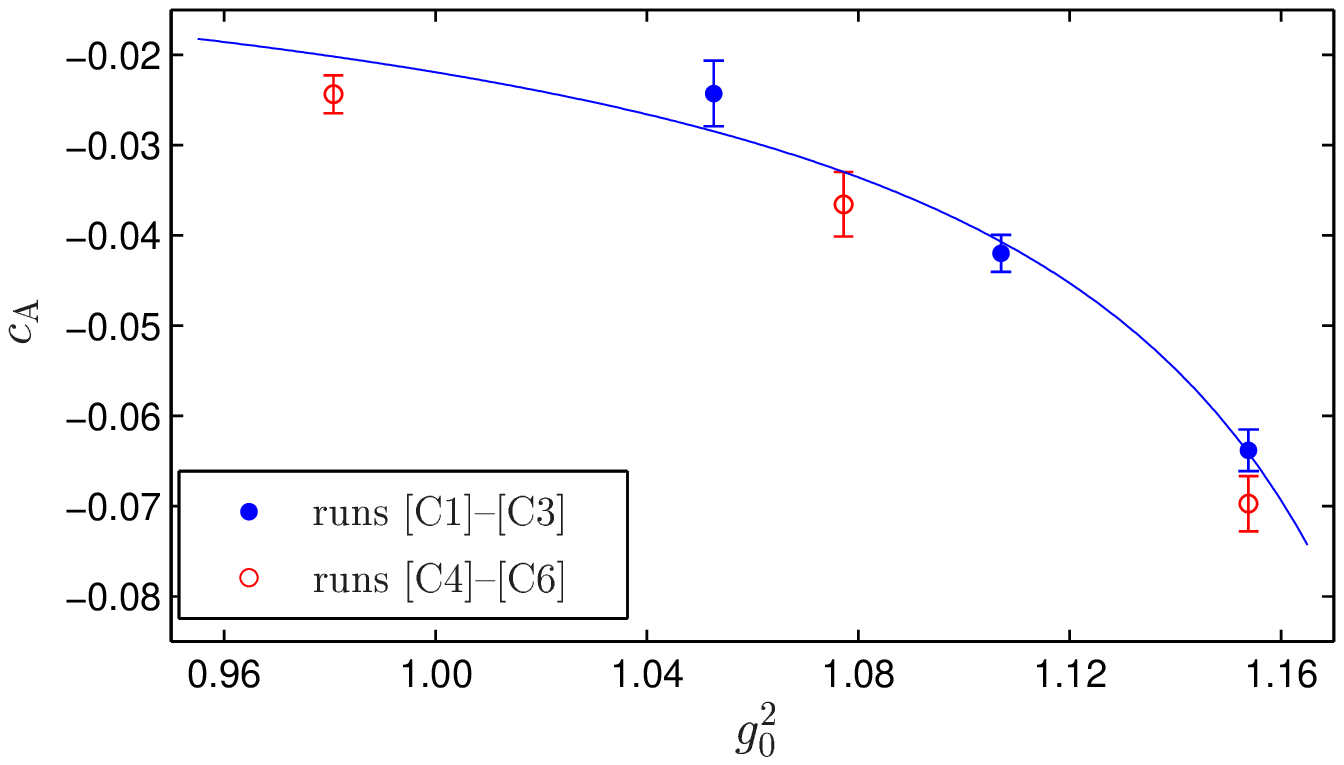,width=12cm}{
Simulation results for $\cA$. The solid line represents a fit
of the data points at constant physics (filled circles).
\label{figpade}}

This non-perturbative result is quite far away form 
1--loop perturbation theory, in particular at the
coarsest lattice spacing. For $\beta=5.2$, the perturbative
value of $\cA=-0.0087$ is almost an order of magnitude smaller.

Of course this has a significant impact if simulation data
are now analyzed using the non--perturbative $\cA$ instead
of its 1--loop value.
Using e.g. data from \cite{Irving:2001vy}, we see that the effect
on the result for the pseudo--scalar decay constant at this lattice
spacing is as large as $10\%$.

\subsection{Uncertainties due to deviations from the\\[-0.5mm]
``constant physics'' condition}

We should check whether the volumes in our runs [C1]-[C3] are scaled
sufficiently precisely or if systematic errors need to
be added to the statistical ones on $\cA$ to cover possible
violations of the constant physics condition.
Table 1 shows that the bare PCAC mass has been
kept constant to within about 10\%. A renormalized quark
mass would differ by the multiplication with a $Z$-factor, which is
a slowly (namely logarithmically) varying function of $a$.
Considering the restricted range
of lattice spacings we cover, such factors can be safely ignored.
Run [C4] is done with a quark mass which is more
than twice as large as the one in run [C1], with otherwise identical
parameters. The barely significant difference in $\cA$ confirms that
the small deviations from the ``constant mass'' condition can be
neglected.

We also need to examine the impact of the uncertainty in $L$ 
due to our perturbative (or asymptotic)
scaling of the lattice spacing. It has been argued in
\sect{sect:two} that the difference to a proper non-perturbative
scaling is rather small. Also, estimating a possible change by
comparing 3--loop to 2--loop and non--perturbative scaling,
gives a deviation in $t(\beta)$ which is smaller than 10\% in
the whole range of \fig{pert_evol} and thus the
same maximum deviation applies to $L/a$. That this is negligible
can be seen from a comparison of run [C5] to the interpolating
formula (\ref{cainterpol}).
In run [C5],
$L/a$ is 20\% lower than the proper value, but $\cA$
does not differ significantly from the fit curve.
As will be discussed in the next section, comparison with
significantly smaller volumes at \emph{smaller} $\beta$ values
can lead to enhanced $\rmO(a^2)$ effects.

Finally, by run [C6] we verify that the dependence of $\cA$ on the
kinematic parameters disappears quickly when going to even larger
values of $\beta$. In this run we used gauge configurations from the
calculation of $Z_{\rm P}$
\cite{DellaMorte:2003jj}. Although those were produced at $m\!=\!0$,
$\theta\!=\!0.5$ and a much smaller volume\footnote
{If one would extend $t(\beta)$ to this value, this would indicate
a required lattice volume of $43^4$.}, the resulting $\cA$ is only
approximately two standard deviations away from our fit.

\section{Discussion}

For the $\rmO(a)$-improved action with non-perturbative $\csw$
\cite{Jansen:1998mx},
we have determined the improvement coefficient $\cA$ 
for $\beta\!\geq\!5.2$, which roughly corresponds to $a\!\leq\!0.1\fm$. 
The improvement condition was evaluated on a line of constant physics,
which is  necessary in the situation when $\rmO(a)$ ambiguities in the
improvement coefficients are not negligible. 

That this is indeed the case here can be seen from an additional run at
$\beta\!=\!5.42$. Decreasing $L/a$ from 16 to 12 at constant
$T/a$ gives $\cA=-0.0559(21)$, which is about 
25\% larger in magnitude than its
value on our line of constant physics. This can be understood
from the fact that the energy levels in the pseudo--scalar channel
increase if the volume is decreased. But in order
to safely exclude large $\rmO(a)$ ambiguities, improvement conditions
should only involve states with energy $E\ll a^{-1}$. On this 
requirement we had to compromise more than we would have liked to do
already in our LCP volume, where the maximum values for $Ea$ are about
$0.7$. With larger energies in smaller volume it is only to be
expected that the value of $\cA$ changes in trying to compensate
the larger $\rmO(a^2)$ effects.

Although one could have improved the situation, i.e. made $Ea$ smaller,
by going to somewhat larger values of $L$ (and $T$), this would
have made the numerical computation much more expensive. 

Large $\rmO(a^2)$ effects have indeed been found in
the $\nf\!=\!2$, $\rmO(a)$-improved theory \cite{Sommer:2003ne}
at $\beta\!=\!5.2$ and these may well
be related to the not so small $\rmO(a)$ ambiguity in $\cA$
that we just mentioned. 

This can only be investigated further by 
studying  the scaling violations
in quantities such as $F_{\pi}r_0$ after improvement and
renormalization.
The next step after successfully implementing non--perturbative
improvement
for the axial current is therefore its (also non--perturbative)
renormalization, which will be discussed in the next chapter.

Clearly, the method described here may also be useful to
compute $\cA$ in the three flavor case, where $\csw$ is known
non--perturbatively with plaquette and Iwasaki gauge actions 
\cite{Aoki:2002vh,Yamada:2004ja,Ishikawa:2003ri}.

\chapter{Axial current renormalization}
\label{chap:renorm}
After the successful non--perturbative improvement of the
axial current, we can now turn to the problem
of its normalization. In the quenched case \cite{Luscher:1996jn},
a non--perturbative renormalization of the isovector axial
current was achieved by requiring certain continuum chiral
Ward identities to hold at finite lattice spacing.
These relate matrix elements of the axial and vector currents
and thus link the normalization of the two.
The local vector current is normalized by enforcing isospin
symmetry, again in the form of an integrated Ward identity.

In fact, the situation is similar to the calculation
of $\cA$. Also here the results (i.e. $\za$ and $\zv$) are
ambiguous due to cutoff effects and in order to obtain
them
as smooth functions of $g_0^2$ we need to evaluate our
normalization conditions on a line of constant physics,
keeping all length scales fixed.
As discussed in \sect{currentnorm}, the normalization
conditions have to be set up at vanishing quark mass since
we want to implement a mass--independent renormalization
scheme.

While this is in principle possible in the
Schr\"odinger functional, we have seen in \chap{chap:algo}
that cutoff effects make simulations at small quark
masses and coarse lattice spacings difficult.
Although this problem is addressed efficiently through
the use of the PHMC algorithm, it is still desirable
to formulate the normalization conditions such that
they have only very little dependence on the quark mass.
This is the point where we improve the method employed
in the quenched case
\cite{Luscher:1996jn} by deriving a normalization condition
from the axial Ward identity including the mass term.
This new normalization condition naturally reduces to
the previously used one
in the chiral limit, but numerical results can now be more
easily extrapolated in the quark mass.

After deriving the normalization conditions for the isovector
currents, the advantage of using the new condition is
demonstrated through its chiral extrapolation
in a quenched example. Then the results
for two dynamical flavors are presented and summarized
in interpolating formulae. We also discuss systematic
errors due to deviations from the line of constant physics
and other variations of the normalization condition.

\section{Continuum Ward identities}

\label{int_ward_later}

In \sect{contward} we considered Ward identities derived from
the flavor chiral symmetry of the continuum QCD Lagrangian.
Here we will generalize these to matrix elements, where
(part of) the operator insertion has support in the region,
where the field transformations are performed. As a result
eqs.~(\ref{cvc}) and (\ref{pcac}) will be modified by
terms containing the variation of this ''internal'' operator.

\subsection{VWI}

Again we consider only the case of two degenerate quarks flavors
and pick a region $\R$ with a smooth boundary $\partial \R$.
Suppose ${\cal O}_{\rm int}$ and ${\cal O}_{\rm ext}$
are polynomials in the basic fields localized
in the interior and exterior of this region, respectively. If we
perform a field transformation with support in $\R$, the variation
of the external operator $\op_{\rm ext}$ vanishes and for the case
of an isovector vector transformation
it follows that
\begin{eqnarray*}
\langle(\dv \op_{\rm int})\op_{\rm ext}\rangle&=&\langle\dv \op_{\rm int}
\op_{\rm ext}\rangle\use{deltaOS}
\langle(\dv S)\op_{\rm int}\op_{\rm ext}\rangle\\
&\use{dVS}&\int_\R\!\!\textstyle\rmd^4x\, \omega^a(x)\,\Big\langle-
\partial_\mu V_\mu^a(x)\op_{\rm int}
\op_{\rm ext}\Big\rangle\;,
\end{eqnarray*}
where $\dv \op_{\rm int}=\omega^a(y)\,\dv^a\op_{\rm int}$
if $y$ denotes the point, where $\op$ lives. We now take $\omega^a$
to be a constant in $\R$. Using Gauss' law the integrated vector current
Ward identity (VWI) is obtained
\begin{eqnarray}
\Big\langle(\dv^a \op_{\rm int})\op_{\rm ext}\Big\rangle&=&-
\int_{\partial \R}\!\!
\textstyle\rmd\sigma_\mu(x)\,
\Big\langle V_\mu^a(x)\op_{\rm int}\op_{\rm ext}\Big\rangle\;.\label{vwa}
\end{eqnarray}

\subsection{AWI}

The same construction is also valid for the axial current variation except
that the mass term in (\ref{dAS}) cannot be written as a
surface integral\footnote
{Due to additive mass renormalization, it is the current quark mass,
e.g. (\ref{massimp}), rather than the bare quark mass, which appears in
the integrated Ward identity.}
\begin{eqnarray}
\Big\langle(\da^a \op_{\rm int})\op_{\rm ext}\Big\rangle&\use{dAS}&-
\int_{\partial \R}\!\!
\textstyle\rmd\sigma_\mu(x)\,
\Big\langle A_\mu^a(x)\op_{\rm int}\op_{\rm ext}\Big\rangle\nonumber\\
&&\qquad\qquad+\displaystyle2m\int_{\R}\!\!\textstyle\rmd^4x\,
\Big\langle P^a(x)\op_{\rm int}\op_{\rm ext}\Big\rangle\;.\label{awa}
\end{eqnarray}
The axial Ward identity (AWI, \ref{awa}) will later be used for massless quarks
with $\op_{\rm int}$ equal to the axial current component $A_\nu^b$ at
some point $y\!\in\!\R$. for this case we have
\begin{eqnarray}
\int_{\partial \R}\!\!\textstyle\rmd\sigma_\mu(x)\,\Big\langle
A_\mu^a(x)A_\nu^b(y)\op_{\rm ext}\Big\rangle
&=&-\Big\langle\Big(\da^a A_\nu^b(y)\Big)\op_{\rm ext}\Big\rangle\nonumber\\
&\use{dA}&i\epsilon^{abc}\Big\langle V_\nu^c(y)\op_{\rm ext}
\Big\rangle\;.\label{2.9}
\end{eqnarray}

\subsection{Euclidean proof of the Goldstone theorem}

As an immediate application of the integrated Ward identities
with operator insertions
we will now show that a non--zero \emph{chiral condensate}
$\ev{\psibar\psi}$ leads to
long--ranged current--density correlation functions
\cite{Luscher:1998pe}, which in
turn implies a massless propagating particle. This is
the basis for the conjectured mechanism of spontaneous
chiral symmetry breaking.

We know from (\ref{pcac}) that for all $x\neq0$ and zero quark mass
the correlation function
$\ev{\partial_\mu A_\mu^a(x)P^a(0)}$ vanishes. From Lorentz invariance
one can then conclude that the integrated version is of the form
\be
\ev{A_\mu^a(x)P^a(0)}=k\frac{x_\mu}{(x^2)^2}\;.\label{euc1}
\ee
The constant $k$ can be fixed by applying (\ref{awa}) with
a sphere of radius $r$ around the origin as the region $\R$,
$m=0$, $\op_{\rm int}
=P^a(0)$ and $\op_{\rm ext}=1$, which gives
\be
\int_{|x|=r}\!\!\!\rmd\sigma_\mu(x)\ev{A_\mu^a(x)P^a(0)}=
-\ev{\da^aP^a(0)}=-{\ts\frac32}\ev{\psibar\psi}\;.\label{euc2}
\ee
In the last step the definition (\ref{varA}) has been applied to
the pseudo--scalar density (\ref{density}).
Since the integrand's divergence is zero everywhere except
at the origin, in an application of Gauss's law the entire contribution
would come from the contact term at the origin. However, we
can simply plug (\ref{euc1}) into (\ref{euc2}) and solve the trivial
integral directly. The result is the surface of the $4$--sphere
times $k$ such that
\be
-{\ts\frac32}\ev{\psibar\psi}=k2\pi^2
\ee
and we finally arrive at
\be
\ev{A_\mu^a(x)P^a(0)}=-\frac3{4\pi^2}\ev{\psibar\psi}\frac{x_\mu}
{(x^2)^2}\;.\label{euc3}
\ee
Thus, if the chiral condensate $\ev{\psibar\psi}$ is non--zero,
the correlation function on the left--hand side of (\ref{euc3}) has
\emph{no} exponential
decay and therefore the energy spectrum of the theory has no
gap. The massless particles propagating in the correlation function
(\ref{euc3}) are the Goldstone bosons of the broken chiral symmetry.
\section{Normalization conditions}

The normalization conditions for the isovector currents are
obtained by deriving an identity between matrix elements
in the continuum from the isovector symmetries of the
QCD action. The renormalized improved currents on the lattice
are then required to satisfy these identities at
finite lattice spacing. This results in conditions
for the renormalization factors $\za$ and $\zv$,
cf. eqs.~(\ref{ar}, \ref{vr}).

In a massless renormalization scheme these conditions
have to be set up a vanishing quark mass. If numerical
simulations at the critical point are not possible,
the normalization factors become the result of a chiral
extrapolation.

We improve the methods employed for the calculation of $\za$
in the quenched theory \cite{Luscher:1996jn} by including
the mass term in the integrated Ward identity (\ref{awa}).
Of course, one still has to extrapolate to the chiral
limit, but we will show that in practice the mass dependence
of this normalization condition is extremely small.

\subsection{The vector current}

Our starting point is the continuum vector Ward identity
(\ref{vwa}) and we take the region $\R$ to be the entire
Euclidean space for times smaller than some positive
$x_0$. As internal
operator we use the
pseudo--scalar density at time zero and as external operator the
pseudo--scalar density at
time $T>x_0$. The surface integral then becomes a spatial
integral over the time component of the vector current. Applying
(\ref{varV}) to (\ref{density}) gives
\begin{equation}
\delta_V^bP^c(x)=-i\epsilon^{bcd}P^d(x)\;.\label{isocharge}
\end{equation}
In this setup the vector Ward identity becomes
\begin{eqnarray*}
\Big\langle(\dv^b P^c(0,\bu))P^a(T,\bv)\Big\rangle&=&-
\int\!\!\rmd^3\bx
\,
\Big\langle V_0^b(x_0,\bx)P^c(0,\bu)P^a(T,\bv)\Big\rangle\nonumber\\
 i\epsilon^{bcd}\Big\langle P^d(0,\bu)P^a(T,\bv)\Big\rangle&=&
\int\!\!\rmd^3\bx
\,
\Big\langle P^a(T,\bv)V_0^b(x_0,\bx)P^c(0,\bu)\Big\rangle.
\end{eqnarray*}
The physical interpretation of this relation is that the
isospin charge $\int\!\!\rmd^3\bx\hspace*{-0.1mm}V_0^b(x)$
generates an infinitesimal
isospin rotation of the state created by $P^c$, which transforms
according to the vector representation of the exact isospin symmetry.

If the open isospin indices on the right--hand side are contracted
in a totally antisymmetric way with $i\epsilon^{abc}$, the result
is
\begin{eqnarray}
-2\Big\langle P^a(T,\bv)P^a(0,\bu)\Big\rangle&=&
i\epsilon^{abc}\int\!\!\rmd^3\bx
\,
\Big\langle P^a(T,\bv)V_0^b(x_0,\bx)P^c(0,\bu)\Big\rangle\;.\qquad\quad
\label{vec_cont}
\end{eqnarray}
On the lattice we now construct these matrix elements in the
framework of the Schr\"odinger functional.
We use the boundary field products (\ref{O}) and (\ref{Oprime})
to create initial and final states that transform according
to the vector representation of the exact isospin symmetry
and insert the renormalized improved vector current (\ref{vr}).
We demand that the correlation function
\begin{equation}
  \fvr(x_0)={a^3\over6L^6}\sum_\bx i\epsilon^{abc}
    \langle\oprime^a(\vr)_0^b(x)\op^c\rangle\;,\label{fvr}
\end{equation}
is equal to
\begin{equation}
f_1=-\frac1{3L^6}\langle\oprime^a\op^a\rangle\label{f1}\;,
\end{equation}
which corresponds to (\ref{vec_cont}) when $\bu$ and $\bv$ are summed over
all of space. In the improved theory it defines the renormalized vector
current up to an $\rmO(a^2)$ uncertainty
\begin{equation}
\fvr(x_0)=f_1+\rmO(a^2)\label{3.11a}\;,
\end{equation}
since both correlation functions approach their (common)
continuum limit with this rate.
Note that we do not need to include the renormalization factors for
the (multiplicatively renormalizable) boundary quark fields here
because they appear on both sides and thus cancel.
With the obvious definition of $\fvi(x_0)$
\be
\fvi(x_0)={a^3\over6L^6}\sum_\bx i\epsilon^{abc}
\langle\oprime^a(\vi)_0^b(x)\op^c\rangle\;,\label{fvi}
\ee
equation (\ref{3.11a}) then
gives
\begin{eqnarray}
\zv(1+b_{\rm V} a\mq)\fvi(x_0)&\!\!\use{vr}\!\!&f_1+\rmO(a^2)\;.\label{fv}
\end{eqnarray}
One can easily evaluate the contribution of the O($a$) counterterm
appearing in the definition
(\ref{VI}) of the improved vector current
to the correlation function
$\fvi(x_0)$.
If we introduce the correlation function for the bare unimproved current,
\be
\fv(x_0)={a^3\over6L^6}\sum_\bx i\epsilon^{abc}
\langle\oprime^aV_0^b(x)\op^c\rangle\;,\label{fvdef}
\ee
it follows that
\bea
\fvi(x_0)-\fv(x_0)&\stackrel{(\ref{VI})}\propto&\sum_\bx\epsilon^{abc}\Big\langle
\oprime^a\drvtilde\nu{} T_{0\nu}^b(x)\op^c\Big\rangle\nonumber\\
&\stackrel{(\ref{Tmunu})}\propto&\sum_\bx\epsilon^{abc}\Big\langle
\oprime^a\drvtilde\nu{} [\psibar(x)\sigma_{0\nu}\tau^b\psi(x)]\op^c\Big\rangle\;.
\eea
Since $\sigma_{0\nu}=\frac i2[\dirac0,\dirac\nu]$ is antisymmetric
in its Dirac indices, we have $\sigma_{00}=0$. The contribution of the
improvement term to the correlation
function will therefore contain the expression
\be
\sum_\bx\drvtilde k{}[\psibar(x)\sigma_{0k}\tau^b\psi(x)]\;,
\ee
which vanishes identically if we impose periodic boundary conditions in
the spatial directions. We can therefore conclude that
$\fvi(x_0)=\fv(x_0)$ and from equation (\ref{fv}) it follows
that
\be
\zv(\tilde g_0^2)(1+b_{\rm V} a\mq)\fv(x_0)=f_1+\rmO(a^2)\;.\label{calculateZV}
\ee
By evaluating the correlation functions $f_1$ and $\fv(x_0)$
through numerical simulation
one is thus able to compute the normalization factor
$\zv(1+b_{\rm V} a\mq)$.
In particular, to calculate $\zv$ it suffices to consider the
theory at vanishing quark mass. Equation (\ref{calculateZV}) also implies
that $\fv(x_0)$ is independent of $x_0$ up to cutoff effects.
In a numerical evaluation one can therefore increase the statistical
accuracy by averaging over a range of $x_0$ values. More details
will be given in \sect{numcomp}.

\subsubsection{Derivation in the operator formalism}

\label{veccurrent}

The vector current normalization condition can also
be derived in the operator formalism
of the Schr\"odinger functional with boundary states $S$ and $S'$.
The fact that the charge is conserved implies that the corresponding
charge operator,
denoted by $\hat Q^b$, commutes with the transfer matrix $\mathbb T$.
The correlation function (\ref{fvr}) now reads
\begin{eqnarray*}
\fvr(x_0)
&=&{1\over6L^6}i\epsilon^{abc}
\langle S'|\hat\oprime^a{\mathbb{T}^{(T-x_0)}}\hat Q^b
\mathbb T^{x_0}\hat\op^c|S\rangle\\
&=&{1\over6L^6}i\epsilon^{abc}
\langle S'|\hat\oprime^a{\mathbb{T}^T}\hat Q^b\hat\op^c|S\rangle\;.
\end{eqnarray*}
Since $\hat Q^b|S\rangle=0$ we can replace $\hat Q^b\hat\op^c$ by
its commutator
and since the boundary fields are isospin vectors we have
$[\hat Q^b,\hat\op^c]=-i\epsilon^{bcd}\hat\op^d$ in the continuum.
The physical interpretation is that the isovector charge generates
an infinitesimal isospin rotation as in (\ref{isocharge}).
In an $\rmO(a)$ improved lattice theory this argumentation holds
up to $\rmO(a^2)$ and therefore
\begin{eqnarray}
  \fvr(x_0)&=&-{1\over6L^6}\underbrace{\epsilon^{abc}
  \epsilon^{bcd}}_{=2\delta^{ad}}
  \langle S'|\hat\oprime^a{\mathbb{T}^T}\hat\op^d|S\rangle+\rmO(a^2)
  \nonumber\\\label{3.11}
  &=&-\frac1{3L^6}\langle\oprime^a\op^a\rangle+\rmO(a^2)\;,
\end{eqnarray}
where in the last line we have switched back to the path integral form
to recover the result (\ref{3.11a}).

\subsection{The axial current}

\label{axnorm}

We will start with a derivation of the normalization
condition used in \cite{Luscher:1996jn}, where the
quark is set to zero in the beginning. This will later
be generalized to $m\neq0$.

\subsubsection{Massless case}

For vanishing quark mass our starting point is the Ward identity
(\ref{2.9}). Contracting the isospin indices gives
\begin{eqnarray}
\int_{\partial \R}\!\!\textstyle\rmd\sigma_\mu(x)\,
\epsilon^{abc}\Big\langle A_\mu^a(x)A_\nu^b(y)\op_{\rm ext}
\Big\rangle&\use{2.9}&i\underbrace{\epsilon^{abc}
\epsilon^{abd}}_{2\delta^{cd}}\Big\langle V_\nu^d(y)\op_{\rm ext}
\Big\rangle\nonumber\\
&=&2i\Big\langle V_\nu^c(y)\op_{\rm ext}\Big\rangle\label{3.14}\;.
\end{eqnarray}
This relation is now transcribed to the $\rmO(a)$ improved lattice
theory with $\R$ being the region between the hyper-planes at
$x_0\!=\!y_0\pm t$ and $\nu\!=\!0$.
With periodic boundary conditions in the spatial directions the surface
integration in (\ref{3.14}) results in the difference between the axial
charge at times $y_0\pm t$, i.e.
\begin{eqnarray}
a^3\sum_\bx
\epsilon^{abc}
\Big\langle
\Big[(\ar)^a_{0}(y_0\!+\!t,\bx)-(\ar)^a_{0}(y_0\!-\!t,\bx)\Big]
(\ar)^b_{0}(y) {\cal O}_{\rm ext}
\Big\rangle\nonumber\\
\quad\quad =2i
\Big\langle
(\vr)^c_{0}(y){\cal O}_{\rm ext}
\Big\rangle
+\rmO(a^2)\;.\label{3.15}
\end{eqnarray}
On-shell improvement is effective in (\ref{3.15}) since the fields in the
correlation functions are localized at non-zero distances from each other.
Equation (\ref{3.15}) is summed over $\bf y$ to obtain the axial charge.
In the form of (\ref{3.15}) the integrated Ward identity requires three
different time--slices, where the axial charge is inserted.

Using the conservation of the (renormalized) axial current the two
insertions at $y_0\mi t$ and $y_0$, associated with the lower
surface of $\R$, can (simultaneously) be shifted
to $y_0$ and $y_0\pl t$.
We thus arrive at
\begin{eqnarray}
\!\!\!\!\!
a^6\sum_{\bf x,\bf y}
\epsilon^{abc}
\Big\langle
(\ar)^a_{0}(x)(\ar)^b_{0}(y)
{\cal O}_{\rm ext}
\Big\rangle\!\!&\!\!=\!\!&\!\!a^3\sum_\by i\Big\langle
(\vr)^c_{0}(y){\cal O}_{\rm ext}
\Big\rangle
+\rmO(a^2)\;,\qquad\label{3.16}
\end{eqnarray}
where $x_0=y_0\pl t$. Since $t$ was arbitrary, the above equation holds for all
insertion points $x_0$ and $y_0$ such that $0<y_0<x_0<T$.

At this point we have a relation that would allow us to calculate the
axial current renormalization from the vector current renormalization.
We can make things even simpler by choosing
the field product $\op_{\rm ext}$ such that the correlation
function $\fvr(y_0)$ appears on the right--hand side of equation (\ref{3.16}).
This amounts to setting
\begin{equation}
\op_{\rm ext}=-\frac1{6L^6}\epsilon^{cde}\oprime^d\op^e\;,\label{oext}
\end{equation}
since with this definition equation (\ref{3.16}) becomes
\pagebreak
\begin{eqnarray}
&&\hspace*{-2cm}-\frac{a^6}{6L^6}\sum_{\bf x,\bf y}
\epsilon^{abc}\epsilon^{cde}
\Big\langle\oprime^d
(\ar)^a_{0}(x)(\ar)^b_{0}(y)
\op^e
\Big\rangle\label{fAAR}\\
&\use{3.16}&-\frac{a^3}{6L^6}\sum_\by i\epsilon^{cde}\Big\langle\oprime^d
(\vr)^c_{0}(y){\cal O}^e
\Big\rangle
+\rmO(a^2)\nonumber\\
&=&\frac{a^3}{6L^6}\sum_\by i\epsilon^{abc}\Big\langle\oprime^a
(\vr)^b_{0}(y){\cal O}^c
\Big\rangle
+\rmO(a^2)\nonumber\\
&\use{fvr}&\fvr(y_0)+\rmO(a^2)\label{calc}\;.
\end{eqnarray}
The normalization condition for the vector current (\ref{3.11a}) allows us to
replace $\fvr(y_0)$ by $f_1$. If we define the unrenormalized version of the
correlator (\ref{fAAR}) as
\begin{equation}
\funci{AA}(x_0,y_0)=-\frac{a^6}{6L^6}\sum_{\bf x,\bf y}
\epsilon^{abc}\epsilon^{cde}
\Big\langle\oprime^d
(\aimpr)^a_{0}(x)(\aimpr)^b_{0}(y)
\op^e
\Big\rangle\;,\label{fAAI}
\end{equation}
one can conclude from (\ref{calc}) that
\begin{equation}
\za^2\funci{AA}(y_0\!+\!t,y_0)=f_1+\rmO(a^2)\label{3.18}
\end{equation}
for all times $t>0$ such that $0<y_0$ and $y_0+t<T$.
As in (\ref{3.11a}) the normalization of the boundary quark fields cancel.
The axial current
normalization constant $\za$ can thus be determined by computing the two
correlation functions $f_1$ and $\funci{AA}(y_0\!+\!t,y_0)$ at vanishing
quark mass.

Numerical simulations (see \sect{quenchedexample}) show that
the above relation has a very pronounced
mass dependence. Since the quark mass was neglected from the very beginning
in its derivation, this should not come as a surprise.
In the following we will derive
a normalization condition from the full PCAC relation. 

\subsubsection{Non-vanishing PCAC mass}

Since we use a mass--independent renormalization scheme the normalization
condition for the axial current can not be set up at finite quark mass.
Instead our goal is to derive a normalization condition, which has a smaller
mass dependence than (\ref{3.18}), such that in practice the chiral extrapolation
is easier. To this end we now use the axial Ward
identity (\ref{awa}) with the same operator
$\op_{\rm int}\!=\!A_\nu^b(y)$, where again the point $y$ is somewhere
in the interior of $\R$. After anti--symmetrically contracting the isospin
indices the result for an arbitrary mass $m$ reads
\begin{eqnarray}
\int_{\partial \R}\!\!\!\textstyle\rmd\sigma_\mu(x)\,\epsilon^{abc}\Big\langle
A_\mu^a(x)A_\nu^b(y)\op_{\rm ext}\Big\rangle\qquad\qquad\nonumber\\\label{firstcontact}
\qquad\qquad-2m\int_\R\!\!\rmd^4x\, \epsilon^{abc}\Big\langle P^a(x)A_\nu^b(y)
\op_{\rm ext}\Big\rangle&\!\!=\!\!&2i\Big\langle V_\nu^c(y)\op_{\rm ext}\Big\rangle\;.
\end{eqnarray}
Already here we note that the volume integral contains a contact--term, i.e.
a contribution from the composite operator $\epsilon^{abc}P^a(x)A_\nu^b(y)$
as $x$ approaches the (fixed) $y$. The operator product expansion
implies that the leading contribution in the $x\!\rightarrow\!y$ limit
comes from $V_\nu^c(x)$. From power counting it then follows that
the composite operator cannot diverge faster
than $|x\mi y|^{-3}$, such that we receive a finite contribution under the
four--dimensional integral over $\R$.

We choose the same region
$\R$ and
set $\nu\!=\!0$ as in the massless case.
Together with an additional spatial integration over
$\by$ the result is
\begin{eqnarray}
&&\!\!\!\!\int\!\!\rmd^3\by\int\!\rmd^3\bx\, \epsilon^{abc}
\left[\Big\langle A_0^a(y_0\!+\!t,\bx)A_0^b(y_0,\by)\op_{\rm ext}\Big\rangle
-\Big\langle A_0^a(y_0\!-\!t,\bx)A_0^b(y_0,\by)\op_{\rm ext}\Big\rangle\right]\nonumber\\\nonumber
&&\quad-2m\int\!\!\rmd^3\by\int\!\rmd^3\bx\int_{y_0-t}^{y_0+t}\!\rmd x_0\,
\epsilon^{abc}\Big\langle P^a(x_0,\bx)A_0^b(y_0,\by)\op_{\rm ext}\Big\rangle\\
&&\qquad\qquad=2i\displaystyle\int\!\!\rmd^3\by\
\Big\langle V_0^c(y_0,\by)\op_{\rm ext}\Big\rangle\;.\label{massive}
\end{eqnarray}
In the massless case the two contributions from the surface integration
were the same due to current conservation and the anti--symmetric
isospin structure.
Here we have to use the partial conservation of the axial
current to transform the second term of the surface integral.
This will cancel the lower part ($y_0\mi t$ to $y_0$) of
the volume integral. The detailed calculation is given in
\app{app:shift} and the result is
\begin{eqnarray}
&&\nonumber
i\displaystyle\int\!\!\rmd^3\by\
\Big\langle V_0^c(y_0,\by)\op_{\rm ext}\Big\rangle=\\&&
\quad\int\!\!\rmd^3\by\int\!\!\rmd^3\bx\, \epsilon^{abc}
\Big\langle A_0^a(y_0\!+\!t,\bx)A_0^b(y_0,\by)\op_{\rm ext}\Big\rangle\nonumber\\
&&\quad\qquad
-2m\int\!\!\rmd^3\by\int\!\!\rmd^3\bx
\int_{y_0}^{y_0+t}\!\!\!\rmd x_0\, \epsilon^{abc}
\Big\langle P^a(x_0,\bx)A_0^b(y_0,\by)\op_{\rm ext}\Big\rangle\;.
\label{aftershift2}
\end{eqnarray}
As before a normalization condition for the axial current on the
lattice is obtained by demanding that eq.~(\ref{aftershift2}) in
terms of the renormalized currents holds at non--zero lattice
spacing.
Inserting the same
external operator (\ref{oext}) as before will again allow us
to replace the matrix--element of the vector current with
the correlation function $f_1$.
As a result of the contact term $\rmO(a)$ improvement fails in the correlator
multiplying the mass term. On the lattice we therefore expect corrections of
order $am$ at finite mass in addition to the overall $\rmO(a^2)$ 
uncertainty and thus have
\begin{eqnarray}
\nonumber&&\hspace*{-8mm}-\frac{a^6}{6L^6}\sum_{\bx,\by}
\epsilon^{abc}\epsilon^{cde}\Big\langle\oprime^d(\ar)^a_{0}(y_0\!+\!t,\bx)
(\ar)^b_{0}(y_0,\by)\op^e\Big\rangle\\
&&\hspace*{-8mm}+\frac{2m_{\rm R}a^7}{6L^6}\!\! \sum_{x_0=y_0}^{y_0+t}w(x_0)
\sum_{\bx,\by}\epsilon^{abc}\epsilon^{cde}\Big\langle\oprime^dP_{\rm R}^a(x)
(\ar)^b_{0}(y)\op^e\Big\rangle\nonumber\\[3mm]
&&\qquad=f_1+\rmO(a^2)+\rmO(am)\;.\qquad\label{massive2}
\end{eqnarray}
The weight factor $w(x_0)$ is needed to implement the trapezoidal
rule in the discretization of the time integral in (\ref{aftershift2}),
i.e. that the boundary terms $x_0=y_0$ and $y_0+t$ should only
contribute with the weight $1/2$.
It is given by
\be
w(x_0)=\left\{
\begin{array}{r@{,\quad}l}
1/2 & x_0\in\{y_0,y_0\pl t\}\\
1	& \textrm{otherwise}\;.
\end{array}\right.
\ee
If we set $m=0$ in (\ref{massive2}) we immediately recover the
normalization condition used in \cite{Luscher:1996jn}, i.e.
(\ref{calc}).
With the definitions (\ref{fAAI}) and (\ref{ar}) the normalization
condition can be written as
\begin{eqnarray*}
\za^2(1\!+\!\ba a\mq)^2\funci{AA}(y_0\!+\!t,y_0)\qquad\qquad\qquad\qquad\qquad&&\\
+\frac{2m_{\rm R}a^7}{6L^6}\!\! \sum_{x_0=y_0}^{y_0+t}
\!w(x_0)\!
\sum_{\bx,\by}\epsilon^{abc}
\epsilon^{cde}\Big\langle\oprime^dP_{\rm R}^a(x)
(\ar)^b_{0}(y)\op^e\Big\rangle\!\!&\!=\!&\!\!f_1+\rmO(a^2)+\rmO(am)\;.
\end{eqnarray*}
Since there is no improvement term for the pseudo-scalar density
we now define the new correlation function
\be
\tfunci{PA}(y_0\!+\!t,y_0)\,=\,-\frac{a^7}{6L^6}\! \sum_{x_0=y_0}^{y_0+t}
\!w(x_0)\!
\sum_{\bx,\by}\epsilon^{abc}\epsilon^{cde}\Big\langle
\oprime^dP^a(x)(\aimpr)_0^b(y)\op^e\Big\rangle\label{ftilde}\;.
\ee
From the PCAC relation we conclude that the product $mP$ renormalizes
with the same factor as $A_\mu$. Thus, the final form for the
normalization condition derived from the Ward identity with mass
term is given by
\begin{eqnarray}
\za^2(1\!+\!\ba a\mq)^2\Big(\funci{AA}(y_0\!+\!t,y_0)\nonumber
-2m\tfunci{PA}(y_0\!+\!t,y_0)\Big)=\qquad\qquad\\
f_1+\rmO(am)+\rmO(a^2)\;.\label{normcon}
\end{eqnarray}

\section{Numerical computation}

\label{numcomp}

Before going into the details of the simulations, we need to
specify our choice of kinematic parameters in the numerical
evaluation of the conditions (\ref{calculateZV}) and (\ref{normcon}).

The spatial boundary conditions are strictly periodic ($\theta_k=0$)
and to accommodate the two insertion points in (\ref{normcon}) in
a symmetric way, we choose a $T=9/4L$ geometry. The background
field ($C_k$ and $C_k'$) is set to zero and the improvement
coefficients have the values specified in Sections~\ref{sec:wilson}
and \ref{sect:latticeSF}. The axial current improvement coefficient
$\cA$ is given by (\ref{cainterpol}).

As first discussed in \cite{Luscher:1996jn} and further
detailed in \cite{Guagnelli:2000jw},
we need to evaluate the normalization conditions
on a line of constant physics, keeping all
length scales fixed.
This ensures
that the $\rmO(a^2)$ ambiguities in the normalization
factors vanish smoothly when the perturbative regime is
approached.
In addition, the normalization conditions have to be
set up at zero quark mass since we are implementing
a mass--independent renormalization scheme
as detailed in \sect{currentnorm}.

To keep
the volume constant, we again employ the approximate scale
setting given by $t(\beta)$, cf. \eq{e:L3l}. We point out that
in the present
case a deviation from the line of constant physics
can change the result by $\rmO(a^2)$ only, whereas in the
computation of $\cA$ such deviations would show up at $\rmO(a)$.
In addition to
three lattice resolutions matched in this way, we simulated at
three larger values of $\beta$ and fixed $L/a=8$, which results
in very small volumes.
This was done in
order to verify that our non--perturbative estimate
smoothly connects to the perturbative predictions,
eqs.~(\ref{pta}) and (\ref{ptv}).

Our ''reference'' volume is an $8^3\times18$ lattice at
$\beta=5.2$, which corresponds to the coarsest lattice
spacing. At all other lattice spacings the systematic error
in $\za$ and $\zv$ due to a mismatch in the volume is estimated
by varying the lattice resolution $L/a$.
To check for smoothness of $\za(g_0^2)$ additional simulations
were performed in an unmatched $8^3\times18$ volume
at $\beta=5.29$ without an estimate of the systematic
error.

These last simulations as well as all but the heaviest run
at $\beta=5.2$ were done with the PHMC algorithm.
All others employ the HMC with two pseudo--fermions
as discussed in \sect{sect:hmc}. For all runs it could
be verified that the problems described in
Sections \ref{instab} and \ref{sect:sampling}
are absent.

For reference we collect again the precise non--perturbative
definitions of $\zv$ and $\za$.
In the simulations both the PCAC mass and the estimate
for $\zv$ are averaged over a few time--slices in the
middle of the lattice. No such time average is performed
for the correlation
functions $\funci{AA}$ and $\tfunci{PA}$.
\bea
\za(g_0^2)&\!\!\!=\!\!\!&\lim_{m\rightarrow 0}\sqrt\frac{f_1}
{\funci{AA}(2T\!/3,T\!/3)-2m\tfunci{PA}(2T\!/3,T\!/3)}\;,\label{za_impl}\\\label{zv_impl}
\zv(g_0^2)&\!\!\!=\!\!\!&\lim_{m\rightarrow 0}\frac1{N_t}\sum_{x_0=t_1}^{t_2}\!\frac{f_1}{\fv(x_0)}\;,\\
m&\!\!\!=\!\!\!&\frac1{N_t}\sum_{x_0=t_1}^{t_2}\!\frac{\tilde\partial_0
\Big[\fa(x_0)\pl\ga(T\mi x_0)\Big]\pl a\cA\partial_0\partial_0^\star
\Big[\fp(x_0)\pl\gp(T\mi x_0)\Big]}{2\Big[\fp(x_0)\pl\gp(T\mi x_0)\Big]}\;,\qquad\quad\label{m_av}
\eea
\be
\textrm{where}\ \ \left\{
\begin{array}{rl}
t_1=\frac {T\mi2}2,\ t_2=\frac {T\pl2}2,\ N_t=3 &  \textrm{ for } T \textrm{ even.}\\
t_1=\frac {T\mi3}2,\ t_2=\frac {T\pl3}2,\ N_t=4 &  \textrm{ for } T \textrm{ odd.}
\end{array}\right.
\ee

\subsection{Implementation notes and quenched example}

\label{quenchedexample}

The starting point in the implementation of the correlation functions
for the axial normalization condition were the APEmille TAO codes
of the ALPHA collaboration. These already include the correlation
functions $\fa$, $\fp$, $\ga$ and $\gp$.

The correlation functions $\fv$, $\funci{AA}$ and $\tfunci{PA}$ have
been implemented in the quenched
code (which uses SSOR preconditioning \cite{Guagnelli:1999nt}) as well as the
even--odd preconditioned HMC and PHMC
codes. An additional version (derived from the quenched one) exists,
which measures the correlation functions on a saved gauge
configuration.
A detailed discussion of all correlation functions and in particular
of $\funci{AA}$ and $\tfunci{PA}$ and their Wick contractions is given
in \app{app:corr}.

The code for $\funci{AA}$ as well as the analysis program
have been checked against those used in \cite{Luscher:1996jn}.
Agreement was found both statistically (i.e. at a given set
of bare parameters) as well as on a given gauge configuration.

As a first test of the new axial current normalization condition,
we consider sets of quenched $8^3\!\times\!18$
gauge configurations at $\beta=8.0$ and various values of the
hopping parameter $\kappa$. The lattice spacing and hence the volume
are extremely small\footnote
{In the quenched approximation $\beta=6.0$ implies $a\simeq0.1\fm$
and hence corresponds to $\beta=5.2$ in the two--flavor case.}
such that the fluctuations of the gauge field are strongly
suppressed and one can (in the Schr\"odinger functional)
easily simulate at the critical point.

The results are shown in \fig{quenched},
\EPSFIGURE[t]{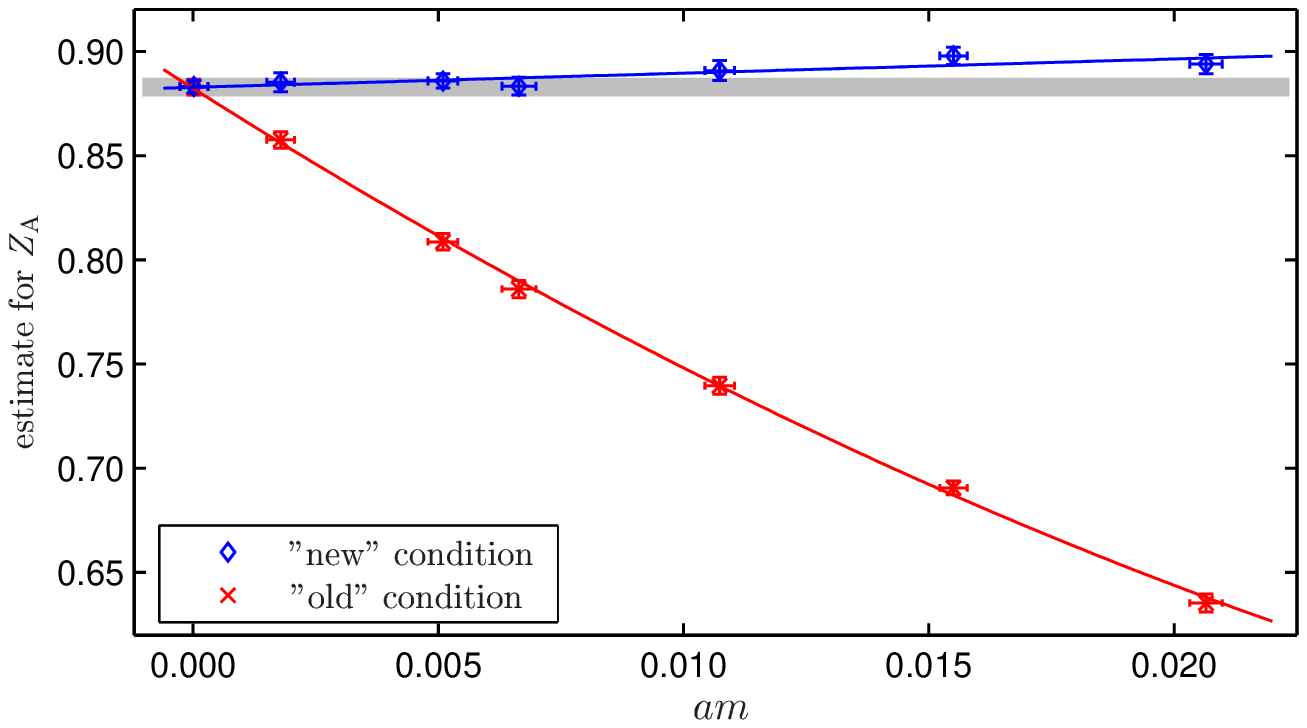,width=13cm}{
Mass dependence of the old and new normalization conditions
for the axial current evaluated in the quenched
approximation. The parameters are $\beta=8.0$,
volume = $8^3\!\times\!18$
and $\kappa\in[0.1325,0.133168]$.
\label{quenched}}
where one can see that the lightest mass (with
$\kappa=0.133168$) is exactly at
the critical point. We show the estimate for $\za$
defined in (\ref{za_impl}) as well as the one from
the old condition \cite{Luscher:1996jn}, where the
term $m\tfunci{PA}$ is neglected. As expected from
the argumentation in the previous section, the new
condition has a significantly smaller mass dependence.
At these parameter values using the new condition
reduces the linear coefficient of a fit in $am$
from $-14.9(8)$ to $0.7(2)$, which is consistent
with our $\rmO(am)$ expectation. With a value of
$0.97(2)$ the slope in the estimate of $\zv$ from
(\ref{zv_impl}), which is not shown in the plot,
is similar to that of $\za$ with the
new condition.

Due to the large slope, an extrapolation of results
from the old normalization condition of the axial current
should take into account the error in the quark mass
as well as the correlation between $am$ and the $\za$
estimate. In contrast, with the small slope from
the new condition the extra-- or interpolation is
essentially flat and uncertainties in the location
of the critical point do not propagate to the
determination of $\za$.

Note that in the shown quark mass range the old
normalization condition already shows non--linear
effects in $am$. Since $a$ is extremely small here,
the physical quark mass is already quite large
and such a behavior is thus not unexpected.

\subsection{Results for the normalization factors}

The results of all simulations are collected in \app{simdata},
while in \tab{t_zasimpar} we show only the results of the chiral
extrapolations (\ref{za_impl}) and (\ref{zv_impl}).

In most cases at least one simulation is very close to the
critical point and in some cases we actually perform
an \emph{inter}polation in $am$. As can be seen from the table,
this results in a very precise estimate of the critical
hopping parameter $\kappa_c$. The latter is obtained by extrapolating
the PCAC mass $am$ linearly in $1/\kappa$.

\TABULAR[hb]{c|rrlll}{
    \hline
$\beta$ &
\multicolumn{1}{c}{$L/a$} &
\multicolumn{1}{c}{$T/a$} &
\multicolumn{1}{c}{$\kappa_c$} &
\multicolumn{1}{c}{$\za$} &
\multicolumn{1}{c}{$\zv$} \\
    \hline
5.200 &  8 & 18& 0.135856(18)& 0.7141(123)    & 0.7397(12)	\\
5.500 & 12 & 27& 0.136733(8) & 0.7882(35)(39) & 0.7645(22)(18)  \\
5.715 & 16 & 36& 0.136688(11)& 0.8037(38)(7)  & 0.7801(15)(27)  \\\hline
5.290 &  8 & 18& 0.136310(22)& 0.7532(79)     & 0.7501(13)    \\\hline
7.200 &  8 & 18& 0.134220(21)& 0.8702(16)(7)  & 0.8563(5)(45)	\\
8.400 &  8 & 18& 0.132584(7) & 0.8991(25)(7)  & 0.8838(13)(45)  \\
9.600 &  8 & 18& 0.131405(3) & 0.9132(11)(7)  & 0.9038(3)(45)	\\\hline
}
{Results of the chiral extrapolations to extract $\za$ (\ref{za_impl})
and $\zv$ (\ref{zv_impl}).\label{t_zasimpar}}

As already mentioned the volumes of the simulations at
$\beta=5.2$, $5.5$ and $5.715$ are
matched in the same way as for the $\cA$ simulations, cf.
\eq{e:L3l}, while for $\beta\geq7.2$ the volumes are much
smaller.\footnote
{The (compared to \chap{chap:impr}) different $\beta$, such that $t(\beta)=0.5$
is due to the use of a version of eq.~(\ref{e:L3l}), which differs by
$\rmO(g_0^4)$ terms.}
The second set of errors for $Z_{\rm A/V}$ represent an estimate of the systematic
uncertainties due a mismatch in the lattice spacing/volume. They will
be discussed in \sect{sect:syst}.
An exception are the simulations at $\beta=5.2$, which is
the reference volume and $\beta=5.29$, which was performed only
to qualitatively
confirm the observed rapid
change of $\za(g_0^2)$ in this region of the coupling.

\EPSFIGURE[p]{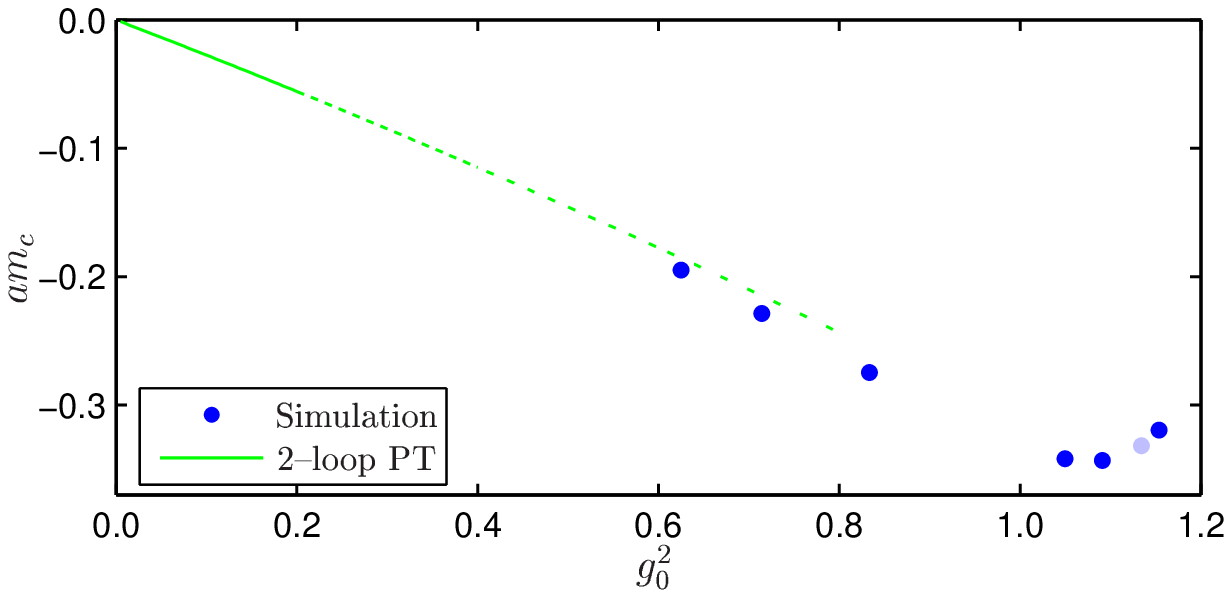,width=127mm}{
The critical mass $a\mc$ (errors are much
smaller than the symbols) as a function of the
bare gauge coupling $g_0^2$. For $g_0^2<0.8$
the non--perturbative data start to approach
the two--loop prediction (\ref{mcrit}) from
\cite{Panagopoulos:2001fn}.\label{fig:kappac}}

\fig{fig:kappac} shows the critical mass $a\mc=(2\kappa_c)^{-1}-4$
as determined in the simulations and compares it to the (continuum)
perturbative prediction.
The non--monotonic behavior of the additive mass renormalization
at $g_0^2>1$ is similar to the one found in the quenched case
(see e.g. \cite{Luscher:1996jn}) and
is clearly a non--perturbative phenomenon.
In contrast, for the largest values of $\beta$ simulated, the numerical
results are already close to the the two--loop
formula (\ref{mcrit}).

\EPSFIGURE[p]{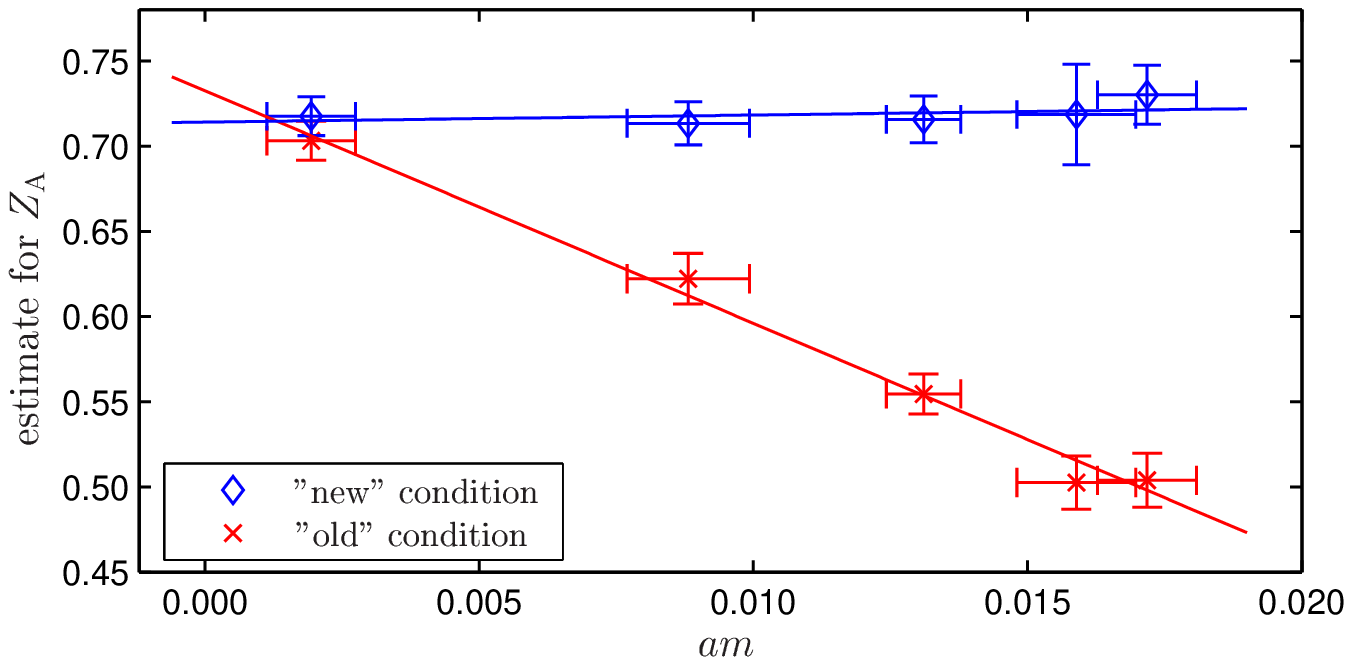,width=130mm}{
The chiral extrapolation of the result from
the axial current normalization conditions
at $\beta=5.2$.\label{fig:extrapol}}

The chiral extrapolation of our estimate for $\za$ at
the coarsest lattice spacing, $\beta\!=\!5.2$, is
plotted in \fig{fig:extrapol}. Apart from the significantly larger
errors the situation is very similar to the
quenched one from \fig{quenched}, showing that also in the
dynamical case the new normalization condition has only a very small
mass dependence. As a result, the errors on $am$ and the correlation of $am$
with the estimate for $\za$ are irrelevant for the
extrapolation. Note however, that the mass
range in \emph{lattice units} is the same in both plots, which
implies that (ignoring renormalization factors of order one) the
physical masses in the dynamical case are much smaller since
$a$ is much larger. As a result,
also for the old normalization condition all mass effects now
show a linear behavior.

The slopes of the two estimate as functions of $am$ are also very similar
to the quenched case. Due to the large statistical errors the slope
for the new condition is consistent with zero (0.4 with an error of 1.2)
and for the old condition we obtain -14(1). Again, the slope of the
estimate for $\zv$ ($0.64(9)$, not shown) is comparable to that from the
new condition for $\za$.
In the quenched case \cite{Luscher:1996jn}
the slope (in $a\mq$) of $f_1/(\zv\fv)$ could be used to determine the $\rmO(a\mq)$ improvement
coefficient $b_{\rm V}$, cf. \eq{calculateZV}.
This was possible because for $\nf=0$ the bare and modified
gauge couplings coincide ($\bg=0$). This is no longer
true in the dynamical case \cite{Luscher:1996sc,Sint:1997jx}
and hence the slope one finds
is not entirely due to $b_{\rm V}$.

In order to keep the discussion transparent,
we will now restrict ourselves to the new normalization condition.
The final results for $\za$ and $\zv$ are shown in \fig{fig:result}
as a function of $g_0^2$.
The errors plotted there are obtained by linearly
summing the statistical and systematic errors. Also shown
are the 1--loop estimates (\ref{pta}) and (\ref{ptv}).

\EPSFIGURE[!t]{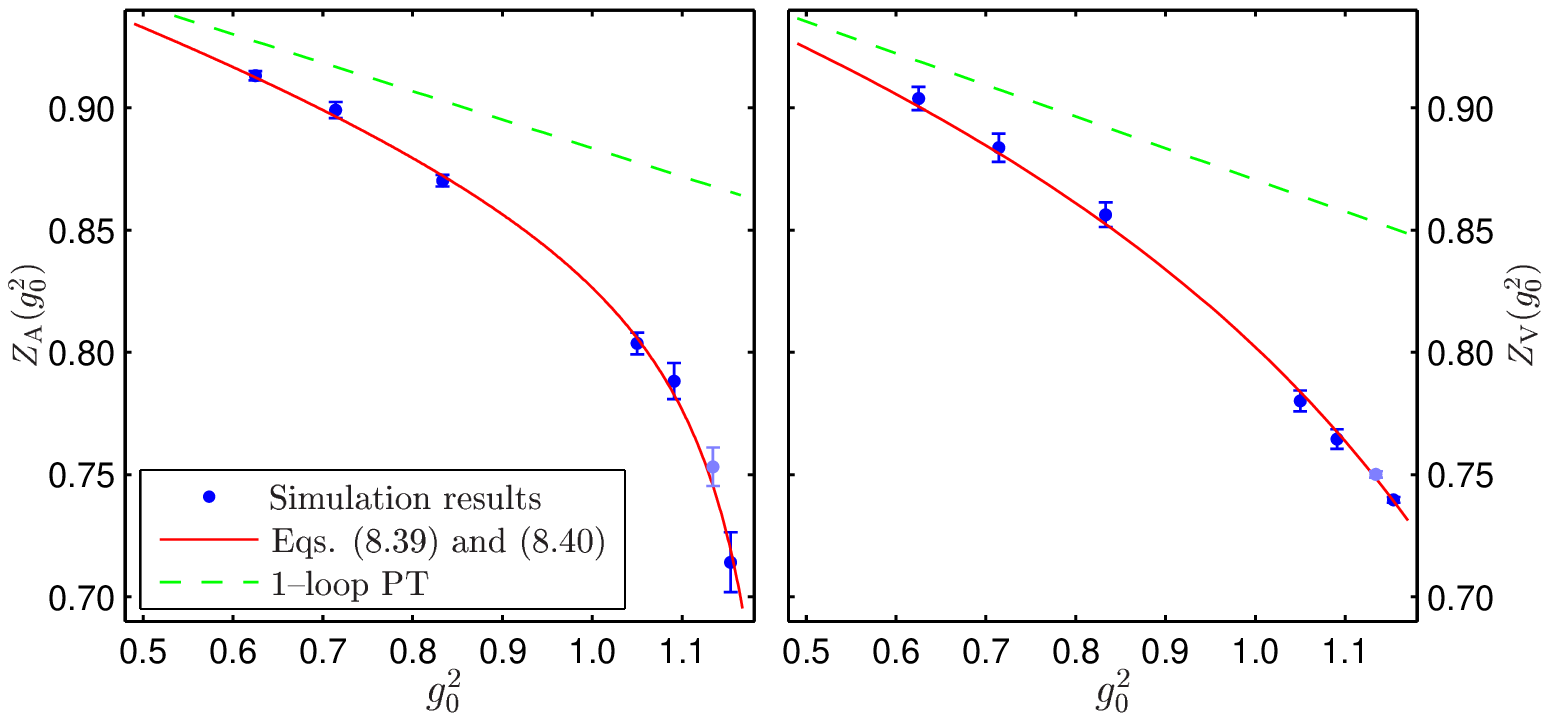,width=144mm}{
The normalization factors $\za$ and $\zv$ as a function of $g_0^2$, together with
the 1--loop results (dashed line) and the Pade interpolating
formulae (solid line).
The data points at $\beta=5.29$ (lighter color) are excluded from
the fit.\label{fig:result}}

One can see that our results for both $\za$ and $\zv$
lie on smooth curves and we can therefore (as for $\cA$)
describe the non--perturbative data 
in form of a rational function of $g_0^2$ (Pade fit). In this
fit the difference of the linear terms of numerator
and denominator is constrained to the 1--loop value, which
ensures a smooth matching with perturbation theory
as $g_0^2$ approaches zero.

The fit is performed with the sum of the errors from
\tab{t_zasimpar} and the data points at $\beta=5.29$
are excluded since no
estimate of the systematics is available in this case.
With the perturbative input (\ref{pta}) and (\ref{ptv})
we thus arrive at the interpolating formulae
\bea
\za(g_0^2)&=&\frac{1 -0.918\,g_0^2+0.062\,g_0^4+0.020\,g_0^6}
{1 -0.8015\,g_0^2}\;,\label{zaint}\\[2mm]
\zv(g_0^2)&=&\frac{1 -0.6715\,g_0^2+0.0388\,g_0^4}
{1 -0.5421\,g_0^2}\;.\label{zvint}
\eea
The low value of $\za$ at $\beta=5.2$ requires that the
$\za$ data are fitted with a polynomial of third degree
in the numerator, while the $\zv$ data can be
well described including only quadratic terms in $g_0^2$
in the numerator. For $\zv$ we find agreement at the $1\%$ level
with the results from \cite{Bakeyev:2003ff}, where isospin
charge conservation is imposed for nucleon matrix elements
of the local vector current in large volume.

The maximal absolute deviations of the fit from the
simulation data are
0.0056 (corresponding to $0.7\,\sigma$ or $0.7\%$)
for $\za$ at
$\beta=5.5$ and 0.004 ($0.9\,\sigma$ or $0.5\%$)
for $\zv$
at $\beta=5.715$. Considering that all the data in one
fit are statistically independent since they come from
different simulations, these fluctuations are
rather small.
For future application we propose to either use
the data from \tab{t_zasimpar} directly or the
interpolating formulae (\ref{zaint}) and (\ref{zvint}).
To $\zv$ we ascribe an absolute error of 0.005 and
for $\za$ the absolute error decreases from
0.01 at $\beta=5.2$ to 0.005 at $\beta=5.7$.

We note that for a lattice spacing of roughly $0.1\fm$, corresponding
to $\beta=5.2$, our non--perturbative estimate of $\za$
is almost $20\%$ smaller than the 1--loop value, while in the quenched
case \cite{Luscher:1996jn} this difference was only
$10\%$ at the same lattice spacing.\footnote
{Although in \cite{Luscher:1996jn} $\za^{\rm con}$ (see next section)
was considered, this comparison is still meaningful since the difference from
our definition of $\za$ was
found to be negligible already for $a\simeq0.1\fm$.}

\subsection{Systematic uncertainties}

\label{sect:syst}
Close to the continuum the dependence of the normalization factors
on the lattice size is expected to be of order $(a/L)^2$ \cite{Luscher:1996jn}
in the improved theory.
This implies that effects in $\za$ and $\zv$ due to deviations from the
line of constant physics, i.e. the constant physical volume
condition, should be strongly suppressed.

To check for these effects, at $\beta=5.5$ and $5.715$ the
simulations closest to the critical point were
repeated on smaller lattices
($L/a=8$ instead of $12$ at $\beta=5.5$ and
$L/a=12$ instead of $16$ at $\beta=5.715$).
From our approximate matching (see \chap{chap:impr})
we estimate the uncertainty in $L$ (measured in units of
$L$ at $\beta\!=\!5.2$) to increase up to 
at most $10\%$ in the range $5.2<\beta<5.715$. We assume that
the uncertainty in $L$ grows linearly in $\beta$
and therefore
assign a $6\%$ error to $L$ at $\beta=5.5$ and $10\%$ at
$\beta=5.715$. Together with the difference in the estimates
of $\za$ and $\zv$ this gives the systematic errors quoted in
\tab{t_zasimpar} through a linear propagation of the error.

Our simulations confirm the expectation of small volume dependence.
The largest difference is seen at $\beta=5.5$ ($L/a=12$ vs. $8$),
where it is $0.022(7)$ corresponding to a $3\%$ effect in $\za$
for a $33\%$ change in $L$.

For the runs with $\beta\geq7.2$ the matched $L/a$ would be extremely
large. On the other hand the volume dependence of the normalization
factors should
become smaller as we are approaching the perturbative regime.
In practice the simulations are thus performed at $L/a=8$ and
the systematic error is
estimated from additional runs at the coarsest of these lattice spacings,
i.e. at $\beta=7.2$, by taking the difference of $\za$ and $\zv$ between
$L/a=8$ and $16$. While for the $\za$ estimate the volume dependence
is hardly visible, in the case of $\zv$ the large volume
($L/a=16$) results in a (statistically) significantly lower value.
This is the reason for the larger systematic error of $\zv$ for
$\beta\geq7.2$.

As already mentioned, the simulations at $\beta=5.29$ (again with
a $8^3\times18$ lattice) were done to check
that the large difference in $\za$ when going from $\beta=5.2$ to $5.5$ is
bridged smoothly. At this value of $\beta$ the matched volume would be $L/a=9$,
which is not accessible with our machine (APE geometry restrictions)
and code (even--odd preconditioning). Although the results were
excluded from the fits for this reason, also the $\beta=5.29$ data
are well reproduced by the interpolating formulae (see \fig{fig:result}),
which implies
that possible systematic effects are not visible at the given statistical
precision.

\subsection[Comparison with alternative normalization conditions]
{Comparison with alternative normalization\\ conditions}

\label{sect:asq}

In this section we consider two variants of the
definition of $\za$ (\ref{za_impl}) and study
the difference in the results for $\za(g_0^2)$.
This provides us with an estimate for the magnitude
of the cutoff effects to be expected
at the considered lattice spacings.

The Wick contractions of $\funci{AA}(x_0,y_0)$ are derived in
\app{app:corr} with the result that for the chosen isospin
assignment only the quark diagrams
shown in \fig{fig:contractions_nonzero} contribute.
\EPSFIGURE[t]{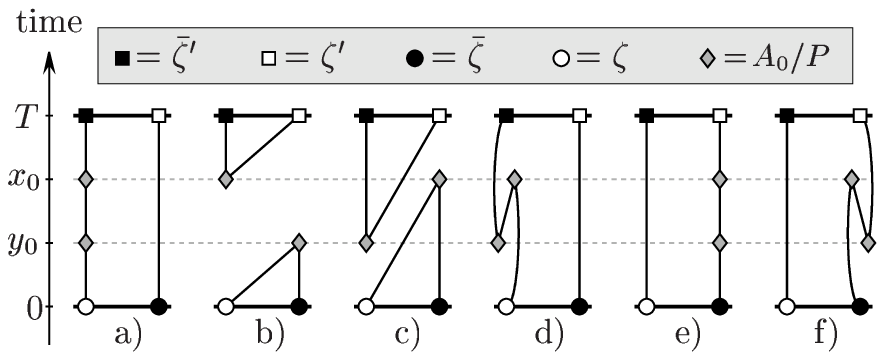,width=114mm}{
Quark diagrams for $\funci{AA}$ with non--vanishing isospin
factors. The gray diamonds indicate the insertions
at times $y_0$ and $x_0$.\label{fig:contractions_nonzero}}
Diagrams related by an exchange $y_0
\leftrightarrow x_0$ have the same isospin factors
with opposite signs. This follows directly from
(\ref{fAAI}), where such an exchange corresponds
to $\epsilon^{abc}\!\leftrightarrow\epsilon^{bac}$.
Among the quark diagrams are the two disconnected diagrams
b) and c), where no quark propagator connects the 
boundaries.

Following \cite{disco}, we now argue that in the massless
continuum limit the
contribution of the disconnected diagrams vanishes.
Thus we consider diagram b) without improvement,
i.e. the current insertions are just axial charges.

In this diagram,
termed $[gf]_{\rm AA}(x_0,y_0)$ in \app{app:simpl}, the spatial insertion
points $\bx$ and $\by$ are integrated over such that it is a function
of $x_0$ and $y_0$ only.
In fact, since we are in the chiral limit, the axial charge is conserved
and hence the diagram is independent of the insertion points in
the two regions $x_0<y_0$ and $x_0>y_0$. If the two insertion meet,
contact terms may arise, which we need to treat separately.

To this end we restrict the spatial integration to $|\bx-\by|>\epsilon$
and let the insertion times approach one another from either region.
No contact terms can appear due to the finite spatial separation.
If the integrand has a divergence
weaker than $|\bx-\by|^{-3}$, the contribution to the spatial integral
from the region
$|\bx-\by|\leq\epsilon$ vanishes as we make it smaller and smaller.
In this case we can take the limit $\epsilon\rightarrow0$ and conclude that
the order of $x_0$ and $y_0$ does not play any r\^ole.
This would imply $[gf]_{\rm AA}(x_0,y_0)=[gf]_{\rm AA}(y_0,x_0)$, i.e.
the diagrams b) and c) have the same value. Since their isospin factors
have opposite signs, their contribution to $\func{AA}$ cancels.

This argumentation shows that in the massless continuum theory the
disconnected quark diagrams b) and c) do not contribute if the composite
field $A_0(x)A_0(y)$ has a divergence weaker than $|\bx\mi\by|^{-3}$.
To decide this we consider the matrix elements of two axial charges
between pseudo--scalar states and assign the flavor quantum numbers
such that 
diagram b) is the only Wick contraction. Using four valence quark species
$u$, $d$, $s$ and $c$ this amounts to
\be
\langle\pi_{ud}|\, (A_0)_{du}(x) (A_0)_{cs}(y) |\pi_{sc}\rangle\;.
\ee
It is now immediately clear that such a flavor assignment excludes
a single quark bilinear as
the leading contribution in the operator product expansion of
$A_0(x)A_0(y)$. Hence the latter has (if any) a divergence weaker
than $|\bx-\by|^{-3}$ in the limit $|\bx-\by|\rightarrow\epsilon$ and
the contribution from the excluded integration region vanishes.

Since the correlation functions approach their continuum value
with a rate proportional to $a^2$, we can conclude that on the
lattice the contribution from the disconnected diagrams is a
cutoff effect of this order.

If, in contrast, we would consider the diagrams a) and d), which
also differ by a sign, they are still independent of the insertion
times in the regions $x_0<y_0$ and $x_0>y_0$. However, in the above
argumentation the flavor structure would be
\be
\langle\pi_{ud}|\, (A_0)_{ds}(x) (A_0)_{sc}(y) |\pi_{cu}\rangle
\ee
and hence the operator product could mix with e.g. $(V_0)_{dc}$ or the
scalar density. This, by dimensional analysis, produces a factor
$|\bx-\by|^{-3}$ in the integrand at leading order of the OPE and thus,
after integrating, a logarithmically divergent contact term.
Therefore, no statement about the contribution of the connected
diagrams can be made from this argument.

\EPSFIGURE[t]{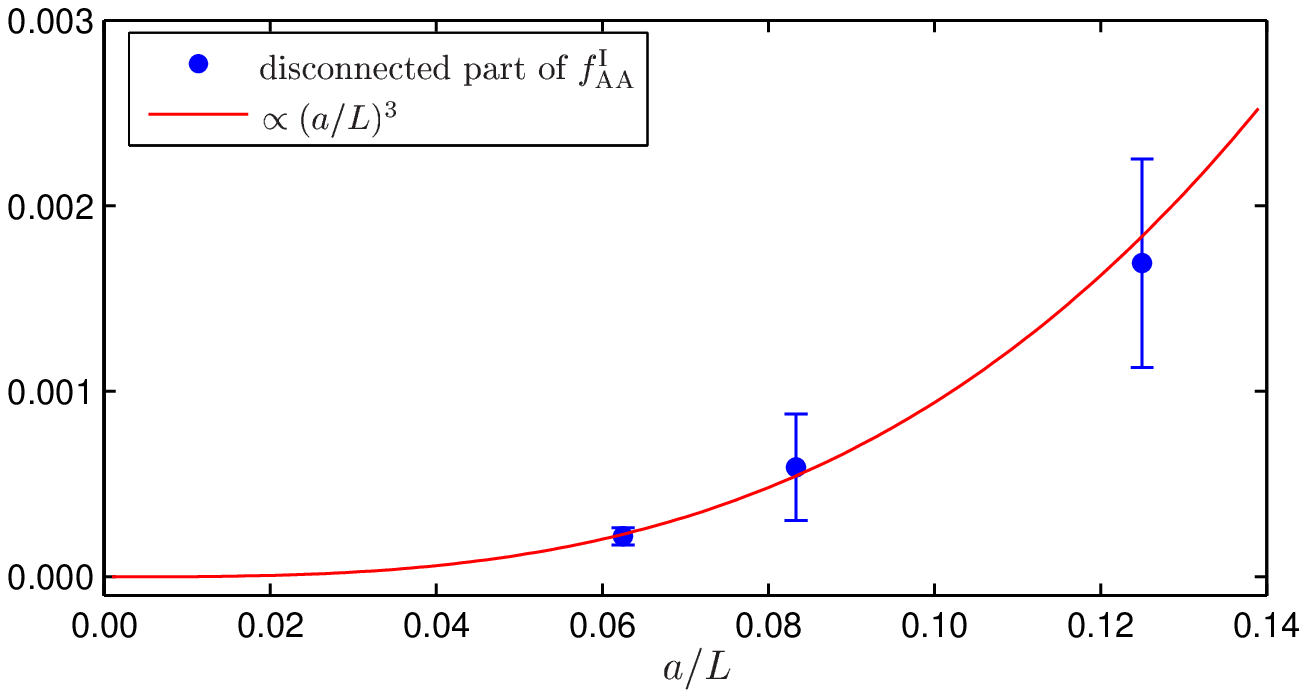,width=114mm}{
Data from the simulations at $\beta=5.2$, $5.5$ and $5.715$ extrapolated
to the chiral limit.
As expected the contributions from the disconnected diagrams vanishes
in the continuum limit with a rate that is faster than linear in $a$.
\label{fig:disco}}

In more physical terms this can be understood directly from the
quark diagrams. Even if the insertion points coincide when we move
them around (on a fixed gauge background), in b) there is no propagator
connecting them and the
axial charges thus don't ''see'' each other. Doing this with diagram d)
changes the propagator $\sum_{\bx,\by}S(x;y)$ and in particular it
gives a contact term at zero separation.

As is shown in \fig{fig:disco}, we can numerically confirm that the
contribution of the disconnected diagrams to $\funci{AA}$ vanishes
in the continuum limit. According to \app{app:simpl} the
contribution is
\bea
\funci{AA}\Big|_{\rm disconnected}\!\!\!\!&=&[gf]_{\rm AA}(x_0,y_0)-[gf]_{\rm AA}
(y_0,x_0)\nonumber\\\nonumber
&&+ac_A\drvtilde0x\,\Big\{-[gf]_{\rm PA}(x_0,y_0)-
[gf]_{\rm AP}(y_0,x_0)\Big\}\\\nonumber
&&+ac_A\drvtilde0y\,\Big\{[gf]_{\rm AP}(x_0,y_0)+[gf]_{\rm PA}(y_0,x_0)
\Big\}\\
    &&+a^2c_A^2
\drvtilde0x\,\drvtilde0y\,\Big\{[gf]_{\rm PP}(y_0,x_0)
-[gf]_{\rm PP}(x_0,y_0)\Big\}\;,\label{discoimp}
\eea
where we have $y_0=T/3$ and $x_0=2\,T/3$. In \fig{fig:disco} we plot data
from the matched simulations and neglect
any systematic effects from volume mismatch.
One can clearly see that the above term approaches zero
\emph{faster} than linear in the lattice spacing and is in fact
compatible with an $a^3$ effect.

The correlation function multiplying the mass term,
$\tfunci{PA}$, only influences the slope of the chiral extrapolation
but not the result for $\za$ in the chiral limit. We are therefore free
to also drop the disconnected diagrams for this correlator.
An estimate of $\za$ using (\ref{za_impl}), where only the connected
part of the correlation functions enter, should therefore
agree with the original definition up to $\rmO(a^2)$. We will denote
this by $\za^{\,\rm con}$.

In the quenched case \cite{Luscher:1996jn} $\za$ was also determined
from the connected quark diagrams only. The authors approach the
above conclusion from a different point. Their argument exploits
the freedom to choose different external operators in the Ward identity
(\ref{3.16}). Instead of (\ref{oext}), one could use
\be
\op_{\rm ext}=\frac{-i}{6L^6}\oprime_i{\ts\frac12}(\tau^c)_{ij}\op_j
+\frac{i}{6L^6}\kprime_i{\ts\frac12}(\tau^c)_{ji}\K_j\;,\label{oextnew}\\[-2mm]
\ee
\bea
\textrm{where }\ \oprime_i  =a^6\sum_{\bu,\bv}\zetabarprime_i(\bu)\dirac5\xiprime(\bv)\;,&&
\op_i                =a^6\sum_{\bu,\bv}\xibar(\bu)\dirac5\zeta_i(\bv)\;,\\
\kprime_i  =a^6\sum_{\bu,\bv}\xibarprime(\bu)\dirac5\zetaprime_i(\bv)\;,&&
\K_i                =a^6\sum_{\bu,\bv}\zetabar_i(\bu)\dirac5\xi(\bv)\;.
\eea
Here $i$ and $j$ are flavor indices and $\xi$ denotes the boundary
field for a third quark species, which is taken to be an isospin singlet.
Therefore no Wick contraction exists that connects it to the axial current
with the result that no disconnected diagrams can appear.
Since the external operator in the Ward identity is arbitrary, the resulting
$\za$ agrees with the original one up to cutoff effects and one thus
again reaches the same conclusion. Note that the first argument takes
place at the level of quark diagrams and is thus valid for any number
of valence and sea quarks, whereas the external operator (\ref{oextnew})
explicitly assumes the existence of three valence quarks flavors.

If the improved axial current is inserted into the correlator
$\funci{AA}$, we obtain expressions proportional to $\cA^2$ as
in the last line of (\ref{discoimp}). Since we implement improvement
only to $\rmO(a)$, we could have decided to drop these as
just another source of $\rmO(a^2)$ ambiguities. Thus, keeping only
improvement terms \emph{lin}ear in $\cA$ provides
us with another estimate of the axial current normalization constant,
which we denote by $\za^{\,\rm lin}$.\footnote{
One might be tempted to also consider $\za^{\,\rm con,lin}$, but
in practice the contribution of the $\cA^2$ term to
the connected part is negligible and
hence $\za^{\,\rm con,lin}\simeq\za^{\,\rm con}$.}

We can now repeat the whole analysis -- chiral limit at
each value of $\beta$ and
estimate of volume dependence -- to extract $\za(g_0^2)$ from the
estimates $\za^{\rm con}$ and $\za^{\rm lin}$.
\EPSFIGURE[t]{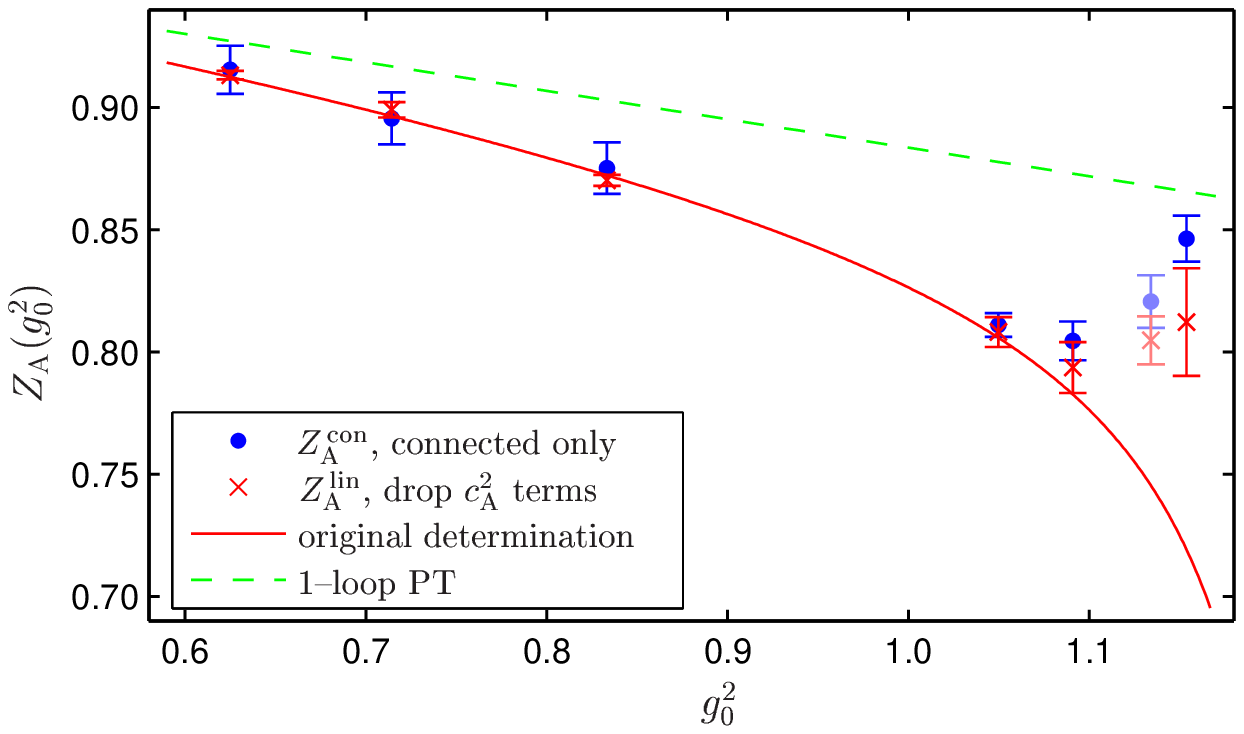,width=122mm}{
Changing the definition of $\za$ at $\rmO(a^2)$ has
significant effect on the result for bare gauge couplings
$g_0^2\gtrsim1.1$.
\label{fig:asq}}
The result of this exercise is shown in \fig{fig:asq}, where it is
compared to the interpolating formula (\ref{zaint}) for the
original estimate.
We note that for $\beta\geq5.7$ ($g_0^2<1.1$) the central values
agree with the interpolating formula (\ref{zaint}).
The increased errors for $\za^{\rm con}$ at $\beta\geq7.2$
are due to the rather large
dependence on $L/a$ we found at $\beta=7.2$ (see \app{simdata}).
 
However, much more noticeable are the $\rmO(a^2)$ effects in the range
of $g_0^2\simeq1.1$ and above, corresponding to $\beta\lesssim5.4$.
The agreement with the original determination in form of the
Pade fit is rapidly lost and at $\beta=5.2$ we see a (statistically
significant) deviation of $15\%$ (for $\za^{\rm con}$).

This might be surprising, given that in \fig{fig:disco} the $\beta=5.2$ estimate
of the disconnected part
is only two standard deviations from zero. It is, 
however, of similar magnitude as the connected part and the small
error in the resulting $\za$ is due to the correlation of
numerator and denominator in (\ref{za_impl}).

From the above observations and \fig{fig:asq} it is clear that it is essentially
the last term in (\ref{discoimp}), which is responsible for
the monotonic behavior in our original estimate. The large
contribution of this term is in turn related to the fact
that in the volumes and time extensions we consider here, the
$\fp$ correlator
still shows a strong exponential decay \emph{even at the critical
point}. This means that it receives large contributions from
excited states in the pseudo--scalar channel \cite{Guagnelli:1999zf}
and therefore its effective mass large.

These deviations
and even the apparent non--monotonic behavior one obtains
from $\za^{\,\rm con}$ and $\za^{\,\rm lin}$ do not signify
any theoretical problems.
It is, however, another sign that even after a successful implementation
of the improvement programme the lattice artifacts in the two--flavor
theory can
be large for $\beta\lesssim5.4$ and a continuum extrapolation remains
necessary to make reliable physical predictions. In particular,
care should be taken
that such extrapolations
are \emph{not} dominated by the data from simulations with
$\beta=5.3$ or lower.

In the quenched case the $\rmO(a^2)$ effects from the disconnected
diagrams are significantly smaller at a lattice spacing $a\simeq0.1\fm$.
In fact, at a statistical precision of the order of $1\%$ these
effects are not visible at all.

\section{Summary}

\label{zasummary}

We have formulated an integrated Ward identity, derived
from the isovector chiral symmetry of the continuum action,
as a normalization condition for the axial current on the
lattice. Through the use of the massive Ward identity we
were able to improve the method from \cite{Luscher:1996jn}
with respect to its mass dependence.

The normalization condition was implemented in terms
of correlation functions in the Schr\"odinger functional
framework and evaluated
on a line of constant physics in order to achieve
a smooth disappearance of the $\rmO(a^2)$ uncertainties.
Through additional simulations at
very small lattice spacings and volumes we verified
that our non--perturbative definition approaches
the perturbative prediction at
small bare gauge coupling.
Simulations were done at or near the critical point
and owing to the new normalization condition, which
keeps track of the mass term in the PCAC relation, any
chiral extrapolations were extremely flat.
Systematic effects due to deviations
from the constant volume condition are
also estimated and turn out to be small.
The results are well described by an interpolating
formula $\za(g_0^2)$ in the range of bare couplings
considered.

Enforcing isospin symmetry with the same programme,
we obtain at the same time a non--perturbative
determination of
the normalization constant $\zv(g_0^2)$ of the
local vector current.

We found large $\rmO(a^2)$ uncertainties in $\za$ at
$\beta\lesssim5.4$ by varying the definition
of $\za$, which indicate that despite improvement
the cutoff effects in this range of lattice spacings can be large.
Together with the algorithmic issues discussed in
\chap{chap:algo} these findings corroborate the worries
expressed in \cite{Sommer:2003ne} about the
status of simulations with improved dynamical Wilson fermions at the
currently accessible lattice spacings.
This merely emphasizes that cutoff effects in physical
quantities can be controlled only if a continuum
extrapolation with several lattice resolutions is
performed.

Finally, the method employed here can also be
used to obtain $\za$ with other actions
formulated in the Schr\"odinger functional. This
includes in particular the $\rmO(a)$ improved
three flavor theory
with either Iwasaki or plaquette gauge action
\cite{Aoki:2002vh,Yamada:2004ja,Ishikawa:2003ri}.

\chapter{Conclusions}

The main result of this work is the non--perturbative determination
of the axial current improvement and normalization constants
for the lattice theory with two flavors of dynamical Wilson fermions.

With this regularization we have thus shown that normalization
conditions can be imposed at the non--perturbative level
such that isovector chiral symmetries are realized
in the continuum limit.
Since we work with an improved theory, chiral
Ward--Takahashi identities are then satisfied up to
$\rmO(a^2)$ at finite lattice spacing.

In the course of the implementation of this programme algorithmic
difficulties were encountered in the numerical simulations.
A more detailed investigation revealed that those are related
to features of the infrared spectrum of the Wilson--Dirac operator.
More precisely, it was found that at a given lattice 
spacing the inclusion of the fermion determinant
enhances the probability of finding very small eigenvalues, which
is contrary to the usual intuition. This effect disappears rapidly
as the lattice spacing is decreased and is thus interpreted
as a lattice artifact.

This spectral distortion is responsible for problems in the
molecular dynamics evolution of the hybrid Monte Carlo algorithm
as well as in the efficient sampling of observables sensitive
to small eigenvalues. One of the important results is that
in this situation it can be advantageous to deviate
from importance sampling. In practice we found the polynomial
hybrid Monte Carlo algorithm with reweighting to be an efficient
tool in the parameter range we considered.

The non--perturbative determination of $\csw$, the improvement
coefficient of the Wilson fermion action, has been known
for some time already \cite{Jansen:1998mx}. However,
for a complete removal of the $\rmO(a)$ lattice artifacts
also the composite fields need to be improved.
We have completed this programme for the axial current,
accomplishing a necessary 
step on the way to a non--perturbative normalization of
the latter.

The non--perturbative determination of the axial current
improvement constant $\cA$ was implemented in the Schr\"odinger
functional framework using spatial
wave functions in the composite boundary fields. In this
way we could obtain a large sensitivity by explicitly
projecting to the ground and first excited state in the
pseudo--scalar channel and requiring the current quark mass
derived from the PCAC relation to be the same in both cases.
With this method the contribution of excited states is
well controlled and we were thus able to ensure that
the unavoidable $\rmO(a)$ ambiguities in $\cA$ remain small.
To achieve a smooth disappearing of the latter
in the perturbative limit we evaluated the improvement condition
on a line of constant physics. Lacking a non--perturbative
estimate of the evolution of the lattice spacing with the
bare gauge coupling, we used (in a small range of bare
gauge couplings) an estimate based on the
three--loop lattice $\beta$--function for improved Wilson
fermions.

In the next step we implemented the non--perturbative
renormalization of the improved axial current.
Similarly to the determination of the improvement coefficient
this is also achieved by enforcing a continuum Ward identity
at finite lattice spacing. We improved the method used
in the quenched case \cite{Luscher:1996jn} to obtain a
normalization condition whose dependence on the quark mass
is only a cutoff effect, thus alleviating the need for
a chiral extrapolation.

Similar to $\cA$, beyond perturbation theory $\za(g_0^2)$ is
affected by an intrinsic $\rmO(a^2)$ ambiguity,
which vanishes smoothly in the continuum limit if 
the normalization condition is evaluated on a line of constant
physics. However, in the case of $\za$ we need to evaluate this
condition at zero quark mass at each lattice spacing and $\za$ is
thus the result of an extra-- or interpolation in the quark mass.
To match the volumes we use the same estimate
of $\partial a/\partial g_0^2$ as in the determination of $\cA$
and with additional simulations at large $\beta$ we verify
that our definition of $\za$ smoothly matches with perturbation
theory.
With little additional effort we also extracted the local
vector current normalization factor $\zv(g_0^2)$ from the same
simulation data by enforcing isospin symmetry at finite
lattice spacing.

The dependence of a non--perturbative definition of $\za$
on the lattice size is expected to be of order $(a/L)^2$
close to the continuum.
In our estimate of the systematic effects in $\za$ due to
deviations from the constant physics condition, we indeed
observed only very little volume dependence.
However, at lattice spacings $a\gtrsim0.07\fm$ large ambiguities
in $\za$ were found by changing the normalization condition
at $\rmO(a^2)$.
Together with the cutoff effects in the spectrum (\chap{chap:algo})
these findings add to the evidence collected in
\cite{Sommer:2003ne} that the lattice artifacts with dynamical
improved Wilson fermions at the current lattice spacings
are much larger than initially expected from quenched
experience. One might even suspect the
proximity of a phase transition in bare parameter space
in analogy to
\cite{Aoki:2001xq}.
Such an extreme example of a lattice artifact would surely
invalidate the Symanzik expansion (\ref{symanzik}).

Setting these worries aside, the results
for $\cA$ and $\za$ can now readily be applied.
The most immediate application would be to $\rmO(a)$ improve
and renormalize the bare pseudo--scalar decay constants
computed in \cite{Allton:2001sk,Irving:2001vy,Aoki:2002uc}.
As already mentioned in \sect{sect:canum}, using the
non--perturbative $\cA$ lowers
the result from \cite{Irving:2001vy} by $10\%$ at $\beta=5.2$.
With a $20\%$ decrease compared to the 1--loop estimate
(again at $\beta=5.2$) the non--perturbative $\za$
will also have a large effect when used to renormalize
existing data.

The methods developed and implemented here can easily be applied
to other actions formulated in the Schr\"odinger functional
framework. This includes improved gauge actions \cite{Iwasaki:1983ck}
as well as
theories with more than two dynamical quark flavors
\cite{Aoki:2002vh,Yamada:2004ja,Ishikawa:2003ri}.

Within the research programme of the ALPHA collaboration,
the results obtained here are an essential step in the
computation of the pseudo--scalar meson decay constant
$F_{\rm PS}$ needed to reliably convert the $\Lambda$ parameter
from \cite{DellaMorte:2004bc} into physical units.
In the short term, together with data from \cite{Knechtli:2002vp}
$\za(g_0^2)$ will be used in a fully
non--perturbative calculation
of the strange quark mass \cite{DellaMorte:2005kg} following the strategy
of \cite{Capitani:1998mq,Garden:1999fg}.

{\renewcommand{\baselinestretch}{0.94}
\small
\bibliography{../refs}
\bibliographystyle{JHEP}}
\appendix

\chapter{Improved action for the Schr\"odinger functional}

\label{app:SF}

The $\rmO(a)$ improved action of the QCD Schr\"odinger functional
\cite{Luscher:1996sc}
is given by (the functional dependence on $U,\psibar$ and $\psi$ is
suppressed)
\be
S_{\rm impr}=S_g+S_f+\delta S_v+\delta S_{f,b}\;.\label{impraction}
\ee
The improvement of the gauge action (\ref{wilsongauge})
is implemented through
\be
S_g[U]=\frac1{g_0^2}\sum_p w(p)\tr\left\{1-U_p\right\}\;,
\ee
where the weight for the oriented plaquette $p$ is given by
\bea
w(p)&\!=\!&\left\{\begin{array}{r@{,\ }l}
1	& \textrm{for $p$ in the interior of the SF}\\
\ct &  \textrm{for $p$ \emph{touching} the $x_0=0$ or $T$ boundaries}\\
\half \cs & \textrm{for $p$ \emph{in} the boundary.}
\end{array}\right.
\eea
A sufficient condition for the $\cs$ term \emph{not} to contribute
are Abelian and spatially homogeneous boundary fields $C_k$
and $C_k'$. This condition is not only fulfilled for vanishing background
field (the setup employed here), but for all boundary gauge fields so far
considered in the SF \cite{Jansen:1998mx,DellaMorte:2004bc}.

The fermionic action is given by (\ref{wilsonaction}) in the interior
of the SF cylinder and the boundary improvement terms (\ref{SFimp1})
and (\ref{SFimp2}) are transcribed on the lattice as
\bea
\delta S_{f,b}&=&a^4\sum_{\bx}\ (\ctildes\mi1)
[\widehat\op_s(\bx)+\widehat\op_s'(\bx)]\ + \
(\ctildet\mi1)
[\widehat\op_t(\bx)-\widehat\op_t'(\bx)]\;,
\eea
with the boundary and near--boundary operators
\bea
\widehat\op_s(\bx)&=&\rhobar(\bx)\dirac k\widetilde\nabla_k\rho(\bx)\;,\\
\widehat\op_s'(\bx)&=&\rhobarprime(\bx)\dirac k\widetilde\nabla_k\rhoprime(\bx)\;,\\
\widehat\op_t(\bx)&=&\Big\{\psibar(x)P_+\nabla_0^\star\psi(x)
+\psibar(x)\lvec\nabla_0^\star P_-\psi(x)\Big\}_{x_0=a}\;,\\
\widehat\op_t'(\bx)&=&\Big\{\psibar(x)P_-\nabla_0\psi(x)
+\psibar(x)\lvec\nabla_0 P_+\psi(x)\Big\}_{x_0=T\mi a}\;.
\eea
While $\widehat\op_t$ and $\widehat\op_t'$ couple to the dynamical variables
close to the boundaries, $\widehat\op_s$ and $\widehat\op_s'$ depend
on the fermionic boundary fields $\rho,\rhobar,\rhoprime$ and $\rhobarprime$
only. Since they are set to zero, a knowledge of $\ctildes$ is not necessary.

The entire fermionic part of (\ref{impraction}) is expressed
in ($\delta D$ summarizes the volume and boundary improvement)
\be
S_{f,\rm impr}=\sum_x\psibar(x)[D+\delta D+m_0]\psi(x)\;,
\ee
and for a given gauge field configuration $U_\mu(x)$ the quark propagator
$S(x;y)$ is thus defined as \cite{Luscher:1996vw}
\be
(D+\delta D+m_0)S(x;y)=a^{-4}\delta_{xy}\;,\ 0<x_0<T\;,\label{fullD}
\ee
and the boundary conditions
\be
P_+S(x;y)\Big|_{x_0=0}=P_-S(x;y)\Big|_{x_0=T}=0\;,
\ee
where color and spin indices have been suppressed.
The basic two-point correlations are obtained \cite{Luscher:1996vw} by
differentiating the generating functional with respect to the source fields.
The result can be expressed in terms
of the quark propagator $S(x;y)$.
\bea
\F{\psi(x)\psibar(y)}&=&S(x;y)\label{eins}\\[2mm]
\F{\psi(x)\zetabar(\by)}&=&\ctildet S(x;a,\by)
U_0(0,\by)^{-1}P_+\label{zwei}\\[2mm]
\F{\psi(x)\zetabarprime(\by)}&=&\ctildet S(x;T\mi a,\by)
U_0(T\mi a,\by)P_-\label{drei}\\[2mm]
\F{\zeta(\bx)\psibar(y)}&=&\ctildet P_-U_0(0,\bx)
S(a,\bx;y)\label{vier}\\[2mm]
\F{\zetaprime(\bx)\psibar(y)}&=&\ctildet P_+U_0(T\mi a,\bx)^{-1}
S(T\mi a,\bx;y)\label{fuenf}\\[2mm]
\F{\zeta(\bx)\zetabarprime(\by)}&=&\ctildet^2P_-U_0(0,\bx)
S(a,\bx;T\mi a,\by)U_0(T\!-\!a,\by)P_-\label{sechs}\\[2mm]
\F{\zetaprime(\bx)\zetabar(\by)}&=&\ctildet^2P_+U_0(T\mi a,\bx)^{-1}
S(T\mi a,\bx;a,\by)U_0(0,\by)^{-1}P_+\qquad\quad\label{sieben}
\eea
The correlators $\F{\zeta\zetabar}$ and $\F{\zetaprime\zetabarprime}$ are
not needed since the isospin factor for such a Wick contraction
vanishes for isovector boundary field of the type (\ref{O}) and (\ref{Oprime}).

\chapter{Correlation functions}

\section{Summed two-point correlators}

\label{app:corr}
We introduce the quark propagator from the lower
boundary to the point $x$
\bea
\label{sbar}   \overline S(x) &=& \frac{\ctildet a^3}{L^{3/2}}\sum_\by S(x;a,\by)U_0(0,\by)^{-1}P_+\\
\nonumber\Rightarrow\quad \overline S(x)^\dagger&=& \frac{\ctildet a^3}{L^{3/2}}\sum_\by P_+U_0(0,\by)
                                                        S(x;a,\by)^{\dagger}\\
                    &=&                             \frac{\ctildet a^3}{L^{3/2}}\sum_\by P_+U_0(0,\by)
                                                    \dirac5 S(a,\by;x)\dirac5\label{sbardag}\;,
\eea
the quark propagator from the upper boundary to the point $x$
\bea
\label{rbar}   \overline R(x) &=& \frac{\ctildet a^3}{L^{3/2}}\sum_\by S(x;T\!-\!a,\by)U_0(T\!-\!a,\by)P_-\\
\Rightarrow\quad \overline R(x)^\dagger&=& \frac{\ctildet a^3}{L^{3/2}}\sum_\by P_-U_0(T\!-\!a,\by)^{-1}
                                                    \dirac5 S(T\!-\!a,\by;x)\dirac5\label{rbardag}\;,
\end{eqnarray}
and the boundary-to-boundary propagator
\bea
\label{sbarT}   \overline S_T &=& \ctildet a^3\sum_\bx-U_0(T\!-\!a,\bx)^{-1}P_+\overline S(T\!-\!a,\bx)\\
\Rightarrow\quad \overline S_T^\dagger&=& \ctildet a^3\sum_\bx-\overline S(T\!-\!a,\bx)^\dagger
                                                        P_+U_0(T\!-\!a,\bx)\label{sbarTdag}\;.
\eea
Of course $\overline S_T$ can also be obtained through a summation of $\overline R(x)$
\begin{eqnarray*}
\dirac5\overline S_T\dirac5 &\use{sbar}& \frac{\ctildet^2 a^6}{L^{3/2}} \sum_{\bx,\by}-U_0(T\!-\!a,\bx)^{-1}P_-
                                \dirac5S(T\!-\!a,\bx;a,\by)\dirac5U_0(0,\by)^{-1}P_-\\
                        &\use{rbardag}& \ctildet a^3 \sum_{\by}-\overline R(a,\by)^\dagger U_0(0,\by)^{-1}P_-\;.
\end{eqnarray*}
With these definitions we can express the basic two--point functions as
\begin{eqnarray}
\frac{a^3}{L^{3/2}}\sum_\by\F{\psi(x)\zetabar(\by)}&\!\!\!\!\use{zwei}\!\!\!&
\frac{a^3\ctildet}{L^{3/2}}\sum_\by
S(x;a,\by)U_0(0,\by)^{-1}P_+
\!\!\use{sbar}\!\overline S(x)\;,\label{use1}\\
\frac{a^3}{L^{3/2}}\sum_\by\F{\psi(x)\zetabarprime(\by)}&\!\!\!\!\use{drei}\!\!\!&
\frac{a^3\ctildet}{L^{3/2}}
\sum_\by S(x;T\!-\!a,\by)
U_0(T\mi a,\by)P_-
\!\!\use{rbar}\!\overline R(x)\label{use2}\;.\ \quad\quad\,
\end{eqnarray}
The $\overline S(x)$ and $\overline R(x)$ propagators inherit the $\dirac5$--Hermiticity
from $S(x;y)$
\begin{eqnarray}
\frac{a^3}{L^{3/2}}\sum_\by\F{\zeta(\by)\psibar(x)}&\use{vier}&\frac{a^3\ctildet}{L^{3/2}}
\sum_\by P_-U_0(0,\by)S(a,\by;x)\nonumber\\
&=&\frac{a^3\ctildet}{L^{3/2}}\sum_\by \Big[S(a,\by;x)^\dagger U_0(0,\by)^{-1}P_-\Big]^\dagger\nonumber\\
&=&\frac{a^3\ctildet}{L^{3/2}}\sum_\by \Big[\dirac5S(a,\by;x)^\dagger \dirac5U_0(0,\by)^{-1}P_+
\dirac5\Big]^\dagger\dirac5\nonumber\\
&\use{sbar}&\dirac5\overline S(x)^\dagger\dirac5\label{use3}\;,\\[1mm]
\frac{a^3}{L^{3/2}}\sum_\by\F{\zetaprime(\by)\psibar(x)}&\use{fuenf}&\frac{a^3\ctildet}{L^{3/2}}
\sum_\by P_+U_0(T\!-\!a,\by)^{-1}S(T\!-\!a,\by;x)\nonumber\\
&=&\frac{a^3\ctildet}{L^{3/2}}\sum_\by \Big[S(T\!-\!a,\by;x)^\dagger U_0(T\!-\!a,\by)P_+\Big]^\dagger\nonumber\\
&=&\frac{a^3\ctildet}{L^{3/2}}\sum_\by \Big[\dirac5S(T\!-\!a,\by;x)^\dagger
\dirac5U_0(T\!-\!a,\by)P_-\dirac5\Big]^\dagger\dirac5\nonumber\\
&\use{rbar}&\dirac5\overline R(x)^\dagger\dirac5\label{use4}\;.
\end{eqnarray}
For the spatially averaged boundary-to-boundary correlators one obtains
\bea
\!
\frac{a^6}{L^{3/2}}\!\sum_{\bx,\by}\!\F{\zeta(\by)\zetabarprime(\bx)}\!\!\!\!&\!\!\use{sechs}
&\!\!\!\!\frac{a^6\ctildet^2}{L^{3/2}}
\sum_{\bx,\by} P_-U_0(0,\by)S(a,\by;T\!-\!a,\bx)U_0(T\!-\!a,\bx)P_-\nonumber\\
&\!\!=&\!\!\!\!\!\!\!\!\frac{a^6\ctildet^2}{L^{3/2}}\sum_{\bx,\by}\!\dirac5 P_+U_0(0,\by)\dirac5
S(a,\by;T\!\!-\!a,\bx)\dirac5U_0(T\!\!-\!a,\bx)P_+\dirac5\nonumber\\
&\!\!\use{sbardag}&\!\!\!\!a^3\ctildet\sum_\bx \dirac5\overline S(T\!-\!a,\bx)^\dagger U_0(T\!-\!a,\bx)P_+\dirac5\nonumber\\
&\!\!\use{sbarTdag}&\!\!\!\!-\dirac5\overline S^\dagger_T\dirac5\label{use5}\;,
\eea

\pagebreak

\bea
\!
\frac{a^6}{L^{3/2}}\!\sum_{\bx,\by}\!\F{\zetaprime(\bx)\zetabar(\by)}\!\!\!&\!\!\use{sieben}
\!\!&\!\!\frac{a^6\ctildet^2}{L^{3/2}}
\sum_{\bx,\by} P_+U_0(T\!-\!a,\bx)^{-1}S(T\!-\!a,\bx;a,\by)U_0(0,\by)^{-1}P_+\nonumber\\
&\!\!\use{sbar}&\!\!\!\!a^3\ctildet\sum_\bx P_+ U_0(T\!-\!a,\bx)^{-1}\overline S(T\!-\!a,\bx)\nonumber\\
&\!\!\use{sbarT}&\!\!\!\!-\overline S_T\label{use6}\;.
\eea

\subsection{2--point functions}
\label{sectionf1}
To derive explicit expressions for $f_1$, $f_{\rm X}$ and $g_{\rm X}$ in terms
of the propagators defined above,
we use $A,B,\ldots$ for the Dirac indices and $i,j,\ldots$
for the indices in $SU(2)$ isospin space.
Inserting equations (\ref{O}, \ref{Oprime}) into (\ref{f1def}) yields
\begin{eqnarray}
\nonumber   f_1 &=& -\frac1{3L^6}\Big\langle\oprime^a\op^a\Big\rangle\\
\nonumber       &=& -\frac{a^{12}}{3L^6}\sum_{\bx,\by, {\bf x'}\!, {\bf y'}}
                \Big\langle(\zetabarprime)_i^A({\bf y'})\dirac5^{AB}{\ts\frac12}\tau^a_{ij}
                (\zetaprime)_j^{B}({\bf x'})
                \zetabar_m^C(\by)\dirac5^{CD}{\ts\frac12}\tau^a_{mn}
                \zeta_n^{D}(\bx)\Big\rangle\;.
\end{eqnarray}
We now perform the Wick contractions of the fields and since the propagator is isospin diagonal,
each contraction requires the corresponding isospin indices to be the same. Due to their
Grassmann nature, the sign changes when the order of fields is changed
\begin{displaymath}
\begin{array}{rclcl}
\zetaprime\zetabarprime, \zeta\zetabar &    \rightarrow & \delta_{ij}\delta_{mn} & \rightarrow &
                        \ts\frac14\tau^a_{ij}\tau^a_{mn}\delta_{ij}\delta_{mn}=\frac14(\tr\tau^a)^2=0\;,\\[1mm]
\zeta\zetabarprime, \zetaprime\zetabar &        \rightarrow & -\delta_{in}\delta_{mj} & \rightarrow &
                                                \ts-\frac14\tau^a_{nj}\tau^a_{jn}=-\frac14\tr(\tau^a)^2=-\frac32\;.
\end{array}
\end{displaymath}
The Dirac indices give the order of the propagators and Dirac--matrices
and we are left with a trace in Dirac and color space
\begin{eqnarray}
f_1 &=& \frac{a^{12}}{2L^6}\sum_{\bx,\by, {\bf x'}\!, {\bf y'}}\Big\langle \tr\Big\{
        \left[\zeta(\bx)\zetabarprime({\bf y'})\right]_{\rm F}\dirac5\left[\zeta'({\bf x'})
        \zetabar(\by)\right]_{\rm F}\dirac5\Big\}\Big\rangle_U\nonumber \\\nonumber
    &\use{use6}& -\frac{a^{6}}{2L^{9/2}}\sum_{\bx, {\bf y'}}\Big\langle\tr\Big\{
\left[\zeta(\bx)\zetabarprime({\bf y'})\right]_{\rm F}\dirac5\overline S_T\dirac5\Big\}\Big\rangle_U\\
&\use{use5}& \frac{1}{2L^3}\Big\langle\tr\Big\{\dirac5\overline S_T^\dagger\dirac5
\dirac5\overline S_T\dirac5\Big\}\Big\rangle_U\ =\ \frac{1}{2L^3}\Big\langle\tr\overline S_T
\overline S_T^\dagger\Big\rangle_U\;.
\label{f1prelast}
\end{eqnarray}
In the same manner we evaluate the correlation function $\fp$ (\ref{fP})
\beann
\fp(x_0)&=&-{a^3\over3L^3}\sum_\bx
    \Big\langle P^a(x)\op^a\Big\rangle\\
    &=& -\frac{a^9}{3L^3}\sum_{\bx,\bu,\bv}
                \Big\langle\psibar_i^A(x)\dirac5^{AB}{\ts\frac12}\tau^a_{ij}
                \psi_j^{B}(x)
                \zetabar_m^C(\bu)\dirac5^{CD}{\ts\frac12}\tau^a_{mn}
                \zeta_n^{D}(\bv)\Big\rangle\;.
\eeann
This is of the same structure as for $f_1$ and hence
the only non--trivial Wick contraction is
$[\psi\zetabar]$, $[\zeta\psibar]$ with an isospin factor of $-3/2$
\bea
\fp(x_0)&=& \frac{a^9}{2L^3}\sum_{\bx,\bu,\bv}\Big\langle
\tr\Big\{\left[\psi(x)
        \zetabar(\bu)\right]_{\rm F}\dirac5\left[\zeta(\bv)
        \psibar(x)\right]_{\rm
        F}\dirac5\Big\}\Big\rangle_U\nonumber\\
&\use{use1}&\frac{a^6}{2L^{3/2}}\sum_{\bx,\bv}\Big\langle
\tr\Big\{\overline S(x)\dirac5\left[\zeta(\bv)
        \psibar(x)\right]_{\rm
        F}\dirac5\Big\}\Big\rangle_U\nonumber\\
&\use{use3}&\frac{a^3}2\sum_{\bx}\Big\langle
\tr\Big\{\overline S(x)\dirac5\dirac5\overline S(x)^\dagger
\dirac5\dirac5\Big\} \Big\rangle_U\nonumber\\
&=&\frac{a^3}2\sum_{\bx}\Big\langle
\tr\Big\{\overline S(x)\overline S(x)^\dagger\Big\}\Big\rangle_U\label{fPlat}\;.
\eea
The correlation function $\fa(x_0)$ (\ref{fA}) differs
from $\fp(x_0)$ only by an additional $\dirac0$ following
$\psibar(x)$. We therefore have
\bea
\fa(x_0)&=&\frac{a^6}{2L^{3/2}}\sum_{\bx,\bv}\Big\langle
\tr\Big\{\overline S(x)\dirac5\left[\zeta(\bv)
        \psibar(x)\right]_{\rm
        F}\dirac0\dirac5\Big\}\Big\rangle_U\nonumber\\
        &\use{use3}&\frac{a^3}2\sum_{\bx}\Big\langle
\tr\Big\{\overline S(x)\dirac5\dirac5\overline S(x)^\dagger
\dirac5\dirac0\dirac5\Big\}\Big\rangle_U\nonumber\\
&=&-\frac{a^3}2\sum_{\bx}\Big\langle
\tr\Big\{\overline S(x)\overline S(x)^\dagger\dirac0\Big\}\Big\rangle_U\;.\label{fAlat}
\eea
These results can also be used to evaluate $\gp(x_0)$ (\ref{gP})
if one makes the replacement
$$[\psibar(x),\psi(x),
                \zetabar(\bu),
                \zeta(\bv)]\quad\longrightarrow\quad[\zetabarprime(\bu),
\zetaprime(\bv),\psibar(x),\psi(x)]$$
in \eq{fPlat}. The result is
\begin{eqnarray}
\gp(x_0)&=& \frac{a^9}{2L^3}\sum_{\bx,\bu,\bv}\Big\langle
\tr\Big\{\left[\zetaprime(\bv)
        \psibar(x)\right]_{\rm F}\dirac5\left[\psi(x)
        \zetabarprime(\bu)\right]_{\rm
        F}\dirac5\Big\}\Big\rangle_U\nonumber\\
&\use{use2}&\frac{a^6}{2L^{3/2}}\sum_{\bx,\bv}\Big\langle
\tr\Big\{\left[\zetaprime(\bv)
        \psibar(x)\right]_{\rm F}\dirac5\overline R(x)
        \dirac5\Big\}\Big\rangle_U\nonumber\\
&\use{use4}&\frac{a^3}2\sum_{\bx}\Big\langle
\tr\Big\{\dirac5\overline R(x)^\dagger\dirac5\dirac5\overline R(x)
        \dirac5\Big\}\Big\rangle_U\nonumber\\
&=&\frac{a^3}2\sum_{\bx}\Big\langle
\tr\Big\{\overline R(x)\overline R(x)^\dagger\Big\}\Big\rangle_U\;.\label{gPlat}
\eea
Again the corresponding axial correlator is obtained by inserting $\dirac0$.
Due to the different sign in (\ref{gA}) we need to put a $\dirac0$ between
$\psibar$ and the following $\dirac5$ to obtain
\bea
\ga(x_0)&=& -\frac{a^6}{2L^{3/2}}\sum_{\bx,\bv}\Big\langle
\tr\Big\{\left[\zetaprime(\bv)
        \psibar(x)\right]_{\rm F}\dirac0\dirac5\overline R(x)
        \dirac5\Big\}\Big\rangle_U\nonumber\\
&\use{use4}&-\frac{a^3}2\sum_{\bx}\Big\langle
\tr\Big\{\dirac5\overline R(x)^\dagger\dirac5\dirac0\dirac5\overline R(x)
        \dirac5\Big\}\Big\rangle_U\nonumber\\
&=&\frac{a^3}2\sum_{\bx}\Big\langle
\tr\Big\{\overline R(x)\overline R^\dagger(x)\dirac0\Big\}\Big\rangle_U\;.
\label{gAlat}
\eea

\subsection{3--point functions}

Inserting equations (\ref{O}, \ref{Oprime}) into (\ref{fvdef}) yields
\bea
\nonumber   \fv(x_0) &=& {a^3\over6L^6}\sum_\bx i\epsilon^{abc}
\langle\oprime^aV_0^b(x)\op^c\rangle\\
\nonumber       &=& \frac{a^{15}}{6L^6}\!\!\sum_{\bx,\bu,\bv,\bw,\bz}\!\!i\epsilon^{abc}
                \Big\langle(\zetabarprime)_i^A(\bu)\dirac5^{AB}{\ts\frac12}\tau^a_{ij}
                (\zetaprime)_j^{B}(\bv)\\[-2mm]
                &&\qquad\quad\qquad\nonumber\qquad\quad\times\, \psibar_k^C(x)\dirac0^{CD}
                {\ts\frac12}\tau^b_{kl}\psi_l^D(x)\zetabar_m^E(\bw)
                \dirac5^{EF}{\ts\frac12}\tau^c_{mn}
                \zeta_n^{F}(\bz)\Big\rangle\;.
\eea
The only contractions with non-vanishing isospin-factors are those where no fields on
the same time-slice are paired, namely
\begin{displaymath}
\begin{array}{rclcl}
\psi\zetabarprime, \zeta\psibar,\zetaprime\zetabar &    \rightarrow & -\delta_{il}\delta_{kn}
\delta_{mj} & \rightarrow &
                        \ts-\frac18\epsilon^{abc}\tau^a_{lj}\tau^b_{nl}\tau^c_{jn}=-\frac18
                        \epsilon^{abc}\tr\tau^a\tau^c\tau^b=\frac32i\;,\\[1mm]
\zeta\zetabarprime, \zetaprime\psibar, \psi\zetabar &        \rightarrow & -\delta_{in}\delta_{kj}
\delta_{ml} & \rightarrow &
                        \ts-\frac18\epsilon^{abc}\tau^a_{nj}\tau^b_{jl}\tau^c_{ln}=-\frac18
                        \epsilon^{abc}\tr\tau^a\tau^b\tau^c=-\frac32i\;.
\end{array}
\end{displaymath}
Again the Dirac indices give the order of the propagators and the Dirac-matrices
\begin{eqnarray}
\nonumber \fv(x_0) \!\!\!\!&\!=\!&\!\!\! \frac{a^{15}}{4L^6}\!\!\!\!\sum_{\begin{array}{cc}\\[-9mm]
\scriptstyle\bx,\bu,\bv\\[-4mm]\scriptstyle
\bw,\bz\end{array}}\!\!\!\!\Big\langle -\tr\Big\{
        \left[\psi(x)\zetabarprime(\bu)\right]_{\rm F}\dirac5\left[\zeta'(\bv)
        \zetabar(\bw)\right]_{\rm F}\dirac5\left[\zeta(\bz)\psibar(x)\right]_{\rm F}
        \dirac0\Big\}\\[-7mm]\nonumber
        &&\hspace*{16.3mm}+\tr\Big\{
        \left[\zeta(\bw)\zetabarprime(\bv)\right]_{\rm F}\dirac5\left[\zeta'(\bu)
        \psibar(x)\right]_{\rm F}\dirac0\left[\psi(x)\zetabar(\bz)\right]_{\rm F}
        \dirac5\Big\}\Big\rangle_U\;,\\\nonumber
\end{eqnarray}
where we have renamed the arguments of the boundary fields ($\bu\leftrightarrow\bv$ and
$\bw\leftrightarrow\bz$) in the second trace under the sum. Using equations (\ref{use2})
and (\ref{use4}) the propagators from the bulk to the upper boundary can be written in
terms of $\overline R$ and $\overline R^\dagger$.
\begin{eqnarray}
\nonumber \fv(x_0) \!\!\!\!&\!=\!&\!\!\! \frac{a^{12}}{4L^{9/2}}\!\!\sum_{\begin{array}{cc}\\[-9mm]
\scriptstyle\bx,\bv\\[-4mm]\scriptstyle
\bw,\bz\end{array}}\!\!\Big\langle \tr\Big\{-
\overline R(x)\dirac5\left[\zeta'(\bv)
\zetabar(\bw)\right]_{\rm F}\dirac5\left[\zeta(\bz)\psibar(x)\right]_{\rm F}
\dirac0\\[-7mm]\nonumber
&&\qquad\quad \qquad\qquad+
\left[\zeta(\bw)\zetabarprime(\bv)\right]_{\rm F}\dirac5\dirac5\overline R(x)^\dagger\dirac5\dirac0
\left[\psi(x)\zetabar(\bz)\right]_{\rm F}
\dirac5\Big\}\Big\rangle_U\;.
\end{eqnarray}
In the next step we insert the expressions for the boundary-to-boundary propagators $\overline S_T$
and $\overline S^\dagger_T$ from equations (\ref{use5}) and (\ref{use6}),
respectively, and then use $\dirac5\dirac5=1$ in both traces
\begin{eqnarray}
\nonumber \fv(x_0) \!\!\!&\!=\!&\!\!\! \frac{a^{6}}{4L^{3}}\sum_{\bx,\bz}\Big\langle \tr\Big\{
\overline R(x)\dirac5\overline S_T\dirac5\left[\zeta(\bz)\psibar(x)\right]_{\rm F}\dirac0\\[-3mm]
\nonumber
&&\qquad\qquad\qquad-\dirac5\overline S_T^\dagger\dirac5\overline R(x)^\dagger\dirac5\dirac0
\left[\psi(x)\zetabar(\bz)\right]_{\rm F}
\dirac5\Big\}\Big\rangle_U\;.\\\nonumber
\end{eqnarray}
Again we use $\dirac5\dirac5=1$ and $\dirac5\dirac0=-\dirac0\dirac5$ as well as the
cyclic property of the trace. After inserting the propagators (\ref{use1}) and (\ref{use3})
we thus end up with
\begin{eqnarray}
\nonumber \fv(x_0) \!\!\!&\!=\!&\!\!\! \frac{a^{3}}{4L^{3/2}}\sum_{\bx}\Big\langle \tr\Big\{
\overline R(x)\dirac5\overline S_T\overline S(x)^\dagger\dirac5\dirac0+
\overline S_T^\dagger\dirac5\overline R(x)^\dagger\dirac0\dirac5\overline S(x)
\Big\}\Big\rangle_U\;.
\end{eqnarray}
We now see that the expression is of the form $\tr(X+X^\dagger)=2\re\tr X$ and the final result is therefore
\be
\nonumber \fv(x_0) = \frac{a^{3}}{2L^{3/2}}\sum_{\bx}\Big\langle \re\tr\Big\{
\overline S(x)^\dagger\dirac5\dirac0 \overline R(x)\dirac5\overline S_T\Big\}\Big\rangle_U\;.
\label{fVprelast}
\ee

\subsection{4--point functions}

\subsubsection{Preliminaries}

We start by
inserting the expression (\ref{AI}) for the improved axial current into (\ref{fAAI})
to arrive at
\begin{eqnarray}
\funci{AA}(x_0,y_0)\!\!\!\!&=&\!\!\!\!-\frac{a^6}{6L^6}\sum_{\bf x,\bf y}
\epsilon^{abc}\epsilon^{cde}
\Big\langle\oprime^d
(\aimpr)^a_{0}(x)(\aimpr)^b_{0}(y)
\op^e
\Big\rangle\nonumber\\
&=&\!\!\!\!-\frac{a^6}{6L^6}\sum_{\bf x,\bf y}
\epsilon^{abc}\epsilon^{cde}
\Big\langle\oprime^d
\Big[A^a_{0}(x)+a\cA\drvtilde0xP^a(x)\Big]\nonumber\\[-1mm]
&&\qquad\qquad\qquad\qquad\qquad\quad \times\Big[A^b_{0}(y)+a\cA\drvtilde0yP^b(y)\Big]
\op^e
\Big\rangle\nonumber\\
&=&\!\!\!\!-\frac{a^6}{6L^6}\sum_{\bf x,\bf y}
\epsilon^{abc}\epsilon^{cde}
\Big\langle\oprime^d
\Big[A^a_{0}(x)A^b_{0}(y)+a\cA\drvtilde0xP^a(x)A^b_{0}(y)\nonumber\\[-1mm]
&&\qquad\qquad\quad\quad +A^a_{0}(x)a\cA\drvtilde0yP^b(y)+
a\cA\drvtilde0xP^a(x)a\cA\drvtilde0yP^b(y)\Big]
\op^e
\Big\rangle\nonumber\\
&=&\!\!\!\!\func{AA}+a\cA[\drvtilde0x\func{PA}+\drvtilde0y\func{AP}]+a^2\cA^2\,
\drvtilde0x\drvtilde0y\func{PP}\;.\label{fAAI_decompose}
\eea
Here we have introduced the notation
\be
\func{XY}(x_0,y_0)\!=\!-\frac{a^6}{6L^6}
\sum_{\bf x,\bf y}
\epsilon^{abc}\epsilon^{cde}
\Big\langle\!\oprime^d
X^a(x)Y^b(y)
\op^e\!
\Big\rangle,\ X,Y\!\in\!\{A_0,P\}\;.\label{fXY}
\ee
In order to evaluate these correlation functions in a closed form we define
the Dirac matrices
\beann
\Xi=\left\{\begin{array}{l@{\ ,\quad}cl}
\dirac0\dirac5&\textrm{if}   &X\!=\!A_0\\[1mm]
\dirac5       &\textrm{if}   &X\!=\!P
\end{array}\right.,\qquad
\Upsilon=\left\{\begin{array}{l@{\ ,\quad}cl}
\dirac0\dirac5 &\textrm{if}   &Y\!=\!A_0\\[1mm]
\dirac5        &\textrm{if}   &Y\!=\!P
\end{array}\right.,\quad\textrm{so that}
\eeann
\be
X^a(x)=\psibar(x)\Xi\ts\frac12\tau^a\psi(x)\ \textrm{ and }\
Y^a(y)=\psibar(y)\Upsilon\ts\frac12\tau^a\psi(y)\;.\label{XYa}
\ee

\subsubsection{Wick contractions for $\func{XY}$}

As before the first step is the insertion of the explicit expressions
for the boundary fields (\ref{O}, \ref{Oprime}) and the currents/densities
(\ref{XYa}) into (\ref{fXY}). We suppress the time arguments of the correlation function
to arrive at
\bea
\nonumber   \func{XY} \!\!\!&=&\!\!\! -\frac{a^{18}}{6L^6}\!\!\!\sum_{\begin{array}{cc}\\[-9mm]
\scriptstyle\bx,\by,\bu\\[-3.5mm]\scriptstyle
\bv,\bw,\bz\end{array}}\!\!\!\epsilon^{abc}\epsilon^{cde}
                \Big\langle(\zetabarprime)_i^A(\bu)\dirac5^{AB}{\ts\frac12}\tau^d_{ij}
                (\zetaprime)_j^{B}(\bv)\psibar_k^C(x)\Xi^{CD}
                {\ts\frac12}\tau^a_{kl}\psi_l^D(x)\\[-7mm]
                &&\qquad\qquad\qquad\nonumber\qquad\quad\times\,
                \psibar_s^E(y)\Upsilon^{EF}{\ts\frac12}\tau^b_{st}\psi_t^F(y)
                \zetabar_m^G(\bw)
                \dirac5^{GH}{\ts\frac12}\tau^e_{mn}
                \zeta_n^{H}(\bz)\Big\rangle\;.
\eea
Since we have two currents/densities in the bulk we need to keep track of
the space--time locations, which we now write as indices.
The number of possible Wick contractions is significantly larger
than for the two-- or three--point
functions.

\EPSFIGURE[!bt]{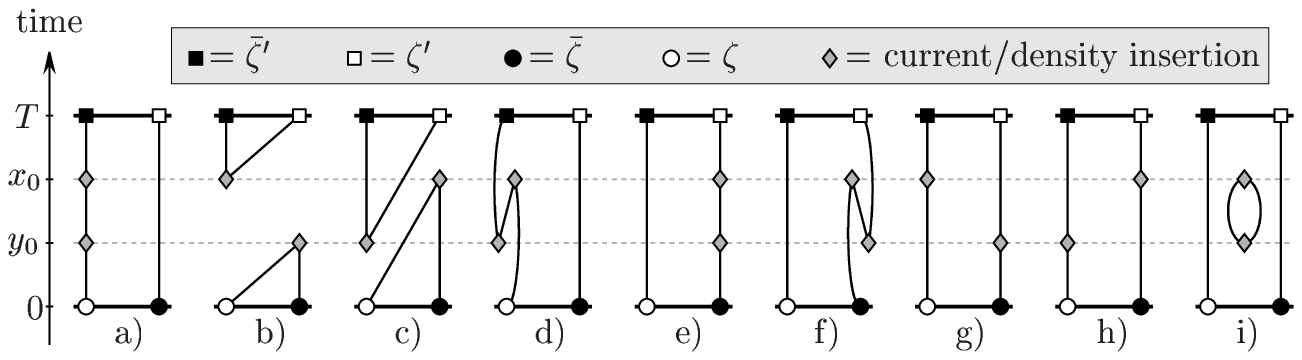,width=13cm}{
Wick contractions of the correlation function
$\func{XY}(x_0,y_0)$. The gray diamonds indicate the insertions
of $Y$ and $X$ at times $y_0$ and $x_0$.
\label{wick_contract}}

\begin{displaymath}
\begin{array}{rr@{,\,}r@{,\,}r@{,\,}r@{\qquad}rrr@{,\,}r@{,\,}r@{,\,}r}
a)& \psi_x\zetabarprime_\bu&\psi_y\psibar_x            &\zeta_\bz\psibar_y     &\zetaprime_\bv\zetabar_\bw&
f)& \zeta_\bz\zetabarprime_\bu &\psi_y\psibar_x        &\zetaprime_\bv\psibar_y&\psi_x\zetabar_\bw\\[1mm]
b)& \psi_x\zetabarprime_\bu&\zetaprime_\bv\psibar_x    &\zeta_\bz\psibar_y     &\psi_y\zetabar_\bw&
g)& \psi_x\zetabarprime_\bu&\zeta_\bz\psibar_x         &\zetaprime_\bv\psibar_y&\psi_y\zetabar_\bw\\[1mm]
c)& \psi_y\zetabarprime_\bu&\zeta_\bz\psibar_x         &\zetaprime_\bv\psibar_y&\psi_x\zetabar_\bw&
h)& \psi_y\zetabarprime_\bu&\zetaprime_\bv\psibar_x    &\zeta_\bz\psibar_y     &\psi_x\zetabar_\bw\\[1mm]
d)& \psi_y\zetabarprime_\bu&\zeta_\bz\psibar_x         &\psi_x\psibar_y        &\zetaprime_\bv\zetabar_\bw&
i)& \zeta_\bz\zetabarprime_\bu &\psi_y\psibar_x        &\psi_x\psibar_y        &\zetaprime_\bv\zetabar_\bw\\[1mm]
e)& \zeta_\bz\zetabarprime_\bu &\zetaprime_\bv\psibar_x&\psi_x\psibar_y        &\psi_y\zetabar_\bw
\end{array}
\end{displaymath}

\subsubsection{Isospin factors}
The calculation of the isospin factors corresponding to the
contractions a) to i) is lengthy but straightforward. With
the notation $\tr(\tau^a\ldots\tau^e)=\tr_{a\ldots e}$ the result is
\begin{displaymath}
\begin{array}{r@{\ }c@{\ \ }l@{\ \ }r@{\,}r@{\,}lr@{\,}r}
a)&-\delta^{il}\delta^{kt}\delta^{sn}\delta^{mj}&\Rightarrow&-\ts\frac1{16}\epsilon^{abc}\epsilon^{cde}
\tau^d_{lj}\tau^a_{tl}\tau^b_{nt}\tau^e_{jn}=&-\ts\frac1{16}\epsilon^{abc}\epsilon^{cde}
&\tr_{deba}&=&-3/2\\[1mm]
b)&+\delta^{il}\delta^{kj}\delta^{sn}\delta^{mt}&\Rightarrow&\ts\frac1{16}\epsilon^{abc}\epsilon^{cde}
\tau^d_{lj}\tau^a_{jl}\tau^b_{nt}\tau^e_{tn}=&\ts\frac1{16}\epsilon^{abc}\epsilon^{cde}
&\tr_{da}\tr_{be}&=&3/2\\[1mm]
c)&+\delta^{it}\delta^{kn}\delta^{sj}\delta^{ml}&\Rightarrow&\ts\frac1{16}\epsilon^{abc}\epsilon^{cde}
\tau^d_{tj}\tau^a_{nl}\tau^b_{jt}\tau^e_{ln}=&\ts\frac1{16}\epsilon^{abc}\epsilon^{cde}
&\tr_{db}\tr_{ae}&=&-3/2\\[1mm]
d)&-\delta^{it}\delta^{kn}\delta^{sl}\delta^{mj}&\Rightarrow&-\ts\frac1{16}\epsilon^{abc}\epsilon^{cde}
\tau^d_{tj}\tau^a_{nl}\tau^b_{lt}\tau^e_{jn}=&-\ts\frac1{16}\epsilon^{abc}\epsilon^{cde}
&\tr_{deab}&=&3/2\\[1mm]
e)&-\delta^{in}\delta^{kj}\delta^{sl}\delta^{mt}&\Rightarrow&-\ts\frac1{16}\epsilon^{abc}\epsilon^{cde}
\tau^d_{nj}\tau^a_{jl}\tau^b_{lt}\tau^e_{tn}=&-\ts\frac1{16}\epsilon^{abc}\epsilon^{cde}
&\tr_{dabe}&=&-3/2\\[1mm]
f)&-\delta^{in}\delta^{kt}\delta^{sj}\delta^{ml}&\Rightarrow&-\ts\frac1{16}\epsilon^{abc}\epsilon^{cde}
\tau^d_{nj}\tau^a_{tl}\tau^b_{jt}\tau^e_{ln}=&-\ts\frac1{16}\epsilon^{abc}\epsilon^{cde}
&\tr_{dbae}&=&3/2\\[1mm]
g)&-\delta^{il}\delta^{kn}\delta^{sj}\delta^{mt}&\Rightarrow&-\ts\frac1{16}\epsilon^{abc}\epsilon^{cde}
\tau^d_{lj}\tau^a_{nl}\tau^b_{jt}\tau^e_{tn}=&-\ts\frac1{16}\epsilon^{abc}\epsilon^{cde}
&\tr_{dbea}&=&0\\[1mm]
h)&-\delta^{it}\delta^{kj}\delta^{sn}\delta^{ml}&\Rightarrow&-\ts\frac1{16}\epsilon^{abc}\epsilon^{cde}
\tau^d_{tj}\tau^a_{jl}\tau^b_{nt}\tau^e_{ln}=&-\ts\frac1{16}\epsilon^{abc}\epsilon^{cde}
&\tr_{daeb}&=&0\\[1mm]
i)&+\delta^{in}\delta^{kt}\delta^{sl}\delta^{mj}&\Rightarrow&\ts\frac1{16}\epsilon^{abc}\epsilon^{cde}
\tau^d_{nj}\tau^a_{tl}\tau^b_{lt}\tau^e_{jn}=&\ts\frac1{16}\epsilon^{abc}\epsilon^{cde}
&\tr_{de}\tr_{ab}&=&0\;.\hspace*{-2.33mm}
\end{array}
\end{displaymath}
Combining this result with the contractions gives the propagators that need to be evaluated.
Keeping track of the Dirac indices gives the order of the propagators in the non--vanishing
contractions
\begin{displaymath}
\begin{array}{rr}
a)& \ts-\frac32\tr\left\{\F{\psi_x\zetabarprime_\bu}\dirac5\F{\zetaprime_\bv\zetabar_\bw}\dirac5
        \F{\zeta_\bz\psibar_y}\Upsilon\F{\psi_y\psibar_x}\Xi\right\}\\[2mm]
b)& \ts\frac32\tr\left\{\F{\psi_x\zetabarprime_\bu}\dirac5\F{\zetaprime_\bv\psibar_x}\Xi\right\}
    \tr\left\{\F{\zeta_\bz\psibar_y}\Upsilon\F{\psi_y\zetabar_\bw}\dirac5\right\}\\[2mm]
c)& \ts-\frac32\tr\left\{\F{\psi_y\zetabarprime_\bu}\dirac5\F{\zetaprime_\bv\psibar_y}\Upsilon\right\}
        \tr\left\{\F{\zeta_\bz\psibar_x}\Xi\F{\psi_x\zetabar_\bw}\dirac5\right\}\\[2mm]
d)& \ts\frac32\tr\left\{\F{\psi_y\zetabarprime_\bu}\dirac5\F{\zetaprime_\bv\zetabar_\bw}\dirac5
        \F{\zeta_\bz\psibar_x}\Xi\F{\psi_x\psibar_y}\Upsilon\right\}\\[2mm]
e)& \ts-\frac32\tr\left\{\F{\zeta_\bz\zetabarprime_\bu}\dirac5\F{\zetaprime_\bv\psibar_x}\Xi
        \F{\psi_x\psibar_y}\Upsilon\F{\psi_y\zetabar_\bw}\dirac5\right\}\\[2mm]
f)& \ts\frac32\tr\left\{\F{\zeta_\bz\zetabarprime_\bu}\dirac5\F{\zetaprime_\bv\psibar_y}\Upsilon
        \F{\psi_y\psibar_x}\Xi\F{\psi_x\zetabar_\bw}\dirac5\right\}\;.
\end{array}
\end{displaymath}

\newpage

\subsubsection{The general correlator $\func{XY}$}

\noindent If we keep the arguments of the fields as indices, the correlation function
$f_{XY}$ can now be written as
\beann
\nonumber   \func{XY} \!\!\!&=&\!\!\! \frac{a^{18}}{4L^6}\!\!\!\sum_{\begin{array}{cc}\\[-9.5mm]
\scriptstyle\bx,\by,\bu\\[-3.5mm]\scriptstyle
\bv,\bw,\bz\end{array}}\!\!\!
\Big\langle\ \ 
\tr\left\{\F{\psi_x\zetabarprime_\bu}\dirac5\F{\zetaprime_\bv\zetabar_\bw}\dirac5
                \F{\zeta_\bz\psibar_y}\Upsilon\F{\psi_y\psibar_x}\Xi\right\}\\[-4.5mm]
&&\quad\qquad\quad\, \ts-\tr\left\{\F{\psi_x\zetabarprime_\bu}\dirac5\F{\zetaprime_\bv\psibar_x}\Xi\right\}
            \tr\left\{\F{\zeta_\bz\psibar_y}\Upsilon\F{\psi_y\zetabar_\bw}\dirac5\right\}\\[2mm]
&&\qquad\quad\,\quad+\tr\left\{\F{\psi_y\zetabarprime_\bu}\dirac5\F{\zetaprime_\bv\psibar_y}\Upsilon\right\}
        \tr\left\{\F{\zeta_\bz\psibar_x}\Xi\F{\psi_x\zetabar_\bw}\dirac5\right\}\\[2mm]
&&\quad\qquad\quad\,-\tr\left\{\F{\psi_y\zetabarprime_\bu}\dirac5\F{\zetaprime_\bv\zetabar_\bw}\dirac5
        \F{\zeta_\bz\psibar_x}\Xi\F{\psi_x\psibar_y}\Upsilon\right\}\\[2mm]
&&\qquad\quad\, \quad+\tr\left\{\F{\zeta_\bz\zetabarprime_\bu}\dirac5\F{\zetaprime_\bv\psibar_x}\Xi
        \F{\psi_x\psibar_y}\Upsilon\F{\psi_y\zetabar_\bw}\dirac5\right\}\\[1mm]
&&\quad\qquad\quad\, -\tr\left\{\F{\zeta_\bz\zetabarprime_\bu}\dirac5\F{\zetaprime_\bv\psibar_y}\Upsilon
        \F{\psi_y\psibar_x}\Xi\F{\psi_x\zetabar_\bw}\dirac5\right\}\Big\rangle_U\;.
\eeann
Using eqs.~(\ref{eins}), (\ref{use1}), (\ref{use4}), (\ref{use5}) and (\ref{use6})
as well as the properties of the
Dirac matrices, one arrives at
\begin{eqnarray*}
\nonumber   \func{XY} \!\!\!&=&\!\!\! \frac{a^{6}}{4L^{3/2}}\sum_{\bx,\by}
\Big\langle
-\tr\left\{\overline R(x)\dirac5\overline S_T
        \overline S(y)^\dagger\dirac5\Upsilon S(y;x)\Xi\right\}\\[-2.5mm]
&&\qquad\qquad\quad\, \ts-L^{3/2}\tr\left\{\overline R(x)\overline R(x)^\dagger\dirac5\Xi\right\}
    \tr\left\{\overline S(y)^\dagger\dirac5\Upsilon\overline S(y)\right\}\\[2mm]
&&\qquad\qquad\,\quad+L^{3/2}\tr\left\{\overline R(y)\overline R(y)^\dagger\dirac5\Upsilon\right\}
    \tr\left\{\overline S(x)^\dagger\dirac5\Xi\overline S(x)\right\}\\[2mm]
&&\qquad\qquad\quad\,+\tr\left\{\overline R(y)\dirac5\overline S_T
    \overline S(x)^\dagger\dirac5\Xi\, S(x;y)\Upsilon\right\}\\[2mm]
&&\qquad\qquad\, \quad-\tr\left\{\overline S_T^\dagger\dirac5\overline R(x)^\dagger\dirac5\Xi
    \,\dirac5S(y;x)^\dagger\dirac5\Upsilon\overline S(y)\right\}\\[1mm]
&&\qquad\qquad\quad\, +\tr\left\{\overline S_T^\dagger\dirac5\overline R(y)^\dagger\dirac5\Upsilon
    \dirac5S(x;y)^\dagger\dirac5\Xi\,\overline S(x)\right\}\Big\rangle_U\;.
\end{eqnarray*}
If we now specialize to the four cases $\func{PP}$, $\func{AA}$, $\func{AP}$
and $\func{PA}$, the
connected contributions will combine to form pairs of Hermitian conjugates.

\vspace*{6mm}

\fbox{$\func{PP}\Rightarrow \Xi=\Upsilon=\dirac5$}
\begin{eqnarray}
\nonumber   \func{PP} \!\!\!&=&\!\!\! \frac{a^{6}}{4L^{3/2}}\sum_{\bx,\by}
\Big\langle
-2\re\tr\left\{\overline R(x)\dirac5\overline S_T
        \overline S(y)^\dagger S(y;x)\dirac5\right\}\\[-2mm]
&&\qquad\qquad\quad\, \ts-L^{3/2}\tr\left\{\overline R(x)\overline R(x)^\dagger\right\}
    \tr\left\{\overline S(y)\overline S(y)^\dagger\right\}\nonumber\\[2mm]
&&\qquad\qquad\,\quad+L^{3/2}\tr\left\{\overline R(y)\overline R(y)^\dagger\right\}
    \tr\left\{\overline S(x)\overline S(x)^\dagger\right\}\nonumber\\[2mm]
&&\qquad\qquad\quad\,+2\re\tr\left\{\overline R(y)\dirac5\overline S_T
    \overline S(x)^\dagger S(x;y)\dirac5\right\}\Big\rangle_U\;.\label{fppfinal}
\end{eqnarray}

\fbox{$\func{AA}\Rightarrow \Xi=\Upsilon=\dirac0\dirac5$}
\begin{eqnarray}
\nonumber   \func{AA} \!\!\!&=&\!\!\! \frac{a^{6}}{4L^{3/2}}\sum_{\bx,\by}
\Big\langle
+2\re\tr\left\{\overline R(x)\dirac5\overline S_T
\overline S(y)^\dagger\dirac0 S(y;x)\dirac0\dirac5\right\}\\[-2mm]
&&\qquad\qquad\quad\, \ts-L^{3/2}\tr\left\{\overline R(x)\overline R(x)^\dagger\dirac0\right\}
\tr\left\{\overline S(y)\overline S(y)^\dagger\dirac0\right\}\nonumber\\[2mm]\nonumber
&&\qquad\qquad\,\quad+L^{3/2}\tr\left\{\overline R(y)\overline R(y)^\dagger\dirac0\right\}
\tr\left\{\overline S(x)\overline S(x)^\dagger\dirac0\right\}\\[2mm]
&&\qquad\qquad\quad\,-2\re\tr\left\{\overline R(y)\dirac5\overline S_T
\overline S(x)^\dagger\dirac0\, S(x;y)\dirac0\dirac5\right\}\Big\rangle_U\;.\label{faafinal}
\end{eqnarray}

\fbox{$\func{AP}\Rightarrow \Xi=\dirac0\dirac5,\, \Upsilon=\dirac5$}
\begin{eqnarray}
\nonumber   \func{AP} \!\!\!&=&\!\!\! \frac{a^{6}}{4L^{3/2}}\sum_{\bx,\by}
\Big\langle
-2\re\tr\left\{\overline R(x)\dirac5\overline S_T
        \overline S(y)^\dagger S(y;x)\dirac0\dirac5\right\}\\[-2mm]\nonumber
&&\qquad\qquad\quad\, \ts+L^{3/2}\tr\left\{\overline R(x)\overline R(x)^\dagger\dirac0\right\}
    \tr\left\{\overline S(y)\overline S(y)^\dagger\right\}\\[2mm]\nonumber
&&\qquad\qquad\,\quad-L^{3/2}\tr\left\{\overline R(y)\overline R(y)^\dagger\right\}
    \tr\left\{\overline S(x)\overline S(x)^\dagger\dirac0\right\}\\[2mm]
&&\qquad\qquad\quad\,-2\re\tr\left\{\overline R(y)\dirac5\overline S_T
    \overline S(x)^\dagger\dirac0 S(x;y)\dirac5\right\}\Big\rangle_U\;.\label{fapfinal}
\end{eqnarray}

\fbox{$\func{PA}\Rightarrow \Xi=\dirac5,\, \Upsilon=\dirac0\dirac5$}
\begin{eqnarray}
\nonumber   \func{PA} \!\!\!&=&\!\!\! \frac{a^{6}}{4L^{3/2}}\sum_{\bx,\by}
\Big\langle
+2\re\tr\left\{\overline R(x)\dirac5\overline S_T
    \overline S(y)^\dagger\dirac0S(y;x)\dirac5\right\}\\[-2mm]\nonumber
&&\qquad\qquad\quad\, \ts+L^{3/2}\tr\left\{\overline R(x)\overline R(x)^\dagger\right\}
    \tr\left\{\overline S(y)\overline S(y)^\dagger\dirac0\right\}\\[2mm]\nonumber
&&\qquad\qquad\,\quad-L^{3/2}\tr\left\{\overline R(y)\overline R(y)^\dagger\dirac0\right\}
    \tr\left\{\overline S(x)\overline S(x)^\dagger\right\}\\[2mm]\label{fpafinal}
&&\qquad\qquad\quad\,+2\re\tr\left\{\overline R(y)\dirac5\overline S_T
    \overline S(x)^\dagger S(x;y)\dirac0\dirac5\right\}\Big\rangle_U\;.
\end{eqnarray}

\subsection*{Evaluation of $\tfunci{PA}$}

Inserting the expression (\ref{AI}) for the improved axial current into (\ref{ftilde}) gives
\begin{eqnarray}
\tfunci{PA}(y_0\pl t,y_0)\!\!\!\!&\!\!\!=\!\!\!&\!\!\!\!-\frac{a^7}{6L^6}\sum_{x_0=y_0}^{y_0+t}
\!w(x_0)\sum_{\bx,\by}\,\epsilon^{abc}
\epsilon^{cde}
\Big\langle\oprime^d
P^a(x)[A_0^b(y)+a\cA\drvtilde0yP^b(y)]
\op^e
\Big\rangle\nonumber\\
&\!\!\!\use{fXY}\!\!\!&\!a\!\sum_{x_0=y_0}^{y_0+t}\!w(x_0)\Big[\func{PA}(x_0,y_0)+a\cA \drvtilde0y\func{PP}(x_0,y_0)\Big]
\label{ftilde_final}
\;.
\end{eqnarray}

\pagebreak
\section{Simplifying the correlation functions}
\label{app:simpl}
We now specify an explicit (chiral) representation of the Euclidean Dirac matrices
\begin{displaymath}
\dirac\mu=\left(\!\!
\begin{array}{cc}
0   \!&\!   e_\mu\\
e_\mu^\dagger   \!&\!   0
\end{array}\!\!\right), \
\textrm{with the }2\!\times\!2\textrm{ matrices }\ e_0=-1,\quad e_k=-i\sigma_k\;,
\end{displaymath}
where $k=1,2,3$ and $\sigma_k$ are the Pauli matrices. With
$\dirac5=\dirac0\dirac1\dirac2\dirac3$ and \linebreak
$P_\pm=\frac12(1\pm\dirac0)$ we then have
$$
\dirac5=\left(\!\!
\begin{array}{rr}
\,1       \!&\!   0\\
\,0   \!&\!   -1
\end{array}\!\!\right)\;,\
P_+=\frac12\left(\!\!
\begin{array}{rr}
1       \!&\!   -1\\
-1   \!&\!   1
\end{array}\!\!\right)\ \textrm{ and }\
P_-=\frac12\left(\!\!
\begin{array}{rr}
1   &   1\\
1   &   1
\end{array}\!\!\right)\;.
$$
If we write the four components explicitly we see that for all matrices $X,Y$
in Dirac space we can conclude that
\bea
0=XP_+&\Rightarrow& \ts0=2(XP_+)_{ A \ast}=(X_{ A 1}\!-\!X_{ A
3},\, X_{ A 2}\!-\!X_{ A 4},\nonumber \, X_{ A 3}\!-\!X_{ A 1}
,\, X_{ A 4}\!-\!X_{ A 2})\\\label{proj1}
&\Rightarrow& \underline{\ts0=X_{ A 1}\!-\!X_{ A 3}=X_{ A 2}\!-\!X_{ A 4}}\;,\\[2mm]
0=XP_-&\Rightarrow& \ts0=2(XP_-)_{ A \ast}=(X_{ A 1}\!+\!X_{ A
3},\, X_{ A 2}\!+\!X_{ A 4},\nonumber \, X_{ A 1}\!+\!X_{ A 3}
,\, X_{ A 2}\!+\!X_{ A 4})\\\label{proj2}
&\Rightarrow& \underline{\ts0=X_{ A 1}\!+\!X_{ A 3}=X_{ A 2}\!+\!X_{ A 4}}\;,\\[2mm]
0=P_+Y&\Rightarrow& \ts0=2(P_+Y)_{\ast B }=(Y_{1 B }\!-\!Y_{3
B },\ Y_{2 B }\!-\!Y_{4 B },\nonumber \ Y_{3 B }\!-\!Y_{1 B }
,\ Y_{4 B }\!-\!Y_{2 B })^T\\\label{proj3}
&\Rightarrow& \underline{\ts0=Y_{1 B }\!-\!Y_{3 B }=Y_{2 B }\!-\!Y_{4 B }}\;,\\[2mm]
0=P_-Y&\Rightarrow& \ts0=2(P_-Y)_{\ast B }=(Y_{1 B }\!+\!Y_{3
B },\ Y_{2 B }\!+\!Y_{4 B },\nonumber \ Y_{1 B }\!+\!Y_{3 B }
,\ Y_{2 B }\!+\!Y_{4 B })^T\\\label{proj4} &\Rightarrow&
\underline{\ts0=Y_{1 B }\!+\!Y_{3 B }=Y_{2 B }\!+\!Y_{4 B }}\;.
\eea
From equations (\ref{sbar}) - (\ref{sbarT}) we can now derive relations between
the Dirac components of the summed
correlators (only Dirac indices are written explicitly)
\begin{eqnarray}
\label{trick1}0\use{sbar}\overline SP_-\ &\ \Rightarrow\ &
    \ts0\use{proj2}\overline S_{ A 1}\!+\!\overline S_{ A 3}=\overline S_{ A 2}\pl
    \overline S_{ A 4}\;,\\[1mm]
\label{trick0}0\use{sbardag}P_-\overline S^\dagger&\ \Rightarrow\ &
    \ts0\use{proj4}(\overline S^\dagger)_{1B}\!+\!(\overline S^\dagger)_{3B}=(\overline S^\dagger)_{2B}\!+
    \!(\overline S^\dagger)_{4B}\;,\\[1mm]
\label{trick2}0\use{rbar}\overline RP_+\,&\ \Rightarrow\ &
    \ts0\use{proj1}\overline R_{ A 1}\!-\!\overline R_{ A 3}=\overline R_{ A 2}\!-\!\overline R_{ A 4}\;,\\[1mm]
\label{trick2a}0\use{rbardag}P_+\overline R^\dagger\!&\ \Rightarrow\ &
    \ts0\use{proj3}(\overline R^\dagger)_{ 1 B}\!-\!(\overline R^\dagger)_{ 3 B}=(\overline R^\dagger)_{ 2 B}\!-
    \!(\overline R^\dagger)_{ 4 B}\;,\\[1mm]
\label{trick3}0\use{sbar}\overline S_TP_-\!\!&\ \Rightarrow\ &
    \ts0\use{proj2}(\overline S_T)_{ A 1}\!+\!(\overline S_T)_{ A 3}=(\overline S_T)_{ A 2}\!+
    \!(\overline S_T)_{ A 4}\;,\\[1mm]
\label{trick4}0\use{sbarT}P_-\overline S_T&\ \Rightarrow\ &
    \ts0\use{proj4}(\overline S_T)_{1 B }\!+\!(\overline S_T)_{3 B }=(\overline S_T)_{2 B }\!+
    \!(\overline S_T)_{4 B }\;,\\[1mm]
0\use{sbarT}P_-\overline S_T&\ \Rightarrow\ &\dirac5P_-\overline S_T=0\ \Rightarrow\ P_+
(\dirac5\overline S_T)=0\nonumber\\[-11mm]\nonumber
\end{eqnarray}
\bea
&\ \Rightarrow\ &\label{trick5}
    \ts0\use{proj3}(\dirac5\overline S_T)_{1 B }\!-\!(\dirac5\overline S_T)_{3 B }=
    (\dirac5\overline S_T)_{2 B }\!-\!(\dirac5\overline S_T)_{4 B }\;.
\eea

\subsection{Explicit form of the correlation functions}

\noindent\fbox{$f_1$}\\[3mm]
We use $A,B,\ldots$ for the Dirac indices and $\alpha,\beta,\ldots$
for color indices.
\bea
\tr \overline S_T\overline S_T^\dagger&=&\sum_{ A , B
=1}^4\sum_{\alpha,\beta=1}^3 \left(\overline S_T\right)^{\alpha \beta}_{ A  B }
    \left(\overline S_T^\dagger\right)^{\beta\alpha}_{ B  A }
=\sum_{ A , B =1}^4\sum_{\alpha,\beta=1}^3 \left(\overline
S_T\right)^{\alpha \beta}_{ A  B }
        \left\{\left(\overline S_T\right)^{\alpha\beta}_{ A  B }\right\}^*\nonumber\\
    &=&\sum_{ A , B =1}^4\sum_{\alpha,\beta=1}^3 \left|\left(\overline S_T\right)^{\alpha \beta}_{ A  B }\right|^2
    \use{trick3}2\sum_{ A =1}^4\sum_{ B =1}^2\sum_{\alpha,\beta=1}^3
    \left|\left(\overline S_T\right)^{\alpha \beta}_{ A  B }\right|^2\nonumber\\
    &\use{trick4}&4\sum_{ A , B =1}^2\sum_{\alpha,\beta=1}^3
    \left|\left(\overline S_T\right)^{\alpha \beta}_{ A  B }\right|^2\;.\label{trick6}
\eea
The final form for the correlation function $f_1$ from (\ref{f1prelast}) is therefore
\bea
f_1&\use{trick6}&\frac2{L^3}\left\langle\sum_{ A , B
=1}^2\sum_{\alpha,\beta=1}^3 \left|\left(\overline S_T\right)^{\alpha\beta}_{ A
 B }\right|^2\right\rangle_U\label{f1last}\;.
\eea
For the more complicated correlation functions it will be useful to introduce
a modified sum convention, where all Dirac indices that are restricted
to $\{1,2\}$ are written in square brackets. In this notation we have
\be
f_1=\frac2{L^3}\left\langle\left(\overline S_T\right)^{\alpha\beta}_{[A][B]}
 \left(\overline S_T^\dagger\right)^{\beta\alpha}_{ [B][A]}
 \right\rangle_U\;.
\ee

\noindent\fbox{$\fv(x_0)$}\\[3mm]
The correlation function $\fv(x_0)$ from ($\ref{fVprelast}$)
contains the trace of $\overline S^\dagger\dirac5\dirac0\overline
R\dirac5\overline S_T$. Using equation (\ref{trick2})
and (\ref{trick5}) one can see that the Dirac index contracting
$\overline R$ and $\dirac5\overline S_T$ can be restricted to
$\{1,2\}$ if a factor of $2$ is included. Since in this representation
$\dirac5$ is equal to the unit matrix if the indices are restricted
to $\{1,2\}$ the $\dirac5$ in front of $\overline S_T$ can be dropped, i.e.
\begin{displaymath}
\ldots\overline R_{AB}(\dirac5\overline S_T)_{BC}\ldots\ =\
\ldots2\overline R_{A[B]} (\overline S_T)_{[B]C}\ldots
\end{displaymath}
Similarly, using equations (\ref{trick0}) and (\ref{trick3}) the
contracting Dirac index between $\overline S_T$ and
$\overline S^\dagger$ can be restricted to $\{1,2\}$ if another
factor of $2$ is included
\begin{displaymath}
\ldots(\overline S_T)_{AB}(\overline S^\dagger)_{BC}\ldots\ =\
\ldots2(\overline S_T)_{A[B]}
(\overline S^\dagger)_{[B]C}\ldots
\end{displaymath}
The result for $\fv(x_0)$ is then
\bea
\fv(x_0)\!\!\!&=&\!\!\!\frac{2a^3}{L^{3/2}}\left\langle\sum_\bx \re
\left[(\overline S(x)^\dagger)_{[A]B}^{\alpha\beta}
(\dirac5\dirac0)_{BC}\overline R(x)^{\beta\gamma}_{C[D]}
(\overline S_T)_{[D][A]}^{\gamma\alpha}\right]\right\rangle_U\nonumber\;.
\eea
Since the combination $\overline S\cdot\overline S_T^\dagger$ will appear
in all $f_{XY}$ as well as $f_V$, we define
\begin{equation}
\overline P(x)^{\beta\gamma}_{B[D]}=\frac12\left[\overline S(x)
\left(\overline S_T\right)^\dagger\right]^{\beta\gamma}_{B[D]}
=\overline S(x)_{B[A]}^{\beta\alpha}\left[\left(\overline S_T
\right)^\dagger\right]^{\alpha\gamma}_{[A][D]}\;.\label{pbar}
\end{equation}
The final result for $\fv(x_0)$ is then
\begin{eqnarray}
\fv(x_0)\!\!\!\!\!&=&\!\!\!\frac{2a^3}{L^{3/2}}\left\langle\sum_\bx \re\left\{
(\overline S_T)_{[D][A]}^{\gamma\alpha}(\overline S(x)^\dagger)_{[A]B}^{\alpha\beta}
(\dirac5\dirac0)_{BC}\overline R(x)^{\beta\gamma}_{C[D]}
\right\}\right\rangle_U\nonumber\\
&\use{pbar}&\!\!\!\frac{2a^3}{L^{3/2}}\left\langle\sum_\bx \re\left\{
\left[\,\overline P(x)^\dagger\right]_{[D]B}^{\gamma\beta}
(\dirac5\dirac0)_{BC}\overline R(x)^{\beta\gamma}_{C[D]}
\right\}\right\rangle_U\;.
\end{eqnarray}

\noindent\fbox{$\fa$, $\fp$, $\ga$, $\gp$}\\[3mm]
In addition to the above simplifications we also have
\begin{eqnarray}
(\ref{trick1}) \wedge (\ref{trick0}) &\rightarrow&  \ldots\overline S_{AB}
(\overline S^\dagger)_{BC}\ldots\ = \
\ldots2\overline S_{A[B]} (\overline S^\dagger)_{[B]C}\ldots\label{newtrick1}\\
(\ref{trick2}) \wedge (\ref{trick2a}) &\rightarrow&  \ldots\overline R_{AB}
(\overline R^\dagger)_{BC}\ldots\ = \
\ldots2\overline R_{A[B]} (\overline R^\dagger)_{[B]C}\ldots\label{newtrick2}
\end{eqnarray}
With these the correlators
$\fp$, $\fa$, $\gp$ and $\ga$ can be written as
\bea
\fp(x_0)&\use{fPlat}&{a^3}\sum_{\bx}\Big\langle
\overline S(x)_{A[B]}^{\alpha\beta}\Big(\overline S(x)^\dagger\Big)_{[B]A}
^{\beta\alpha}\Big\rangle_U\label{fPsimp}\\[1mm]
\fa(x_0)&\use{fAlat}&-{a^3}\sum_{\bx}\Big\langle
\overline S(x)_{A[B]}^{\alpha\beta}\Big(\overline S(x)^\dagger\Big)_{[B]C}
^{\beta\alpha}(\dirac0)_{CA}
\Big\rangle_U\label{fAsimp}\\[1mm]
\gp(x_0)&\use{gPlat}&{a^3}\sum_{\bx}\Big\langle
\overline R(x)_{A[B]}^{\alpha\beta}\Big(\overline R(x)^\dagger\Big)_{[B]A}
^{\beta\alpha}\Big\rangle_U\label{gPsimp}\\[1mm]
\ga(x_0)&\use{gAlat}&{a^3}\sum_{\bx}\Big\langle
\overline R(x)_{A[B]}^{\alpha\beta}\Big(\overline R(x)^\dagger\Big)_{[B]C}
^{\beta\alpha}(\dirac0)_{CA}\Big\rangle_U\label{gAsimp}\;.
\eea

\noindent\fbox{$\func{XY}(x_0,y_0)$}\\[3mm]
The expressions for $\func{XY}$ can be written in the same form as $\fv$
with the definitions
\begin{eqnarray}
\overline {N_5}(x,y_0)^{\gamma\alpha}_{C[A]}&=&a^3\sum_\by
\left[ S(x;y)\dirac5\overline R(y)\right]_{C[A]}^{\gamma\alpha}
\label{n5}\;,\\
\overline {N_{05}}(x,y_0)^{\gamma\alpha}_{C[A]}
&=&a^3\sum_\by
\left[ S(x;y)\dirac0\dirac5\overline R(y)\right]_{C[A]}^{\gamma\alpha}
\label{n05}\;.
\end{eqnarray}
Using these, the expression (\ref{fppfinal}) for $\func{PP}$ becomes

\pagebreak

\begin{eqnarray*}
\nonumber   \func{PP} \!\!\!&=&\!\!\!
\frac{a^{6}}{L^{3/2}}\sum_{\bx,\by} \Big\langle
L^{3/2}\left\{\overline R(y)_{A[B]}^{\alpha\beta}
(\overline R(y)^\dagger)_{[B]A}^{\beta\alpha}\right\}
    \left\{\overline S(x)_{D[E]}^{\gamma\delta}
    (\overline S(x)^\dagger)_{[E]D}^{\delta\gamma}\right\}\\[-1mm]
&&\qquad\quad\quad\,+2\re\left\{(\overline S_T)_{[A][B]}^{\alpha\beta}
        \left(\overline S(x)^\dagger\right)_{[B]C}^{\beta\gamma}
        \left( S(x;y)\dirac5\overline R(y)
        \right)_{C[A]}^{\gamma\alpha}\right\}\\[2mm]
&&\qquad\quad\quad\,-\{x\longleftrightarrow y\}\Big\rangle_U\\[1mm]
&\use{pbar}&\!\!\!
\frac{a^{6}}{L^{3/2}}\sum_{\bx,\by} \Big\langle
L^{3/2}\left\{\overline R(y)_{A[B]}^{\alpha\beta}(\overline
R(y)^\dagger)_{[B]A}^{\beta\alpha}\right\}
    \left\{\overline S(x)_{D[E]}^{\gamma\delta}(\overline S(x)^\dagger)_{[E]D}^{\delta\gamma}\right\}\\[-1mm]
&&\qquad\quad\quad\,+2\re\left\{\left(\overline P(x)_{C[A]}^{\gamma\alpha}\right)^*
    \left( S(x;y)\dirac5\overline R(y)\right)_{C[A]}^{\gamma\alpha}\right\}\\[1mm]
&&\qquad\quad\quad\,-\{x\longleftrightarrow y\}\Big\rangle_U\\[2mm]
&\use{n5}&\!\!\!
\frac{a^{6}}{L^{3/2}} \Big\langle
L^{3/2}\sum_{\bx,\by}\left\{\overline R(y)_{A[B]}^{\alpha\beta}(\overline
R(y)^\dagger)_{[B]A}^{\beta\alpha}\right\}
    \left\{\overline S(x)_{D[E]}^{\gamma\delta}(\overline S(x)^\dagger)_{[E]D}^{\delta\gamma}\right\}\\[0mm]
&&\quad\quad\quad+\frac2{a^3}\re\sum_{\bx}\left\{\left(\overline P(x)_{C[A]}^{\gamma\alpha}\right)^*
    \overline{N_{5}}(x,y_0)_{C[A]}^{\gamma\alpha}\right\}\\[1mm]
&&\qquad\quad\quad\,-\{x\longleftrightarrow y\}\Big\rangle_U\;.
\end{eqnarray*}
The disconnected parts are built from the propagator products,
which also appear in $\fp$ and $\gp$. However, this is done
on \emph{each} gauge configuration and we therefore have to define
new correlation functions
\begin{eqnarray*}
[gf]_{\rm PP}(x_0,y_0)&\!\!=\!\!&{a^6}\!\sum_{\bx,\by}\!\Big\langle\!\!
\left\{\overline R(x)_{A[B]}^{\alpha\beta}(\overline
R(x)^\dagger)_{[B]A}^{\beta\alpha}\right\}\!
    \left\{\overline S(y)_{D[E]}^{\gamma\delta}(\overline S(y)^\dagger)_{[E]D}^{\delta\gamma}\right\}\!\!
    \Big\rangle_{\!U}\\[2mm]
[gf]_{\rm AA}(x_0,y_0)&\!\!=\!\!&-{a^6}\!\sum_{\bx,\by}\!\Big\langle\!\!
\left\{\overline R(x)_{A[B]}^{\alpha\beta}(\overline
R(x)^\dagger)_{[B]C}^{\beta\alpha}(\dirac0)_{CA}\right\}\\[-1mm]&&\qquad\qquad\qquad\times
    \left\{\overline S(y)_{D[E]}^{\gamma\delta}(\overline S(y)^\dagger)_{[E]F}^{\delta\gamma}
    (\dirac0)_{FD}\right\}\!\!\Big\rangle_{\!U}\\[2mm]
[gf]_{\rm AP}(x_0,y_0)&\!\!=\!\!&{a^6}\!\sum_{\bx,\by}\!\Big\langle\!\!
\left\{\overline R(x)_{A[B]}^{\alpha\beta}(\overline
R(x)^\dagger)_{[B]C}^{\beta\alpha}(\dirac0)_{CA}\right\}\!\!
    \left\{\overline S(y)_{D[E]}^{\gamma\delta}(\overline S(y)^\dagger)_{[E]D}^{\delta\gamma}
    \right\}\!\!\Big\rangle_{\!U}\\[2mm]
[gf]_{\rm PA}(x_0,y_0)&\!\!=\!\!&\!-{a^6}\!\sum_{\bx,\by}\!\Big\langle\!\!
\left\{\overline R(x)_{A[B]}^{\alpha\beta}(\overline
R(x)^\dagger)_{[B]A}^{\beta\alpha}\right\}\!\!
    \left\{\overline S(y)_{D[E]}^{\gamma\delta}(\overline S(y)^\dagger)_{[E]F}^{\delta\gamma}
    (\dirac0)_{FD}\right\}\!\!\Big\rangle_{\!\!U}\;.
\end{eqnarray*}
Without any additional effort these correlation functions
can be constructed from the estimates for $f_{\rm X}$ and $g_{\rm X}$
in the analysis program. Inserting them into the last expression for $\func{PP}$ gives
\pagebreak
\begin{eqnarray}
\nonumber   \func{PP} \!\!\!&=&\!\!\!
\frac{2a^3}{L^{3/2}} \Big\langle\nonumber
-\re\sum_{\by}\left\{\left(\overline P(y)_{C[A]}^{\gamma\alpha}\right)^*
    \overline{N_{5}}(y,x_0)_{C[A]}^{\gamma\alpha}\right\}\\[0mm]\nonumber
&&\quad\quad\quad+\re\sum_{\bx}\left\{\left(\overline P(x)_{C[A]}^{\gamma\alpha}\right)^*
    \overline{N_{5}}(x,y_0)_{C[A]}^{\gamma\alpha}\right\}\Big\rangle_U\\
    &&-{[gf]_{\rm PP}(x_0,y_0)}+{[gf]_{\rm PP}(y_0,x_0)}\label{fPP_decompose}\;.
\end{eqnarray}
In the expression (\ref{faafinal}) for $\func{AA}$ again two indices in each line can be
restricted, thus giving an overall factor of 4
\begin{eqnarray}
\nonumber   \func{AA} \!\!\!&=&\!\!\! \frac{a^{6}}{L^{3/2}}\sum_{\bx,\by}
\Big\langle
\pl2\re\left\{(\overline S_T)_{[A][B]}^{\alpha\beta}
        \left(\overline S(y)^\dagger\right)_{[B]C}^{\beta\gamma}(\dirac0)_{CD}\left( S(y;x)\dirac0\dirac5
        \overline R(x)\right)_{D[A]}^{\gamma\alpha}\right\}\\[-2mm]\nonumber
&&\qquad\quad\,\quad\pl L^{3/2}\left\{\overline R(y)_{A[B]}^{\alpha\beta}(\overline
R(y)^\dagger)_{[B]C}^{\beta\alpha}(\dirac0)_{CA}\right\}\\
&&\qquad\qquad\qquad\qquad\qquad\qquad\times
    \left\{\overline S(x)_{D[E]}^{\gamma\delta}(\overline S(x)^\dagger)_{[E]F}^{\delta\gamma}
    (\dirac0)_{FD}\right\}\Big\rangle_U\nonumber\\[0mm]
&&\qquad\quad\quad\,-\{x\longleftrightarrow y\}\Big\rangle_U\nonumber\\[3mm]\nonumber
&\use{pbar}&\!\!\! \frac{a^{6}}{L^{3/2}}\sum_{\bx,\by}
        \Big\langle
        +2\re\left\{\left(\overline P(y)_{C[A]}^{\gamma\alpha}\right)^*(\dirac0)_{CD}\left( S(y;x)\dirac0\dirac5
        \overline R(x)\right)_{D[A]}^{\gamma\alpha}\right\}\\[-2mm]\nonumber
&&\qquad\quad\,\quad+L^{3/2}\left\{\overline R(y)_{A[B]}^{\alpha\beta}(\overline
R(y)^\dagger)_{[B]C}^{\beta\alpha}(\dirac0)_{CA}\right\}\\
&&\qquad\qquad\qquad\qquad\qquad\qquad\times\nonumber
    \left\{\overline S(x)_{D[E]}^{\gamma\delta}(\overline S(x)^\dagger)_{[E]F}^{\delta\gamma}
    (\dirac0)_{FD}\right\}\\[2mm]\nonumber\\[-1mm]
&&\qquad\quad\quad\,-\{x\longleftrightarrow y\}\Big\rangle_U\nonumber\\[3mm]\nonumber
&\use{n05}&\!\!\! \frac{2a^3}{L^{3/2}}
        \Big\langle
        +\re\sum_{\by}\left\{\left(\overline P(y)_{C[A]}^{\gamma\alpha}\right)^*(\dirac0)_{CD}
        \overline{N_{05}}(y,x_0)_{D[A]}^{\gamma\alpha}\right\}\\[0mm]
&&\qquad\;\,-\re\sum_{\bx}\left\{\left(\overline
P(x)_{C[A]}^{\gamma\alpha}\right)^*(\dirac0)_{CD}
\overline{N_{05}}(x,y_0)_{D[A]}^{\gamma\alpha}\right\}\Big\rangle_U\nonumber\\
&&+{[gf]_{\rm AA}(x_0,y_0)}-{[gf]_{\rm AA}(y_0,x_0)}\label{fAA_decompose}\;.
\end{eqnarray}
The evaluation of $\func{AP}$ and $\func{PA}$ from equations (\ref{fapfinal}) and (\ref{fpafinal})
is now entirely straightforward. However, the antisymmetry under the exchange $x\leftrightarrow y$
is obtained only when the two are combined.

\pagebreak

\begin{eqnarray}
\nonumber   \func{AP} \!\!\!\!\!\!&=&\!\!\!\!\!\!\! \frac{a^{6}}{L^{3/2}}\sum_{\bx,\by}
\Big\langle
-2\re\left\{(\overline S_T)_{[A][B]}^{\alpha\beta}
        \left(\overline S(y)^\dagger\right)_{[B]C}^{\beta\gamma}\left( S(y;x)\dirac0\dirac5
        \overline R(x)\right)_{C[A]}^{\gamma\alpha}\right\}\\[-2.5mm]
\nonumber&&\qquad\quad\quad\, \ts+L^{3/2}\left\{\overline R(x)_{A[B]}^{\alpha\beta}(\overline
R(x)^\dagger)_{[B]C}^{\beta\alpha}(\dirac0)_{CA}\right\}
    \left\{\overline S(y)_{D[E]}^{\gamma\delta}(\overline S(y)^\dagger)_{[E]D}^{\delta\gamma}\right\}\\[1mm]
\nonumber&&\qquad\quad\,\quad-L^{3/2}\left\{\overline R(y)_{A[B]}^{\alpha\beta}(\overline
R(y)^\dagger)_{[B]A}^{\beta\alpha}\right\}
    \left\{\overline S(x)_{D[E]}^{\gamma\delta}(\overline S(x)^\dagger)_{[E]F}^{\delta\gamma}(\dirac0)_{FD}
    \right\}\\[1mm]
\nonumber&&\qquad\quad\quad\,-2\re\left\{(\overline S_T)_{[A][B]}^{\alpha\beta}
        \left(\overline S(x)^\dagger\right)_{[B]C}^{\beta\gamma}(\dirac0)_{CD}\left( S(x;y)\dirac5
        \overline R(y)\right)_{D[A]}^{\gamma\alpha}\right\}\Big\rangle_U\nonumber\\[2mm]
\nonumber&\use{pbar}&\!\!\! \frac{a^{6}}{L^{3/2}}\sum_{\bx,\by}
        \Big\langle
        -2\re\left\{\left(\overline P(y)_{C[A]}^{\gamma\alpha}\right)^*
        \left( S(y;x)\dirac0\dirac5\overline R(x)\right)_{C[A]}^{\gamma\alpha}\right\}\\[-2.5mm]
\nonumber&&\qquad\quad\quad\, \ts+L^{3/2}\left\{\overline R(x)_{A[B]}^{\alpha\beta}(\overline
        R(x)^\dagger)_{[B]C}^{\beta\alpha}(\dirac0)_{CA}\right\}
        \left\{\overline S(y)_{D[E]}^{\gamma\delta}(\overline S(y)^\dagger)_{[E]D}^{\delta\gamma}\right\}\\[1mm]
\nonumber&&\qquad\quad\,\quad-L^{3/2}\left\{\overline R(y)_{A[B]}^{\alpha\beta}(\overline
        R(y)^\dagger)_{[B]A}^{\beta\alpha}\right\}
        \left\{\overline S(x)_{D[E]}^{\gamma\delta}(\overline S(x)^\dagger)_{[E]F}^{\delta\gamma}(\dirac0)_{FD}
        \right\}\\[1mm]
\nonumber&&\qquad\quad\quad\,-2\re\left\{\left(\overline P(x)_{C[A]}^{\gamma\alpha}\right)^*
(\dirac0)_{CD}\left( S(x;y)\dirac5
        \overline R(y)\right)_{D[A]}^{\gamma\alpha}\right\}\Big\rangle_U\nonumber\\[1mm]
&=&\!\!\! \frac{2a^3}{L^{3/2}}
        \Big\langle
        -\re\sum_{\by}\left\{\left(\overline P(y)_{C[A]}^{\gamma\alpha}\right)^*\nonumber
        \overline{N_{05}}(y,x_0)_{C[A]}^{\gamma\alpha}\right\}\\[0mm]
&&\qquad\quad-\re\sum_{\bx}\left\{\left(\overline P(x)_{C[A]}^{\gamma\alpha}\right)^*
(\dirac0)_{CD}\overline{N_5}(x,y_0)_{D[A]}^{\gamma\alpha}\right\}\Big\rangle_U\nonumber\\
&&+{[gf]_{\rm AP}(x_0,y_0)}+{[gf]_{\rm PA}(y_0,x_0)}\label{fAP_decompose}\;.
\end{eqnarray}

\begin{eqnarray}
\nonumber   \func{PA} \!\!\!&=&\!\!\! \frac{a^{6}}{L^{3/2}}\sum_{\bx,\by}
\Big\langle
+2\re\left\{(\overline S_T)_{[A][B]}^{\alpha\beta}
        \left(\overline S(y)^\dagger\right)_{[B]C}^{\beta\gamma}(\dirac0)_{CD}\left( S(y;x)\dirac5
        \overline R(x)\right)_{D[A]}^{\gamma\alpha}\right\}\\[-2.5mm]
\nonumber&&\qquad\quad\quad\, \ts+L^{3/2}\left\{\overline R(x)_{A[B]}^{\alpha\beta}(\overline
R(x)^\dagger)_{[B]A}^{\beta\alpha}\right\}
    \left\{\overline S(y)_{D[E]}^{\gamma\delta}(\overline S(y)^\dagger)_{[E]F}^{\delta\gamma}(\dirac0)_{FD}
    \right\}\\[1mm]
\nonumber&&\qquad\quad\,\quad-L^{3/2}\left\{\overline R(y)_{A[B]}^{\alpha\beta}(\overline
R(y)^\dagger)_{[B]C}^{\beta\alpha}(\dirac0)_{CA}\right\}
    \left\{\overline S(x)_{D[E]}^{\gamma\delta}(\overline S(x)^\dagger)_{[E]D}^{\delta\gamma}\right\}\\[1mm]
\nonumber&&\qquad\quad\quad\,+2\re\left\{(\overline S_T)_{[A][B]}^{\alpha\beta}
        \left(\overline S(x)^\dagger\right)_{[B]C}^{\beta\gamma}
        \left( S(x;y)\dirac0\dirac5\overline R(y)\right)_{C[A]}^{\gamma\alpha}\right\}\Big\rangle_U\nonumber\\[2mm]
\nonumber&\use{pbar}&\!\!\! \frac{a^{6}}{L^{3/2}}\sum_{\bx,\by}
\Big\langle
+2\re\left\{\left(\overline P(y)_{C[A]}^{\gamma\alpha}\right)^*(\dirac0)_{CD}\left( S(y;x)\dirac5
        \overline R(x)\right)_{D[A]}^{\gamma\alpha}\right\}\\[-2.5mm]
\nonumber&&\qquad\quad\quad\, \ts+L^{3/2}\left\{\overline R(x)_{A[B]}^{\alpha\beta}(\overline
        R(x)^\dagger)_{[B]A}^{\beta\alpha}\right\}
        \left\{\overline S(y)_{D[E]}^{\gamma\delta}(\overline S(y)^\dagger)_{[E]F}^{\delta\gamma}(\dirac0)_{FD}
        \right\}\\[1mm]
\nonumber&&\qquad\quad\,\quad-L^{3/2}\left\{\overline R(y)_{A[B]}^{\alpha\beta}(\overline
        R(y)^\dagger)_{[B]C}^{\beta\alpha}(\dirac0)_{CA}\right\}
        \left\{\overline S(x)_{D[E]}^{\gamma\delta}(\overline S(x)^\dagger)_{[E]D}^{\delta\gamma}\right\}\\[1mm]
\nonumber&&\qquad\quad\quad\,+2\re\left\{\left(\overline P(x)_{C[A]}^{\gamma\alpha}\right)^*
        \left( S(x;y)\dirac0\dirac5\overline R(y)\right)_{C[A]}^{\gamma\alpha}\right\}\Big\rangle_U\nonumber\\[1mm]
\nonumber&=&\!\!\! \frac{2a^3}{L^{3/2}}
        \Big\langle
        +\re\sum_{\by}\left\{\left(\overline P(y)_{C[A]}^{\gamma\alpha}\right)^*(\dirac0)_{CD}
        \overline{N_5}(y,x_0)_{D[A]}^{\gamma\alpha}\right\}\\[0mm]
&&\qquad\quad+\re\sum_{\bx}\left\{\left(\overline P(x)_{C[A]}^{\gamma\alpha}\right)^*
\overline{N_{05}}(x,y_0)_{C[A]}^{\gamma\alpha}\right\}\Big\rangle_U\nonumber\\
&&-{[gf]_{\rm PA}(x_0,y_0)}-{[gf]_{\rm AP}(y_0,x_0)}\label{fPA_decompose}\;.
\end{eqnarray}
We can now plug the expressions (\ref{fPP_decompose}) to
(\ref{fPA_decompose}) back into (\ref{fAAI_decompose})
to obtain
\beann
\funci{AA}(x_0,y_0)\!\!\!\!
&=&\!\!\!\!\func{AA}(x_0,y_0)+ac_A[\drvtilde0x\func{PA}(x_0,y_0)+\drvtilde0y\func{AP}(x_0,y_0)]\\[1mm]
\nonumber&&\!\!\!\!+a^2c_A^2
\drvtilde0x\drvtilde0y\func{PP}(x_0,y_0)
\eeann
\bea
&=\!\!\!\!&\frac{2a^3}{L^{3/2}}
        \Big\langle\nonumber
        \re\sum_{\by}\left\{\left(\overline P(y)_{C[A]}^{\gamma\alpha}\right)^*(\dirac0)_{CD}\Big[
        \overline{N_{05}}(y,x_0)+ac_A\drvtilde0x\overline{N_5}(y,x_0)\Big]_{D[A]}^{\gamma\alpha}\right\}
        \Big\rangle_U\\[0mm]
&&-ac_A\drvtilde0y\Bigg[
\frac{2a^3}{L^{3/2}}
        \Big\langle
        \re\sum_{\by}\left\{\left(\overline P(y)_{C[A]}^{\gamma\alpha}\right)^*\nonumber
        \Big[\overline{N_{05}}(y,x_0)+ac_A
\drvtilde0x\overline{N_5}(y,x_0)\Big]_{C[A]}^{\gamma\alpha}\right\}\Bigg]\\[0mm]
&&-\frac{2a^3}{L^{3/2}}\Big\langle\re\sum_{\bx}\left\{\left(\overline P(x)_{C[A]}^{\gamma\alpha}\right)^*
(\dirac0)_{CD}\Big[
\overline{N_{05}}(x,y_0)+ac_A\drvtilde0y\overline{N_5}(x,y_0)\Big]_{D[A]}^{\gamma\alpha}\right\}
\Big\rangle_U\nonumber\\
&&+ac_A\drvtilde0x\Bigg[
\frac{2a^3}{L^{3/2}}
        \Big\langle\re\sum_{\bx}\left\{\left(\overline P(x)_{C[A]}^{\gamma\alpha}\right)^*\Big[
        \overline{N_{05}}(x,y_0)+ac_A
        \drvtilde0y\overline{N_5}(x,y_0)\Big]_{C[A]}^{\gamma\alpha}\right\}\Big\rangle_U\nonumber\Bigg]\\\nonumber
    &&+{[gf]_{\rm AA}(x_0,y_0)}-{[gf]_{\rm AA}(y_0,x_0)}
    +ac_A\drvtilde0x\Big[-{[gf]_{\rm PA}(x_0,y_0)}-{[gf]_{\rm AP}(y_0,x_0)}\Big]\\\nonumber
&&+ac_A\drvtilde0y\Big[{[gf]_{\rm AP}(x_0,y_0)}+{[gf]_{\rm PA}(y_0,x_0)}
\Big]\\
    &&+a^2c_A^2
\drvtilde0x\drvtilde0y\Big[{[gf]_{\rm PP}(y_0,x_0)}-{[gf]_{\rm PP}(x_0,y_0)}\Big]\;.
\label{fexplicit}
\eea
The expressions in square brackets, which multiply the $\overline P$ propagator,
can be obtained with one additional inversion for $x_0$ and $y_0$
and each combination of external indices $[A]$ and $\alpha$. More details about
the necessary inversions are given in Sections~\ref{sources} and \ref{countinv}.
If it is necessary to keep $\funci{AA}$ as an explicit function of $c_A$ more inversions are
needed.

The correlation function $\tfunci{PA}(y_0+t,y_0)$ (\ref{ftilde_final})
multiplying the mass term in the integrated Ward identity can also be expanded in the
same way.
\begin{eqnarray}\nonumber
&&\!\!\!\!\!\!\!\!\!\!\!\!\!\tfunci{PA}(y_0+t,y_0)=
a\!\!\sum_{x_0=y_0}^{y_0+t}\!w(x_0)
\Big[\func{PA}(x_0,y_0)+c_Aa\drvtilde0y\func{PP}(x_0,y_0)\Big]\qquad\qquad\\
&=&\!\!\!\!\frac{2a^4}{L^{3/2}}
        \Big\langle\re \!\!\sum_{x_0=y_0}^{y_0+t}\!w(x_0)\!\sum_{\bx}\left\{\left(\overline P(x)_{C[A]}^{\gamma\alpha}\right)^*
\Big[\overline{N_{05}}(x,y_0)\Big]_{C[A]}^{\gamma\alpha}\right\}\Big\rangle_U\nonumber\\
&&\!\!\!\!+ac_A\Bigg[\frac{2a^4}{L^{3/2}}
        \Big\langle\re \!\!\sum_{x_0=y_0}^{y_0+t}\!w(x_0)\sum_{\bx}\left\{\left(\overline P(x)_{C[A]}^{\gamma\alpha}\right)^*
\Big[\drvtilde0y\overline{N_5}(x,y_0)\Big]_{C[A]}^{\gamma\alpha}\right\}\Big\rangle_U\nonumber\Bigg]\\
&&\!\!\!\!+\frac{2a^4}{L^{3/2}}
\Big\langle
\re \!\sum_{\by}\left\{\left(\overline P(y)_{C[A]}^{\gamma\alpha}\right)^*(\dirac0)_{CD}
\Bigg[\!\sum_{x_0=y_0}^{y_0+t}\!w(x_0)\overline{N_5}(y,x_0)_{D[A]}^{\gamma\alpha}\Bigg]
\right\}\Big\rangle_U\nonumber\\[0mm]
&&\!\!\!\!-c_Aa\Bigg[
\frac{2a^4}{L^{3/2}}\sum_{x_0=y_0}^{y_0+t}\!w(x_0)\,\drvtilde0y\Big\langle\nonumber
\re \!\sum_{\by}\left\{\left(\overline P(y)_{C[A]}^{\gamma\alpha}\right)^*
\Big[\overline{N_5}(y,x_0)_{C[A]}^{\gamma\alpha}\Big]\right\}\Big\rangle_U\Bigg]\\
&&\!\!\!\!+c_Aa^2\sum_{x_0=y_0}^{y_0+t}\!\!w(x_0)\Bigg[\drvtilde0y
\Big(-[gf]_{\rm PP}(x_0,y_0)+[gf]_{\rm PP}(y_0,x_0)\Big)\Bigg]\nonumber\\
&&\!\!\!\!+a\!\!\sum_{x_0=y_0}^{y_0+t}\!\!w(x_0)\Big(-[gf]_{\rm PA}(x_0,y_0)-[gf]_{\rm AP}(y_0,x_0)\Big)\;.
    \label{ftildeexplicit}
\end{eqnarray}

\subsection{Sources for the inversion of the Dirac operator}

\label{sources}

With the definition of the propagator $S(x;y)$ as the inverse of the Dirac operator\footnote
{Here $D$ stands symbolically for $D\pl\delta D\pl m_0$ from (\ref{fullD}).}
\begin{equation}
D(x,y)_{AB}^{\alpha\beta}\cdot S(y;z)_{BC}^{\beta\gamma}=a^{-4}\delta_{xz}\delta_{\alpha\gamma}
\delta_{AC}\label{propdef}\;,
\end{equation}
one can easily calculate the action of $D$ on the summed correlators.
\begin{eqnarray}
D(x,y)_{AB}^{\alpha\beta}\overline S(y)_{B[C]}^{\beta\gamma}&\use{sbar}&
\frac{\ctildet a^3}{L^{3/2}}\sum_\bz D(x,y)_{AB}^{\alpha\beta} S(y;a,\bz)_{BD}^{\beta\delta}
[U_0(0,\bz)^{-1}]_{\delta\gamma}[P_+]_{D[C]}\nonumber\\
&\use{propdef}&
\frac{\ctildet a^3}{L^{3/2}}\sum_\bz a^{-4}\delta_{x_0,a}\delta^{(3)}_{\bx\bz}
\delta_{\alpha\delta}\delta_{AD}\left[U_0(0,\bz)^{-1}\right]_{\delta\gamma}[P_+]_{D[C]}\nonumber\\
&=&\frac{\ctildet}{aL^{3/2}}\delta_{a,x_0}\left[U_0(0,\bx)^{-1}\right]_{\alpha\gamma}[P_+]_{A[C]}
\label{sourceS}\;,\\[2mm]
D(x,y)_{AB}^{\alpha\beta}\overline R(y)_{B[C]}^{\beta\gamma}&\use{rbar}&
\frac{\ctildet}{aL^{3/2}}\delta_{T\!-\!a,x_0}\left[U_0(T\!-\!a,\bx)\right]_{\alpha\gamma}
[P_-]_{A[C]}\label{sourceR}\;.
\end{eqnarray}
The right--hand sides of equations (\ref{sourceS}) and (\ref{sourceR}) are therefore the sources
to be used in the calculation of $\overline S(x)$ and $\overline R(x)$. For all index combinations
of
these propagators that were used, six inversions, corresponding to the possible combinations
of $\gamma$ and $[C]$, are needed.
Similarly we obtain for
the expressions containing the full propagator $S(x;y)$
\begin{eqnarray}
D(z,y)_{BC}^{\beta\gamma}\, a^3\!\!\sum_\bx\left[S(y;x)\Gamma\overline R(x)\right]_{C[A]}^{\gamma\alpha}\nonumber
&\!\!\!=\!\!\!&D(z,y)_{BC}^{\beta\gamma}\, a^3\sum_\bx S (y,x)_{CD}^{\gamma\delta}\Gamma_{DE}
\left[\,\overline R(x)\right]_{E[A]}^{\delta\alpha}\nonumber\\
&\!\!\!=\!\!\!&\sum_\bx a^{-1}\delta_{zx}\delta_{BD}\delta_{\beta\delta}\Gamma_{DE}
\left[\,\overline R(x)\right]_{E[A]}^{\delta\alpha}\nonumber\\
&\!\!\!=\!\!\!&a^{-1}\delta_{z_0x_0}\Gamma_{BE}\left[\,\overline R(z)\right]_{E[A]}^{\beta\alpha}\label{source_insert}\;.
\end{eqnarray}
Thus, to calculate $\overline N_5$ and $\overline N_{05}$, the knowledge of
the full propagator $S(x;y)$ is not necessary. Instead, one uses $\overline R$
as the source
for an inversion to obtain the desired product of propagators. When
$\overline N_5$ is used with a time derive, this is included
in the source, such that e.g. $\drvtilde0x\overline{N_5}(y,x_0)$ also requires
only one inversion. The same applies to the correlator
$$\sum_{x_0=y_0}^{y_0+t}\!w(x_0)\overline{N_5}(y,x_0)_{D[A]}^{\gamma\alpha}\;,$$
appearing in both the bare and the improvement term of $\tfunci{PA}$.

\subsection{Counting inversions}

\label{countinv}

Here we give the structure of the program that calculates the correlation function
necessary for the estimate of $\za$. We write $\funci{AA}$ as
\verb|fAA1|$+\cA$\verb|fAA2|$+\cA^2$\verb|fAA3| and
$\tfunci{PA}$ as \verb|fPA1|$+\cA$\verb|fPA2|. On a given gauge configuration the
program proceeds according to

\vspace*{5mm}

\begin{tabular}{rl}
for all $\gamma$ and $[C]$\bigg\{
&Solve for $\overline S(\star)_{\star [C]}^{\star\gamma}$ and\\[-4mm]
& accumulate result into $\fp$, $\fa$, $\overline S_T$, $f_1$ and $\overline P$.\bigg\}\\[2mm]
for all $\gamma$ and $[C]$\bigg\{
&Solve for $\overline R(\star)_{\star [C]}^{\star\gamma}$ and\\[-3mm]
& accumulate result into $\gp$, $\ga$, $\fv$.\\[2mm]
&Solve for $\overline N_{05}(\star,x_0)_{\star [C]}^{\star\gamma}$ and\\[-1mm]
& accumulate result into \verb|fAA1| and \verb|fAA2|.\\[2mm]
&Solve for $\drvtilde0x\overline N_{5}(\star,x_0)_{\star [C]}^{\star\gamma}$ and\\[-1mm]
& accumulate result into \verb|fAA2| and \verb|fAA3|.\\[2mm]
&Solve for $\overline N_{05}(\star,y_0)_{\star [C]}^{\star\gamma}$ and\\[-1mm]
& accumulate result into \verb|fAA1|, \verb|fAA2| and \verb|fPA1|.\\[2mm]
&Solve for $\drvtilde0y\overline N_{5}(\star,y_0)_{\star [C]}^{\star\gamma}$ and\\[-1mm]
& accumulate result into \verb|fAA2|, \verb|fAA3| and \verb|fPA2|.\\[2mm]
&Solve for $\sum_{x_0=y_0}^{y_0+t}\!\overline{N_5}(\star,x_0)_{\star[C]}^{\star\gamma}$ and\\[-2.5mm]
& accumulate result into \verb|fPA1| and \verb|fPA2|.
\bigg\}
\end{tabular}\\[4mm]
The ''accumulate into\ldots'' corresponds to the trace over color and (restricted) Dirac
indices in the correlation functions. For the connected correlators this is done by taking the
scalar product of $\overline P$
with the result of the inversion, taking into account temporal derivatives or sums and
the correct combination of Dirac indices as given by (\ref{fexplicit}) and
(\ref{ftildeexplicit}).

Thus, a calculation of $\fa$, $\fp$ and $f_1$ needs six inversions on
each gauge configuration, while six more are required for $\ga$, $\gp$
and $\fv$. A total of 36 is needed for the correlation function $\funci{AA}$
and only six more for the volume correlator $\tfunci{PA}$. However, even
this is negligible compared to the cost of the inversions in the molecular dynamics.
One should note that if the value of $\cA$ were fixed at run--time, two of
the additional inversions could be saved by using the linearity of the Dirac
equation, e.g. by solving
\beann
&&D(z,y)_{BC}^{\beta\gamma}\Big[
        \overline{N_{05}}(y,x_0)+ac_A\drvtilde0x\overline{N_{5}}(y,x_0)\Big]_{C[A]}^{\gamma\alpha}\nonumber
\\[2mm]&&\qquad\qquad\qquad\use{source_insert}a^{-1}\delta_{z_0x_0}(\dirac0\dirac5)_{BE}
\left[\,\overline R(z)\right]_{E[A]}^{\beta\alpha}\\
  &&\qquad\qquad\qquad\qquad +\frac{c_A}{2a}\delta_{z_0,x_0+a}(\dirac5)_{BE}
  \left[\,\overline R(z)\right]_{E[A]}^{\beta\alpha}\\
  &&\qquad\qquad\qquad\qquad -\frac{c_A}{2a}\delta_{z_0,x_0-a}(\dirac5)_{BE}
  \left[\,\overline R(z)\right]_{E[A]}^{\beta\alpha}\;,
\eeann
thus reducing the total number of inversions from 42 to 30.
An overview is given in table \tab{tab:countinv}.

\vspace*{1cm}

\TABULAR[hb]{|l|c|}{\hline
correlation functions	&  \# inversions    \\\hline
$\fa$, $\fp$, $f_1$               						& 6\\
$\fa$, $\fp$, $f_1$, $\ga$, $\gp$, $\fv$				& 12\\\hline
$\fa$, $\fp$, $f_1$, $\ga$, $\gp$, $\fv$, $\funci{AA}$	& 36 (24)\\
$\fa$, $\fp$, $f_1$, $\ga$, $\gp$, $\fv$, $\funci{AA}$, $\tfunci{PA}$	& 42 (30)\\\hline}
{Number of inversions needed to compute the correlation functions.
The numbers in parenthesis refer to the case, where the value of $\cA$ is fixed
at run--time.\label{tab:countinv}}

\chapter{Transforming the integrated Ward identity}
\label{app:shift}

We start by isolating
the contact term in the volume integration in (\ref{massive})
\begin{eqnarray}
&&\hspace*{-12mm}-2m\int\!\!\rmd^3\by\int\!\rmd^3\bx\int_{y_0-t}^{y_0+t}\!\rmd x_0\,
\epsilon^{abc}\Big\langle P^a(x_0,\bx)A_0^b(y_0,\by)\op_{\rm ext}\Big\rangle\nonumber\\
&=&-2m\int\!\!\rmd^3\by\int\!\rmd^3\bx\int_{y_0+\epsilon}^{y_0+t}\!\rmd x_0\,
\epsilon^{abc}\Big\langle P^a(x_0,\bx)A_0^b(y_0,\by)\op_{\rm ext}\Big\rangle\nonumber\\
&&-2m\int\!\!\rmd^3\by\int\!\rmd^3\bx\int_{y_0-\epsilon}^{y_0+\epsilon}\!\rmd x_0\,
\epsilon^{abc}\Big\langle P^a(x_0,\bx)A_0^b(y_0,\by)\op_{\rm ext}\Big\rangle\nonumber\\
&&-2m\int\!\!\rmd^3\by\int\!\rmd^3\bx\int_{y_0-t}^{y_0-\epsilon}\!\rmd x_0\,
\epsilon^{abc}\Big\langle P^a(x_0,\bx)A_0^b(y_0,\by)\op_{\rm ext}\Big\rangle\;.
\label{volumeterm}
\end{eqnarray}
We will now use the
partial conservation of the axial current to relate the two contributions
from the surface integral in eq.(\ref{massive}).
Using $\int\!\rmd^3\bx\,\partial_kf_k(x)=0$ (for periodic spatial
boundary conditions) the partial conservation reads as an operator identity
\begin{eqnarray}
\int\!\rmd^3\by\,\partial_0 A_0^b(y)&\!\!=\!\!&\int\!\rmd^3\by\, \partial_\mu A_\mu^b(y)=
2m\int\!\rmd^3\by P^b(y)\nonumber\qquad\Rightarrow\\[2mm]
\int\!\rmd^3\by A_0^b(y_0\!-\!t,\by)&\!\!=\!\!&\int\!\rmd^3
\by A_0^b(y_0\mi\epsilon,\by)-2m\int_{y_0-t}^{y_0-\epsilon}\!\!\rmd x_0
\int\!\rmd^3\by P^b(x_0,\by)\;.\qquad\label{shift1}
\end{eqnarray}
We will also use (\ref{shift1}) in the form
\begin{equation}
\int\!\rmd^3\bx A_0^a(y_0,\bx)=\int\!\rmd^3\bx A_0^a(y_0\!+\!t,\bx)-2m\int_{y_0}^{y_0+t}\!\!\rmd x_0
\int\!\rmd^3\bx\, P^a(x_0,\bx)\label{shift2}\;.
\end{equation}
These relations can be
used only in matrix elements with fields
that are not located in the integration region since otherwise the axial variation
of the latter will appear as an additional term. We therefore
have to be careful how far we can shift the
current insertions.
Due to the antisymmetry of
$\epsilon^{abc}$ the second expression in the first line of
equation (\ref{massive}) can be rewritten as
\begin{eqnarray}
&& \hspace*{-12mm}\int\!\!\rmd^3\by\int\!\!\rmd^3\bx\, \epsilon^{abc}\nonumber
\Big\langle A_0^a(y_0,\bx)A_0^b(y_0\!-\!t,\by)\op_{\rm ext}\Big\rangle\\[2mm]
& \use{shift1}&\int\!\!\rmd^3\by\int\!\!\rmd^3\bx\,\nonumber
\epsilon^{abc}\Big\langle A_0^a(y_0,\bx)A_0^b(y_0\!-\!\epsilon,\by)\op_{\rm ext}\Big\rangle\\
&&\quad -2m\int\!\!\rmd^3\by\int\!\!\rmd^3\bx
\int_{y_0-t}^{y_0-\epsilon}\!\!\!\rmd x_0\, \epsilon^{abc}\Big\langle A_0^a(y_0,\bx)P^b(x_0,\by)\nonumber
\op_{\rm ext}\Big\rangle\\[2mm]
& \use{shift2}&\int\!\!\rmd^3\by\int\!\!\rmd^3\bx\,\nonumber
\epsilon^{abc}\Big\langle A_0^a(y_0\!+\!t,\bx)A_0^b(y_0\!-\!\epsilon,\by)\op_{\rm ext}\Big\rangle\\
&&\quad -2m\int\!\!\rmd^3\by\int\!\!\rmd^3\bx
\int_{y_0}^{y_0+t}\!\!\!\rmd x_0\, \epsilon^{abc}\nonumber
\Big\langle P^a(x_0,\bx)A_0^b(y_0\!-\!\epsilon,\by)\op_{\rm ext}\Big\rangle\\
&&\quad +2m\int\!\!\rmd^3\by\int\!\!\rmd^3\bx
\int_{y_0-t}^{y_0-\epsilon}\!\!\!\rmd x_0\, \epsilon^{abc}\Big\langle P^a(x_0,\by)A_0^b(y_0,\bx)
\op_{\rm ext}\Big\rangle\;.\label{shift}
\end{eqnarray}
The last term cancels the last integral in the split up integration
 (\ref{volumeterm}) and in the limit $\epsilon\rightarrow0$ the first
combines with the result from the integration over the upper surface.
Thus, inserting (\ref{shift}) and (\ref{volumeterm}) into equation
 (\ref{massive}) yields
\begin{eqnarray}&&\nonumber
2i\displaystyle\int\!\!\rmd^3\by\
\Big\langle V_0^c(y_0,\by)\op_{\rm ext}\Big\rangle=\\&&\nonumber
{\displaystyle\quad\lim_{\epsilon\rightarrow0}}\ \ 
2\int\!\!\rmd^3\by\int\!\!\rmd^3\bx\, \epsilon^{abc}
\Big\langle A_0^a(y_0\!+\!t,\bx)A_0^b(y_0,\by)\op_{\rm ext}\Big\rangle\\
&&\ \ \quad\qquad-2m\int\!\!\rmd^3\by\int\!\!\rmd^3\bx
\int_{y_0}^{y_0+t}\!\!\!\rmd x_0\, \epsilon^{abc}\nonumber
\Big\langle P^a(x_0,\bx)A_0^b(y_0\!-\!\epsilon,\by)\op_{\rm ext}\Big\rangle\\
&&\ \ \quad\qquad-2m\int\!\!\rmd^3\by\int\!\rmd^3\bx\int_{y_0+\epsilon}^{y_0+t}\!\rmd x_0\,
\epsilon^{abc}\Big\langle P^a(x_0,\bx)A_0^b(y_0,\by)\op_{\rm ext}\Big\rangle\nonumber\\
&&\ \ \quad\qquad-2m\int\!\!\rmd^3\by\int\!\rmd^3\bx\int_{y_0-\epsilon}^{y_0+\epsilon}
\!\rmd x_0\,
\epsilon^{abc}\Big\langle P^a(x_0,\bx)A_0^b(y_0,\by)\op_{\rm ext}
\Big\rangle\;.
\end{eqnarray}
In the third line we can replace $A_0^b(y_0\mi\epsilon)$ by $A_0^b(y_0)$
if we start the time integration at $y_0\pl\epsilon$. This preserves the order
of the operator insertions and allows us to combine two of the time integrations
with the result given by
\pagebreak
\begin{eqnarray}&&\nonumber
i\displaystyle\int\!\!\rmd^3\by\
\Big\langle V_0^c(y_0,\by)\op_{\rm ext}\Big\rangle=\\&&
{\displaystyle
\quad\lim_{\epsilon\rightarrow0}}
\int\!\!\rmd^3\by\int\!\!\rmd^3\bx\, \epsilon^{abc}
\Big\langle A_0^a(y_0\!+\!t,\bx)A_0^b(y_0,\by)\op_{\rm ext}\Big\rangle\nonumber\\
&&\quad\qquad\nonumber
-2m\int\!\!\rmd^3\by\int\!\!\rmd^3\bx
\int_{y_0+\epsilon}^{y_0+t}\!\!\!\rmd x_0\, \epsilon^{abc}\nonumber
\Big\langle P^a(x_0,\bx)A_0^b(y_0,\by)\op_{\rm ext}\Big\rangle\\
&&\quad\qquad-m\int\!\!\rmd^3\by\int\!\rmd^3\bx\int_{y_0-\epsilon}^{y_0+\epsilon}
\!\rmd x_0\,
\epsilon^{abc}\Big\langle P^a(x_0,\bx)A_0^b(y_0,\by)\op_{\rm ext}
\Big\rangle\;.\label{aftershift}
\end{eqnarray}
The contact terms are the same as those appearing in (\ref{firstcontact})
and thus integrable under a four--dimensional integration. The limit of
$\epsilon\rightarrow0$ can hence be performed and the final result is
\begin{eqnarray}&&\nonumber
i\displaystyle\int\!\!\rmd^3\by\
\Big\langle V_0^c(y_0,\by)\op_{\rm ext}\Big\rangle=\\&&
\quad\int\!\!\rmd^3\by\int\!\!\rmd^3\bx\, \epsilon^{abc}
\Big\langle A_0^a(y_0\!+\!t,\bx)A_0^b(y_0,\by)\op_{\rm ext}\Big\rangle\nonumber\\
&&\quad\qquad\nonumber
-2m\int\!\!\rmd^3\by\int\!\!\rmd^3\bx
\int_{y_0}^{y_0+t}\!\!\!\rmd x_0\, \epsilon^{abc}\nonumber
\Big\langle P^a(x_0,\bx)A_0^b(y_0,\by)\op_{\rm ext}\Big\rangle\;.
\end{eqnarray}
\chapter{List of simulation parameters and results}
\label{simdata}

In \tab{longtable} the simulation results for $\za$ and $\zv$
are collected. The algorithm is specified in the same manner as in
\tab{t_simpar_algo}. The number of measurements $N_{\rm meas}$
is multiplied with the number
of replica and as before $\tau_{\rm meas}$ is the molecular dynamics time
between consecutive measurements.
The PCAC mass $am$ is defined through (\ref{m_av})
and the column $\zv$ refers to the definition (\ref{zv_impl}) before
taking the chiral limit.

For completeness the results for $\za$ are also given for the estimate using
only connected diagrams in the evaluation of the correlation functions $\func{XY}$,
see \sect{sect:asq}. This is denoted by $\za^{\,\rm con}$.
An (additional) superscript ''old'' refers to the estimate from the
massless condition from ref.~\cite{Luscher:1996jn}, where the $m\tfunci{PA}$ term is neglected.

\begin{sidewaystable}
\centering{\small
\begin{tabular}{|lllrrr@{$\cdot$}lc|llllll|}
\hline\hline
\multicolumn{1}{|c}{algo$\!\!$}  & 
\multicolumn{1}{c}{$\beta$}  & 
\multicolumn{1}{c}{$\kappa$}  &
\multicolumn{1}{c}{$L$}  &
\multicolumn{1}{c}{$T$}  &
\multicolumn{2}{c}{$N_{\rm meas}$}   &
\multicolumn{1}{c|}{$\!\!\!\!\tau_{\rm meas}\!\!$}   &
\multicolumn{1}{c}{$am$}  &
\multicolumn{1}{c}{$\!\!\!\!\za$}  &
\multicolumn{1}{c}{$\!\!\!\!\za^{\,\rm con}$}  &
\multicolumn{1}{c}{$\!\!\!\!\za^{\,\rm old}$}  &
\multicolumn{1}{c}{$\!\!\!\!\za^{\,\rm con,old}$}  &
\multicolumn{1}{c|}{$\!\!\!\!Z_{\rm V}$}
\\\hline
$\rm H_2\!\!$  &5.200 & 0.13550 & 8 & 18 & 16 & 200 & 4 &$\phantom{-} 0.01718(90)$  & $\!\!\!\!$  0.7301(173)
  & $\!\!\!\!$  0.8411(80)  & $\!\!\!\!$  0.5039(159)  & $\!\!\!\!$  1.0124(224)  & $\!\!\!\!$  0.7509(6)\\[-0.6mm]
$\rm P_{140}\!\!$  &5.200 & 0.13550 & 8 & 18 & 16 & 40 & 10 &$\phantom{-} 0.0159(11)$  & $\!\!\!\!$  0.7186(295)
  & $\!\!\!\!$  0.8455(108)  & $\!\!\!\!$  0.5026(156)  & $\!\!\!\!$  1.0174(227)  & $\!\!\!\!$  0.7497(14)\\[-0.6mm]
$\rm P_{140}\!\!$  &5.200 & 0.13560 & 8 & 18 & 16 & 225 & 3 &$\phantom{-} 0.01310(68)$  & $\!\!\!\!$  0.7157(137)
  & $\!\!\!\!$  0.8212(96)  & $\!\!\!\!$  0.5546(117)  & $\!\!\!\!$  0.9627(123)  & $\!\!\!\!$  0.7471(7)\\[-0.6mm]
$\rm P_{140}\!\!$  &5.200 & 0.13570 & 8 & 18 & 16 & 230 & 2 &$\phantom{-} 0.0088(11)$  & $\!\!\!\!$  0.7134(126)
  & $\!\!\!\!$  0.8302(70)  & $\!\!\!\!$  0.6222(149)  & $\!\!\!\!$  0.9114(113)  & $\!\!\!\!$  0.7447(8)\\[-0.6mm]
$\rm P_{140}\!\!$  &5.200 & 0.13580 & 8 & 18 & 16 & 230 & 2 &$\phantom{-} 0.00194(81)$  & $\!\!\!\!$  0.7176(114)
  & $\!\!\!\!$  0.8588(99)  & $\!\!\!\!$  0.7032(115)  & $\!\!\!\!$  0.8773(104)  & $\!\!\!\!$  0.7424(14)\\[-0.2mm]\hline
$\rm P_{140}\!\!$  &5.290 & 0.13625 & 8 & 18 & 16 & 50 & 2 &$\phantom{-} 0.0031(18)$  & $\!\!\!\!$  0.7527(102)
  & $\!\!\!\!$  0.8103(167)  & $\!\!\!\!$  0.7391(108)  & $\!\!\!\!$  0.8437(179)  & $\!\!\!\!$  0.7507(19)\\[-0.6mm]
$\rm P_{140}\!\!$  &5.290 & 0.13641 & 8 & 18 & 16 & 120 & 2 &$-0.00512(61)$  & $\!\!\!\!$  0.7540(124)
  & $\!\!\!\!$  0.8378(73)  & $\!\!\!\!$  0.7421(120)  & $\!\!\!\!$  0.8081(62)  & $\!\!\!\!$  0.7490(12)\\[-0.2mm]\hline
$\rm H_2\!\!$  &5.500 & 0.13606 & 12 & 27 & 16 & 25 & 6 &$\phantom{-} 0.02254(26)$  & $\!\!\!\!$  0.8417(222)
  & $\!\!\!\!$  0.8077(26)  & $\!\!\!\!$  0.3918(137)  & $\!\!\!\!$  1.0732(261)  & $\!\!\!\!$  0.7853(14)\\[-0.6mm]
$\rm H_2\!\!$  &5.500 & 0.13650 & 12 & 27 & 16 & 44 & 3 &$\phantom{-} 0.00758(27)$  & $\!\!\!\!$  0.7987(153)
  & $\!\!\!\!$  0.8100(45)  & $\!\!\!\!$  0.6063(143)  & $\!\!\!\!$  0.8630(57)  & $\!\!\!\!$  0.7738(8)\\[-0.6mm]
$\rm H_2\!\!$  &5.500 & 0.13672 & 12 & 27 & 16 & 80 & 3 &$\phantom{-} 0.00041(25)$  & $\!\!\!\!$  0.7888(32)
  & $\!\!\!\!$  0.8048(54)  & $\!\!\!\!$  0.7861(33)  & $\!\!\!\!$  0.8096(56)  & $\!\!\!\!$  0.7650(21)\\[-0.6mm]
$\rm H_2\!\!$  &5.500 & 0.13672 & 8 & 18 & 1 & 318 & 4 &$-0.00168(62)$  & $\!\!\!\!$  0.8105(64)
  & $\!\!\!\!$  0.8168(38)  & $\!\!\!\!$  0.8148(58)  & $\!\!\!\!$  0.8111(42)  & $\!\!\!\!$  0.7750(45)\\[-0.2mm]\hline
$\rm H_2\!\!$  &5.715 & 0.13665 & 16 & 36 & 1 & 106 & 2 &$\phantom{-} 0.00194(57)$  & $\!\!\!\!$  0.8142(135)
  & $\!\!\!\!$  0.8079(31)  & $\!\!\!\!$  0.7811(60)  & $\!\!\!\!$  0.8199(20)  & $\!\!\!\!$  0.7827(11)\\[-0.6mm]
$\rm H_2\!\!$  &5.715 & 0.13670 & 16 & 36 & 1 & 54 & 2 &$-0.00060(69)$  & $\!\!\!\!$  0.8004(26)
  & $\!\!\!\!$  0.8120(30)  & $\!\!\!\!$  0.8014(23)  & $\!\!\!\!$  0.8098(28)  & $\!\!\!\!$  0.7793(20)\\[-0.6mm]
$\rm H_2\!\!$  &5.715 & 0.13670 & 12 & 27 & 4 & 62 & 2 &$-0.00100(34)$  & $\!\!\!\!$  0.8021(38)
  & $\!\!\!\!$  0.8182(18)  & $\!\!\!\!$  0.8071(34)  & $\!\!\!\!$  0.8138(19)  & $\!\!\!\!$  0.7861(18)\\[-0.2mm]\hline
$\rm H_2\!\!$  &7.200 & 0.13420 & 8 & 18 & 1 & 220 & 2 &$\phantom{-} 0.00029(45)$  & $\!\!\!\!$  0.8721(24)
  & $\!\!\!\!$  0.8772(18)  & $\!\!\!\!$  0.8699(22)  & $\!\!\!\!$  0.8787(18)  & $\!\!\!\!$  0.8573(9)\\[-0.6mm]
$\rm H_2\!\!$  &7.200 & 0.13424 & 8 & 18 & 1 & 164 & 2 &$-0.00028(42)$  & $\!\!\!\!$  0.8683(22)
  & $\!\!\!\!$  0.8732(18)  & $\!\!\!\!$  0.8707(36)  & $\!\!\!\!$  0.8716(28)  & $\!\!\!\!$  0.8553(6)\\[-0.6mm]
$\rm H_2\!\!$  &7.200 & 0.13424 & 12 & 27 & 16 & 50 & 2 &$-0.00049(15)$  & $\!\!\!\!$  0.8685(23)
  & $\!\!\!\!$  0.8717(8)  & $\!\!\!\!$  0.8717(23)  & $\!\!\!\!$  0.8697(7)  & $\!\!\!\!$  0.8543(18)\\[-0.6mm]
$\rm H_2\!\!$  &7.200 & 0.13424 & 16 & 36 & 1 & 80 & 2 &$-0.00023(41)$  & $\!\!\!\!$  0.8678(18)
  & $\!\!\!\!$  0.8670(17)  & $\!\!\!\!$  0.8694(14)  & $\!\!\!\!$  0.8652(11)  & $\!\!\!\!$  0.8508(18)\\[-0.2mm]\hline
$\rm H_2\!\!$  &8.400 & 0.13258 & 8 & 18 & 4 & 40 & 2 &$\phantom{-} 0.00023(40)$  & $\!\!\!\!$  0.8990(28)
  & $\!\!\!\!$  0.8956(16)  & $\!\!\!\!$  0.8962(26)  & $\!\!\!\!$  0.8973(19)  & $\!\!\!\!$  0.8839(15)\\[-0.6mm]
$\rm H_2\!\!$  &8.400 & 0.13262 & 8 & 18 & 4 & 45 & 2 &$-0.00183(42)$  & $\!\!\!\!$  0.8998(25)
  & $\!\!\!\!$  0.8953(13)  & $\!\!\!\!$  0.9184(39)  & $\!\!\!\!$  0.8845(15)  & $\!\!\!\!$  0.8826(7)\\[-0.2mm]\hline
$\rm H_2\!\!$  &9.600 & 0.13140 & 8 & 18 & 4 & 100 & 2 &$\phantom{-} 0.00021(15)$  & $\!\!\!\!$  0.9137(14)
  & $\!\!\!\!$  0.9154(7)  & $\!\!\!\!$  0.9110(18)  & $\!\!\!\!$  0.9166(8)  & $\!\!\!\!$  0.9040(4)\\[-0.6mm]
$\rm H_2\!\!$  &9.600 & 0.13142 & 8 & 18 & 4 & 125 & 2 &$-0.00059(15)$  & $\!\!\!\!$  0.9118(12)
  & $\!\!\!\!$  0.9155(7)  & $\!\!\!\!$  0.9188(18)  & $\!\!\!\!$  0.9121(10)  & $\!\!\!\!$  0.9034(4)\\[-0.2mm]\hline
\hline
\end{tabular}}
\caption{Summary of simulation parameters and results for $\za$ and $\zv$.
\label{longtable}}\end{sidewaystable}

\chapter*{Lebenslauf}

\begin{tabular}{ll}

Name: & \dcauthorname  \dcauthorsurname \\[3mm]
1996              & Abitur am Ernst-Mach-Gymnasium Haar\\
10/1997 - 8/2002  & Studium an der Universit\"at Regensburg\\[-1mm]
                  & in der Fachrichtung Physik\\
8/2000 - 7/2001   & Studium der Physik an der\\[-1mm]
                  & University of Colorado\\
10/2002 - 8/2005  & Promotion an der\\[-1mm]
                  & Humboldt-Universit\"at zu Berlin,\\[-1mm]
                  & Lehrstuhl Prof. Dr. U. Wolff, \\[-1mm]
                  & Institut f\"ur Physik\\

\end{tabular}                                                                      

\chapter*{Publications}

\begin{itemize}

\item A.~Hasenfratz, R.~Hoffmann and F.~Knechtli,
{\it The static potential with hypercubic blocking},
Presented at LATTICE 2001,
Berlin, Germany, 19-24 Aug 2001
{\em Nucl. Phys. Proc. Suppl.} {\bf 106} (2002) 418,
arXiv: hep-lat/0110168.

\item C.~Gattringer, R.~Hoffmann and S.~Schaefer,
{\it Setting the scale for the L\"u\-scher--Weisz action},
{\em Phys. Rev. D}{\bf 65} (2002) 094503,\\
arXiv: hep-lat/0112024.

\item C.~Gattringer, R.~Hoffmann and S.~Schaefer,
{\it The topological susceptibility of $SU(3)$ gauge theory near $T_C$},
{\em Phys. Lett. B}{\bf 535} (2002) 358,
arXiv: hep-lat/0203013.

\item R.~Hoffmann, F.~Knechtli, J.~Rolf, R.~Sommer, U.~Wolff,
{\it Non--perturbative renormalization of the axial current with
                  improved Wilson quarks},
Presented at LATTICE 2003,
Tsukuba, Ibaraki, Japan, 15-19 Jul 2003, {\em Nucl. Phys. Proc. Suppl.}
{\bf 129} (2004) 423,
arXiv: hep-lat/0309071.

\item R. Sommer et al.,
{\it Large cutoff effects of dynamical Wilson fermions},
{\em Nucl. Phys. Proc. Suppl.} {\bf 129} (2004) 405,
arXiv: hep-lat/0309171.

\item M. Della Morte, R. Hoffmann, F. Knechtli, U. Wolff,
{\it Impact of large cutoff--effects on algorithms for improved
Wilson fermions},
{\em Comput. Phys. Commun.} {\bf 165} (2005) 49,
arXiv: hep-lat/0405017.

\item M. Della Morte, R. Hoffmann, F. Knechtli, U. Wolff,
{\it Cutoff effects in the spectrum of dynamical Wilson
                  fermions},
Presented at LATTICE 2004,
Fermilab, Batavia, USA, 15-19 Jul 2004,
{\em Nucl. Phys. Proc. Suppl.} {\bf 129} (2004) 423,
arXiv: hep-lat/0409005.

\item M. Della Morte, R. Hoffmann, R. Sommer,
{\it Non--perturbative improvement of the axial current for
                  dynamical Wilson fermions},
{\em JHEP} {\bf 03} (2005) 029,
arXiv: hep-lat/0503003.
   
\end{itemize}

\chapter*{Acknowledgments}

Here is a (most likely incomplete) list of people
whose help, knowledge, patience and support were
indispensable for a successful completion of this work.

\begin{itemize}
\item I would like to thank Ulli Wolff for giving me the opportunity
to work in his group and the encouragement throughout my time here.

\item Thanks to the ALPHA people for a fruitful and interesting
collaboration. In particular to Rainer Sommer for the many
interesting discussions about our work.

\item Special thanks to all the members of the COM
group in the last three years, especially Francesco,
Andreas, Michele, Bj\"orn, Tom, Magdalena and Oliver.
I think we had a great time.

\item I am especially indebted to Francesco Knechtli for
recruiting me to Berlin and to Andreas J\"uttner for
sharing the COM-PhD experience!

\item Michele not only showed amazing knowledge and patience in
the discussion of physical questions, his advice was also essential
for bringing my thesis into its final form. Thanks!

\item I am grateful to Martin L\"uscher and Stefan Sint for illuminating discussions
and Hartmut Wittig for communicating details about the simulations and the TAO code
from the quenched $\za$ project.

\item I thank Stephan D\"urr for his valuable
work in the early stages of the $\cA$ project
and Heiko Molke for contributing the implementation of the spatial
wave functions.

\item Thanks to the DFG for the scholarship within the Graduiertenkolleg 271
(Strukturuntersuchungen, Pr\"azisionstests und Erweiterungen des Standardmodells der Elementarteilchenphysik).

\item I am also indebted to my family, who provided continuous support and
encouragement for my scientific endeavors.

\end{itemize}
\selectlanguage{german}


\chapter*{Selbst\"andigkeitserkl\"arung}

\noindent Hiermit erkl\"are ich, die vorliegende Arbeit
selbst\"andig ohne fremde Hilfe verfa{\ss}t
und nur die angegebene Literatur und Hilfsmittel verwendet zu haben.

\vspace{55mm}
\noindent\dcauthorname \dcauthorsurname \\  
\dcdatesubmitted

\end{document}